%% file: CompleteWhitepaper.tex
\RequirePackage{lineno}
\documentclass[12pt,tightenlines,superscriptaddress,floatfix,linenumbers]{revtex4PATCH}
\pdfoutput=1

\usepackage{enumitem}
\usepackage{latexsym}
\usepackage{graphicx}
\usepackage{url}
\usepackage{feynman}
\usepackage{mathrsfs,amssymb,amsmath,amsfonts,amstext,graphics, multirow}
\usepackage[export]{adjustbox}
\usepackage{epstopdf}
\usepackage{url}
\usepackage{appendix}
\usepackage{color,slashed}
\usepackage{tikz}
\usepackage{slashed}
\usepackage{cancel}
\usepackage{mciteplus}
\usepackage{verbatim}
\usetikzlibrary{arrows,shapes}
\usetikzlibrary{trees}
\usetikzlibrary{matrix,arrows} 
\usepackage{pdflscape}
\usepackage{xspace}
\usepackage{color}
\usepackage{wrapfig}
\usepackage{rotating}
\usepackage{dcolumn}% Align table columns on decimal point
\usepackage{bm}% bold math
\usepackage{longtable}% Include figure files
\usepackage{changepage} % for expanding margins to accomodate wide table
\usepackage{rotating} % for rotating table 90deg
\usepackage[colorlinks=false]{hyperref}

\usepackage{xcolor}
\usepackage{graphicx}
\usepackage{lipsum}

\usepackage[utf8]{inputenc}											
\usepackage{relsize}

\usepackage[printwatermark]{xwatermark}
\newcommand{\sauth}[1]{\noindent {\small{Section Editors: #1}} \vspace{-0.2in}}

\newcommand{\comments}[1]{} 

\newcommand{\TeV}{\,\mathrm{TeV}}
\newcommand{\GeV}{\,\mathrm{GeV}}
\newcommand{\MeV}{\,\mathrm{MeV}}
\newcommand{\keV}{\,\mathrm{keV}}
\newcommand{\eV}{\,\mathrm{eV}}

\newcommand{\gev}{{\rm GeV}}

\newcommand{\lsim}{\lesssim}

\newcommand{\bef}{\begin{figure}[htbp]\begin{center}}
\newcommand{\eef}{\end{center}\end{figure}}
\newcommand{\bea}{\begin{eqnarray}}
\newcommand{\eea}{\end{eqnarray}}
\def\beq{\begin{equation}}
\def\eeq#1{\label{#1}\end{equation}}
\def\eeqn{\end{equation}}
\def\beqa{\begin{eqnarray}}
\def\eeqa#1{\label{#1}\end{eqnarray}}
\def\eeqan{\end{eqnarray}}

\newcommand{\benum}{\begin{enumerate}}
\newcommand{\eenum}{\end{enumerate}}
\newcommand{\bi}{\begin{itemize}}
\newcommand{\ei}{\end{itemize}}
\newcommand{\ee}{\ensuremath{e^+e^-}\xspace}

\newcommand{\apr}{A^{\prime}}

\def\babar{\mbox{\sl B\hspace{-0.4em} {\small\sl A}\hspace{-0.37em} \sl B\hspace{-0.4em} {\small\sl A\hspace{-0.02em}R}}}

\newcommand{\be}{\begin{eqnarray}}

\newcommand{\secref}[1]{Sec.~\ref{WG4sec:#1}}
\newcommand{\ssection}[1]{\vspace{.1in} \noindent {\em #1.}}
\makeatletter
\def\l@subsubsection#1#2{}
\def\l@paragraph#1#2{}
\makeatother

\newcommand*{\nl}{\newline}

\input{institutionAliases.tex}

\begin{document}

\pagestyle{plain}
\title{\Large US Cosmic Visions: New Ideas in Dark Matter 2017 : Community Report}

\input{authorlist.tex}

\date{\today}

\maketitle
%%%%%%%%%%%%%%%%%%%%%%%%%%%%%%%%%%%%%%%%%%
\newpage

\tableofcontents

\newpage
%\linenumbers

\input{abstract.tex}

\newpage

\input{exec.tex}

\newpage

\input{introduction.tex}

\newpage

\input{ScienceCase.tex}

\newpage
\sauth{Rouven Essig, Juan Estrada, Dan McKinsey}
\input{WG1.tex}

\newpage
\sauth{Aaron Chou, Peter Graham}
\input{WG2.tex}

\newpage
\sauth{Bertrand Echenard and Eder Izaguirre}
\input{WG3.tex}

\clearpage

\sauth{Jonathan Feng and Patrick Fox}
\input{WG4.tex}

\newpage

\input{conclusions.tex}

\newpage
\bibliography{figs/fullBib_May16}

\end{document}

%% file: institutionAliases.tex
\newcommand{\Amherst}{University of Massachusetts, Amherst, MA 01003 USA}
\newcommand{\ASU}{University of Arizona, Tucson, AZ  85721}
\newcommand{\Weizmann}{Weizmann Institute of Science, 76100 Rehovot, Israel}
\newcommand{\BNL}{Brookhaven National Laboratory, Upton, NY 11973}
\newcommand{\BU}{Boston University, Boston, MA 02215}
\newcommand{\CNYang}{Stony Brook, NY 11794}
\newcommand{\Caltech}{California Institute of Technology, Pasadena, CA 91125}
\newcommand{\CERN}{CERN, Geneva, Switzerland}
\newcommand{\Mines}{Colorado School of Mines, Golden, CO 80401}
\newcommand{\Columbia}{Columbia University, New York, NY 10027}
\newcommand{\Cornell}{Cornell University, Ithaca, NY 14853}
\newcommand{\NCU}{National Central University, Taoyuan City 32001, Taiwan (R.O.C.)}
\newcommand{\Purdue}{Purdue University, West Lafayette, IN 47907}
\newcommand{\Bub}{Boston University, Boston, MA 02215}
\newcommand{\Northwestern}{Northwestern University, Evanston, IL 60201}
\newcommand{\Tamu}{Texas AandM University, College Station, TX 77843 }
\newcommand{\UCI}{University of California, Irvine, CA 92697}
\newcommand{\Hawaii}{University of Hawaii, Honolulu, HI 96822}
\newcommand{\UKY}{University of Kentucky, Lexington, KY 40506}
\newcommand{\Carleton}{Carleton University, K1S 5B6 Ottawa, Canada}
\newcommand{\houston}{University of Houston, Houston, TX 77204}
\newcommand{\UW}{University of Washington, Seattle 98195}
\newcommand{\UCLA}{University of California at Los Angeles, Los Angeles,  CA 90095}
\newcommand{\wiscIce}{University of Wisconsin, Madison, WI 53706}
\newcommand{\ggali}{Universit\'a di Firenze, 50019 Firenze, Italy}
\newcommand{\drexel}{Drexel University, Philadelphia, PA 19104}
\newcommand{\Duke}{Duke University and Triangle Universitites Nuclear Laboratory, Durham, NC 27708}
\newcommand{\ETH}{ETH Zurich, Institute for Particle Physics, 8093 Zurich, Switzerland}
\newcommand{\FNAL}{Fermi National Accelerator Laboratory, Batavia, IL 60510}
\newcommand{\NIU}{Northern Illinois University, DeKalb, Illinois 60115}
\newcommand{\FSU}{Florida State University,Tallahassee, FL 32306}
\newcommand{\GWU}{George Washington University, Washington, DC 20052}
\newcommand{\Hampton}{Hampton University, Hampton, VA 23668}
\newcommand{\IBS}{Institute for Basic Science (IBS), Daejeon 34051, Korea}
\newcommand{\Indiana}{Indiana University, Bloomington, IN 47405}
\newcommand{\INFNG}{Istituto Nazionale di Fisica Nucleare, Sezione di Genova, 16146 Genova, Italy}
\newcommand{\INFNLNS}{Istituto Nazionale di Fisica Nucleare, Laboratori Nazionali del Sud, 95125 Catania, Italy}
\newcommand{\INFNCat}{Istituto Nazionale di Fisica Nucleare, Sezione di Catania, 95125 Catania, Italy}
\newcommand{\INFNCag}{Istituto Nazionale di Fisica Nucleare, Sezione di Cagliari,  09126 Cagliari, Italy}
\newcommand{\INFNN}{Istituto Nazionale di Fisica Nucleare, Sezione di Napoli, 80125 Napoli, Italy }
\newcommand{\UFN}{Universit\`a Federico II di Napoli, 80125 Napoli, Italy}
\newcommand{\KICP}{The University of Chicago,Chicago, IL 60637}
\newcommand{\KPH}{Johannes Gutenberg University, 55128 Mainz, Germany}
\newcommand{\Lafayette}{Lafayette College, Easton, PA 18042}
\newcommand{\LANL}{Los Alamos National Laboratory, Los Alamos, NM 87545}
\newcommand{\LBL}{Lawrence Berkeley National Laboratory, Berkeley, CA 94720}
\newcommand{\UCB}{University of California at Berkeley, Berkeley, CA 94720}
\newcommand{\LLNL}{Lawrence Livermore National Laboratory, Livermore, CA, 94550}
\newcommand{\MIT}{Massachusetts Institute of Technology, Cambridge, MA 02139}
\newcommand{\NMSU}{New Mexico State University, Las Cruces, NM 88003}
\newcommand{\NYU}{New York University, New York, NY 10003}
\newcommand{\NWU}{Northwestern University, Evanston, IL 60208}
\newcommand{\ORNL}{Oak Ridge National Laboratory, Oak Ridge, TN 37831}
\newcommand{\Oxy}{Occidental College, Los Angeles, CA 90041}
\newcommand{\PNNL}{Pacific Northwest National Laboratory ,Richland, WA 99352}
\newcommand{\PI}{Perimeter Institute,  N2L 2Y5 Waterloo, Canada}
\newcommand{\Brown}{Brown University, Providence, RI 02912}
\newcommand{\Queens}{Queen's University , K7L 3N6 Kingston, Canada}
\newcommand{\Princeton}{Princeton University, Princeton, NJ 08544}
\newcommand{\TelAviv}{Tel-Aviv University,  69978 Tel-Aviv, Israel}
\newcommand{\Sanford}{Sanford Underground Research Facility, Lead, SD 57754}
\newcommand{\SCIPP}{University of California at Santa Cruz, Santa Cruz, CA 95064}
\newcommand{\RomaS}{Sapienza Universit\`a di Roma, 00185 Roma, Italy}
\newcommand{\SLAC}{SLAC National Accelerator Laboratory, Menlo Park, CA 94205}
\newcommand{\SNOLAB}{SNOLAB, Lively, ON P3Y 1N2, Canada}
\newcommand{\Stanford}{Stanford University, Stanford, 94305}
\newcommand{\TUM}{Technical University of Munich,  80333 Munich, Germany}
\newcommand{\Techsource}{Techsource Incorporated, Los Alamos, NM 87544}
\newcommand{\DOE}{US. Department of Energy, Germantown, MD 20874}
\newcommand{\Temple}{Temple University, Philadelphia, PA 19122}
\newcommand{\OSU}{The Ohio State University, Columbus, OH 43212}
\newcommand{\USD}{The University of South Dakota, Vermillion, SD 57069}
\newcommand{\Utenn}{The University of Tennessee, Knoxville, TN 37996}
\newcommand{\UBC}{University of British Columbia, BC V8P 5C2 Victoria, Canada}
\newcommand{\IPP}{Institute for Particle Physics, BC V8W 3P6 Victoria, Canada}
\newcommand{\UCD}{University of California at Davis, Davis, CA    95616}
\newcommand{\UCSB}{University of California at Santa Barbara, Santa Barbara, CA, 93106}
\newcommand{\SUNYA}{University at Albany SUNY, Albany, NY 12222}
\newcommand{\UCL}{University College London, WC1E 6BT London, United Kingdom}
\newcommand{\Liverpool}{University of Liverpool,  L69 7ZE Liverpool , United Kingdom}
\newcommand{\DESY}{DESY,  22607 Hamburg, Germany}
\newcommand{\UA}{University of Alabama, Tuscaloosa, AL 35487}
\newcommand{\UCR}{University of California at Riverside, Riverside, CA 92521}
\newcommand{\UCSC}{University of California at Santa Cruz, Santa Cruz, CA 95064}
\newcommand{\UCSD}{University of California San Diego, La Jolla, CA 92093}
\newcommand{\UFL}{University of Florida, Gainesville, FL 32611}
\newcommand{\Umich}{University of Michigan, Ann Arbor, MI 48109}
\newcommand{\UMN}{University of Minnesota, Minneapolis, MN 55455}
\newcommand{\UNV}{University of Nevada, Reno, NV 89557}
\newcommand{\UNM}{University of New Mexico,vAlbuquerque, USA, 87131}
\newcommand{\ND}{University of Notre Dame,vNotre Dame, IN 46556}
\newcommand{\OK}{ University of Oklahoma, Norman, OK 73019}
\newcommand{\Sassari}{Universit\`a di Sassari, 07100 Sassari,  Italy}
\newcommand{\Sheffield}{University of Sheffield, S3 7RH Sheffield, United Kingdom}
\newcommand{\Uvic}{University of Victoria, BC V8P 5C2 Victoria, Canada}
\newcommand{\UVA}{University of Virginia, Charlottesville, VA 22903}
\newcommand{\VUU}{Virginia Union University, Richmond, Virginia, 23220}
\newcommand{\WUSL}{Washington University in St Louis, St. Louis, MO 63130}
\newcommand{\Wellesley}{Wellesley College, Wellesley, MA 02481}
\newcommand{\WM}{College of William and Mary, Newport News, VA 23606}
\newcommand{\UMD}{University of Maryland, College Park, MD 20742}
\newcommand{\JLAB}{Thomas Jefferson National Laboratory, Newport News, VA 23606}
\newcommand{\Jerusalem}{Hebrew University of Jerusalem, 91904 Jerusalem, Israel}
\newcommand{\PNPI}{Petersburg Nuclear Physics Institute, 188300 Gatchina, Russia}
\newcommand{\INFNLNF}{Istituto Nazionale di Fisica Nucleare, Laboratori Nazionali di Frascati, 00044 Frascati, Italy}
\newcommand{\INFNRM}{Istituto Nazionale di Fisica Nucleare, Sezione di Roma, 00185 Roma, Italy}
\newcommand{\INFNT}{Istituto Nazionale di Fisica Nucleare, Sezione di Torino, 10125, Italy }
\newcommand{\UNC}{University of North Carolina at Chapel Hill, Chapel Hill, NC 27599}
\newcommand{\UNH}{University of New Hampshire, Durham, NH 03824}
\newcommand{\CPthree}{CP3-Origins, 5230 Odense, Denmark}
\newcommand{\KPMU}{University of Tokyo, 277-8583  Kashiwa , Japan}
\newcommand{\KINGS}{King's College London, WC2R 2LS London, United Kingdom}
\newcommand{\Glasgow}{University of Glasgow, G12 8QQ Glasgow, United Kingdom}
\newcommand{\Kobe}{Kobe University, 657-8501 Kobe, Japan}
\newcommand{\ED}{University of Edinburgh, EH8 9YL Edinburgh, United Kingdom}
\newcommand{\Nottingham}{University of Nottingham, NG7 2RD Nottingham, United Kingdom}
\newcommand{\IHEP}{Institute of High Energy Physics, Austrian Academy of Sciences, 1050 Vienna, Austria}
\newcommand{\ICRR}{Institute for Cosmic Ray Resaerch, The University of Tokyo, 456 Higashi-Mozumi, Kamioka, Hida, Gifu 506-1205, Japan}

%% file: authorlist.tex
\author{Marco Battaglieri (SAC co-chair)}
 %\email{battaglieri@ge.infn.it}
\affiliation{\INFNG}

\author{Alberto Belloni (Coordinator)}
 %\email{abelloni@mail.cern.ch}
\affiliation{\UMD}

\author{Aaron Chou (WG2 Convener)}
 %\email{achou@fnal.gov}
\affiliation{\FNAL}

\author{Priscilla Cushman (Coordinator)}
 %\email{prisca@physics.umn.edu}
\affiliation{\UMN}

\author{Bertrand Echenard (WG3 Convener)}
 %\email{echenard@hep.caltech.edu}
\affiliation{\Caltech}

\author{Rouven Essig (WG1 Convener)}
 %\email{rouven.essig@stonybrook.edu}
\affiliation{\CNYang}

\author{Juan Estrada (WG1 Convener)}
 %\email{estrada@fnal.gov}
\affiliation{\FNAL}

\author{Jonathan L.~Feng (WG4 Convener)}
 %\email{jlf@uci.edu}
\affiliation{\UCI}
 
\author{Brenna Flaugher (Coordinator)}
 %\email{brenna@fnal.gov}
\affiliation{\FNAL}

\author{Patrick J. Fox (WG4 Convener)}
 %\email{pjfox@fnal.gov}
\affiliation{\FNAL}

% ...

%% file: abstract.tex
%\noindent \textbf{Preamble}

\section*{Preamble}

\noindent
This white paper summarizes the workshop ``U.S. Cosmic Visions: New Ideas in Dark Matter'' held at University of Maryland from March 23-25. The flagships of the US Dark Matter search program are the G2 experiments ADMX, LZ, and SuperCDMS, which will cover well-motivated axion and WIMP dark matter over a range of masses.  The workshop assumes that a complete exploration of this parameter space remains the highest priority of the dark matter community, and focuses instead on the science case for additional new small-scale projects in dark matter science that complement the G2 program (and other ongoing projects worldwide).  It therefore concentrates on exploring distinct, well-motivated parameter space that will not be covered by the existing program; on surveying ideas for such projects (i.e. projects costing $\sim$\$10M or less); and on placing these ideas in a global context.  The workshop included over 100 presentations of new ideas, proposals and recent science and R\&D results from the US and international scientific community.

%% file: exec.tex
\section*{Executive Summary}

Deciphering the fundamental nature of dark matter---its cosmological origin, its constituents, and their interactions---is one of the foremost open questions in fundamental science today with tremendous potential to deepen our understanding of the laws of Nature.  The existing dark matter experimental program is focused primarily on weakly-interacting massive particles (WIMPs), which remain of great interest.  At the same time, given the importance of dark matter, there is strong motivation to explore a broader set of dark matter candidates.  Indeed, the 2014 P5 report calls out the importance of ``search[ing] for dark matter along every feasible avenue.''  

In recent years, the field of dark matter has been characterized by the blossoming of many innovative ideas.  New dark matter candidates have emerged that, like previous candidates, are highly motivated by beautiful theoretical results or experimental data, but are qualitatively different in their experimental implications. Most notably, some of these candidates can be explored by small experiments with short timescales, where a modest investment can have an outsize impact.  
Two broad classes of dark matter models stand out as ripe for exploration: 

\noindent {\em Hidden-Sector Dark Matter} candidates are completely neutral under Standard Model forces, but interact through a new force.  The low-mass (sub-GeV) parameter space for hidden-sector dark matter is both important and beyond the reach of the existing program.  Particularly well-motivated milestones in parameter space are derived from general production mechanisms, theory, and current experimental anomalies. Novel small-scale direct detection experiments, fixed-target experiments, and even astrophysical, nuclear, and atomic probes each have unique sensitivities to fully explore these milestones.  

\noindent {\em Ultra-Light Dark Matter} candidates have masses from $10^{-22}$ eV to about a keV, and that can be produced during inflation or phase transitions in the very early Universe.  A particularly motivated case is QCD axion dark matter, predicted by the axion solution to the strong CP problem, which defines an important milestone in coupling sensitivity as a function of mass.  Much of this parameter space, including low-mass QCD axions, was thought to be completely inaccessible several years ago, but can now be explored by a suite of new experimental approaches.

The community has presented a diverse and innovative set of experimental proposals, including potential game-changers in the search for their respective dark matter candidates.  Many exploit unique US-based facilities and/or expertise, and represent natural opportunities for US leadership in the field.  
The proposals described in this document demonstrate the vibrancy of the dark matter community in universities and labs in the United States and around the world.  Many of the new ideas presented here were spawned not by a programmatic approach to the dark matter problem, but by small groups developing ideas and technologies to tackle a variety of fundamental questions. In many cases, these proposals are the result of close collaboration between experimentalists and theorists and include researchers from disciplines outside of high energy physics, such as nuclear, atomic, and condensed matter physics.  

We highlight five important directions (\emph{not} ordered by priority) for a small experiments program and continued innovation in dark matter (DM) physics: 
\begin{itemize}
\item Low-threshold direct detection is an active field with a wealth of low-cost new and innovative ideas that can probe a variety of highly motivated Hidden-Sector and Ultralight DM candidates, and affords the only prospect to begin testing the tiny couplings associated with hidden-sector freeze-in through an ultralight mediator.  It is important to pursue both DM-electron and DM-nuclear scattering experiments, as they have complementary sensitivities.  Some proposals are ready for small-project-scale funding now, while several will be ready for small-project-scale funding within the next 1 to 2 years.  In addition, the potential to lower energy thresholds by orders of magnitude motivates continued technology R\&D. 
\item A suite of experiments using multiple technologies are required to explore the wide parameter space of light new-force carriers, and in particular the full mass range for QCD axions.  The ADMX G2 experiment is currently exploring an exciting region of QCD axion mass range, and many new experimental approaches are in pilot phases now.  Together with ADMX, next-generation experiments at the small-project scale can explore much of the highly motivated QCD axion parameter space over the next decade.  
\item Accelerator experiments can both produce and detect new particles, such as dark matter and the particles mediating new interactions. This unique ability has enabled beam dump, missing mass/energy, and visible mediator search experiments to achieve world-leading sensitivity to highly sought-after dark matter scenarios. Building on these proven techniques and exploiting existing US accelerator facilities, a small number of fixed-target experiments can broadly explore sub-GeV dark matter and associated forces with sufficient sensitivity to test all predictive thermal DM scenarios.  This focused effort is based on established detector technology, with a number of modest-cost proposals ready for funding now to achieve significant science in the next few years.
\item Existing data may already be pointing to dark sector physics.  Anomalies in $(g-2)$ of the muon and in the properties of beryllium-8 nuclei provide tentative evidence for a new boson at the 10 MeV-scale that can be tested by nuclear and atomic spectroscopy experiments.  The small-scale structure of dark matter halo distributions may be explained by dark matter self-interactions with 1-100 MeV mediators.  LIGO's discovery of colliding black holes motivates micro-lensing probes of solar mass black hole dark matter.  These puzzles each define sharp, highly-motivated targets that can be resolved by small investments in experiment, simulations, and theory.  Typical timescales are 1 to 2 years and budgets are a small fraction of the small projects threshold. 
%There are many more low-cost opportunities that can make a major impact on dark matter science.  These include modest investments in experiments and theory to investigate current anomalies in muon $(g-2)$, beryllium-8, and small-scale structure in dark matter halo distributions; probes of new forces from macroscopic to sub-nuclear length-scales; and a diverse set of proposed searches for motivated DM candidates beyond those focused on above.  
\item Progress in theory has been the driving force behind recent developments in dark matter, particularly proposals for small-scale experiments and innovative connections to other subfields.  Additional investments in theory are essential to exploit cosmological and astrophysical data to improve measurements of dark matter's particle properties and to develop the novel connections to nuclear, atomic, and condensed matter physics that have already been identified.
\end{itemize}
All of these directions are scientifically important, and they motivate a portfolio of multiple small experiments in dark matter, including experiments in direct detection, accelerator-based searches, searches for coherent-field dark matter, and broad investigations of dark-sector properties, as well as targeted investments in theory. A healthy dark matter research program should include both large- and small-scale efforts.  This document illustrates both the breadth and the promise of small-scale opportunities in dark matter science, any one of which may lead to a breakthrough and transform our understanding of the cosmos.

%% file: introduction.tex
\section{Introduction}
The evidence for dark matter comes from cosmological and astrophysical measurements in many different contexts and over a wide range of scales --- from the shape of the cosmic microwave background (CMB) power spectrum to cluster and galactic rotation curves and gravitational lensing.  Yet all of these data are essentially gravitational, and therefore tell us little directly about the particle nature of dark matter.  In particular, constituents of dark matter could be as light as $10^{-22} \eV$ or as heavy as $100 M_\odot$, and still be consistent with these observations.  Deciphering the fundamental nature of the dark matter --- its cosmological origin, its constituents, and their interactions --- is one of the foremost open questions in basic science today.  Answering this question involves synergies across multiple levels: between experimentalists and theorist and between high energy physics and other disciplines, such as nuclear, atomic, and condensed matter physics.

The search for dark matter can be focused by putting it in the context of known cosmology and particle physics: how our Universe's cosmic history gives rise to the dark matter abundance, and how the Standard Model both informs and restricts the possibilities for dark matter particles' interactions.  That these guiding questions have many possible answers is part of the reason why uncovering the particle nature of dark matter is so important, and so challenging; it also necessitates a multi-faceted program with different techniques optimized for different dark matter mass ranges and interactions.    

The 2014 P5 report has called out the importance of a broad dark matter search program: ``It is imperative to search for dark matter along every feasible avenue,'' and the breadth of ``well-motivated ideas for what dark matter could be, [which] include weakly interacting massive particles (WIMPs), gravitinos, axions, sterile neutrinos, asymmetric dark matter, and hidden sector dark matter'' \cite{P5}.  Some of these scenarios -- including (with some notable exceptions) WIMP, gravitino, and sterile neutrino DM --- are the purview of larger experiments, as reviewed for example in \cite{Feng:2014uja}.
%\cite{Cushman:2013zza,Buckley:2013bha,Adhikari:2016bei}.  
But much of the well-motivated parameter space for dark matter \textbf{can be explored by small experiments in the near future}.  

\begin{figure}[tbp]
\begin{center}
\includegraphics[width=0.98\textwidth]{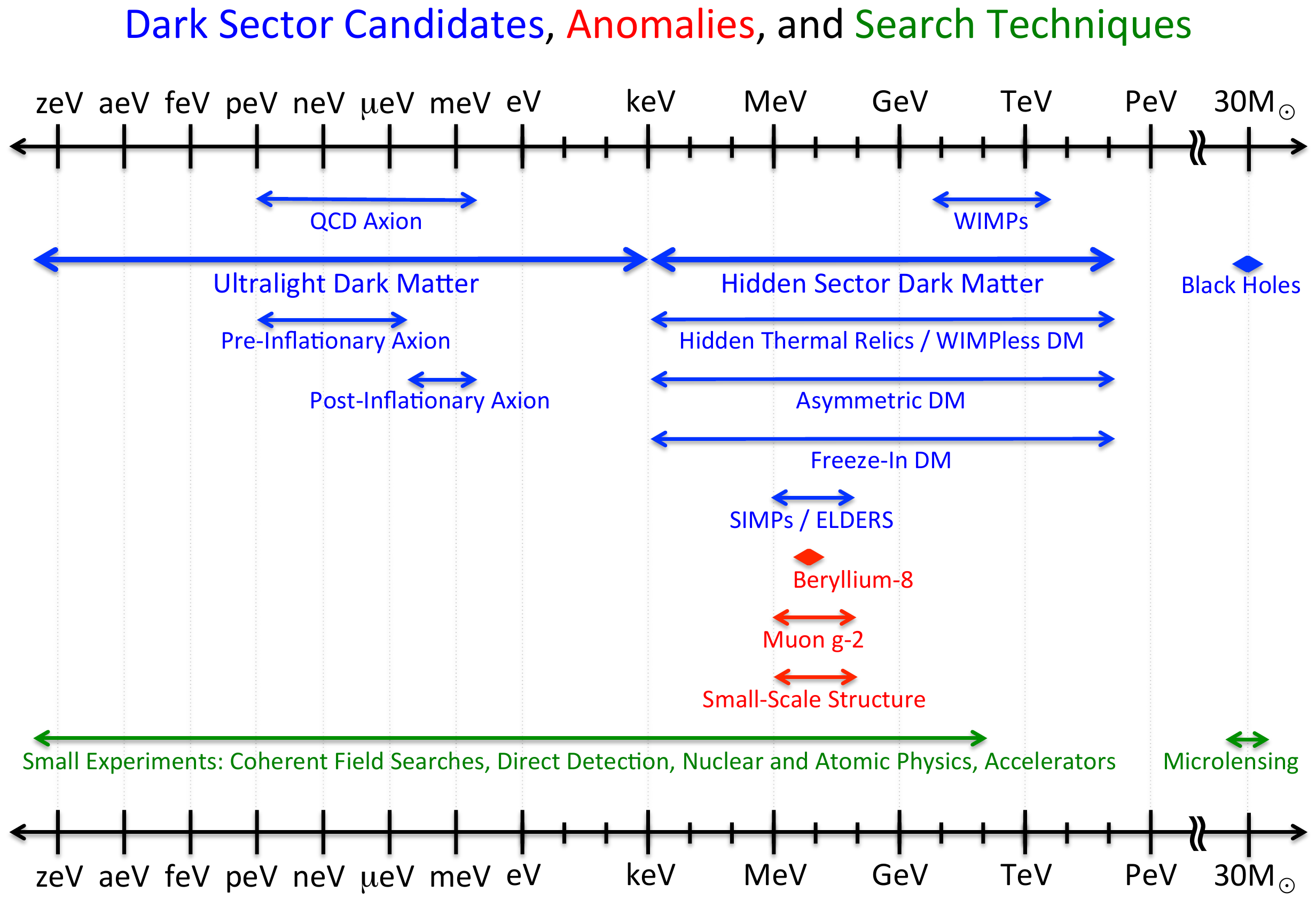}
\caption{Mass ranges for dark matter and mediator particle candidates, experimental anomalies, and search techniques described in this document.  All mass ranges are merely representative; for details, see the text.  The QCD axion mass upper bound is set by supernova constraints, and may be significantly raised by astrophysical uncertainties. Axion-like dark matter may also have lower masses than depicted.  Ultralight Dark Matter and Hidden Sector Dark Matter are broad frameworks.  Mass ranges corresponding to various production mechanisms within each framework are shown and are discussed in Sec.~\ref{sec:sciencecase}. The Beryllium-8, muon $(g-2)$, and small-scale structure anomalies are described in \ref{WG4sec:WG4}.  The search techniques of Coherent Field Searches, Direct Detection, and Accelerators are described in Secs.~\ref{sec:WG2experiments}, \ref{sec:WG1experiments}, and \ref{sec:WG3experiments}, respectively, and Nuclear and Atomic Physics and Microlensing searches are described in Sec.~\ref{WG4sec:WG4}.}
\label{fig:SummaryPlot}
\end{center}
\end{figure}

Two broad classes of dark matter model stand out as strongly motivated possibilities where small experiments can have an outsized impact: 
\begin{itemize}
%\item{\bf Dark matter in the vicinity of Standard Model scales} includes WIMPs, which interact through SM forces, and hidden-sector DM --- dark matter that interacts through a new force (sometimes called a ``dark sector'').  In both cases, the thermal history of the Universe and the couplings to familiar matter play key roles in generating the observed DM abundance.  Small experiments can probe new parameter space for low-mass WIMPs, and are particularly powerful in exploring low-mass hidden-sector DM.  For example, accelerator-based experiments offer a robust probe of sub-GeV hidden-sector DM produced through thermal freeze-out, fully exploring the most predictive models, and low-threshold direct detection experiments can exploit a unique kinematic enhancement at low mediator masses to start exploring the very weakly coupled ``freeze-in'' scenarios.
\item Dark matter in the vicinity of Standard Model scales includes WIMPs, which interact through SM forces, and also {\bf hidden-sector DM} --- dark matter that interacts through a new force (sometimes called a ``dark sector'').  They share common motivations --- in both cases, the thermal history of the Universe and the couplings to familiar matter play key roles in generating the observed DM abundance.  Hidden-sector DM is viable over a wider mass range than WIMPs.  The parameter space below the GeV scale is largely invisible to WIMP searches, but presents multiple opportunities for new experiments. For example, low-threshold direct detection experiments can explore well-motivated parameter space even with gram-scale target masses, and exploit a unique kinematic enhancement at low mediator masses to start exploring the very weakly coupled ``freeze-in'' scenarios. Accelerator-based experiments offer a robust probe of sub-GeV hidden-sector DM produced through thermal freeze-out, fully exploring the most predictive models.  Moreover, the DM physics in these hidden-sector models also encompasses other effects of the new force, such as DM self-interactions and new reactions of ordinary matter.
\item{\bf Ultra-light dark matter}, in the mass range from $10^{-22}$ to about a keV, includes scalar, pseudo-scalar, and vector boson DM that are produced during inflation or a high-temperature phase transition.  In most of this parameter space, the DM acts as an oscillating classical field, whose coupling to matter can be detected by a variety of precision experiments.  The QCD axion solution to the strong CP problem motivates axion dark matter, and provides a well defined target in the coupling sensitivity-dark matter mass parameter space.  Much of the axion parameter space was believed to be inaccessible several years ago, but the development of several new experimental techniques and advances in detector capabilities opens the possibility of searching a large portion of this space in the near future.
\end{itemize}
These two frameworks, as well as specific production mechanisms within each framework, experimental anomalies, and search techniques, are shown in Fig.~\ref{fig:SummaryPlot}.

%The search for dark matter is of paramount importance for two reasons.  The first is evident: We know that there is an undiscovered form of matter, which in fact dominates the Universe, yet we know almost nothing about it.  Identifying the nature of this matter is  scientifically important in its own right, and could lead us in scientific directions that are not yet imagined.  A second, less obvious but crucial motivation to search for dark matter is the promise that a discovery will teach us deep lessons about the laws of Nature and change the course for future exploration.  For example, the discovery of hidden-sector dark matter would demonstrate that the forces and particles of the Standard Model are just one corner of the theory of Nature.  This would strongly motivate an exploration of other physics weakly coupled to us.  Axion dark matter would explain the long-standing puzzle of CP symmetry in the strong interactions, while any very light dark matter particle would provide a glimpse of the earliest moments of our Universe's history.  

%%%%%%%%%%%%%%%%%%%%%%%%%%%%%%%%%
%%%%%%%%%%%%%%%%%%%%%%%%%%%%%%%%%%%

This white paper summarizes the workshop ``U.S. Cosmic Visions: New Ideas in Dark Matter'' held at University of Maryland from March 23-25.  The workshop focused on the science case for new small-scale projects in dark matter science,  on surveying ideas for such projects (i.e.~projects costing $\sim$ \$10M or less) in the U.S. Dark Matter search program, and on placing these ideas in a global context.  The workshop included over 100 presentations of new proposals and recent results from the US and international scientific community. 

This report parallels the structure of the workshop.  Section \ref{sec:sciencecase} summarizes the science case for exploring dark matter parameter space, as well as high-value parameter-space targets that were highlighted at the workshop and the prospects for testing them. The following sections are organized, parallel to the structure of the workshop, around four working groups:
\begin{itemize}
\item {\bf New Avenues in Direct Detection} (Section \ref{sec:WG1experiments}), including spin-dependent WIMP scattering and many new ideas aimed at direct detection of dark matter lighter than traditional WIMPs, from several GeV down to the meV scale,
\item {\bf Detection of Ultra-Light (sub-eV) Dark Matter } (Section \ref{sec:WG2experiments}), exploiting several physical effects to search for coherent-field effects of dark matter ranging from $10^{-22}-1$ eV, including QCD axions,
\item {\bf Dark Matter Production at Fixed Target and Collider Experiments} (Section \ref{sec:WG3experiments}), searching for sub-GeV dark matter (and related new forces) with small-scale fixed-target experiments to a level of sensitivity motivated by models of light thermal Dark Matter, and
\item {\bf New Candidates, Targets, and Complementarity} (Section \ref{WG4sec:WG4}), surveying new theoretical models, high-value regions of parameter space motivated by existing experimental anomalies and theoretical ideas, the interplay of dark matter searches from different subfields and the complementarity of proposed small-scale experiments with the data expected from the existing program. 
\end{itemize}

There is a broad and active community of physicists pursuing these new directions, and developing experiments that, taken together, cover broad parameter regions with great sensitivity and decisively explore several high-priority targets.  
The experimental approaches presented at the workshop are highly complementary --- each of the working groups has identified models for which particular techniques are uniquely sensitive, while in many other cases a combination of different experimental approaches is required to move from discovery to a physical understanding of dark matter.

% INTRO TENTATIVE OUTLINE: 
% Dark matter is important
% Many possibilities for its nature and origin.  
% Workshop intended to address the science case for Dark Matter searches beyond  the current program, and in particular the opportunities for small experiments to play an important role in Dark Matter science. 
% The workshop was organized in four working groups...
% Summary of science case  -- broad regions & sharp targets
% One-paragraph summary of key messages from each working group. 
% Conclusion: not only are there opportunities -- including broad regions of interest that can be explored by small experiments, and sharp targets that can be tested decisively -- there are also a wealth of 
% Outline of the whitepaper

%% file: ScienceCase.tex
% !TEX root = ScienceCase_wrapper.tex
\section{Science Case for a Program of Small Experiments}\label{sec:sciencecase}
% SCIENCE CASE
%The evidence for dark matter comes from cosmological and astrophysical measurements in many different contexts and over a wide range of scales --- from the shape of the cosmic microwave background (CMB) power spectrum to cluster and galactic rotation curves and gravitational lensing.  Yet all of these data are essentially gravitational, and therefore tell us little directly about the particle nature of dark matter.  In particular, constituents of dark matter could be as light as $10^{-22} \eV$ or as heavy as $100$ solar masses, and still be consistent with these observations.

Given the wide range of possible dark matter candidates, it is useful to focus the search for dark matter by putting it in the context of what is known about our cosmological history and the interactions of the Standard Model, by posing questions like: What is the (particle physics) origin of the dark matter particles' mass? What is the (cosmological) origin of the abundance of dark matter seen today?  How do dark matter particles interact, both with one another and with the constituents of familiar matter? And what other observable consequences might we expect from this physics, in addition to the existence of dark matter?  Might \emph{existing} observations or theoretical puzzles be closely tied to the physics of dark matter?  These questions have many possible answers --- indeed, this is one reason why understanding the particle nature of dark matter is so important --- with each case pointing to a different range of dark matter properties  and hence motivating different techniques to search for dark matter particles.  
%
%The 2014 P5 report has called out the importance of a broad dark matter search program: ``It is imperative to search for dark matter along every feasible avenue,'' and the breadth of ``well-motivated ideas for what dark matter could be, [which] include weakly interacting massive particles (WIMPs), gravitinos, axions, sterile neutrinos, asymmetric dark matter, and hidden sector dark matter'' \cite{P5}.  Some of this vast landscape -- including (with some notable exceptions) WIMP, gravitino, and sterile neutrino DM --- is the purview of larger experiments, as reviewed for example in \cite{Cushman:2013zza,Buckley:2013bha,Adhikari:2016bei}.  But much of the well-motivated phase space for dark matter \textbf{can be explored by small experiments in the near future}.  These are the regions that we focus on in this report.

The WIMP hypothesis that has dominated the DM search program to date offers a good example of many of these motivations: new, weakly interacting particles at or above the weak scale are predicted in many models that address the electroweak hierarchy problem.  In this mass range, the weak interactions with familiar matter can naturally explain their abundance.  While WIMP dark matter remains highly motivated, several factors motivate a broader approach to the dark matter question.  Significant parameter space for both WIMPs and the supersymmetric models that realize  WIMP dark matter have been explored by recent advances in direct detection, indirect detection, and LHC searches.  New experimental ideas have opened prospects to explore the full viable mass range for axion DM, another long-standing DM candidate motivated by the strong CP problem.  Meanwhile, experimental anomalies are suggestive of substantial new interactions between DM particles and of new forces very weakly coupled to ordinary matter.  In parallel with (and often spurred by) these experimental developments, theoretical progress has underscored that both WIMP and axion dark matter are special cases of broader theoretical frameworks that have many of the same attractive features. 
%Arguably, too much weight has been given to the idea that dark matter be a byproduct of models motivated by other Standard Model puzzles, when simple solutions to the Dark Matter problem itself should be motivation enough to explore these general ideas.  
That these general frameworks can be very effectively explored by small experiments makes them particularly exciting. 

\subsection{Broad Frameworks for Dark Matter Motivating Small Experiments}
The dark matter candidates motivating small experiments can be organized into two broad classes, within which we highlight sharp, well-motivated targets that can be explored or robustly tested by a program of small experiments:

{\bf Hidden-sector Dark Matter} is a natural generalization of the WIMP idea  to include interactions through a new force rather than just SM forces.  These two scenarios are closely related: both suggest DM near Standard Model mass-scales, and in both cases the thermal history of the Universe and coupling to familiar matter play key roles in generating the observed DM abundance --- whether or not the DM ever reached thermal equilibrium with familiar matter.  However, the hidden-sector case --- in particular, the parameter space with GeV-scale or lighter DM and/or mediators --- opens up qualitatively new directions for experiment.  Cosmological DM production mechanisms, theoretical ideas, and observations point to parameter-space milestones of particular interest.  {\bf Proposed small experiments are poised to conclusively test the most predictive possibility for thermal freeze-out of hidden-sector DM, where DM annihilates directly into SM particles, over most of the sub-GeV mass range. They will also explore parameter space for many other production mechanisms, including thermal DM with ``secluded'' annihilation, asymmetric DM, and very weakly coupled DM that ``freezes in'' without reaching equilibrium.}  Any new interaction between dark and familiar matter necessarily has consequences in the self-interactions of DM and of familiar matter, as well.  Intriguingly, several anomalies in data point to possible new physics, weakly coupled to familiar matter, in the 1-100 MeV scale, while a suppression of cosmological small-scale structure may be explained by DM self-scattering through a mediator in the same mass range.
%In addition to thermal freeze-out, several other mechanisms for the cosmological origin of DM can be naturally realized in a hidden sector.  These offer additional targets that can be explored \emph{along with} the thermal scenario, by the same experiments.  At the same time, the rich possibilities opened up by hidden-sector dark matter are still active areas of theoretical study, and recent efforts continue to motivate new parameter regions for exploration. 

{\bf  Ultralight dark matter}, bosonic particles with sub-keV mass include the QCD axion and generic light scalar, pseudo-scalar, and vector bosons coupled linearly to familiar matter.  Such particles can arise from string theory or other high-energy phenomena, with their masses protected by symmetries that generically also lead to exponentially small couplings to matter.  The observed DM abundance can be produced during an initial inflationary phase of cosmology or a high-temperature phase transition.  A distinctive feature of these models is that for sub-meV mass, the DM mode occupation numbers are high, so that ultralight DM can lead to classical oscillating field signals that in much of the parameter space offer the most promising path to detection.  The ``invisible'' QCD axion, long proposed as a solution to the strong CP problem, is an attractive dark matter candidate; while the viable mass range spans eight orders of magnitude, sharp predictions can be made for QCD axion couplings as a function of mass.  {\bf Small experiments can explore an enormous amount of ultralight boson dark matter parameter space, with several techniques capable of reaching sensitivity to the QCD-axion.}

\subsection{The Need for a Multi-Experiment Program}
\input{multiExp}

%\subsection{High-Impact Parameter-Space Targets}
%\input{figs/wg-targets}

\newpage

\section{Theory Overview and Motivations}\label{sec:theory-overview}
This section further explains the motivations for DM candidates noted in the Science Case, and defines several sharp parameter-space targets and broad regions of interest within their parameter spaces.  

\subsection{WIMPs}
WIMP dark matter --- composed of particles that interact through the SM weak interactions, and usually assumed to be produced through thermal freeze-out --- has long been an important benchmark model.  Indeed, most of the effort in direct and indirect DM detection, including the G2 program in the US, is motivated by the WIMP hypothesis.  There is scientific value to exploring WIMP parameter space beyond the G2 program, but this area is typically the purview of large-scale experiments and is beyond the scope  of this report.  We do note, however that there are some aspects of WIMP physics complementary to the G2 program where small experiments have an important role to play.  In particular, the workshop included discussions of small experiments searching for spin-dependent WIMP interactions and the ``low-mass WIMP" parameter space from $\sim1-10$ GeV.  The latter is, in fact, mostly below the mass range for conventional thermal WIMPs, and scientifically motivated by hidden-sector dark matter, discussed below.

Both top-down and bottom-up considerations motivate multi-GeV to TeV-scale WIMP masses.  These are the natural mass scales for any particle involved in solving the hierarchy problem, or for a particle whose mass shares a common origin with the Standard Model Higgs.  A similar mass range is singled out for annihilation through weak interactions to give rise to the observed DM abundance -- at masses much higher than a TeV or lower than several GeV (the Lee-Weinberg bound), the DM annihilation cross-sections are too small and therefore an overabundance of thermal DM would be expected.  

\subsection{Hidden Sector DM}
\input{hiddenSector}

\subsection{Ultralight Dark Matter}\label{sec:Science-ultralight}
\input{lightBoson}

%% file: multiExp.tex
We emphasize that a comprehensive exploration of this science requires multiple experiments with complementary sensitivity.  This is perhaps most obvious in the case of ultralight DM, where different experimental techniques cover different ranges of dark matter mass.  Though many different probes of hidden-sector DM explore the keV-to-GeV mass ranges, they do so while achieving substantially different science --- not only do they present distinct discovery opportunities, they also provide qualitatively different information about the DM.

\subsubsection{Ultralight DM}
In the case of ultra-light bosonic dark matter, the QCD axion remains one of the best motivated dark matter candidates.  While axions provide perhaps the simplest solution to the strong-CP problem, these models also inevitably produce dark matter via release of their initial potential energy density.   Axion dark matter searches and cosmic microwave background experiments provide complementary probes of inflation;  a measurement of the axion mass by detection of the dark matter beam would also immediately determine or constrain the energy scale of cosmic inflation;  Recent phenomenology has also indicated a possibly rich interplay between QCD axions and the electroweak hierarchy problem.

\begin{figure}[htbp]
\includegraphics[width=\textwidth]{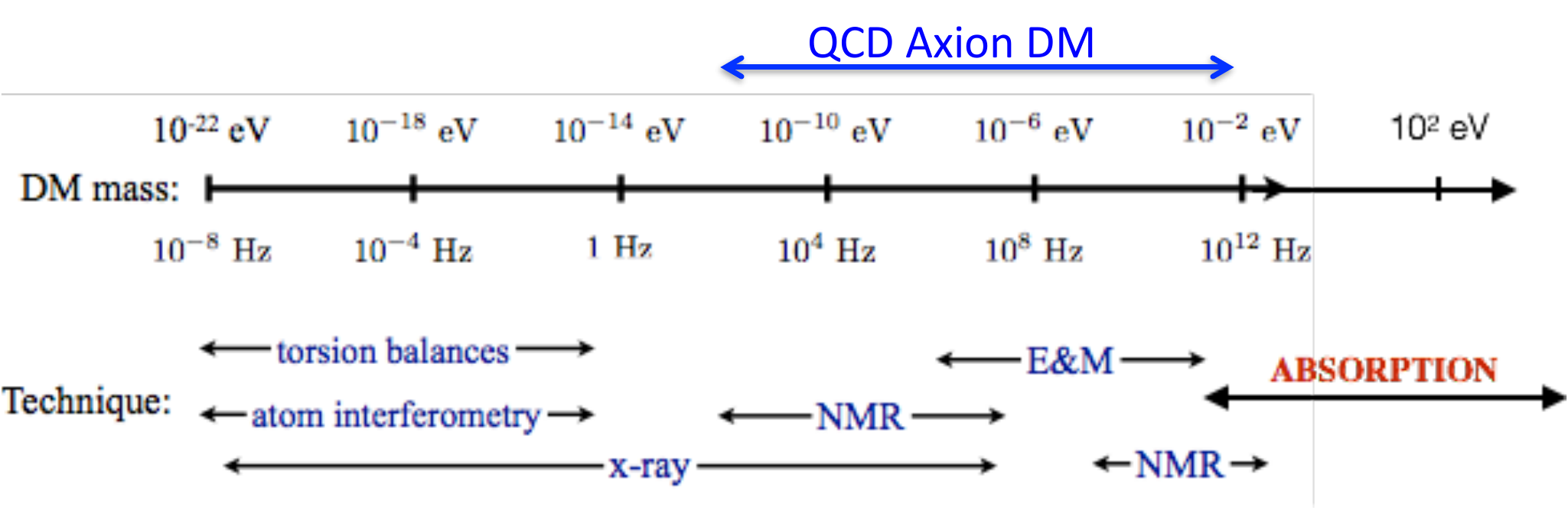}
\caption{\label{fig:ULcomplementarity}Schematic illustration of the complementarity of different types of experiments in exploring QCD axion DM and ultralight DM more generally.  The horizontal axis illustrates the observationally allowed mass range for ultralight DM, with an arrow highlighting the viable mass range for the QCD axion specifically.  Indicative ranges of sensitivity for different techniques are illustrated by dark blue arrows for coherent field, new-force, and X-ray helioscope techniques (see Sec.~\ref{sec:WG2experiments}, while a red arrow indicates the range of DM masses that can be explored by absorption in direct-detection experiments (see Sec.~\ref{sec:WG1experiments}).}
\end{figure}

Several distinct techniques are proposed to search for the QCD axion (see Sec.~\ref{sec:WG2experiments}), many of which search for a coherent  field induced by the axion DM, which oscillates at a frequency $\omega = m_a /\hbar$ for axion mass $m_a$.  The large range of viable QCD axion masses, from $10^{-12}$ to $10^{-2}$ eV, implies a correspondingly large range of frequencies for possible DM signals.  No one technique can cover this wide range of frequencies. Searching for QCD axion DM over their full parameter space requires a suite of techniques, such as cavity resonators (including ADMX G2) at high frequencies, lumped element resonators at medium frequencies, and nuclear magnetic resonance.  

Searches for the QCD axion, including continuation of the current ADMX generation 2 experiment should be given high priority in any future dark matter program.  However, it should be noted that  more general scalar, pseudoscalar, or vector dark matter models over an even wider range of masses (from $10^{-22}$ eV to $10^3$ eV) are also well motivated.  Similarly cost-effective experimental techniques have been identified that would cover these possibilities.  Many  (see Sec.~\ref{sec:WG2experiments}) import non-traditional detector technology that has nonetheless been well developed in other fields of physics, including atom interferometry, nuclear magnetic resonance, and fifth force measurements.  Large improvements in sensitivity to low mass bosonic dark matter can and should be quickly obtained by engaging these other communities in cross-disciplinary collaborations.   Meanwhile, the same direct detection experiments that can search for sub-GeV hidden-sector DM (see Sec.~\ref{sec:WG1experiments}) can also be used to search for absorption of ultralight DM particles in the heavier part of its allowed mass range, from meV to keV scales, where THz-scale frequencies for the oscillating DM field make field coherence harder to exploit.  The importance of a multi-experiment program to explore these models comprehensively is illustrated in Figure \ref{fig:ULcomplementarity}.

\subsubsection{Hidden-Sector DM}
\begin{figure}[htbp]
\includegraphics[width=0.8\textwidth]{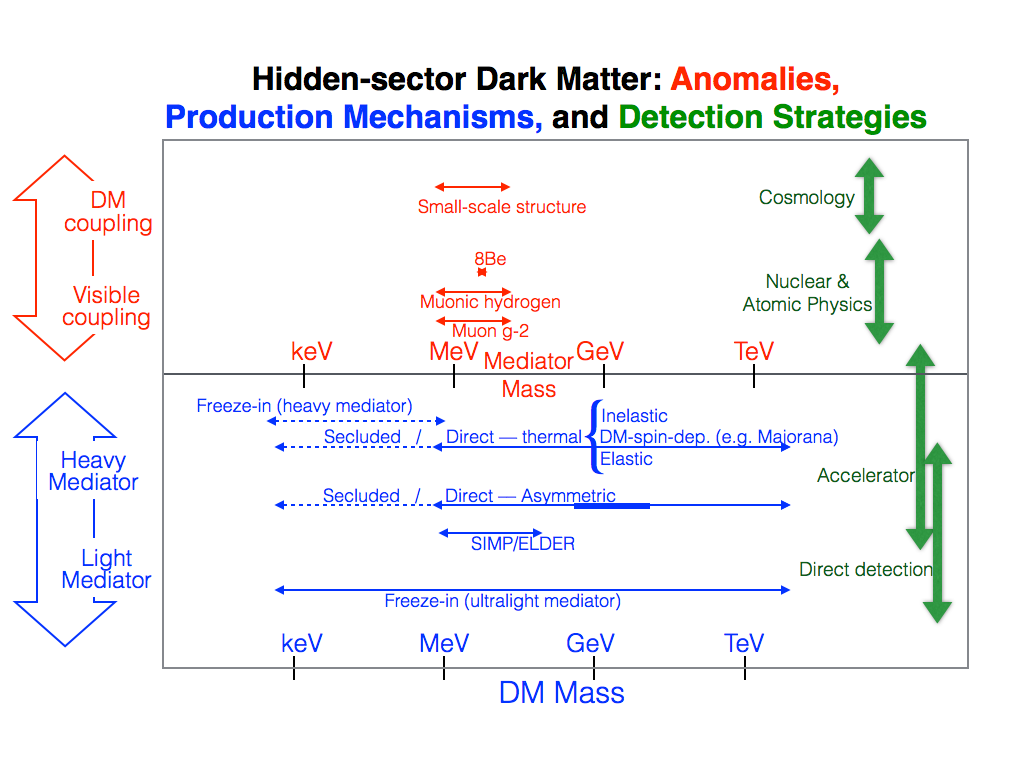}
\caption{\label{fig:hiddenComplementarity}Schematic illustration of the complementarity of different types of experiments in exploring sharp targets and general regions of interest for hidden-sector DM.  Anomalies in data (see Section \ref{sssec:HS-opportunities}) highlight regions of interest in mediator mass and/or coupling to visible or dark matter; the red arrows highlight the suggested regions of mediator mass.  Blue horizontal arrows for production mechanisms (see Sections \ref{sssec:HS-thermal}-\ref{sssec:HS-freezein}) indicate the parameter regions over which they are viable (dashed), regions in which they motivate a sharp parameter-space target (solid arrow), and, in the case of asymmetric DM, a ``natural'' range where the DM and baryon number densities are comparable (thick band).  Blue and red vertical arrows highlight directions in ``theory space'' that have significant impact on detection strategies, while the green vertical arrows indicate the models to which different experimental approaches are most sensitive.  Direct detection is discussed in Section \ref{sec:WG1experiments}, accelerator-based experiments in Section \ref{sec:WG3experiments}, and cosmology and nuclear and atomic physics probes in Section \ref{WG4sec:WG4}.}
\end{figure}

In contrast, searches for hidden-sector DM are primarily exploring a more focused DM mass range from a keV to several GeV.  However, there are very important differences between the sensitivities of different experiments.  These differences can be loosely classified along three directions: the difference between relativistic and non-relativistic probes of DM interactions, the importance of probing DM-DM and SM-SM (in addition to DM-SM) interactions, and experimental signals' dependence on the precise nature of mediator interactions with familiar matter. We discuss each of these points in turn below and illustrate them graphically in Figure \ref{fig:hiddenComplementarity}. 

In general, accelerator searches (Sec.~\ref{sec:WG3experiments}) explore the relativistic production and/or interactions of DM candidates, while direct detection experiments (Sec.~\ref{sec:WG1experiments}) search for the scattering of DM in the Milky Way halo off matter, with relative velocity $\sim 10^{-3} c$.  The effect of this kinematic difference is that well-motivated scenarios 10--20 orders of magnitude beyond the reach of one technique are accessible to the other.  For example, thermal freeze-out of hidden sector DM via a mediator coupled to familiar matter (the ``direct'' channel) represents a precise target of interest.  For elastically scattering scalar DM charged under a new force, most of the sub-GeV parameter space for this scenario can be explored by the next generation of both accelerator and direct detection experiments.  If instead the DM is axially coupled (as a Majorana fermion must be) or scatters inelastically, then direct detection rates are suppressed by anywhere from 6 to 18 orders of magnitude, while accelerator production rates are within one to two decades.  Therefore, while both techniques can explore this possibility, only accelerators are able to do so robustly.  The converse is true if the mediator of DM-SM scattering is much lighter than the DM itself.  In this case, direct detection rates are parametrically enhanced by up to 12 orders of magnitude, because of their low momentum transfer.  This opens the possibility of testing the idea that the DM abundance ``freezes in'' through DM and SM interactions with a very light  mediator, which would be too weakly coupled to be seen at accelerators.  

It may be that the new force at the heart of hidden sector DM is most readily explored, not through DM-SM interactions as discussed above, but through the new physics it induces within the DM or visible sector.  These possibilities imply considerable synergy with disciplines including astrophysics, cosmology, and nuclear, atomic, and condensed matter physics.  Indeed, there are existing experimental anomalies that may be pointing towards dark sector physics; testing these is a crucial piece of the search for dark matter and one of the recurring themes of Sec.~\ref{WG4sec:WG4}.  For example, dark matter self-interactions have been suggested as an explanation of puzzles in small-scale cosmological structure.  Small investments in simulations and astroparticle theory can leverage the enormous amount of cosmological data already being collected to shed light on these puzzles, providing not just {\em constraints} on dark matter candidates, but {\em measurements} of the properties of dark matter.  Another sharp motivation for small experiments is the  $^8$Be anomaly, a possible signal of a new force interacting with nuclei and electrons.  The $^8$Be anomaly strongly motivates proposed followup nuclear experiments that are fast (under 2 years) and cheap (a small fraction of the small projects threshold), as well as isotope shift spectroscopy experiments and accelerator searches for new bosons with masses $\sim 10$ MeV and electron couplings $\varepsilon \sim 10^{-4} - 10^{-3}$.  It is quite intriguing that new models, astrophysical observations, and existing experimental anomalies point to the 1 to 100 MeV mass scale as a high-value target region for dark matter and dark mediator searches.  

We also highlight the value of exploring both electron and nucleon couplings to DM and dark forces.  While the most widely used benchmark model --- a kinetically mixed dark photon --- couples equally to electrons and protons, other possibilities would have one of these couplings much larger than the other (for example, a scalar mediator with mass-proportional couplings or a vector mediator coupled primarily to baryon number, lepton number, or another combination of charges).  This provides motivation for a program of small direct-detection experiments that includes both proton and electron recoil experiments, and for an accelerator-based program that includes both electron- and proton-beam experiments to maximize sensitivity.   Similarly, ultralight DM can have a variety of couplings and there is good motivation to search for all of them, even if experiments cover overlapping mass ranges.  

It is just as important to emphasize that, even when they probe the same DM candidates, various kinds of experiments offer different information about the DM.  For example, a discovery of a new particle at an underground direct-detection experiment would constitute strong evidence that such a particle constitutes all or at least part of the DM, but cannot disentangle the particle's couplings from its abundance.  In contrast, accelerator-based experiments provide a clear probe of particle properties, but not its stability on cosmological timescales.  

\subsubsection{Further Motivations and Opportunities}
While the science case for many of these opportunities falls nicely into either ultralight or hidden-sector DM, these are certainly not the limits of well-motivated possibilities.  For example, the LIGO discovery of gravitational waves from colliding black holes has renewed interest in multi-solar-mass primordial black hole DM.  The LIGO observation sharply motivates a proposed microlensing search (see Sec.~\ref{WG4sec:WG4}) that can confirm or exclude the possibility of intermediate mass black hole dark matter using existing facilities with minimal funding.

The hunt for dark matter now crosses multiple frontiers and benefits from vibrant communication between many subdisciplines of physics, including astrophysics, cosmology, and nuclear, atomic, and condensed matter physics.  Innovations springing from these collaborations have created a wealth of new ideas that can be explored by inexpensive experiments.  
 Healthy support for theory is essential to maintaining the flow of creative and cross-disciplinary ideas that have been seen in recent years, and which may finally unmask the particle identity of dark matter.

%% file: hiddenSector.tex
The absence of sizable DM interactions with ordinary matter motivates the simple hypothesis that it consists of particles neutral under SM forces, but perhaps charged under new forces that have not yet been discovered.  Such hidden sectors have been considered as a possible origin for dark matter for decades \cite{Kobzarev:1966qya,Blinnikov:1982eh,Foot:1991bp,Hodges:1993yb,Berezhiani:1995am}.  Hidden-sector DM arises in Hidden Valleys \cite{Strassler:2006im} and their explicit top-down string constructions \cite{Cvetic:2012kj}, can live alongside TeV-scale Standard Model extensions, including supersymmetry \cite{Strassler:2006qa,Hooper:2008im,Feng:2008ya} and composite Higgs sectors, and is largely unconstrained by current data.  

While hidden-sector dark matter could logically have \emph{no} interactions with the Standard Model, there are several strong motivations to look for such interactions.  First, general symmetry arguments allow several types of ``portal'' interaction between generic hidden sectors and the Standard Model, which can be generated by radiative corrections.  Second, these modest couplings can play a key role in realizing the dark matter abundance --- for example, determining the DM abundance via thermal freeze-out (like in the standard WIMP paradigm), depleting a thermal component in Asymmetric DM \cite{Kaplan:2009ag}, mediating the production of DM from a bath of SM particles in freeze-in scenarios \cite{Moroi:1993mb,Asaka:2005cn,Shaposhnikov:2006xi,Kusenko:2006rh,Hall:2009bx}, or maintaining kinetic equilibrium while hidden-sector dynamics depletes the DM number density (SIMP/ELDER scenarios) \cite{Dolgov:1980uu,Carlson:1992fn,Hochberg:2014dra,Kuflik:2015isi}.  Each of these mechanisms implies sharp targets in coupling space, which are strikingly compatible with the (broad) expectations for radiatively generated portal couplings, and in many cases, experimentally accessible.

The natural mass range for hidden-sector DM is broader than for WIMPs, but still in the vicinity of Standard Model mass scales --- 
from about 100 TeV down to keV masses or perhaps even lower.  If the physics that generates the weak scale couples to the hidden sector only through the portal interaction, it is natural for hidden sector matter to be parametrically lighter than the weak scale \cite{Hooper:2008im,Feng:2008ya,ArkaniHamed:2008qp,Morrissey:2009ur}.  Alternately, the hidden sector mass scale may arise from confinement of a hidden gauge group.  It has long been argued that supersymmetry breaking (or other mechanism responsible for generating weak-scale masses) could also generate  masses for many hidden-sector particles, triggering a confining phase transition in the broad vicinity of the weak scale.  

The high-mass parameter space for hidden-sector DM, above several GeV, overlaps WIMP parameter space and has similar phenomenology.  In contrast, the low-mass parameter space for hidden-sector DM, below a few GeV DM mass, is \emph{not} well explored by traditional WIMP searches and motivates new experimental strategies for detection.  This low-mass region also opens up the possibility of cosmologically significant DM self-interactions, and also enables new mechanisms for quasi-thermal DM production.

%\draftnote{ADD: summary of the sub-sections?}

\input{HS-benchmarks}

\subsubsection{Thermal Relic Targets} \label{sssec:HS-thermal}
\input{HS-thermal}

\subsubsection{Targets from quasi-thermal DM production}

\input{HS-quasithermal}

\subsubsection{Light mediators and Freeze-in}\label{sssec:HS-freezein}
\input{HS-freezein}
\subsubsection{Further Opportunities in hidden-sector physics}\label{sssec:HS-opportunities}
\input{HS-opportunities}

%% file: HS-benchmarks.tex
% !TEX root = ScienceCase_wrapper.tex
\subsubsection{Benchmark Models of Hidden-Sector DM} 
The observable signatures of hidden sector DM are dictated by the type of new force coupling the DM to familiar matter, and the nature of the DM coupling to this force. 

\paragraph{Mediators and their SM Couplings}
A new force can be mediated by a vector or scalar boson, which may couple to the SM in a variety of ways. A useful characterization of these interactions is by the following simplified models:
\begin{eqnarray}
{\cal L}_V & \supset & V_\mu \bar f (g^V_f \gamma^\mu + a^V_f \gamma^\mu \gamma^5) f \label{vectorSimplified}  \\ 
{\cal L}_S & \supset & \bar f (g^S_f + \gamma_5 a^S_f) f \phi \label{scalarSimplified}
\end{eqnarray}
for (axial) vector mediator $V_\mu$ or (pseudo)-scalar mediator $\phi$.  

The structure of the couplings $g_f$ and $a_f$ depends on how the mediator coupling to familiar matter arises.  Two important special cases are the ``horizontal portals'' --- the unique renormalizable interactions of an SM-neutral boson compatible with all SM symmetries are \cite{Galison:1983pa,Holdom:1985ag,Patt:2006fw}: %ields renormalizable interactions of hidden-sector matter with 
% that the Standard Model Two important benchmark models are  important benchmark scenarios can arise from mixing of hidden-sector states with Standard Model fields in an \ important benchmark models arise from mixing of hidden-sector states directly with SM fields,  couplings of An important special case is  case is  of these Demanding that the hidden-sector coupling be fully compatible with SM symmetries singles out two ``portals'' \cite{Galison:1983pa,Holdom:1985ag,scalar-refs}: 
\bea
\mathcal{L} \supset 
\begin{cases}
-\tfrac{\epsilon}{2\cos\theta_W}\, B_{\mu\nu} F'^{\mu\nu} & \textrm{ vector portal} \quad \Rightarrow \quad g^V_f \approx \epsilon e q_f  \label{vectorportal} \\ 
%A'^\mu J^G_{\mu} &  \textrm{ global portal} \quad \Rightarrow \quad g^V_f = g_SM Q^G_f  \label{globalportal} \\
(\mu \phi + \lambda \phi^2) H^\dagger H &  \textrm{ Higgs portal}\quad \Rightarrow \quad g^S_f = \mu m_f/m_h^2, \label{higgsportal}
\end{cases}
\eea
where $B_{\mu \nu},~F'_{\mu \nu}\equiv \partial_\mu A'_{\nu}- \partial_\nu A'_{\mu}$ are the hypercharge and dark $U(1)_D$ vector boson field strengths, $e q_f$ the electric charge of each SM particle, %$J^G_\mu$ and $Q^G$ are the current and charges associated with a global symmetry of the SM (such as baryon number, $Q^{B}_{u,d}=1/3,\,Q^{B}_{e,\nu}=0$, or lepton number $Q^{L}_{u,d}=0,\, Q^{L}_{e,\nu}=1$), 
$H$ is the Higgs doublet,  $m_{f}$ the mass of fundamental fermion $f$, and $m_h$ the SM Higgs mass.   

While these are justifiably emphasized as benchmark models, high-energy extensions of the Standard Model readily open up the more general parameter space of \eqref{vectorSimplified} and \eqref{scalarSimplified} --- for example, vector couplings to anomalous global symmetries of the SM like baryon or lepton number; chiral couplings with non-zero $a^V$ from $Z$-mixing or ``effective $Z^\prime$'' models; and pseudo-scalar couplings or enhanced first-generation scalar couplings from an extended Higgs sector.  

It is natural for any of these couplings to be small enough to have escaped detection thus far, yet large enough to explain the primordial generation of dark matter.  For example, loops of heavy particles of mass $M$ charged under both $U(1)_Y$ and the new $U(1)_D$ gauge group generate mixing at the level of $\epsilon \sim g' g_D/16\pi^2 \log(M/\Lambda)$, where $g'$ and $g_D$ are the $U(1)_Y$ and $U(1)_D$ charges respectively of the heavy particle, and $\Lambda$ is an ultraviolet cutoff.  Assuming an $O(1)$ log and $g_D \sim g$ suggests $\epsilon \sim 10^{-3}-10^{-2}$.  Enhanced symmetry of the fundamental theory (e.g. grand unification of SM forces) leads to an approximate cancellation so that the effective log is itself loop-suppressed, suggesting $\epsilon \sim 10^{-5}-10^{-3}$.   Such couplings are in the natural ballpark suggested by thermal or quasi-thermal DM generation mechanisms.  Even smaller couplings, as needed for DM freeze-in, can easily be generated, for example if  $U(1)_D$ is also embedded in a non-Abelian group or is weakly coupled, if the coupling to ordinary matter is suppressed by an additional small mixing angle, or by non-perturbative effects.

We comment briefly on the status of model-independent constraints on the portal couplings:
\begin{itemize}
\item  The \emph{Vector portal} is most constrained by muon and electron magnetic dipole moments for sub-GeV mediators \cite{Endo:2012hp,Davoudiasl:2012ig}, and by precision electroweak physics \cite{Hook:2010tw} for heavier mediators.  These model-independent constraints are generally (and especially at low mediator masses) surpassed by those arising from searches visible or invisible mediator decays, or from DM physics.  
\item The proportionality of \emph{Higgs portal} couplings to particle masses implies strong constraints on these models from heavy meson decays, although some new territory can nonetheless be explored by proposed dark matter experiments (see e.g. \cite{Krnjaic:2015mbs}).  It is also worth emphasizing that these constraints are very specific to the minimal model --- scalar portal mixing with a minimal SM Higgs --- and constraints directly on the first-generation couplings of \eqref{scalarSimplified} are many orders of magnitude weaker.
\item Another simple benchmark is the coupling to an SM global symmetry like baryon or lepton number.   The resulting interactions of electrically neutral particles lead to additional constraints --- in particular, limits on $e-\nu$ scattering \cite{Bellini:2011rx,Izaguirre:2014dua} and low-energy neutron scattering data \cite{Barbieri:1975xy,Tulin:2013teo} set the tightest constraints on new bosons coupled to lepton and baryon number, respectively.  Even so, searches for DM-electron and DM-hadron interactions explore regions allowed by these constraints over most of the relevant DM mass range.   
\end{itemize}
In summary, the next generation of searches for DM interactions will probe viable and motivated parameter space for all the portal interactions.  The viability of global symmetry couplings underscores the importance of separately exploring DM couplings to leptons and hadrons.

\paragraph{Coupling to the Dark Sector: a vector portal case-study}

Turning our attention to the dark sector, it is once again useful to introduce a simplified model.  Focusing for concreteness on vector mediators (though analogous phenomenology arises for scalar mediators),  we consider dark sector matter with mass structure
\bea   
 \hspace{-0.2cm }-{\cal L} \supset m_D\eta \xi +\frac{m_\eta}{2} \eta \eta \! + \frac{m_\xi}{2} \xi\xi   + \mathrm{h.c.}~ (fermion),\\
 \hspace{-0.2cm }-{\cal L} \supset  \mu^2 \varphi^* \varphi  + \frac{1}{2} \rho^2 \varphi \varphi +\mathrm{h.c.}~(scalar).
\eea 	
where $\eta$ and $\xi$ are Weyl fermions with $U(1)_D$ charge $\pm g_D$ and $\varphi$ a complex scalar with $U(1)_D$ charge $g_D$.  
While dark sectors can have much richer structure ---  including for example confined or Higgsed non-Abelian gauge groups, or multiple kinematically accessible matter species (see Section \ref{WG4sec:models}) --- these simplified models encapsulate much of the phenomenology of the DM state itself.  

In the above, $m_D$ and $\mu$ are $U(1)_D$-preserving mass terms and $m_\eta$, $m_\xi$, and $\rho$ are $U(1)_D$ breaking mass terms.   Since $m_{\apr} \neq 0$ breaks the $U(1)_D$ symmetry, it is reasonable for all mass terms to be present giving rise to two dark-sector mass eigenstates with the $\apr$ primarily mediating an \emph{inelastic} transition between them.  Alternately, residual discrete symmetries can lead to symmetry limits where the $\apr$-mediated transition is mass-diagonal:  Dirac fermions ($m_{\eta,\xi}=0$) or complex scalars ($\rho=0$) charged under $U(1)_D$ or an axially coupled Majorana fermion ($m_D=0$).   These distinctions significantly affect the DM phenomenology, especially at the low velocities relevant for direct and indirect detection --- for example, Majorana fermions have $p$-wave annihilation and direct detection cross-sections suppressed by a factor of $(q/m_\chi)^2$, where $q$ is the momentum transfer, relative to Dirac fermions or elastically scattering scalars.  A detailed classification, including scalar mediators, can be found in e.g. \cite{Berlin:2015ymu}.

Accelerator experiments searching for DM production are particularly robust probes of models with significant $U(1)_D$-breaking masses, which generally suppress direct detection cross-sections by either velocity factors or higher powers of $\epsilon$.  On the other hand, models with small $U(1)_D$-breaking so that $m_{\apr} \ll m_{DM}$ can have dramatically enhanced direct detection cross-sections due to low momentum transfer.   Therefore, a broad experimental program is required to search for hidden-sector DM.

%% file: HS-thermal.tex
\label{ssec:thermal}
If DM couples sufficiently to ordinary matter that it reached thermal equilibrium with the Standard Model in the early Universe, there \emph{must} be some interaction that depletes its abundance.  The simplest possibility is that the abundance is depleted by DM annihilation arising from the portal interactions noted above.  

Dark matter coupled to a dark vector or scalar mediator can annihilate in two qualitatively different ways, depending on the DM-dark photon mass hierarchy:
\begin{itemize}
\item ``Secluded'' annihilation to pairs of mediators (e.g. $\chi \chi$ $\to$ $\rm{MED}~\rm{MED}$), followed by mediator decays to SM particles%\cite{Pospelov:2007mp}
,  is allowed for  $m_\chi > m_{\rm{MED}}$.  The annihilation cross-section
$\langle \sigma v \rangle \propto \frac{  g_D^4 }{  m_\chi^2  } $, where $g_D$ is the DM-mediator coupling,
is independent of the mediator-Standard Model coupling, and so this process does not imply a thermal target for the latter. 
For vector mediators, this process leads to unsuppressed annihilations down to low temperatures and is therefore excluded by CMB data for sub-GeV dark matter, as discussed below.  This argues for the regime in which $m_{\chi} < m_{A'}$, where a different channel must dominate DM annihilation. Secluded annihilation into scalar mediators is phenomenologically viable, provided the DM Yukawa couplings are suitably small (e.g. $\sim 3\cdot 10^{-5}-3\cdot 10^{-3}$ for MeV--GeV dark matter) to achieve the thermal relic cross-section. 

\item ``Direct" annihilation into Standard Model fermions through an $s$-channel mediator has an annihilation cross-section scaling as 
\bea
\langle \sigma v \rangle   \sim     \frac{      g_{D}^2   g_{\rm SM}^2  m_{\chi}^2 }{   m_{\rm MED}^4 } \label{directAnn}
\eea
for a vector mediator, where $g_{\rm SM}$ is the SM-mediator coupling.
This process offers a clear, predictive target for discovery or falsifiability, since the dark coupling $g_{D}$  and mass ratio $m_{{\chi}}/m_{A'}$ are 
at most ${\cal O}(1)$, so there is a minimum SM-mediator coupling $g_{SM}$ compatible with a thermal history. This mixing target for the vector portal, at the level of 
$\epsilon \sim 10^{-7} m_{A^\prime}^2/(m_\chi \MeV \sqrt{\alpha_D})$ with $\alpha_D = g_D^2/4\pi$ (and therefore quite compatible with the level of mixing expected from one- or two-loop effects), is an important benchmark for both mediator and dark matter searches.   Direct annihilation of sub-GeV DM through a scalar mediator requires fairly large scalar mixing to compensate for the small Yukawa couplings of SM annihilation products, and is excluded by meson decay constraints \cite{Krnjaic:2015mbs}. 
\end{itemize}

An important constraint on low-mass thermal DM comes from the effect of late-time DM annihilation on the CMB power spectrum~\cite{Chen:2003gz,Galli:2009zc,Padmanabhan:2005es,Slatyer:2009yq,Finkbeiner:2011dx}.  Planck data constrains the power injected by DM annihilation at $\sim$ eV temperatures \cite{Ade:2015xua}:
\bea
p_{ann}  & = &  f_{\rm{eff}} \langle\sigma v\rangle_{T\sim \eV}/m_{DM}  < 3.5\times 10^{-11} \GeV^{-3}%\\
%&  \sim & f_{\rm{eff}} \left(\frac{\langle\sigma v\rangle_{T\sim \eV}}{3\cdot 10^{-26} \cm^3/\s}\right) \left(\frac{70 \GeV}{m_{DM}}\right),
\eea
where $f_{\rm{eff}}\sim 0.15-1$ for most annihilation modes (see e.g. \cite{Slatyer:2009yq}), so that the annihilation rate of sub-GeV thermal dark matter at eV-scale temperatures must be suppressed by 1--5 orders of magnitude relative to the annihilation rate at $T\sim m_{DM}/20 $ relevant for DM annihilation, for DM in the MeV-GeV regime.  

This constraint rules out secluded annihilation into vectors  and direct annihilation of Dirac fermions through the vector portal, but many of the generic DM models presented above experience suppressed annihilation at low temperatures, due to one of three effects:
\begin{itemize}
\item Velocity-suppression, for example from $p$-wave annihilation processes with $\sigma v \propto v^2$ (as in direct annihilation of scalar or Majorana fermion through a vector mediator, or secluded annihilation to scalars).
\item Population suppression, if the leading annihilation process involves an excited state that decays or is thermally depopulated in between freeze-out and recombination eras (as in direct annihilation of pseudo-Dirac or inelastic scalar DM through a vector mediator). 
\item Particle-anti-particle asymmetry, if annihilation in the early universe is sufficiently effective to cosmologically deplete the anti-particle; note in this case, cosmological constraints imply a bound on the {\em minimum} annihilation cross-section \cite{Lin:2011gj}.
\end{itemize}

%
%A consequence of this constraint is that any thermal DM annihilating through the vector portal \emph{must annihilate through the predictive ``direct annihilation'' mode} (and cannot be a Dirac fermion).  Thus, exploring dark matter scattering and/or production at the level of sensitivity needed to test this possibility --- shown in Figure \ref{fig:thermal} ---  implies a broad constraint on light vector-portal DM. 

%\begin{figure}
%\caption{\label{fig:thermal} Thermal freeze-out targets for vector-portal DM with each of the allowed spin and mass structures, in the parameter spaces natural for direct detection (left) and accelerator production (right).
%Also shown are the targets associated with two quasi-thermal production mechanisms: Asymmetric DM (the indicated line is a \emph{minimum} annihilation cross-section) and SIMP/ELDER DM, whose abundance is depleted by dark sector self-interactions (the indicated line corresponds to the ELDER scenario; SIMPs are allowed anywhere between the ELDER and scalar freeze-out lines).}
%\end{figure}

In summary, the paradigm of hidden-sector DM that was in thermal equilibrium with the Standard Model in the early universe features viable models that evade existing constraints.Moreover, the subset of models where DM annihilates directly into the SM are of particular interest to the community, as these offer a predictive and bounded target that new direct detection and accelerator probes can aim to robustly discover or falsify.  

The mapping of these thermal targets onto direct detection and accelerator observables are described in more detail in Sections \ref{sec:WG1experiments} and \ref{sec:WG3experiments}, respectively.  Broadly speaking, for each case of DM spin and mass structure (which, together with mediator spin, dictate the velocity-dependence of both annihilation and scattering cross-sections) thermal freeze-out implies a \emph{minimum} production rate at accelerators and a precise prediction for the direct detection cross-section.  Accelerator yields vary by 2--3 orders of magnitude depending on these spins, with the variation arising entirely from the velocity-dependence of annihilation cross-sections and the resulting spread in the coupling constants implied by \eqref{directAnn}; direct detection yields vary more dramatically because of the low velocity of DM in the local Halo --- particularly when annihilation proceeds through a vector mediator, in which case the leading contribution to elastic scattering  may be a one-loop diagram. 

% Of course, variations of the thermal freeze-out scenario exist that can evade this constraint.  Two examples are secluded annihilation to a dark-sector Higgs boson (with a small Yukawa coupling, as noted above), or ``forbidden annihilation'' (where $m_{A'}$ is slightly greater than $m_{DM}$, but secluded-channel annihilation proceeds off the Boltzmann tail of the DM distribution \cite{DAgnolo:2015ujb}.  But it is also notable that a wide variety of quasi-thermal scenarios, where annihilation does \emph{not} dictate the abundance of DM, still have targets relatively close to the thermal line highlighted above.
 
%\draftnote{Consider moving quasi-thermal discussion here?}

%% file: HS-quasithermal.tex
Asymmetric dark matter (ADM) is a paradigm where the DM relic abundance is set by a primordial asymmetry, similar to the baryons, rather than by a thermal freeze-out process.  In a hidden sector model of ADM, where the dark matter sector is in chemical equilibrium with the standard model early in the universe, the DM and baryon abundances are naturally related, so that 
${\Omega_X} \sim r {\Omega_b} m_X/m_p$  where $r$ is an ${\cal O}(1)$ number that depends on the nature of the operator maintaining chemical equilibrium  \cite{Kaplan:2009ag,Zurek:2013wia}.  This, combined with the observed ratio $\Omega_{DM} \sim 5 \Omega_b$, motivates DM masses of several GeV.  Models where the DM is produced from an out-of-equilibrium decay have $r \ll 1$, so that the DM particle $X$ can be much lighter than the proton. 

The interaction between the dark sector and the Standard Model that generates a DM asymmetry may not be detectable, but ADM also requires an annihilation process that depopulates the symmetric component of DM.  If $X$ is a fermion, this symmetric component must be substantially depopulated to evade the CMB constraints discussed in \ref{ssec:thermal}.  As for thermal freeze-out, suitably large annihilation cross-sections require a new force.  Assuming the annihilation proceeds through the direct channel (i.e.~$m_{MED}>m_X$), the minimum annihilation cross-section implies a minimum coupling and hence target for direct detection and accelerator production that can be explored by near-future experiments \cite{Lin:2011gj}.  Secluded annihilation (to vector or scalar mediators) is also possible \cite{Kaplan:2009ag}, and allows viable ADM models with scattering or production cross-sections below this milestone.  

Another intriguing possibility arises for dark matter particles $\chi$ with mass near the QCD confinement scale, $\Lambda_{\rm QCD}\sim 100$ MeV, which could arise as mesons or baryons of a hidden-sector ''mirror copy'' of QCD \cite{Mohapatra:2000qx,Strassler:2006im}. 
For example, number-changing process $3\chi\leftrightarrow 2\chi$ can deplete the $\chi$ abundance~\cite{Dolgov:1980uu,Carlson:1992fn}, naturally achieving the correct relic density~\cite{Hochberg:2014dra,Hochberg:2014kqa,Hochberg:2015vrg}.  These ``Strongly Interacting Massive Particle" (SIMP) models require elastic scattering between the SIMP and SM particles to keep the two sectors in kinetic equilibrium until the $3\to 2$ scattering freezes out --- so, even though DM annihilation occurs within the hidden sector, there is a robust lower bound on the SIMP-electron elastic scattering cross-section (saturating the bound realizes the ''Elastically Decoupling Relic'' (ELDER) scenario~\cite{Kuflik:2015isi}), which can naturally be realized by a dark photon.  The resulting predictions for DM direct detection cross-sections (assuming elastic DM coupling to the dark photon) and accelerator production are shown in Sections \ref{sec:WG1experiments} and \ref{sec:WG3experiments} 
%Figure \ref{fig:thermal} 
alongside the thermal targets.  The allowed mass range for SIMP or ELDER DM is restricted to $5 \MeV \lsim m_\chi\lsim 200$ MeV, with the lower bound arising from CMB measurements and the upper bound from unitarity of $\chi$ self-scattering.  These models are discussed further in Section \ref{WG4sec:models}.

%% file: HS-freezein.tex
%In models where the DM relic abundance is set through freeze-in, the DM is so weakly coupled to the Standard Model that it never thermalizes.  Instead, the DM abundance is gradually populated through very rare interactions at low temperatures, implying no sensitivity to initial conditions.  If these interactions are with the electron or proton, it gives rise to a prediction for the scattering cross-section in a direct detection experiment.  For example, a process for producing the dark matter through a massless dark photon via $e^+ e^-$ annihilation gives rise to a relic abundance $Y_X \sim \epsilon^2 g_D^2/m_X$ if $m_X > m_e$ and $Y_X \sim \epsilon^2 g_D^2/m_e$ if $m_X < m_e$.  If the direct detection process happens through that same massless dark photon, the scattering cross-section similarly scales with $\epsilon g_D$, fixing $\bar \sigma_e$ for a given $m_X$.  For example, at $m_X = 10 \mbox{ MeV}$, $\epsilon g_D \simeq 6 \times 10^{-12}$ and $\bar \sigma_e \simeq 4 \times 10^{-37} \mbox{ cm}^2$.  This benchmark is shown in Fig.~XX

If the DM is very weakly coupled to the SM and never thermalizes, its abundance can ``freeze in'' through very rare interactions at temperatures near the DM mass --- a mechanism first noted in the contexts of gravitino~\cite{Moroi:1993mb}, sneutrino~\cite{Asaka:2005cn}, and sterile neutrino~\cite{Shaposhnikov:2006xi,Kusenko:2006rh} DM, and subsequently generalized by~\cite{Hall:2009bx}.  Freeze-in can be realized by hidden-sector DM with very weak mixing.  In this case, it implies a prediction for the DM production/annihilation cross-section that is orders of magnitude below the thermal freeze-out level, and correspondingly low predictions for couplings.  For example, $\alpha_D \epsilon^2 \sim 10^{-22} m_{\chi}/\GeV$ for freeze-in via a light vector-portal mediator \cite{Essig:2011nj,Chu:2011be,Essig:2015cda}, compared to $\alpha_D \epsilon^2 \sim 10^{-8} m_{\chi}^2/\GeV^2$ for the Dirac fermion direct annihilation benchmark of Section \ref{ssec:thermal}.  Thus, even if freeze-in arises through interactions with the electron or proton, much of the relevant parameter space is beyond the reach of laboratory experiments.  

It is, indeed, remarkable that \emph{any} of this parameter space can be explored by laboratory experiments.  
In particular, if the particle mediating DM-SM interactions is much lighter than the DM itself, the DM-SM scattering cross section is enhanced, scaling (assuming a velocity-independent elastic interaction) as $\mu_{\chi,T}^2 \alpha_D \epsilon^2/m_{\apr}^4$, where $\mu_{\chi,T}$ is the DM-target reduced mass, at scattering momentum transfers $q$ with $q^2\ll m_{\apr}^2$, saturating at $\mu_{\chi,T}^2 \alpha_D \epsilon^2/q^4$ for sufficiently light mediators.  Because the typical momentum transfer in DM-electron scattering is $O(\keV)$, this enhancement can be dramatic enough to compensate for the small couplings predicted in the freeze-in scenario.
Such a large mass hierarchy between the mediator and DM masses can arise from an enhanced symmetry limit.  Indeed, for a vector mediator the approximate $U(1)_D$-symmetric limit realizes both $m_\chi \gg m_{\apr}$ and the Dirac (or elastic scalar) mass structure for which velocity-independent elastic scattering dominates, as assumed above. 
%Figure \ref{fig:HS-freezein} illustrates the potential sensitivity of low-threshold direct detection experiments to freeze-in via the vector portal, in the case of an ultralight (sub-keV-mass) mediator and elastic velocity-independent DM-mediator couplings, over a wide range of DM masses.  
Freeze-in via heavier mediators would lead to a smaller predicted scattering cross-section; significant phase space can be explored for mediator masses $\lesssim 10$ keV.  

%% file: HS-opportunities.tex
The primary focus of the preceding discussion is on the role of new forces in the dark sector in explaining the cosmological abundance of DM.  But any such force can also have important consequences for the physics of familiar matter and of dark matter, separately.  This motivates searches for new bosons with weak coupling to the Standard Model, and for effects of dark matter self-interactions on cosmological structure formation. Intriguingly, there are hints of both kinds of effect, summarized further in Sections \ref{WG4sec:anomalies} and \ref{WG4sec:astrophysics} respectively.  We summarize these briefly here, and note the parameter regions suggested by each anomaly. 

A number of curious experimental results may point indirectly to the existence of light bosons with weak 
couplings to the Standard Model.
The most famous example is the measurement of anomalous magnetic moment of the muon, $(g-2)_\mu$, whose
experimental determination \cite{Bennett:2006fi} famously shows a $\sim 3\sigma$ discrepancy with the most sophisticated 
available theoretical predictions \cite{Jegerlehner:2009ry}.  As constraints from the null results of LHC searches
on the masses of electroweakly charged particles push into the several hundred GeV regime, a light weakly coupled
boson with mass of order $10-100$ MeV and coupling to the muon around $10^{-3}$
remains as a leading new physics explanation \cite{Pospelov:2008zw}.
In addition, the radius of the proton as measured by the $2S-2P$ transition frequency
in muonic hydrogen is famously discrepant compared with the value extracted from $e$-$p$ scattering
\cite{Antognini:1900ns}.  The tension between the two measurements can be alleviated by positing
a $\sim 10$~MeV mass spin 1 boson with muon-specific couplings around $10^{-3}$ \cite{Karshenboim:2014tka}.
Finally,
the rate of $\pi^0 \rightarrow e^+ e^-$ measured by KTeV \cite{Abouzaid:2006kk} shows a $2-3\sigma$ deviation
compared to Standard Model expectations \cite{Dorokhov:2007bd} and can also be explained by
the existence of a vector particle of mass around $10$ MeV with axial couplings to both first generation quarks and electrons
on the order of $10^{-3}$ \cite{Kahn:2007ru}.

A direct search for light weakly coupled bosons produced in nuclear transitions and decaying into $e^+ e^-$ by
the ATOMKI group \cite{Krasznahorkay:2015iga}
finds an excess with a high statistical significance in the transition of the $1^+$ 18.15 MeV excited
state of Beryllium-8 to its $0^+$ ground state at a particular $e^+ e^-$ opening angle and invariant mass consistent
with an intermediate boson of mass $\sim 16.5$ MeV and couplings of order $10^{-4}-10^{-3}$ to quarks
and leptons \cite{Feng:2016jff}.  The parameter space is roughly consistent with regions explaining
the $(g-2)_\mu$ and KTeV experimental results \cite{Feng:2016ysn,Kahn:2016vjr}

On the astrophysical side, numerical simulations of galaxy formation, while widely successful at the largest
scales, show discrepancies when confronted with observations of smaller scale structures.  While it may
be possible that these discrepancies represent poor modeling of the baryonic components of galaxies,
much of the tension between simulation and observation can be removed by postulating that the dark matter
is self-interacting.  Combined fits to observations of dwarf galaxies, low surface brightness spiral galaxies, and
galaxy clusters points to a parameter space where dark matter whose mass is $10-100$~GeV interacts
via exchange of a dark force carrier whose mass is $10-20$~MeV \cite{Kaplinghat:2015aga}.

%% file: lightBoson.tex
%\input{figs/preamble.tex}   
%\begin{document}

The size of dwarf galaxies constrains the nature of sub-keV dark matter to be bosonic because Fermi degeneracy pressure would prevent the formation of galactic substructure at this scale from gravitationally clumped fermionic dark matter.   Moreover, the mass of this light bosonic dark matter, whether a scalar, pseudoscalar, or vector, should be greater than $10^{-22}$ eV to avoid having its Compton wavelength exceed the size of the observed dwarf galaxies.   Due to their low mass, bosons lighter than about a keV can couple linearly to Standard Model fields, and nonetheless be cosmologically long-lived dark matter candidates.  Due to this linear coupling, these bosons are also mediators of long-range spin-dependent or spin-independent forces.  As with neutrinos, the smallness of the boson mass suggests a connection to UV physics at high scales via a see-saw mechanism which generates the mass.  Various viable cosmological production mechanisms have been proposed for ultralight dark matter, thus offering interesting probes of this new UV physics.

% below eV mass, this bosonic DM acts as a coherent, oscillating background field, while eV-to-keV-mass bosons can be detected via the energy they deposit when absorbed on matter.  In general, any new scalar or vector particle will be produced with some abundance in the early universe, allowing it to be a good dark matter candidate. 
\subsubsection{The QCD Axion}
The best known example of light bosonic DM is the QCD axion, a well-motivated candidate because it can solve the long-standing strong CP problem~\cite{Peccei:1977hh, Peccei:1977ur,Weinberg:1977ma, Wilczek:1977pj}, explaining the puzzle of the vanishing neutron electric dipole moment.  It also simultaneously guarantees the production of dark matter at some abundance through a natural production mechanism of vacuum relaxation~\cite{Preskill:1982cy, Dine:1982ah, Abbott:1982af}.  Axions and axion-like particles are generic in many UV theories (see for example \cite{Svrcek:2006yi}) and they may also be related to the electroweak hierarchy problem \cite{Graham:2015cka}. 

The QCD axion model is quite economical, requiring only a single parameter  --  a high mass scale $f_a> 10^9$~GeV at which a postulated new global U(1) ``Peccei-Quinn" symmetry is broken, resulting in a massless Nambu-Goldstone boson -- the axion --  living in the trough of a Mexican Hat potential.  During the QCD phase transition, the defining axion-gluon coupling causes the trough of the potential to become tilted by an amount of potential energy density of approximately $\Lambda_{QCD}^4$ when the QCD instantons condense to define the QCD vacuum.  The axion field rolls to the bottom of the tilted potential and zeroes out any pre-existing QCD CP-violating ``theta" angle.  Simultaneously, the initial potential energy density is released as ultracold dark matter -- excitations about the new potential minimum whose second derivative determines the tiny axion mass $m_a \approx \Lambda_{QCD}^2/f_a$.  Meanwhile, all couplings to standard model particles are suppressed by $f_a$ and are determined up to a constant factor of order unity.  Axion search experiments typically use the axion mass $m_a$ as the single free model parameter, and aim to cover a range between the ``KSVZ" coupling strength \cite{Kim:1979if, Shifman:1979if} and the ``DFSZ" coupling strength~\cite{Dine:1981rt, Zhitnitsky:1980he} which is around a factor of 3 weaker.  

The QCD axion is allowed to lie in the mass range of roughly $10^{-12}$~eV to $10^{-2}$~eV (corresponding to experimental frequencies 250~Hz - 2.5~THz).  The lower bound arises from requiring $f_a$ not exceed the Planck scale.  The upper bound comes from the neutrino pulse observed from SN1987A having a duration consistent with supernova cooling primarily via neutrino emission, thus placing a bound on the axion-nucleon coupling and immediately constraining all phenomenological features of the single-parameter QCD axion model.  However, given astrophysical uncertainties as well as the limited statistics of a single supernova event, one may also obtain a more conservative upper bound of 1~eV axion mass from hot dark matter limits.

The QCD axion model has an intricate interplay with models of cosmic inflation, and discoveries in either field immediately inform the physics of the other.  For example, the amount of initial potential energy density $\Lambda_{QCD}^4 \sin^2{(\theta_0/2)}$ to be released as axion dark matter depends on the random initial value $\theta_0$ of the strong CP-violating angle to be zeroed out by the rolling axion field. In models in which the Peccei-Quinn phase transition occurs after cosmic inflation, many topological domains of different $\theta_0$ form and are contained within our cosmological horizon.  The average energy density released as dark matter, averaged over all domains is then well-determined --  $\Lambda_{QCD}^4 \times 1/2$.  The axion vacuum relaxes to its minimum at cosmological time $1/3H \approx 1/m_a$, during the radiation-dominated era when the photon density is rapidly red-shifting away.  Since vacuum energy does not red-shift, small values of $m_a$ would delay too long the release of this energy as dark matter, giving too large a axion/photon ratio and thus overproducing dark matter.  Another complication is that topological features such as domain walls and cosmic strings can form, thus temporarily stabilizing the vacuum energy density and further delaying its release.  Assuming equal contributions to dark matter production from vacuum relaxation and  from topological defect decay, recent lattice calculations estimate that $m_a \lesssim 10-50 \ \mu$eV would not be compatible with this post-inflationary axion scenario~\cite{Dine:2017swf,Berkowitz:2015aua,diCortona:2015ldu,Borsanyi:2016ksw}.

The alternative pre-inflationary scenario is one in which the Peccei-Quinn symmetry is broken prior to inflation so that the initial $\theta_0$ is single-valued throughout our cosmological horizon and nothing then disallows a small value of $\sin^2{(\theta_0/2)} \ll 1/2$ to occur by chance.  The much smaller amount of initial vacuum energy could then be released later in cosmic time without overproducing dark matter, thus allowing lower axion masses.  This scenario includes the parameter space at large $f_a$ near the GUT or Planck scale which is preferred by string theory~\cite{Svrcek:2006yi}.  Cosmic inflation also sources a spectrum of axionic excitations resulting in a potentially observable CMB isocurvature power spectral density scaling as $(H_I/f_a)^2$ where $H_I$ is the Hubble scale during inflation.  Constraints on isocurvature then constrain this ratio of inflation and Peccei-Quinn scales~\cite{Fox:2004kb,Hertzberg:2008wr,Hamann:2009yf}.

If a low mass axion with $m_a \lesssim 10-50 \ \mu$eV is discovered, then this immediately implies the pre-inflationary axion scenario.  The upper bounds on isocurvature then constrain $H_I$ to a scale too low to produce any potentially observable CMB B-modes (with primordial gravitational wave spectral density $(H_I/M_p)^2$) and dark matter axion studies would become the primary tool to probe cosmic inflation.  Conversely, if CMB B-modes are discovered first, then only the post-inflationary axion scenario remains viable, the low axion mass window is closed and dark matter axion searches should be focused on higher masses.

%It has a fairly sharply defined relationship between its mass and the strengths of its couplings, and so it lies on a relatively narrow line in the mass-coupling plane. 

 %The most important goal would be to cover the entire QCD axion parameter space.

\subsubsection{General phenomenology of sub-meV mass bosonic dark matter (including axions)}
Since we do not know the nature of dark matter, it is  important to look broadly to cover all candidates in this entire mass range from $10^{-22}$ eV to 1 keV.  
%When the dark matter is lighter than a meV, the dark matter
While non-relativistic cold dark matter of any form has very small kinetic broadening, low mass bosonic dark matter particles act collectively as a coherently oscillating semiclassical wave with high mode occupation number.  For masses less than a few milli-eV corresponding to signal frequencies less than THz, this property can be used to detect bosonic dark matter via novel experimental techniques targeting continuous wave signals rather than impulse detection.  In many cases, these experimental techniques have been well-developed in other fields of physics and had not previously been applied to the problem of dark matter detection.  Cost-effective experiments are therefore possible which can quickly explore new parameter space.
  
These direct detection experiments rely on coupling the coherently oscillating dark matter field to Standard Model (SM) particles via four basic types of operators:
\begin{enumerate}
\item {\bf Electromagnetism}:  This coupling allows  dark matter to transfer energy into electromagnetic fields to be detected via photon, voltage, or flux sensors.  For example, the well-established haloscope technique~\cite{Sikivie:1983ip} uses a microwave cavity to resonantly enhance the transfer of power from the incoming axion or hidden photon dark matter beam into electromagnetic modes.
\item{\bf QCD}:  This gluon coupling gives a time-oscillating electric dipole moment (EDM) for nucleons which can be detected via nuclear magnetic resonance (NMR) techniques. 
\item {\bf Spins of fermions} (either electrons or nucleons):  These couplings cause the spins of electrons or nucleons to precess which can again be detected via NMR or electron spin resonance.
\item {\bf Scalar couplings}:  These couplings can give a force directly on SM particles, or can affect fundamental constants such as SM particle masses or charges.  For example these are couplings to a fermion's mass (without a $\gamma_5$) or a coupling to a gauge boson's kinetic term.  Any precision measurement sensitive to small forces can potentially be modified to search for anomalous AC signal modulations.

\end{enumerate}
%{\bf Rather than the following commentary, it would be better to explain the physics of how different kinds of experiments probe these different operators.}  
It is desirable to have a variety of experiments to probe all of these possible couplings.  First, using different couplings of the dark matter leads to a very different and highly complementary detection techniques that can together allow searches through much more of parameter space than would otherwise be possible.  Second, we do not know what couplings the dark matter has so it is important to probe all possibilities as broadly as possible.  Third, if dark matter is discovered in one of these experiments it will be extremely important to detect it with a different technique both for confirmation and because it is crucial to measure as many couplings as possible in order to learn as much as possible about the dark matter model.  Such light dark matter often arises from physics at very high energies.  Measuring the mass and couplings of this particle would be in many cases the only way to study such high energy scales experimentally;  interesting scales such as the Planck, GUT, or string scales, are far beyond what can be accessed in a collider.  Finally, as with any type of dark matter, it is of critical importance to follow-up the direct detection signal with an accelerator or fifth-force type experiment to directly measure the couplings in order to disentangle them from the uncertainty in the intensity of the dark matter flux.  These laboratory experiments would presumably be easier to design, once armed with knowledge of the dark matter mass and coupling scale.

The high temporal and spatial coherence of the collectively oscillating modes of bosonic dark matter also leads immediately to some interesting follow-up studies.  Because the experiments often rely on narrowband resonant detectors which must be tuned to the signal, they are usually designed to be able to reproduce the signal on very short time scales of minutes to hours.  So blind analyses need not be used since a new, independent dataset can be immediately acquired.  Moreover, by simply integrating longer before Fourier transforming the time-series signal, the energy spectrum of the dark matter can be measured with higher frequency resolution.  This allows the substructure of the dark matter velocity distribution to also be quickly measured with the same detector, as well as its annual modulation.  Finally, the high spatial coherence of the bosonic wave on scales of order the deBroglie wavelength allow the use of multiple identical but spatially separated detectors to map out the local wavefront of the dark matter and hence to determine its local phase space distribution.  These studies can indicate whether the dark matter is fully virialized or if there is substructure due to recent galaxy merger activity.

Furthermore, all these light fields are produced or influenced by cosmic inflation and their discovery can provide valuable information on the inflationary sector and hence the earliest times in the universe.  As discussed above, a measurement of the axion mass can provide critical information on the scale of inflation.  As another example, vectors dark matter (hidden or dark photons) are directly produced through quantum fluctuations during inflation and for high scale inflation would naturally be predicted to have a mass in the range that can be searched for in many of these experiments~\cite{Graham:2015rva}.  If this vector dark matter is detected, then its power spectrum can be measured, giving a confirmation of this production mechanism and a determination of the scale of cosmic inflation.  

Detectors for this ultralight dark matter often rely on very high precision experimental techniques that have a wide range of broader impacts.  On the fundamental physics side other applications for these sensors include searching for new forces of nature, violations of the equivalence principle, and detecting gravitational waves.  There are also more practical applications including geological mapping, inertial navigation, and a connection with quantum information.

There are a variety of these high precision sensor technologies that are complementary including probing different couplings and complementary mass ranges.  Excitingly, experiments now appear able to cover this entire mass range, as discussed in Section~\ref{sec:WG2experiments}.

\subsubsection{Bosonic dark matter from meV-keV}
The same considerations as above apply to meV - keV bosonic dark matter but in this mass range, even the fastest THz electronics cannot resolve the collectively oscillating signal, and micro-calorimetric techniques must be used for detection of individual particle scattering processes.   For this ultra-low threshold impulse detection, it has been shown that coherent modes in the detector target material ({\em i.e.} phonons) can be utilized.  Bosonic dark matter may be absorbed on a target electron in a superconductor through single phonon emission \cite{Hochberg:2016ajh}, or in a semiconductor through single \cite{Hochberg:2016sqx,Bloch:2016sjj} or multiple \cite{Hochberg:2016sqx} phonon emission (see Section~\ref{sec:WG1experiments}). %Fig.~\ref{fig:absorption}).  

%\begin{figure}
%\includegraphics[width=0.4\textwidth]{figs/electron-phonon-photon}
%\caption{Dark matter ($X$) absorption process on an electron, made possible by emission of a phonon $\Phi$.  Figure from Ref.~\cite{Hochberg:2016ajh}. \label{fig:absorption}}
%\end{figure}

The advantage here is that the bosonic dark matter particle can be absorbed onto a fermion line and transfer energy equal to its entire mass, whereas the same microcalorimeter detecting fermionic dark matter can only absorb the recoil kinetic energy which is at most $10^{-6}$ of the dark matter rest mass.  The experiments capable of detecting dark matter through absorption over the meV - keV mass range are the same as those searching for keV - GeV mass dark matter via scattering discussed in Section~\ref{sec:WG1experiments}.

%% file: WG1.tex
% !TEX root = WG1_wrapper.tex
\section{New Avenues in Direct Detection }
\label{sec:WG1experiments}
%%%%%%%%%%%%%%%%%%%%%%%%%%%%%%%%%%%%%%%%%%%%%%%%%%%%%%%%%%%%%%%%%%%%%
%%%%%%%%%%%%%%%%%%%%%%%%%%%%%%%%%%%%%%%%%%%%%%%%%%%%%%%%%%%%%%%%%%%%%
\subsection{Introduction} 

Dark matter (DM) direct-detection experiments are an essential laboratory tool in our quest to identify DM.  Their goal is to search for DM particles in our Milky-Way halo that scatter or absorb in a detector target material.  
The last few decades have seen enormous advances in designing and building direct-detection experiments that has led to a many orders of magnitude improvement in searches for $\sim$10 keV scale nuclear recoils that are characteristic of spin-independent scattering of Weakly Interacting Massive Particles (WIMPs) with masses $> 10$~GeV.
The next generation ``G2'' LZ experiment is poised to probe a large fraction of the remaining theoretically well-motivated parameter space 
for this mass range over the next few years.  
Another exciting possibility is that DM has a mass in the $\mathcal{O}$(GeV) 
range, and   
SuperCDMS, the second ``G2'' direct-detection experiment, is poised to probe this mass range to unprecedented sensitivity. 

As described in Part I of this white paper and summarized below, there are several scientifically well-motivated DM candidates 
that will not be probed by either LZ or SuperCDMS.  
The  ``New Avenues In Direct Detection'' working group has identified the following four additional areas in which novel theoretical ideas and impressive experimental advances enable new small projects that can probe orders of magnitude of previously unexplored DM parameter space:
\begin{enumerate}
\item {\bf Sub-GeV Dark Matter (Electron Interactions)} 
\item {\bf Sub-GeV Dark Matter (Nucleon Interactions)}
\item {\bf Searches down to the  Neutrino Floor  for $\mathcal{O}$(GeV) Dark Matter}
\item {\bf WIMP Spin-Dependent Interactions (Proton)}  
\end{enumerate}

A fifth area of parameter space --- high-mass WIMPs ($m_{\rm DM}\gtrsim 10$~GeV) --- was also identified as 
scientifically well-motivated.  However, to probe this region beyond the projected LZ sensitivity will require 
experiments with very large target masses and significant funds ($\gtrsim$ 10 million dollars).  Consequently, this parameter space falls outside of the scope of the workshop and will not be discussed further in this white paper. 

%%%%%%%%%%%%%%%%%%%%%%%%%%%%%%%%%%%%%%%%%%%%%%%%%%%%%%%%%%%%%%%%%%%%%
%%%%%%%%%%%%%%%%%%%%%%%%%%%%%%%%%%%%%%%%%%%%%%%%%%%%%%%%%%%%%%%%%%%%%
\subsection{Summary of Science Case for New Small-Scale Direct-Detection Experiments}\label{subsec:science-DD}

\begin{figure}[t]
\includegraphics[width=0.8\textwidth]{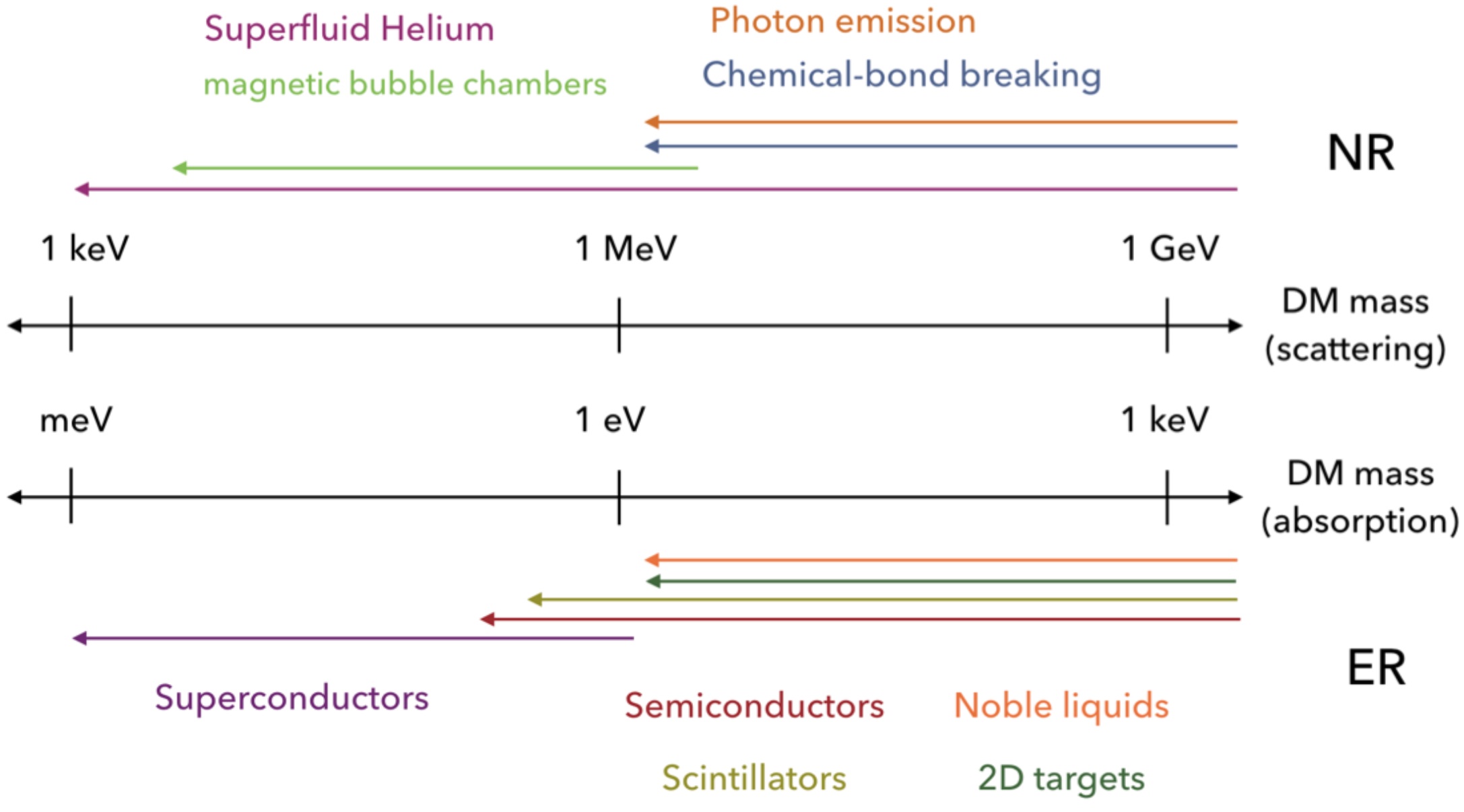}
\caption{Ideas to probe low-mass DM via scattering off, or absorption by, nuclei (NR) or electrons (ER). 
}
\label{fig:overview}
\end{figure}

Direct-detection experiments play a unique and essential role in our quest to identity the DM.  Several proposals and ideas exist for new experiments that present a low-cost opportunity --- well within the ``small-project'' scale --- to {\bf probe DM with masses between the meV to GeV scale}, many orders of magnitude in mass below the planned searches by the G2 experiments LZ and SuperCDMS (see Fig.~\ref{fig:overview} for a schematic overview).  
In fact, the working group recognizes that recent advances in theory and experiment means that {\it now} is the right time for targeted investments to bring to fruition several recent new ideas and proposals and develop them into real experiments.  

Several well-motivated DM candidates can be probed. 
In several cases, {\it sharp theory targets} in parameter spaces can be identified, which can be probed by first-generation, low-cost experiments with target exposures of as little as 100 gram-days.  
These sharp targets have been discussed in Section~III.  They assume that the basic interaction between the DM 
and SM particles are through a dark photon, which allows the DM to couple to all electrically charged particles: 
\begin{itemize}
\item {\bf Elastic Scalar} -- a (complex) scalar particle, $\chi$, can obtain the observed relic abundance from thermal freeze-out of  
the ``direct-annihilation'' process $\chi + \chi^* \leftrightarrow A'^* \to {\rm SM}+{\rm SM}$, where $A'$ is the dark photon~\cite{Boehm:2003hm}.  
The annihilation cross section, $\sigma_{\rm ann}$ is proportional to $\alpha_D \epsilon^2 \mu_{\chi,e}/m_{A'}^4$, and has precisely the same 
dependence as the direct-detection cross section, $\sigma_{\rm DD}$ does on the fundamental parameters, $m_{A'}$ (the dark-photon mass), 
$\epsilon$ (the kinetic mixing), and $\alpha_D$ (the ``fine-structure constant'' of the dark U(1))~\cite{Essig:2015cda} 
($\mu_{\chi,e}$ is the DM-electron reduced).  
In fact, since the final DM relic abundance, $n_\chi$, is proportional to $1/\sigma_{\rm ann}$, the direct-detection rate 
is proportional to $n_\chi \sigma_{\rm DD} \sim \sigma_{\rm DD}/\sigma_{\rm ann}$, which is a constant for a given $m_\chi$.  
So even if $\chi$ constitutes only a subdominant component of the entire DM, the ``target'' cross section on the $\sigma_{\rm DD}-m_\chi$ plane is a fixed line.  
\item {\bf Asymmetric Fermion} -- a Dirac fermion can obtain the correct relic abundance from an initial asymmetry and provides an ``asymmetric'' DM candidate~\cite{Kaplan:2009ag}.  However, 
direct annihilation between DM and SM particles from $\chi + \bar\chi \to A'^* \leftrightarrow {\rm SM}+{\rm SM}$ 
produces also a symmetric component, whose abundance is smaller for larger annihilation cross sections~\cite{Lin:2011gj}.  
The symmetric component can annihilate and, if its abundance is too large, distort the Cosmic Microwave Background power spectrum.  
The CMB thus sets a lower bound on the annihilation cross section and, therefore, on $\sigma_{\rm DD}$~\cite{Essig:2015cda}.  
\item   {\bf ELDER} -- An ``elastically decoupling relic'' (ELDER) has its relic abundance set by its elastic scattering off 
SM particles through $A'$ exchange (as opposed to annihilation into SM particles as in the thermal freeze-out scenario)~\cite{Kuflik:2015isi}.  
This again predicts a specific line in the $\sigma_{\rm DD}-m_\chi$ plane.  
\item  {\bf SIMP} -- 
A strongly interacting massive particle (SIMP) obtains the correct relic abundance from $3\to2$ DM to DM annihilations 
while remaining at the same temperature as SM sector due to its elastic scattering off SM particles~\cite{Hochberg:2014dra,Hochberg:2014kqa}.  This defines a region in the $\sigma_{\rm DD}-m_\chi$ plane, with the lower boundary being set by the ELDER line mentioned previously, and the upper boundary being set by the $2\to2$ DM to SM thermal relic line (the elastic scalar line mentioned above).  
\item  {\bf Majorana} -- A Majorana fermion can obtain the observed 
relic abundance through thermal freeze-out, but has a velocity suppressed scattering cross section off SM particles.  
This scenario again predicts a line in the $\sigma_{\rm DD}-m_\chi$ plane, but lies at lower cross sections than the targets mentioned above, due to the low velocity of the DM in the Milky-Way halo. 
The Majorana line is given in terms of the elastic scalar freeze-out line mentioned above, but multiplied by a factor of 
$2 (\mu_{\chi, e/n}^2/m_\chi^2) v_\chi^2$, where $\mu_{\chi, e/n}$ is the DM-electron or DM-nucleon reduced mass (as applicable) and 
$v_\chi$ is the DM halo velocity.  
\item {\bf Freeze-in} -- An initially empty hidden sector, which remains thermally decoupled from the SM sector, 
can be populated by SM particles annihilating to hidden-sector 
DM particles through the process ${\rm SM}+{\rm SM} \to A'^*\to \chi + \bar\chi$ (we assume $m_{A'}\ll\ $~keV).  
We say that the abundance is obtained through ``freeze-in''~\cite{Hall:2009bx}.  
This again fixes the model parameters and predicts a line in the 
$\sigma_{\rm DD}-m_\chi$ plane~\cite{Essig:2011nj,Chu:2011be,Essig:2015cda}.  
\end{itemize}

We note that other sharp targets exist, so the above list is not complete; for example, 
the DM could also interact only with baryons, or only with leptons.  
This emphasizes the need for experiments that probe DM couplings to electrons as well as experiments that probe 
DM couplings to nuclei.  
Beyond the above sharp theory targets, new direct-detection experiments can probe orders of magnitude of DM parameter space that is well-motivated but does not have a sharp target.  
This includes the scenario in which DM obtains its relic abundance from thermal freeze-out by annihilating into a hidden 
sector (the ``secluded annihilation DM scenario''), from the misalignment mechanism, and others.  
It includes well-motivated bosonic DM candidates such as axion-like particles and dark photon DM.  
We emphasize that the {\it same} experiment can probe {\it many} DM candidates.  
Indeed, several of the above theory targets lie close to each other in the direct-detection parameter space.  
Moreover, an experiment sensitive, for example, to keV DM scattering off electrons will simultaneously also be sensitive to absorption of a 
meV DM particle by an electron (see Fig.~\ref{fig:overview}). 

We emphasize that direct detection provides the {\it only} possibility to test the above freeze-in scenario (with $m_{A'}\ll$~keV).  
This is perhaps surprising:  
since the DM is never in thermal equilibrium with ordinary matter in the early Universe, the interactions between DM and SM particles 
are necessarily tiny.  Nevertheless, if the mediator is ultralight ($\ll\ $keV), there is a large enhancement of the direct-detection cross section at low momentum transfers, which for the above freeze-in scenario is given by 
\begin{equation}
\sigma_{\rm DD} \sim 4 \pi \alpha_D \epsilon^2 \alpha \frac{\mu_{\chi,e}^2}{q^4}\,,
\end{equation}
where the momentum transfer $q$ is at most $q_{\rm max} \sim \mu_{\chi,e} v_\chi$.  Because this momentum transfer is so small in direct-detection experiments, the direct-detection experiments receive a parametric enhancement relative to higher energy experiments, allowing 
new low-threshold experiments to probe couplings much smaller than accessible through other types of experiments.  

A discovery of a new particle at an underground direct-detection experiment would constitute strong evidence that such a 
particle constitutes all or at least part of the DM.  
For some models of DM, new direct-detection and accelerator-based experiments can cover overlapping parameter space.  
This is very exciting, as it allows for testing a potential DM signal by using entirely different approaches.  
However, we note that there are also models that can be probed either by accelerators alone or by direct detection alone.  
For example, models in which the DM scatters inelastically, or a Majorana DM particle that has a velocity-suppressed scattering cross section 
off SM particles, are best probed with accelerator-based experiments 
(due to the DM's non-relativistic velocity in the Milky-Way halo), while models for which the mediator is ultralight (e.g.~axion-like or dark-photon DM) or some models of freeze-in are best probed by direct-detection experiments. 
This emphasizes that a small-scale program will be most successful if it contains a multitude of approaches to probe DM.  

A key point emphasized by the working group is that by leveraging new theoretical ideas together with technological advances that allow 
for the detection of low-threshold signals, vast regions of DM parameter space can be explored by small detectors that are only a fraction 
of the cost of the G2 experiments.  
The close collaboration between theorists and experimentalists has been essential in developing these new ideas, which are 
now ripe for implementation.

\begin{figure}[t]
\includegraphics[width=0.175\textwidth]{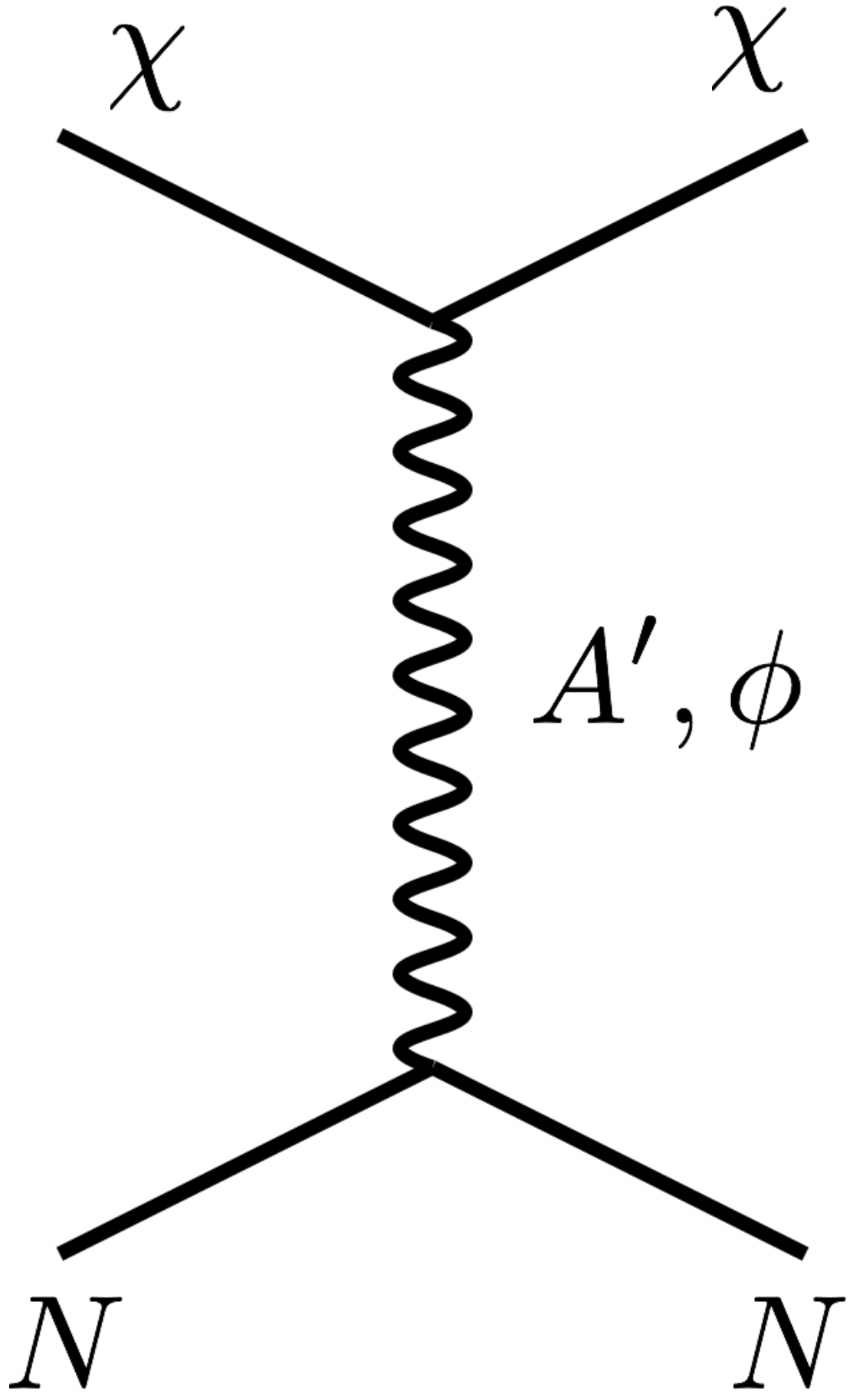}
~~~~~~\includegraphics[width=0.185\textwidth]{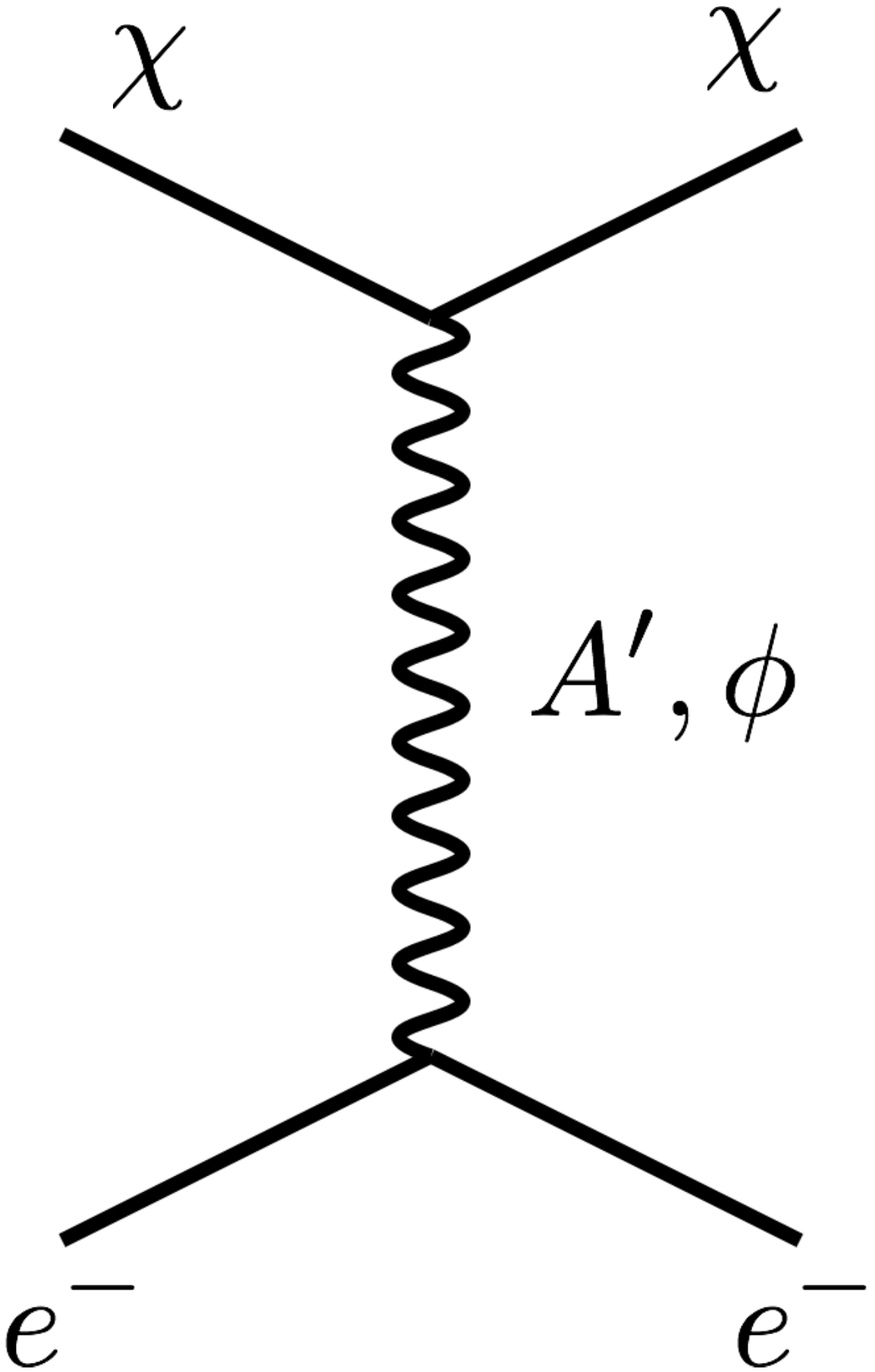}
~~~~~~\includegraphics[width=0.225\textwidth]{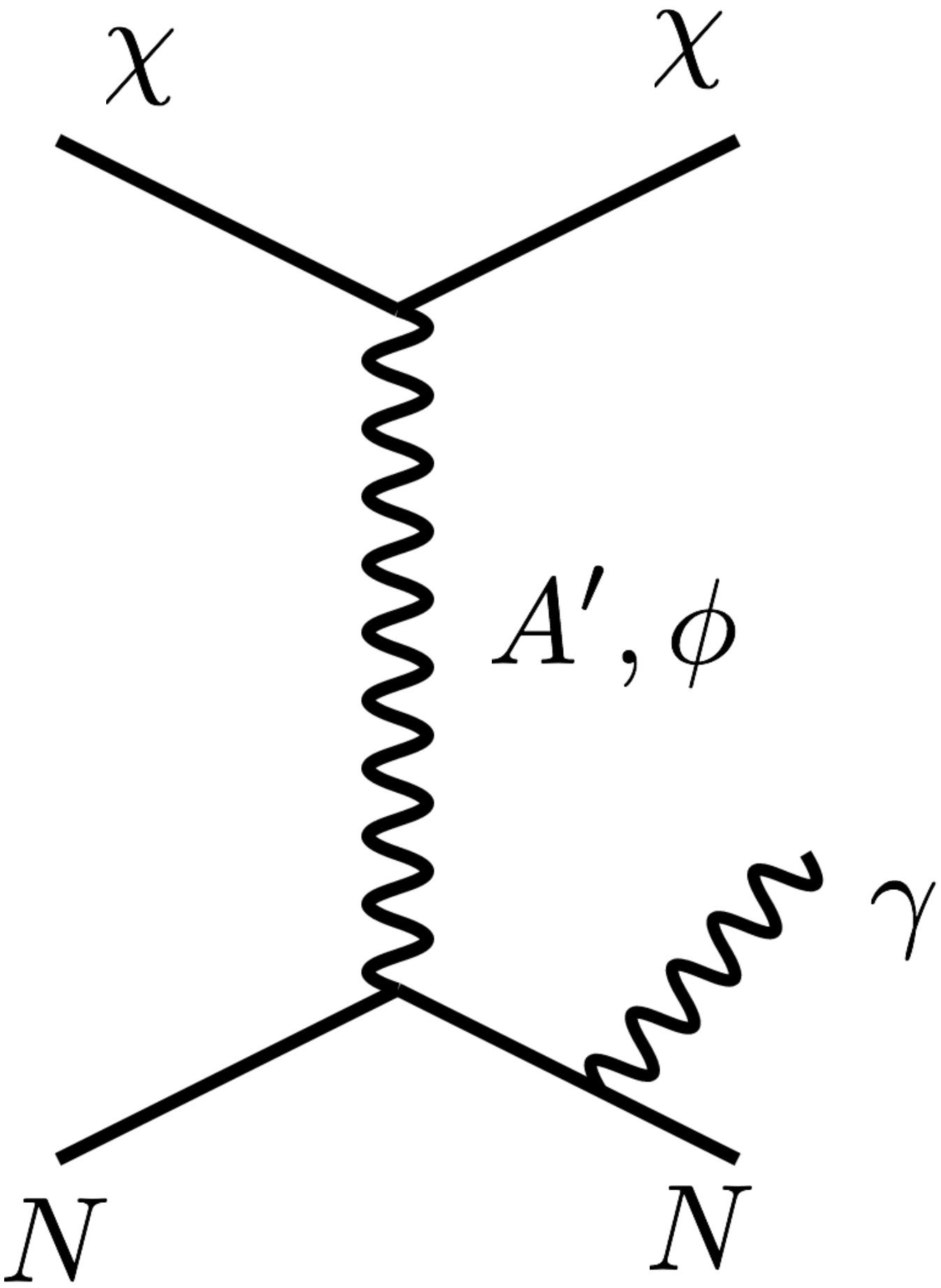}
\\ \vskip 1.0cm
\includegraphics[width=0.23\textwidth]{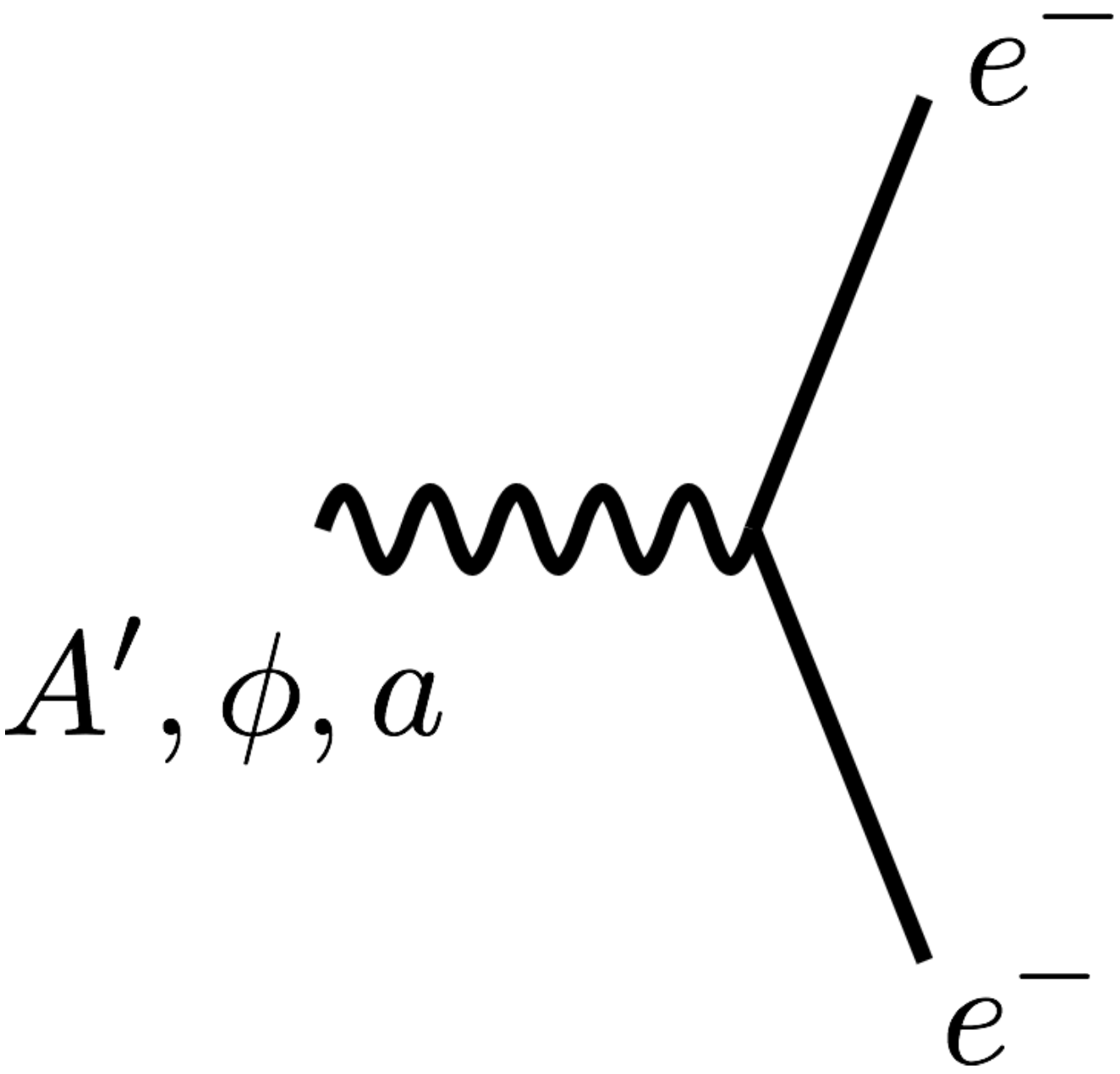}
~~~~~~\includegraphics[width=0.25\textwidth]{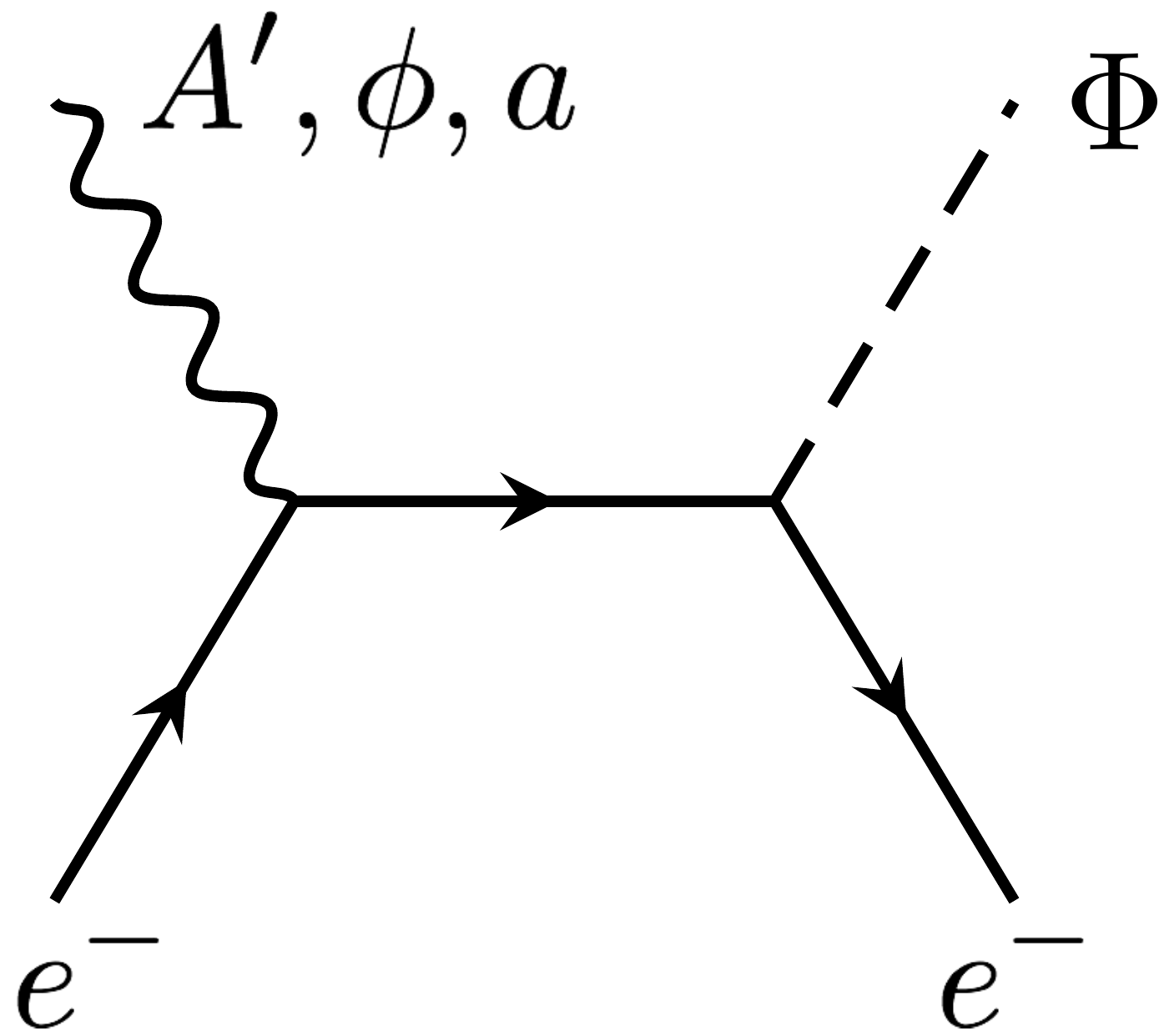}
~~~~~~\includegraphics[width=0.27\textwidth,height=0.23\textwidth]{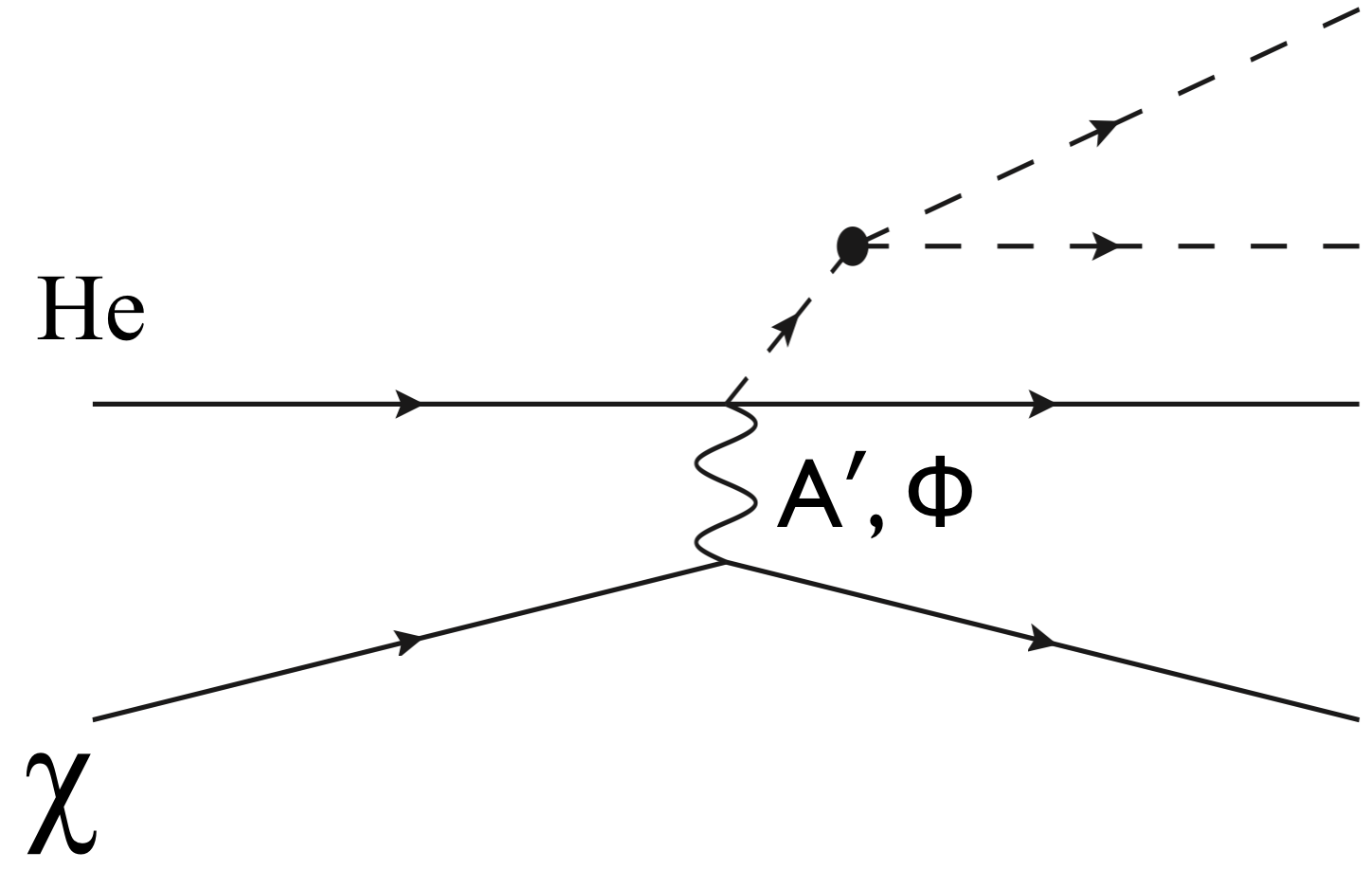}
\caption{Sample processes considered in this section to detect DM, $\chi$.  
{\it Top left:} DM-nucleus scattering.  
{\it Top middle:} DM-electron scattering.  
{\it Top right:} DM-nucleus scattering with emission of a photon.  
{\it Bottom left:} Absorption by an electron of a bosonic DM particle (a vector $A'$, scalar $\phi$, or pseudoscalar $a$).  
{\it Bottom middle:} Absorption by an electron of a bosonic DM particle, made possible by emission of a phonon $\Phi$. 
{\it Bottom right:} Emission of multiple phonons in DM scattering off helium. 
}
\label{fig:feynman}
\end{figure}
%

%%%%%%%%%%%%%%%%%%%%%%%%%%%%%%%%%%%%%%%%%%%%%%%%%%%%%%%%%%%%%%%%%%%%%
%%%%%%%%%%%%%%%%%%%%%%%%%%%%%%%%%%%%%%%%%%%%%%%%%%%%%%%%%%%%%%%%%%%%%
\subsection{New Directions for Low-Mass Dark Matter Searches}
\subsubsection{Energy Threshold}

The fundamental technical challenge in searching for sub-GeV DM is simply the size of the detectable signal. This is because the velocity of bound DM within the Milky Way galaxy, $v_{\chi}$, is non-relativistic and limited by the galactic escape velocity ($\sim 10^{-3}c$), and thus the maximum possible energy transfer to the detector decreases as the DM mass, $m_{\chi}$, is lowered.  

For the traditional nuclear recoil signals from DM scattering elastically off nuclei (Fig.~\ref{fig:feynman}, top left), the need to conserve both momentum and energy suppresses the recoil energy even further for sub-GeV masses.  
In particular, the nuclear recoil energy is given by 
\begin{equation}
E_{ \rm NR}=\frac{q^2}{2 m_N} \le \frac{2\mu_{\chi N}^2 v_\chi^2}{m_N} \lesssim 190 \mathrm{~eV} \times\left(\frac{m_\chi}{500\mathrm{~MeV}}\right)^2\left(\frac{16\mathrm{~GeV}}{m_N}\right) \,,
\label{eq:E_NR}
\end{equation}
In the latter inequality we take the DM speed to be the galactic escape velocity plus the Earth velocity, $v_\chi \simeq (544+220)$~km/s,   
to estimate the maximum nuclear recoil energy.  
We see that the energy transfer to a nucleus from an elastic DM scatter is inefficient, decreasing as $m_\chi^2$ as the DM mass is lowered 
below the GeV scale, and quickly falls below the threshold of the most sensitive current generation DM experiments.

\subsubsection{Ideas to Probe Low-Mass Dark Matter}

Over the past decade, several strategies have been proposed that maximize the energy transfer to the target. In some cases this is at the expense of a modest rate suppression, but this is at least 
partially offset by the larger DM particle flux expected as $m_\chi$ is lowered. These interactions include:

\begin{itemize}
\item {\bf DM-Electron Scattering (1~keV -- 1~GeV):}  
For low-mass DM elastic scattering (Fig.~\ref{fig:feynman}, top middle), the DM energy is transferred far more efficiently to an electron than to a nucleus~\cite{Essig:2011nj}.   If the DM is heavier than the electron, the maximum energy 
transfer is equal to the DM kinetic energy, 
\begin{equation}
\label{eq:maxkin}
E_e \le \frac{1}{2} m_\chi v_\chi^2 \lesssim 3~{\rm eV} \left(\frac{m_\chi}{{\rm MeV}}\right)\,. 
\end{equation}
Bound electrons with binding energy $\Delta E_B$ can thus in principle produce a measurable signal for 
\begin{equation}
m_\chi \gtrsim 0.3~{\rm MeV} \times \frac{\Delta E_B}{1~{\rm eV}}\,. 
\end{equation}
This allows low-mass DM to produce ionized excitations in drift chambers  ($\Delta E_B \sim 10$~eV) for $m_\chi \gtrsim 3$~MeV~\cite{Essig:2011nj,Essig:2012yx,Essig:2017kqs}, to promote electrons from the valence band 
to the conduction band of semiconductors producing ionized excitations (Ge, Si)~\cite{Essig:2011nj,Graham:2012su,Lee:2015qva,Essig:2015cda} or scintillation photons (GaAs, NaI, CsI)~\cite{Derenzo:2016fse}  ($\Delta E_B \sim 1-5$~eV) for $m_\chi \gtrsim$~0.3~MeV, and to eject an electron from a two-dimensional material such as graphene~\cite{Hochberg:2016ntt}.
DM-electron scattering searches have already illustrated their potential, probing down to $m_\chi \sim 5$~MeV~\cite{Essig:2012yx,Essig:2017kqs} using XENON10 
data~\cite{Angle:2011th} sensitive to single electrons and down to $m_\chi \sim 35$~MeV~\cite{Essig:2017kqs} using XENON100 data~\cite{Aprile:2016wwo}.  

We note that the {\it typical} recoil energy of an electron in a crystal or in the outer atomic shells of an atom is a few eV, 
and while larger recoils are possible, they are suppressed by an atomic or crystal form factor~\cite{Essig:2015cda}.  
However, this implies that as new experiments lower their thresholds, an enormous increase in the DM-electron scattering rate will lead to 
a much larger sensitivity than might naively be expected.  
Several proposals summarized below are thus expected to significantly improve on both the mass threshold and the cross-section sensitivity 
beyond the current constraints.

If the DM is lighter than the electron, the target electron velocity, $v_T$, becomes essential for extracting all of the available DM kinetic energy \cite{Hochberg:2015fth}:
\begin{equation}
E_e \simeq \frac{1}{2}\frac{{\bf q}^2}{m_e} + {\bf q} \cdot {\bf v}_T + \Delta E_B,
\end{equation}
where ${\bf q}$ is the momentum transfer in the scattering.
The target electron velocity is important in a proposal to utilize superconductors to detect DM \cite{Hochberg:2015pha,Hochberg:2015fth}, which is possible as long as the DM kinetic energy is larger than the quasi-particle binding energy ($\Delta E_B \sim $~few meV), allowing superconductors to probe DM as light as $m_\chi \gtrsim$~1~keV. 

\item {\bf DM Absorption on Electrons (1 meV -- 1 keV):}  Besides DM {\it scattering} off electrons, bosonic DM can also be {\it absorbed} 
by an electron in an atom (e.g.~in xenon~\cite{An:2014twa}) (Fig.~\ref{fig:feynman}, bottom left), in a 
superconductor through single phonon emission~\cite{Hochberg:2016ajh}, or in a semiconductor through emission of one~\cite{Hochberg:2016sqx,Bloch:2016sjj} or more~\cite{Hochberg:2016sqx} phonons (Fig.~\ref{fig:feynman}, bottom right) (the phonon emission ensures momentum conservation).  
Both the recoiling electron and the emitted phonons from the absorption can be detected in principle. 
The resulting signal is the same in both cases, but in the case of absorption the electron recoil energy is simply given by the DM mass.  
This means that bound electrons can produce a measurable signal for 
\begin{equation}
m_\chi \ge 1~{\rm eV} \times \frac{\Delta E_B}{1~{\rm eV}}\,. 
\end{equation}
Using the same target materials as described above, this allows bosonic DM to be probed down to $m_\chi \gtrsim 1\ $meV.

\item {\bf DM-low-$Z$ elastic nucleus interactions (1 MeV -- 10 GeV):}   By switching to smaller nuclei (H, He, O), kinematic matching is improved and the characteristic nuclear recoil energy is boosted by an order of magnitude from those for larger nuclei used in traditional WIMP searches (Fig.~\ref{fig:feynman}, top left).  If paired with  roton (He) or athermal phonon (O) excitation readout with 100 meV energy threshold, such experiments can probe DM masses as low as $\sim 10-100$~MeV while in the ultimate limit of single roton ($\sim$2 meV) sensitivity, such experiments would have sensitivity down to 1 MeV DM~\cite{Schutz:2016tid}. 
  
Experiments based on ionization readout of low-$Z$ nuclear recoils in gaseous drift chambers have also been proposed. Due to ionization production thresholds, such experiments would be sensitive to DM throughout the $1-10$~GeV mass range.

\item {\bf DM-off-shell nuclear interactions (1 keV -- 1 MeV):} 

In 3-body scatters, all kinematic constraints disappear and thus the entire DM kinetic energy can be transferred to the target. Specifically, a scatter that produces 2 nearly back-to-back nuclear excitations can transfer all the kinetic energy of the DM to the target nuclei while conserving total momentum \cite{Schutz:2016tid,Knapen:2016cue}. Of course, one must pay a penalty factor in the expected rate since the process is 
off-shell, but even so, a He detector sensitive to two rotons ($\sim$4 meV recoils) would probe decades of unexplored parameter space DM down to the warm DM limit of $\mathcal{O}$(keV). 

\item {\bf Bremsstrahlung in inelastic DM-nucleus scattering (10 MeV -- 1 GeV):} 

The emission of a photon when DM scatters off a nucleus (Fig.~\ref{fig:feynman}, top right) can produce a measurable signal in a detector well below the threshold for detecting an elastic DM-nucleus scattering event~\cite{Kouvaris:2016afs}.  
Since the emitted photon will typically produce an ionization signal, this signal is similar to an electron recoil signal, but originates from a DM interaction with a nucleus.  Constraints from XENON10, XENON100, and LUX already exist, and improvements are expected from upcoming experiments~\cite{Kouvaris:2016afs,McCabe:2017rln}. 

\item {\bf DM-induced Chemical-Bond Breaking (10 MeV -- 10 GeV):} 

DM scattering off nuclei can break chemical bonds between atoms, which includes the dissociation of molecules and the creation of defects in a lattice such as color centers~\cite{Essig:2016crl,Budnik:2017sbu}.  With thresholds of a few to 10's of eV, such an experiment could probe the nuclear couplings of DM particles as light as a $\mathcal{O}$(MeV).  This requires the ability to detect single defects in a macroscopic bulk of material.  

\item {\bf  DM-induced spin-flip avalanches (10 keV -- 10 MeV)}

Single molecule magnets are crystals in which the molecular spins act as independent nano-magnets. A crystal can be prepared with spins in a meta-stable state, such that localized heat generated by DM-nuclei inelastic scattering can cause the spins in that region to flip and release their stored (Zeeman) energy~\cite{Bunting:2017net}. This constitutes a positive feed-back loop, which results in a spin-flip ``magnetic bubble'' avalanche that generates a measurable magnetic flux change. The avalanche threshold can be tuned analogously to the tuning of a conventional bubble chamber, and can range from a few 10s of eV down to a few meV. 

\end{itemize}

\subsubsection{Backgrounds and Exposure}

For high mass WIMPs ($>$10 GeV), the rarity of the expected interaction requires that the active mass of  experiments be quite large, $O$(10 tons). Secondly, any backgrounds that are indistinguishable from the DM signal must be strictly controlled. Thus, the experiment must be located underground to suppress cosmogenic backgrounds and  be constructed from materials with excellent radiopurity. Furthermore, since common radioactive backgrounds such as beta decays and compton scattering produce electronic recoils with characteristic energies of $O$(100 keV) that significantly overlaps the expected WIMP-nucleus recoil signal spectrum, the capability to distinguish between nuclear and electron recoils has been found to be essential. 

Coherent scattering of solar neutrinos off nuclei will also soon become an important background that mimics the 
DM signal.  For example, LZ will likely be sensitive to $^8$B solar neutrinos that will limit their sensitivity to 5--10~GeV WIMP masses and will approach the atmospheric neutrino floor for higher masses.

Due to the dearth of experimental constraints on sub-GeV DM particles as well as the fact that the number density and thus the flux of DM varies inversely with $m_{\chi}$, the requirements on active target mass and background rejection to probe unexplored parameter space are significantly relaxed for a low-mass DM search (of course, a detector needs to have a sufficiently low energy threshold to see the DM signal in the first place).  
Sub-GeV DM experiments, for example, need only have e.g.~a mass of 100~g and run for less than $\mathcal{O}$(minutes) to probe unexplored parameter space, while those experiments in the 1--6~GeV range require O(50 kg) active mass to reach the $^{8}$B neutrino floor.

Searches for lighter mass DM also have to contend with radioactive and neutrino backgrounds,  in addition to controlling new backgrounds that exist only at low energy.  
However, the small recoil energies of these interactions means that there is very little overlap with the flat Compton and beta backgrounds with characteristic energies of $O$(100 keV). Thus, underground operation and use of radiopure materials developed for high mass WIMP searches alone should largely be sufficient to guarantee subdominant radioactive backgrounds for ``first-generation'' sub-GeV searches, while 1--10 GeV searches that reach the neutrino floor will still require either some level of electron/nuclear recoil discrimination capability, or achieve significant reduction in the total radioactive-background rate.  

Here we list and expand on the possible backgrounds: 
\begin{itemize}

\item {\bf Radioactivity.} Unlike traditional WIMP searches, in the search for MeV-GeV mass DM, radioactive backgrounds are not expected to be important for exposures $\lesssim 1\ $kg-year, given shielding and radioactivity levels comparable to those in existing experiments.  Experiments typically achieve radioactive background rates of 
$\lesssim 1~{\rm dru} \simeq 0.4$~event/kg/year/eV through 
a combination of high target-material purity and shielding of the detectors.  
These backgrounds have been measured down to 50~eV~\cite{Agnese:2015nto,Aguilar-Arevalo:2016ndq,Ramanathan:2017dfn}, 
and are expected to be approximately flat at lower energies.  
Electron recoils from sub-GeV DM scattering off electrons in a semiconductor or scintillator target 
have typical recoil energies of a few eV.     

\item {\bf Dark  Counts.}  Thermal fluctuations or other detector-specific processes can mimic the DM signal and constitute perhaps 
the most significant background challenge in probing sub-GeV DM.  
For example, the current XENON10 limit for $m_\chi \gtrsim 5$~MeV is limited by a dark-count rate, 
a significant fraction of which is likely due to ionized electrons, originally created by highly ionizing background events outside of 
the DM scattering region of interest, that become trapped at the liquid-gas interface and are released spontaneously at a later 
time~\cite{Sorensen:2017ymt}.  In general, systems that are maintained out of equilibrium with respect to a signal of interest can be expected to have dark count rates. For example, photomultipliers have dark current because their photocathodes are subject to electric fields, and cathodic surfaces under high field can be expected to emit electrons.
\item {\bf Vibrations.} The energy sensitivity of these detectors can be significantly degraded by environmental noise induced by vibrations, for example from cryocoolers \cite{Agnese:2014aze}. For experiments using thermal readout, vibration noise can arise from frictional slipping between mechanical support structures and the detector.
\item {\bf Electromagnetic Interference.} Spurious low frequency noise can be induced by external electronics, if there is not sufficient filtering at electrical feedthroughs. For experiments seeking very low energy thresholds, protections must be taken to minimize electromagnetic environmental interference from coupling to the detector.
\item {\bf Solar neutrinos.} For electron-recoil searches for sub-GeV DM, coherent solar neutrino-nucleus scattering will only 
be a background for exposures of $\gtrsim 1\ $kg-year~\cite{Essig:2011nj,Hochberg:2015fth}.  
For nuclear recoil searches for sub-GeV DM, coherent neutrino-nucleus scattering is a less significant background than in the 1--10 GeV range. This is because the DM signal becomes concentrated in a smaller and smaller energy range with decreasing mass, the solar neutrinos have characteristic energies in the hundreds of keV to several MeV with much less flux below these energies, and because the neutrino scattering cross-section scales as $E^2$ and thus decreases significantly at low energy.
\item {\bf Coherent photon background for sub-MeV DM searches.} It is common for DM direct detection experiments to have a significant background from Compton scattering of gamma rays, included above under {\bf Radioactivity}. But for experiments designed to reach extremely low energy thresholds, one must also take into account the coherent scattering of these gamma rays from atoms in the target material. The cross-section for coherent scattering is large, scaling as $Z^2$ of the target material, and this scattering can be significant in the energy regime less than $\sim$ 100 meV. Consequently, active Compton vetoes must be considered for beyond pathfinder 
experiments~\cite{Robinson:2017prd}.

\end{itemize}

\subsection{New Directions for Spin-Dependent (Proton) Interaction Searches}

Xenon contains two spin isotopes that have an unpaired neutron, $^{129}$Xe (spin-1/2) and $^{131}$Xe (spin-3/2) with an abundance 
of about $\mathcal{O}(20-25\%)$ each. XENON100 and LUX have thus set the best constraints on spin-dependent DM-neutron 
interactions~\cite{Aprile:2013doa,Akerib:2016lao}, and LZ is expected to provide the best constraint in future. 
However, the G2 experiments will not probe spin-dependent DM-proton couplings as effectively as an experiment using a target material with unpaired protons. The strongest constraint currently comes from PICO-60, using C$_3$F$_8$ in which $^{19}$F contains an unpaired 
proton~\cite{Amole:2017dex}.  
Additional experiments using C$_3$F$_8$ or other appropriate target nuclei have been proposed and will be summarized below.

%%%%%%%%%%%%%%%%%%%%%%%%%%%%%%%%%%%%%%%%%%%%%%%%%%%%%%%%%%%%%%%%%%%%%
%%%%%%%%%%%%%%%%%%%%%%%%%%%%%%%%%%%%%%%%%%%%%%%%%%%%%%%%%%%%%%%%%%%%%
\subsection{Brief Descriptions of Experimental Efforts}

In this subsection, we summarize various R\&D and other experimental efforts.  
One-page summaries for some of these efforts can be found in~\cite{WG1-one-page-summaries}.
A summary of these efforts appears in Table~\ref{tab:all-experiments}, together with an estimated cost and timescale.  
{\it We emphasize that several proposals can probe more than one science target, but we have grouped each idea into only one (primary) 
science target.} 
\subsubsection{Sub-GeV Dark Matter (Electron Interactions)}
\begin{itemize}
\item 
{\bf SENSEI:} 
SENSEI will use a recently demonstrated technological breakthrough to search for DM-electron scattering interactions to explore a wide range 
of currently unconstrained DM candidates with masses in the 1~eV -- few~GeV range.  
This project would use a thick fully depleted silicon CCD in the far sub-electron regime ($\sim 0.05$~rms/pix) using a new generation of Skipper-CCDs designed by the LBL Micro Systems Lab. 
For the first time, it has been demonstrated that the charge in each pixel of a CCD --- in a detector consisting of millions of pixels --- can be measured with sub-electron noise~\cite{Tiffenberg:2017aac}. 
A 1-gram detector is already operating in the NUMI access tunnel.  
A larger project (100 grams) can be deployed at a deeper site on a timescale of $\sim 1-2$ years  
if funding is obtained (the required funding is well within the small-project scale).  
A $\mathcal{O}(100)$-gram detector running for one year is expected to be essential free of radioactive backgrounds, assuming a 
background level of $\approx 5$~dru, which has already been demonstrated by the current DAMIC detector operating at SNOLAB.  
Moreover, dark counts are expected to be negligible for a threshold of two or three electrons, allowing SENSEI 
to achieve unprecedented sensitivity to Hidden-Sector and Ultralight DM.  
\item  
{\bf DAMIC-1K:}  DAMIC-1K is a low-background ($\approx$ 0.1 dru), low-threshold
($2 e^-$) experiment with a detector mass of $\approx 1$~kg. It builds on
the success of the DAMIC experiment at SNOLAB, which employs
high-resistivity, thick CCDs to detect sub-keV energy deposits in the bulk
silicon. The technology to fabricate DAMIC-1K CCDs is already proven, with
modest increase in area and thickness of the DAMIC detectors. Skipper
design --- developed, tested, and implemented by the SENSEI collaboration --- 
will be used to reach sub-electron noise, combined with digital
filtering for fast readout. Improvements in the design of the shielding,
in the selection of materials, and in handling procedures will be
implemented to reach a radiogenic background of $\approx 0.1$~dru.
DAMIC-1K will search for low-mass DM in a broad range from 1~eV -- few GeV
with unprecedented sensitivity to DM-electron scattering and hidden-photon
DM. Also, DAMIC-1K will demonstrate the rejection of cosmogenic $^{32}$Si
--- the dominant background for SuperCDMS Si-HV --- through spatial
correlation of candidate events with the decay of the $^{32}$P daughter,
providing a path to the exploration of low-mass DM interactions down to
the Neutrino Floor.
\item  
{\bf UA$'$(1):} Direct detection of dark sector DM via counting single to few electron ionization events in a liquid xenon target.  A primary goal of this experiment will be to understand and mitigate the electron backgrounds in a two-phase xenon detector. Such mitigation R\&D needs to happen in a small (10 kg scale) target and flexible test bed. Studies ultimately need to be carried out underground due to the long lifetime of trapped electrons at the liquid xenon surface. While this experiment is expected to be sensitive to new parameter space, success in mitigation of electron backgrounds would be a great success on its own, because it could enable much larger detectors (e.g. LZ) to perform far more sensitive searches for this class of DM.
\item
{\bf Cryogenic GaAs(Si,B) scintillator for transition edge sensor readout:} 
In~\cite{Derenzo:2016fse}, n-type GaAs was suggested as a promising target material for sub-GeV DM detection due to its commercial availability in high purity and large sizes (15~cm), and its known fluorescent properties at cryogenic temperatures. GaAs has a direct gap of 1.52~eV, and thus a DM particle can scatter off a valence-band electron exciting it into the conduction band. Doping with Si (n-type donor) and boron (p-type) creates trapping sites that scintillate, producing 1.33~eV photons with a measured scintillation yield of 30~photons/keV in crystals with non-optimized dopant densities.

The production of scintillation photons at long time scales after a particle interaction (``afterglow") has been seen in e.g.~NaI and CsI, and is a primary background concern. Recent measurements by the scintillation research group at LBNL, however, have seen no thermally stimulated emission after cryogenic x-ray bombardment.  This suggests that highly-doped n-type GaAs has no afterglow, probably because it has delocalized electrons that can easily annihilate any metastable radiative states. Radioactive backgrounds are also expected to be non-limiting, since $^{3}$H and other cosmogenic spallation contamination can be minimized by limiting surface exposure following crystal production, since no sensor fabrication occurs on the GaAs crystal itself (unlike with SuperCDMS). Furthermore, U/Th-chain radioactive backgrounds are expected to be sub-dominant since commercial GaAs is highly purified. 

This project plans to find dopant concentrations that optimize scintillation performance to hopefully approach the theoretical limit of 200~photons/keV, optimize surface roughness / use of anti-reflective coatings to improve transmission, as well as to develop large-area detectors sensitive to single optical photons within the next 2 years. A 10~kg pathfinder experiment could then be run in 2019 at the CUTE facility.

\item 
{\bf NICE:} The intrinsic light yields of pure NaI/CsI at 77~K have been found to be about twice higher than those of thallium-doped NaI/CsI at room temperature. Integrated with light sensors working at cryogenic temperatures, those pure crystals can be used for various rare event detections. In a phased approach, the first step would be to use cylindrical crystals (about 1~kg) wrapped with PTFE tape, watched by 2 photomultipliers from the ends, cooled by liquid nitrogen or argon, with a background measurement down to 0.2~keV$_{ee}$. In a second step, the system could be switched to SiPM readout for higher quantum efficiency and to explore an active veto with liquid argon and neon. Finally, the project could move to transition edge tensor readout for 100\% quantum efficiency, single photon trigger and an accompanying phonon signal. 
\item 
{\bf Germanium Detector with Avalanche Ionization Amplification:} 
We propose to develop ionization amplification technology for Ge in which very large localized E-fields are used to accelerate ionized excitations produced by particle interaction to kinetic energies larger than the Ge bandgap at which point they can create additional 
$e^{-}/h^{+}$ pairs, producing intrinsic amplification. This amplified charge signal could then be readout with standard high-impedance JFET- or HEMT-based charge amplifiers. Such a system would potentially be sensitive to single ionized excitations produced by DM interactions with both nuclei and electrons. In addition, purposeful doping of the Ge could lower the ionization threshold by $\sim \times$10 ($\sim\ $100~meV), making the detector sensitive to 100~keV DM via electron recoils.

A 3 year R\&D program could develop both the avalanche ionization amplification and impurity ionization technology, after which a 10~kg pathfinder experiment could be constructed in 2 years.
\item 
{\bf PTOLEMY-G$^{3}$:} In the PTOLEMY-G$^{3}$ experiment, graphene field-effect transistors (G-FETs) arranged into a fiducialized volume of stacked planar arrays, called a graphene
cube (G3), would be used to search for MeV DM scattering events that liberate an electron from the graphene target. A narrow, vacuum-separated front-gate of
the G-FET imposes a kinematic discrimination on the maximum electron recoil energy, and the FET-to-FET hopping trajectory of an ejected electron indicates the scattering direction, shown to be correlated to the DM wind. High radio-purity wafer-level fabrication, ultra-low ratio 14C/C graphene growth, a cryogenic fiducialized volume, and the coincidence of the FET-to-FET trajectories of electron recoils would provide the conditions for a low background observatory of MeV DM interactions. The evaluation of the G3
active target and low background methods are an important step for the PTOLEMY project whose long-term goal is the direct detection of the cosmic neutrino background. PTOLEMY-G3 is the only proposed experiment with direct directional detection capability for MeV DM.
\item 
{\bf Superconducting aluminum:} Superconducting detectors can be sensitive to O(meV) electron recoils from DM-electron scattering, using the superconducting gap (e.g in aluminum this is 0.6 meV). Such devices could detect DM as light as a few keV. The use of superconductors as DM targets would be a natural extension of the TES-based DM detection program, as TES resolution reaches the meV scale.
\end{itemize}
\subsubsection{Sub-GeV Dark Matter (Nucleon Interactions)}
\begin{itemize}
\item 
{\bf Superfluid helium with transition edge sensor readout:} Superfluid helium is an extremely pure material with no intrinsic radioactivity and little coupling of vibrations from surrounding solid materials, allowing substantial background suppression. In addition, while superfluid helium produces electronic excitations like the other noble liquids, it is also amenable to calorimetric readout since it remains a liquid at extremely low temperatures. In this detector concept, superfluid helium scintillation light and triplet excimers are detected using athermal, cryogenic sensors with transition edge sensor readout \cite{Guo:2013,Car:2017}.  Rotons and phonons are detected using quantum evaporation, using a bolometer array suspended above the superfluid helium. Background rejection efficiency has been estimated, using the ratio of scintillation light to heat in the form of phonons and rotons. Such detectors should enable DM searches sensitive to extremely low energy deposition, as the phonon/roton signal is amplified through the helium atom desorption/adsorption process. Very low-mass DM candidates might be detected using multi-excitation processes. in which back-to-back phonons or rotons are produced, enabling extraction of the DM kinetic energy while conserving energy and momentum. Assuming gamma ray backgrounds comparable to those projected for SuperCDMS-SNOLab, existing transition edge sensor technology, coupled to a $\sim 1$~kg superfluid helium target, would allow sensitivity to DM candidates with mass as low as 10-30 MeV.  
\item 
{\bf Evaporation and detection of helium atoms by field ionization:} In this variation on a superfluid helium-based DM detector, nuclear recoils would be detected using a scheme based on quantum evaporation of helium atoms followed by field ionization~\cite{Maris:2017xvi}. WIMP scattering events from nuclei with recoil energy of the order of 1 meV produce quasiparticle excitations (phonons and rotons) which can desorb a helium atom from the surface of superfluid helium or other crystalline target materials. The ability to detect single helium atoms in the gas by field ionization thus obtains a threshold energy sensitivity low enough to search for ~ 1 MeV DM particles. A helium atom becomes field-ionized when one of its electrons tunnels into a positively charged metal tip through a field-distorted barrier. The helium ion then accelerates from a high potential, typically several keV, to a cathode which can be a calorimeter, a channeltron, or a microchannel plate. The impact of a single ion is easy to detect. Given that the field-ionization approach could be applied to a range of target materials, from superfluid helium to solids with a long mean free path for phonons, a dedicated research effort extending over several years is required to develop a scalable fabrication process, establish the quantum detection efficiency, and investigate the possibility of dark counts.
\item 
{\bf Color centers:} This experimental initiative involves using defects in crystals created by nuclear recoils with energy of the order of 10 eV. This probe for light DM elastic scattering is in principle sensitive down to DM masses of $\sim$100 MeV. On top of that, sensitivity to solar neutrinos is reached with exposures of about 100 kg year. The defects live practically forever, and in many cases are spectroscopically active. The concept is to look at a bulk of these and count extra defects as they form. Challenges are many; to list the most important: Finding a handle of the optimal signal, rejecting backgrounds, removing existing defects (production, annealing), as well as calculations of rates, branching ratios and response.
\item
{\bf Magnetic bubble chambers:} A proof-of-concept magnetic bubble chamber~\cite{Bunting:2017net} with a $\sim\,$eV energy threshold is currently under development. This prototype will aim to demonstrate stability of the proposed detector; a neutron beam will further be used to demonstrate and perform calibration of the spin avalanche mechanism. The initial design is based on a (1\,cm$^2)\times(\sim\,$few mm) powder sample of compound 3 of~\cite{doi:10.1021/ja068961m} placed in a 50 mK fridge, and shielded with four inches of low activity, cadmium-lined lead. The use of single molecule magnet crystals with lower ($\sim\,$meV) energy thresholds would follow successful experimental demonstration. \end{itemize}
\subsubsection{Searches down to the  Neutrino Floor  for $\mathcal{O}$(GeV) Dark Matter}
\begin{itemize}

\item
{\bf SuperCDMS SNOLAB G2+:} The currently funded G2 experiment SuperCDMS SNOLAB can probe large areas of unexplored DM parameter space in the DM mass range 0.5 -- 6 GeV/c$^{2}$ 
with an ultimate sensitivity that is expected to be limited by $^{3}$H $\beta$ decays produced by cosmogenic spallation during detector fabrication at $\sim \times$20 the neutrino floor. Thus, developing new detector technology with 1:20 electronic recoil / nuclear recoil background discrimination for sub-keV recoils would allow a subsequent upgrade to reach the neutrino floor.  The athermal phonon sensor technology and Luke-Neganov phonon amplification techniques developed by SuperCDMS lead to two natural detector concept evolutions that could achieve this capability:
\begin{enumerate}
\item Encoding 3D position information and ionization yield into the Luke-Neganov phonon signal: SuperCDMS HV detectors currently use Luke-Neganov phonons produced during the drifting of e$^{-}$/h$^{+}$ across a planar electro-static potential in a semi-conductor (Luke-Neganov gain), to lower the energy threshold so as to be sensitive to recoils from very light mass DM. 
We propose to develop a  high voltage detector with 2 interdigitated phonon sensors that replicate the E-field pattern found in the High Mass SuperCDMS iZIP detector design, thereby regaining the $z$-position and electronic/nuclear recoil discrimination capabilities seen in the phonon pulse shape and energy partition.
\item Encoding ionization yield into Luke-Neganov quantization offsets: If the sensor resolution of the SuperCDMS detectors can be decreased significantly below the drift voltage across the crystal, then the total phonon energy spectrum will resolve into quantized peaks depending upon the number of e$^{-}$/h pairs generated by the interaction. Since the average recoil energy to produce a given number of e$^{-}$/h is vastly different for electronic and nuclear recoils, the quantized peaks for nuclear recoils will be offset from that for electronic recoils, and consequently recoil type discrimination should be possible.
\end{enumerate}

\item 
{\bf NEWS-G:} The goal of the NEWS-G (New Experiments with Spheres - filled with Gas) collaboration is to search for galactic DM particles in the 0.1 to few GeV mass region. Detectors are constituted of spherical metallic vessels, each equipped with a small ball sensor set at high voltage at the center of the sphere. Each sphere is filled with a noble gas mixture (Ar, Ne, He, H), operated in proportional mode at pressure up to 10 bar.
\item 
{\bf NEWS-dm:} NEWSdm is meant to be the first experiment with a solid target for directional DM searches: the use of a nuclear emulsion based detector, acting both as target and tracking device, would allow to extend DM searches beyond the neutrino floor and provide an unambiguous signature of the detection of Galactic DM. The novel emulsion technology, based on the use of nuclear emulsion films with nanometric AgBr crystals (NIT), makes it possible to record the sub-micrometric tracks produced by the WIMP scattering off a target nucleus.  In March 2017 the NEWSdm Collaboration has installed an experimental setup for the exposure of a ~10g detector at the Gran Sasso INFN Underground Laboratories.  This test aims at measuring the detectable background from environmental and intrinsic sources and to validate estimates from measurements and simulations. The confirmation of a negligible background will pave the way for the construction of a pilot experiment with an exposure on the $\sim 10$~kg year scale. This pilot experiment will act as a demonstrator to further extend the sensitivity towards the neutrino floor. 

\item
{\bf CYGNUS HD-10:} This directional DM experiment would be a 10~m$^{3}$ gas target time projection chamber with a He:SF$_6$ gas mixture. The SF$_6$ component enables negative ion drift (for reduced diffusion) and 3D fiducialization via minority carriers. 
High resolution charge readout, via resistive Micromegas, will be used to image ionization from nuclear recoils in 3D (``charge cloud tomography'').
This is expected to enable excellent electron-event rejection, fiducialization techniques via transverse diffusion of drift charge, 3D-directionality for unambiguous WIMP discovery, and penetration of the neutrino floor. 
A first, 10~m$^3$ CYGNUS HD-10 detector is expected to have sensitivity competitive with the G2 experiments, to  
both SD and SI interactions, with improved electron rejection for low WIMP masses. The proposed He:SF$_6$ gas mixture is a starting point, and could be optimized to target primarily SD or (at low masses) SI interactions with further improvements in sensitivity. Detailed imaging of ionization allows sensitivity to DM models with multiple-particle final states. CYGNUS HD-10 would be a first step towards a large-scale CYNUS directional detector capable of unambiguously demonstrating the cosmic origin of a WIMP signal, penetration of the neutrino floor, and eventually, WIMP astronomy.
\item 
{\bf Scintillating bubble chambers:} These detectors combine the extremely effective electron rejection and simple instrumentation of a bubble chamber with the event-by-event energy resolution of a liquid scintillator. Recently simultaneous scintillation and bubble nucleation by low-energy nuclear recoils in superheated xenon has been demonstrated. Superheated water bubble chambers are also being pursued to take advantage of advances in water-based scintillators. The goal of the scintillating bubble chamber effort is the development of detectors with the scalability, target flexibility, and background discrimination needed to push WIMP sensitivity towards the neutrino floor (or follow-up a new signal) after the G2 program. The energy information provided through scintillation will allow reduction of backgrounds from high-light-output alpha particles, as well as non-scintillating backgrounds like dust particulates. In addition, scintillating bubble chambers will have even higher rejection of minimum-ionizing backgrounds than non-scintillating bubble chambers, by virtue of having a new energy loss mechanism for particle interactions. The scintillating bubble chamber technique is now established in liquid xenon at the 30-gram scale, and the key next step is the construction of an O(10)-kg scale xenon bubble chamber to demonstrate the scalability of these detectors~\cite{Baxter:2017ozv}. 
\end{itemize}
\subsubsection{Spin-Dependent (Proton) Interactions}
\begin{itemize}
\item 
{\bf PICO:} The PICO bubble chamber detectors can be made very large, have extremely low backgrounds, and work with diverse target nuclei. Most important recent scientific impacts have come from $\rm C_{3}F_{8}$ targets, where the $\rm ^{19}F$ nucleus gives unique sensitivity to spin-dependent WIMP couplings to the proton. Due to coherent enhancement of the background neutrino rate, the ultimate background from atmospheric and solar neutrinos is expected to be two orders of magnitude lower for $\rm C_{3}F_{8}$ than for xenon, when cast in terms of spin-dependent sensitivity. In addition to the $\rm C_{3}F_{8}$ program, the PICO Collaboration is investigating alternative targets for future searches in PICO-40L, PICO-500, or an array of PICO-500 detectors. These include hydrocarbons for low-mass WIMP searches, $\rm CF_{3}I$ to search for coupling to proton orbital angular momentum in iodine or as follow-up to a spin-independent signal in xenon, and superheated nobles (argon, xenon) to take advantage of the extra discrimination and event-by-event energy information provided by the scintillation signal.
\end{itemize}

%%%%%%%%%%%%%%%%%%%%%%%%%%%%%%%%%%%%%%%%%%%%%%%%%%%%
\begin{table}[htbp]
\footnotesize
\begin{center}
    \begin{tabular}{ | p{3.3cm} | p{2.9cm} | p{2.8cm} | p{2.5cm} | p{4.1cm} |}
    \hline
   Main Science Goal    &      Experiment 			& Target 		& Readout	& Estimated Timeline	 \\ 
    \hline
    \hline
    \multirow{ 7}{3cm}[-1cm]{Sub-GeV Dark Matter (Electron Interactions)}  
        & SENSEI \nl & Si & charge &  ready to start project \nl (2 yr to deploy 100g)  \\
    \cline{2-5}
    &  DAMIC-1K  \nl & Si & charge & ongoing R\&D \nl 2018 ready to start project \nl (2 yr to deploy 1 kg) \\  
    \cline{2-5}
	&   UA$'$(1) \nl liquid Xe TPC   & Xe & charge & ready to start project \nl (2 yr to  deploy 10kg) \\
    \cline{2-5}
     & Scintillator w/ TES readout & GaAs(Si,B) & light  &  2 yr R\&D  \nl 2020 in sCDMS cryostat   \\
    \cline{2-5}
    & NICE; NaI/CsI cooled crystals & NaI \nl CsI & light &   3 yr R\&D \nl 2020 ready to start project  \\ 
    \cline{2-5}
    &  Ge Detector w/ Avalanche Ionization Amplification & Ge & charge &  3 yr  R\&D \nl 1 yr 10kg detector \nl 1 yr 100kg detector \\ 
    \cline{2-5}
    & PTOLEMY-G3, 2d graphene & graphene & charge \nl directionality &  1 yr fab prototype \nl 1 yr data   \\
    \cline{2-5}
    & supercond.~Al cube & Al & heat  & 10+ yr program  \\ 
   \hline
   \hline
        \multirow{ 5}{3cm}[-0.7cm]{Sub-GeV Dark Matter (Nucleon Interactions)} 
    & Superfluid helium with TES readout  & He & heat, light &  1 yr R\&D; 2018 ready to start project; 2022 run  \\
     \cline{2-5}
   & Evaporation \& detection of He-atoms by field ionization  & superfluid helium, crystals with long phonon mean free path (e.g. Si, Ge) & heat & 3 yr R\&D; 2020 ready to start project  R\&D  \\
   \cline{2-5}
    & color centers & crystals (CaF) & light &   R\&D effort ongoing  \\
    \cline{2-5}
    & Magnetic bubble chamber & Single molecule magnet crystals & Spin-avalanche (Magnetic flux) & R\&D effort ongoing \\ 
   \hline
   \hline
           \multirow{ 5}{3cm}[-0.7cm]{Searches down to Neutrino Floor for $\mathcal{O}$(GeV) Dark Matter}   
           &  SuperCDMS-G2+  & Ge & heat, ionization &  3 yr R\&D; 1 yr fabrication; 2022 start running  \\
    \cline{2-5}
      & NEWS-G \nl  & H, He & charge & 140cm sphere installed at SNOLAB in 2018 \\
    \cline{2-5}
    & NEWS-dm \nl emulsions & Si, Br, I, C, O, N, H, S &  charge \nl directionality & R\&D  phase complete. \nl Now technical test \\
    \cline{2-5}
     & CYGNUS  HD-10 	&  SF$_6$, He \nl flexible  	& 	charge \nl directionality 	&  1 yr R\&D;   1 yr  1 m$^3$;  \nl 2 yr 10 m$^3$   \\ 
    \cline{2-5}
     & Scintillating  bubble chamber 	& Xe, Ar \nl C$_6$F$_6$, H$_2$0 & light \nl heat(bubble) &   2 yr program; test 10kg Xe chamber with CENNS \\
    \hline
    \hline
   Spin-Dependent \nl (Proton) Interactions  &     PICO \nl bubble chambers  &  wide range  & heat(bubble) &  40 l chamber now \nl PICO 500 l next \\
     \hline

      \end{tabular}
      \caption{\label{tab:all-experiments}
{\footnotesize Proposals and ideas for new experiments, grouped according to their main science target as identified in Working Group 1: 1) Sub-GeV DM (Electron Interactions), 2) Sub-GeV DM (Nucleon Interactions), 3) Searches down to the Neutrino Floor for $\mathcal{O}$(GeV) Dark Matter, and 4) Spin-dependent (Proton) Interactions. {\it Note that several proposals can probe more than one science target.}  
Within each category, the proposal/idea is ordered roughly according to the timescale needed to start the project.  The target material and main readout channel are also listed.}
}
\end{center}
\end{table}

%%%%%%%%%%%%%%%%%%%%%%%%%%%%%%%%%%%%%%%%%%%%%%%%%%%%%%%%%%%%%%%%%%%%%
%%%%%%%%%%%%%%%%%%%%%%%%%%%%%%%%%%%%%%%%%%%%%%%%%%%%%%%%%%%%%%%%%%%%%
\subsection{Facilities}

The capability of the direct DM search community to develop the next generation of detectors
depends in great manner of the availability of facilities for detector R\&D, calibration, and early science
tests. Some of the facilities recently developed, or planned for the near future, are discussed here. The existence
of these facilities reduces significantly the time and cost from detector idea to early science, and also calibration.

\subsubsection{Nuclear Recoil Calibration Facility at TUNL}

The community needs precision measurements of the quenching factor for
several detector technologies in order to perform direct DM searches 
with nuclear recoils. A facility to perform these measurements has been established at
TUNL (Triangle Universities Nuclear Laboratory).  This facility is set to produce pulsed,
tunable, and quasi-mono-energetic neutron beams, with very flexible beam configurations.
Because of the dedicated space available (3 target areas), it is possible to do calibrations
requiring long setup times. 

The TUNL facility uses a 10 MV Tandem accelerator, with bunching and chopping 
capabilities. The facility can operate with various ion sources ($^1$H,$^2$H,$^3$He,$^4$He),
a maximum current of 1 $\mu$A, and 70~keV to 15~MeV neutron energies.
The backing neutron detectors needed for scattering experiments are available at the facility. Several
experiments have already used, or a planning to make use of TUNL for their quenching factor
measurements.

\subsubsection{Northwestern Experimental Underground Site at Fermilab (NEXUS)}

A clean, low-background, testing facility with convenient access for prototyping and testing the next-generation 
cryogenic detectors is being set at the NuMI access tunnel at Fermilab (300 m.w.e.). The facility  is being established
by the Northwestern SuperCDMS group, and will also be available for other users 30\% of the time. This  depth 
gives a muon rate of 3.4 muons/cm$^2$/day, ideal for long term detector testing without 
the risk of cosmogenic activation. A dilution fridge with a 10~mK base temperature will be available at this facility 
with large experimental volume (33~cm diameter $\times$ 53~cm height), and 150~$\mu$W of cooling power at 100~mK. 
Background is expected to be 100~dru, and a  D-D neutron generator will be available for
calibration purposes. This facility is expected to be online early in 2018.

The University of Minnesota group is developing a very low energy nuclear recoil calibration technique 
that could be implemented at NEXUS. This technique is based on thermal neutron capture and 
would produce calibrations for recoils with energies in the 100~eV -- 400~eV range. It is expected that the first measurement 
of these recoils in solids will take place at the University of Minnesota in the Summer of 2017. Improvements to the 
technique are hoped to allow the method to be employed at planned calibration facilities discussed here (NEXUS, CUTE, TUNL).

\subsubsection{Cryogenic Test Underground Facility (CUTE) at  SNOLAB }

This facility operated by the Queen's SuperCDMS group provides a very low background cryogenic test stand
at SNOLAB. CUTE is set to perform low background studies, and can also be used as a science platform. A dilution fridge
similar to the NEXUS (same vendor) will be available, with lower background of 3--30~dru. The fridge will be 
installed inside a water tank for shielding.

\subsubsection{SuperCDMS cryogenic facility at SNOLAB}

The SuperCDMS project is building the SNOBOX facility at SNOLAB.  This is the dilution fridge to be used for the G2 program with
5~$\mu$W cooling power at 15~mK. 
SNOBOX will have a very large volume available, capable of holding 31 SuperCDMS towers, with an expected  
background that is 30 times lower than CUTE. The SuperCDMS G2 program is scheduled to start science operations 
in 2020 with a 4 tower payload. The additional space available
is the ultimate location for either a large payload or a very-low background measurement. The additional space
available in SNOBOX offers an opportunity for operating other low background cryogenic detectors.

%%%%%%%%%%%%%%%%%%%%%%%%%%%%%%%%%%%%%%%%%%%%%%%%%%%%%%%%%%%%%%%%%%%%%
%%%%%%%%%%%%%%%%%%%%%%%%%%%%%%%%%%%%%%%%%%%%%%%%%%%%%%%%%%%%%%%%%%%%%
\subsection{Projected Sensitivities and Yield Estimates}

In this section, we summarize the sensitivities for the experimental ideas mentioned above.  
We also show a few benchmark targets that were discussed in Sec.~III and summarized in Sec.\ref{subsec:science-DD}.  
These benchmark targets assume that the DM scattering is mediated through a dark photon.  
For this case, we define the DM-electron scattering cross section $\bar\sigma_e$ and the DM form factor $F_{\rm DM}(q)$ 
as~\cite{Essig:2011nj,Essig:2015cda} 
\bea
\overline\sigma_e = \frac{16\pi\mu^2_{\chi e} \alpha \epsilon^2\alpha_D}{(m_{A'}^2+\alpha^2 m_e^2)^2}
\simeq
\begin{cases}
\frac{16 \pi \mu_{\chi e}^2 \alpha \epsilon^2 \alpha_D}{m_{A'}^4}\,, & m_{A'} \gg \alpha m_e \\
\frac{16 \pi \mu_{\chi e}^2 \alpha \epsilon^2 \alpha_D}{(\alpha \, m_e)^4}\,, & m_{A'} \ll \alpha m_e
\end{cases}\,, \\
 F_{DM}(q) = \frac{m_{A'}^2+\alpha^2m_e^2}{m_{A'}^2+q^2} \simeq
\begin{cases}
1\,, & m_{A'} \gg \alpha m_e \\
\frac{\alpha^2 m_e^2}{q^2}\,, & m_{A'} \ll \alpha m_e
\end{cases}\,.
\eea
Here $\alpha_D\equiv g_D^2/4\pi$, $\mu_{\chi e}$ is the DM-electron reduced mass, and $q$ is the momentum transfer between the 
DM and electron.

\begin{figure}[!t]
\includegraphics[width=0.48\textwidth]{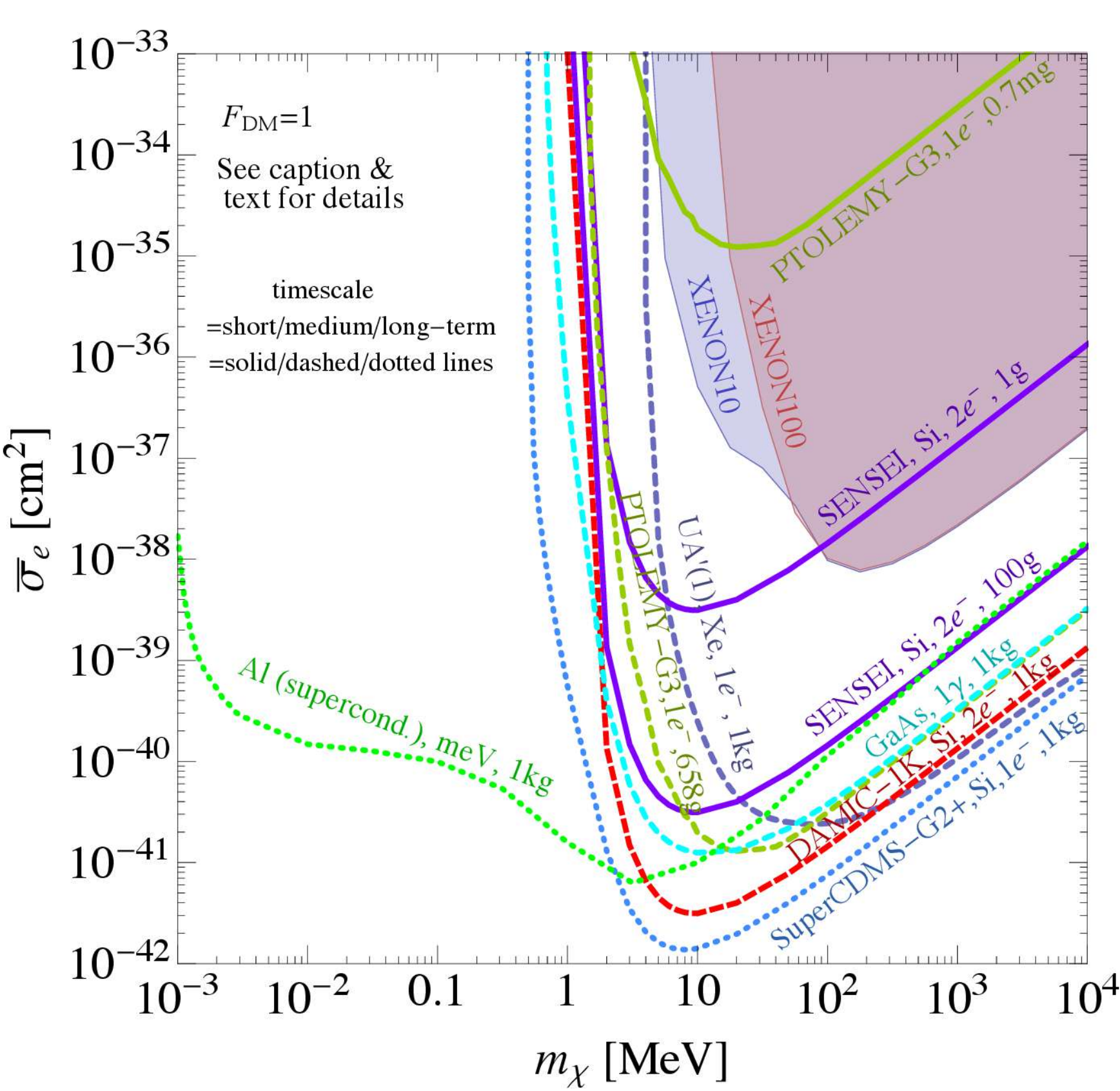}
~~
\includegraphics[width=0.48\textwidth]{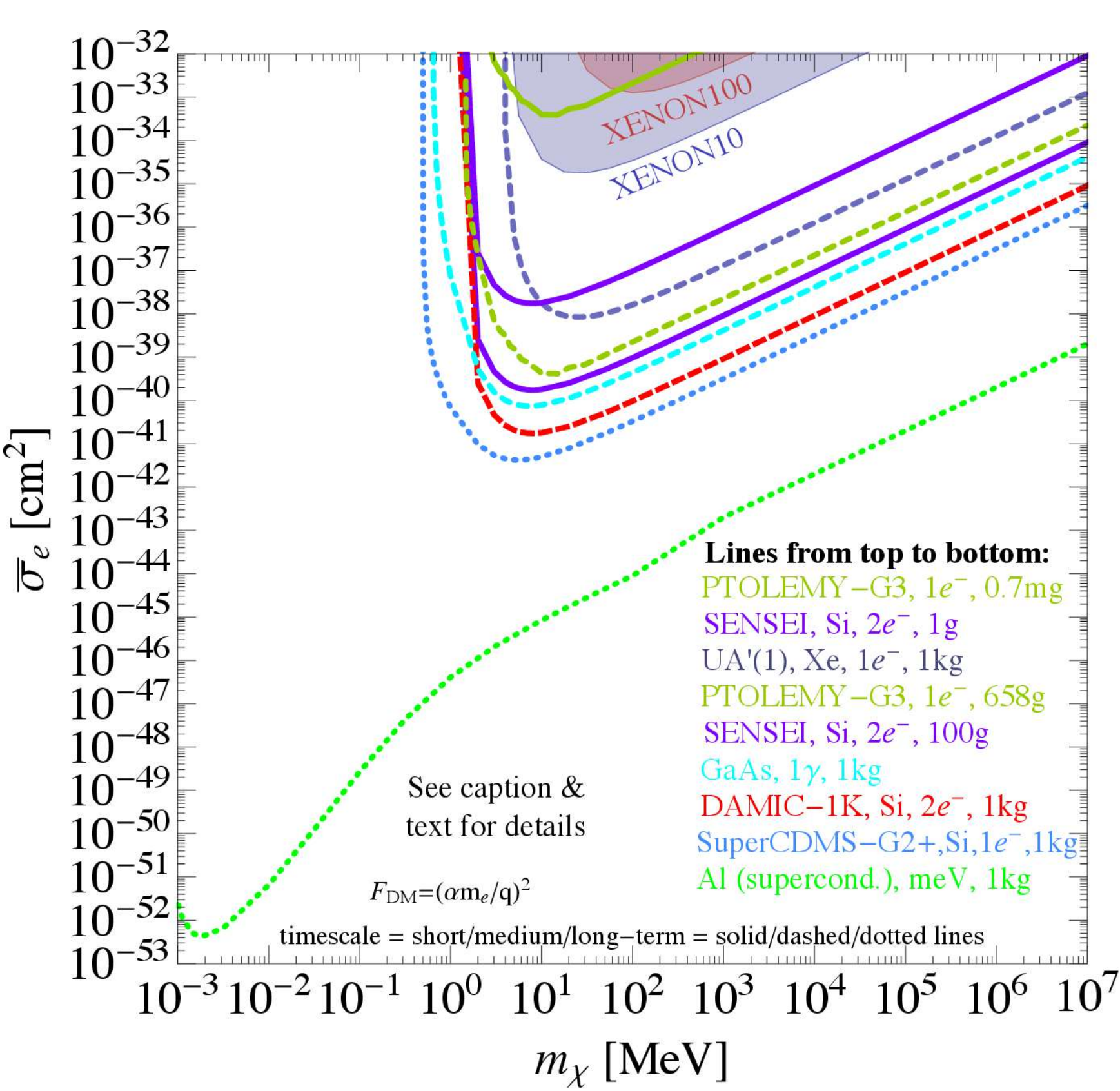} \\
\includegraphics[width=0.48\textwidth]{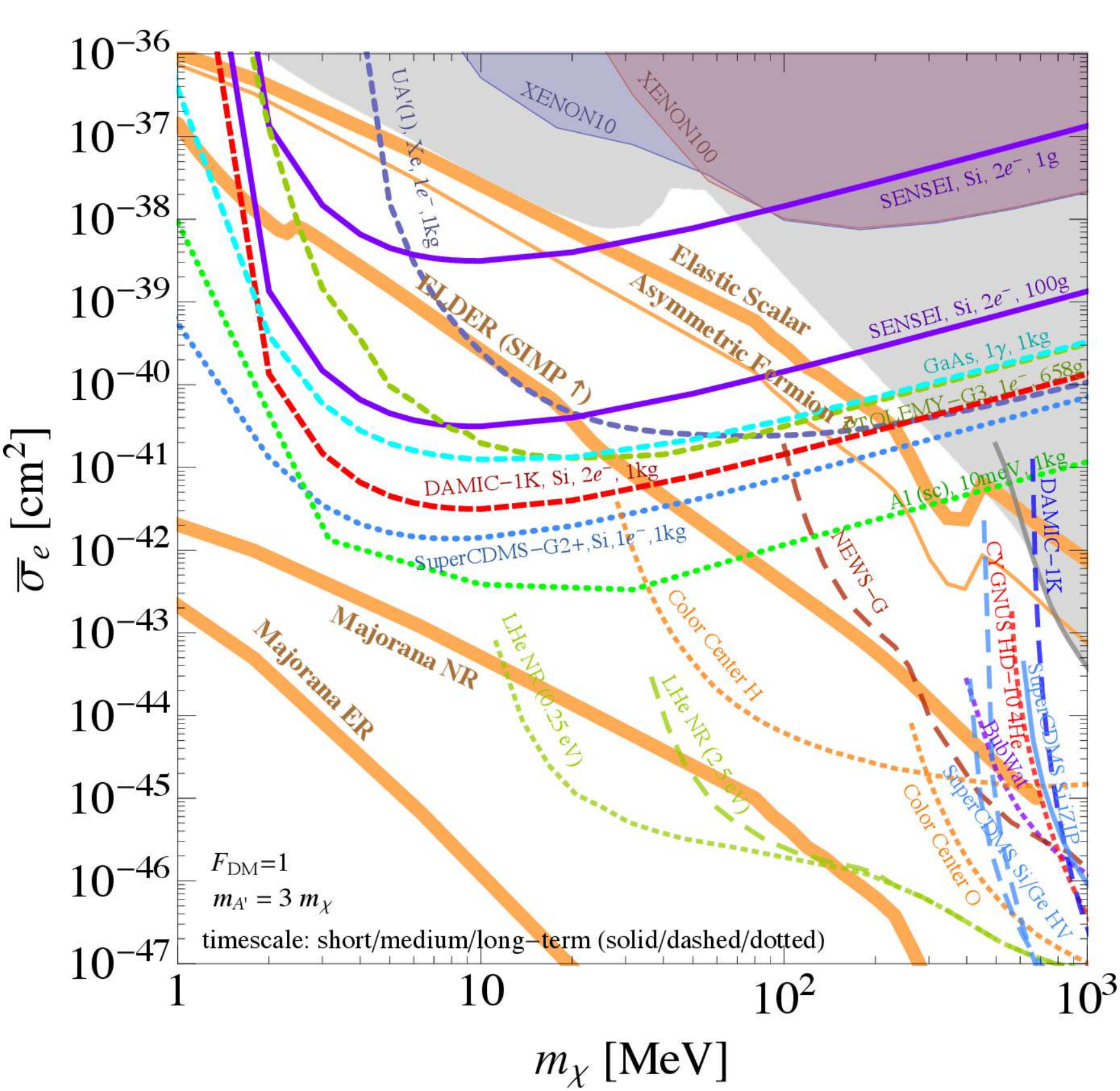}
~~
\includegraphics[width=0.48\textwidth]{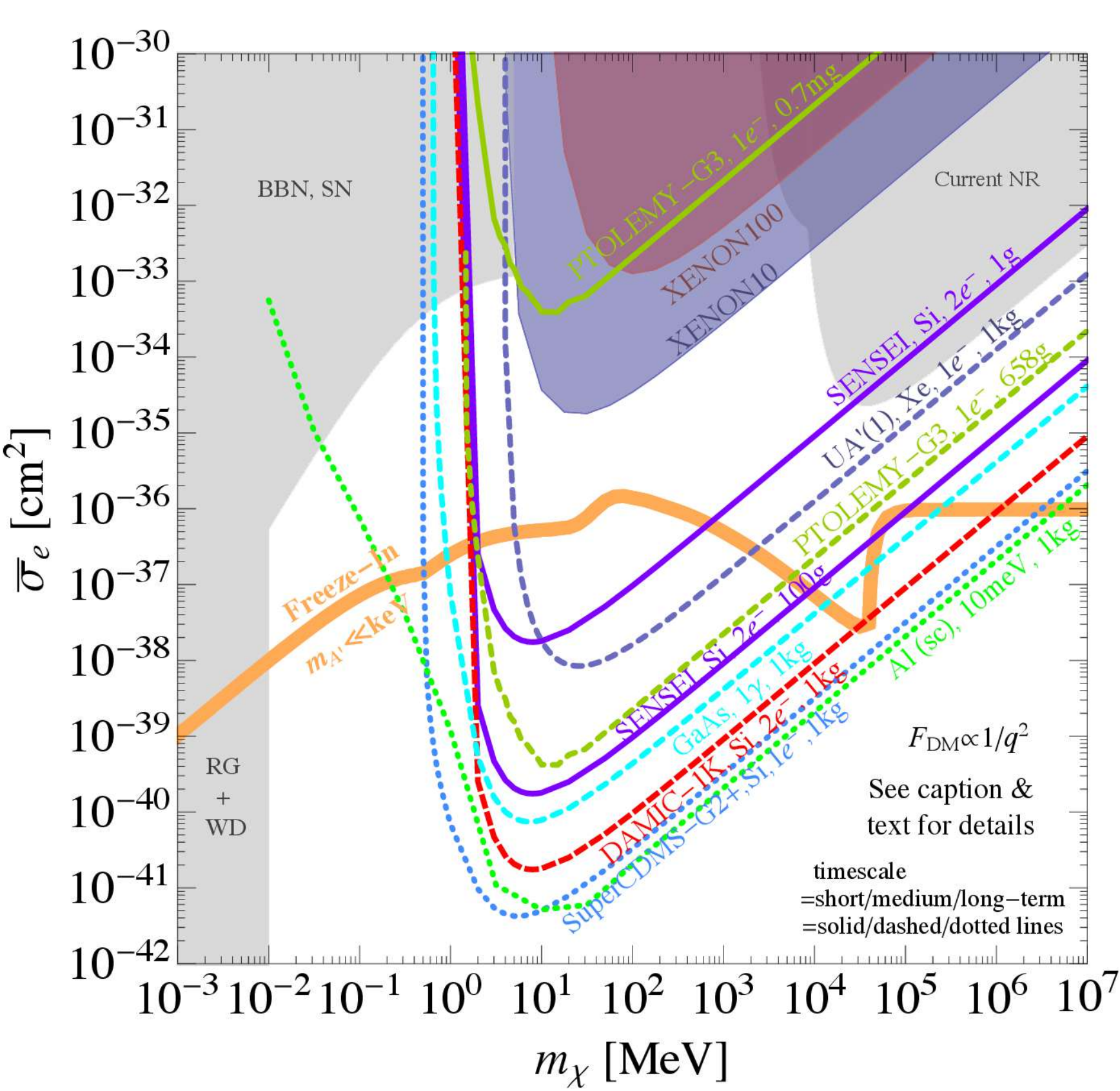}
\caption{
\footnotesize{Constraints and projections for the {\bf DM-electron scattering} cross section $\bar{\sigma_e}$. 
The left (right) plots assume a momentum-independent (dependent) interaction, $F_{\rm DM}=1$ ($F_{\rm DM}=(\alpha m_e/q)^2$).  
Existing constraints from XENON10 (XENON100)~\cite{Essig:2012yx,Essig:2017kqs} are shown in the blue (red) shaded regions. 
Projections show 3 events for a 1-year exposure~\cite{Essig:2015cda,Essig:2012yx,Hochberg:2016ntt,Hochberg:2015pha,Hochberg:2015fth,Derenzo:2016fse}; the label includes the threshold (in terms of number of electrons, photons, or the electron recoil energy)  
and target mass. 
Solid/dashed/dotted lines indicate an estimate of the time to start taking data, corresponding roughly to a 
short/medium/long timescale, respectively.  
A solid line indicates a mature technology: data taking can begin in $\lesssim 2$ years 
and a zero background (radioactivity or dark currents) is reasonable for the indicated thresholds.  
A dashed line indicates more R\&D is required and, if successful, data taking could start in $\sim 2-5$~years; 
the projected sensitivity assumes that backgrounds can be controlled. 
A dotted line indicates longer-term R\&D efforts.  
{\bf Bottom left} plot assumes {\bf DM scatters through an $A'$ with $m_{A'}= 3 m_\chi$}.  
Five theory targets are shown as explained in Section~\ref{subsec:science-DD}. 
In addition to electron-recoil experiments, we show projections from nuclear-recoil experiments (from Fig.~\ref{fig:scattering-NR}). 
Gray shaded regions are constraints from LSND, E137, BaBar, and current WIMP nuclear-recoil searches~\cite{Essig:2015cda}. 
{\bf Bottom right} plot assumes {\bf DM scatters through an $A'$ with $m_{A'} \ll\ $keV}; a freeze-in target is shown. 
Shaded gray regions are bounds from WIMP nuclear-recoil searches, stellar, and BBN 
constraints~\cite{Essig:2015cda}. 
The superconductor projection in bottom plots include in-medium effects for an $A'$ and assume a dynamic range of 10~meV--10~eV.}
}
\label{fig:scattering-ER}
\end{figure}

The following pages contain several representative figures, showcasing that orders of magnitude of new parameter space can 
be probed beyond existing constraints with first-generation small-scale experiments.  
{\it We emphasize that these plots are meant to be illustrative of the enormous parameter space 
that could be covered in the next few years by several small projects.  Not all theoretical ideas for experiments, and proposed 
experimental projects, appear on a plot.  Moreover, additional motivated DM candidates exist that are not represented with a plot in this 
white paper.} 
\begin{itemize}
\item Fig.~\ref{fig:scattering-ER}: Four plots show projections for DM scattering off an electron through a mediator 
with mass $\gg\ $keV ({\bf left two plots}) and mass $\ll\ $keV ({\bf right two plots}), 
leading to a momentum-independent scattering ($F_{\rm DM}(q)=1$) and momentum-dependent scattering ($F_{\rm DM}=(\alpha m_e/q)^2$), 
respectively. 
The {\bf bottom left} plot shows several model scenarios in which the scattering occurs through a dark photon with $m_{A'}= 3 m_\chi$.  
Five theory targets are shown, four of which can be probed by first-generation small-scale experiments sensitive to 
electron recoils.  
Additional projections from experiments sensitive to nuclear recoils are also included.  This assumes that the mediator interacts with 
both electrons {\it and} nuclei, as is the case for a dark photon mediator but not necessarily the case for other types of mediators. 
The nuclear-recoil projections have been converted to electron-recoil 
projections using 
\begin{equation}\label{eq:NR-to-ER}
\bar\sigma_e = 4 \ \frac{\mu_{\chi,e}^2}{\mu_{\chi,N}^2}\ \sigma_{\rm N}\,.
\end{equation}
We show two Majorana targets, since the Majorana target for DM-nucleus scattering differs by a factor of $\mu_{\chi,n}^2/\mu_{\chi,e}^2$ from the 
Majorana target for DM-electron scattering.  
It is exciting to note that new accelerator-based probes (not shown) can cover similar parameter space, and a detection of 
a new particle with both probes would provide compelling evidence for the properties of the DM particle. 
The {\bf bottom right} plot shows a model scenario in which the scattering occurs through an ultralight dark-photon mediator ($m_{A'}\ll\ $keV).  
A benchmark target (freeze-in) is shown, which can be probed by first-generation small-scale experiments sensitive to electron recoils.  
Direct-detection experiments are essential to probe this benchmark target, since they are sensitive to low momentum transfers (accelerator-based probes have large momentum transfers).  
Note that proposed nuclear-recoil searches, shown on bottom left plot, have sensitivity to this dark-photon mediator scenario; 
these projections are left to future work. 
\begin{figure}[t]
\includegraphics[width=0.49\textwidth]{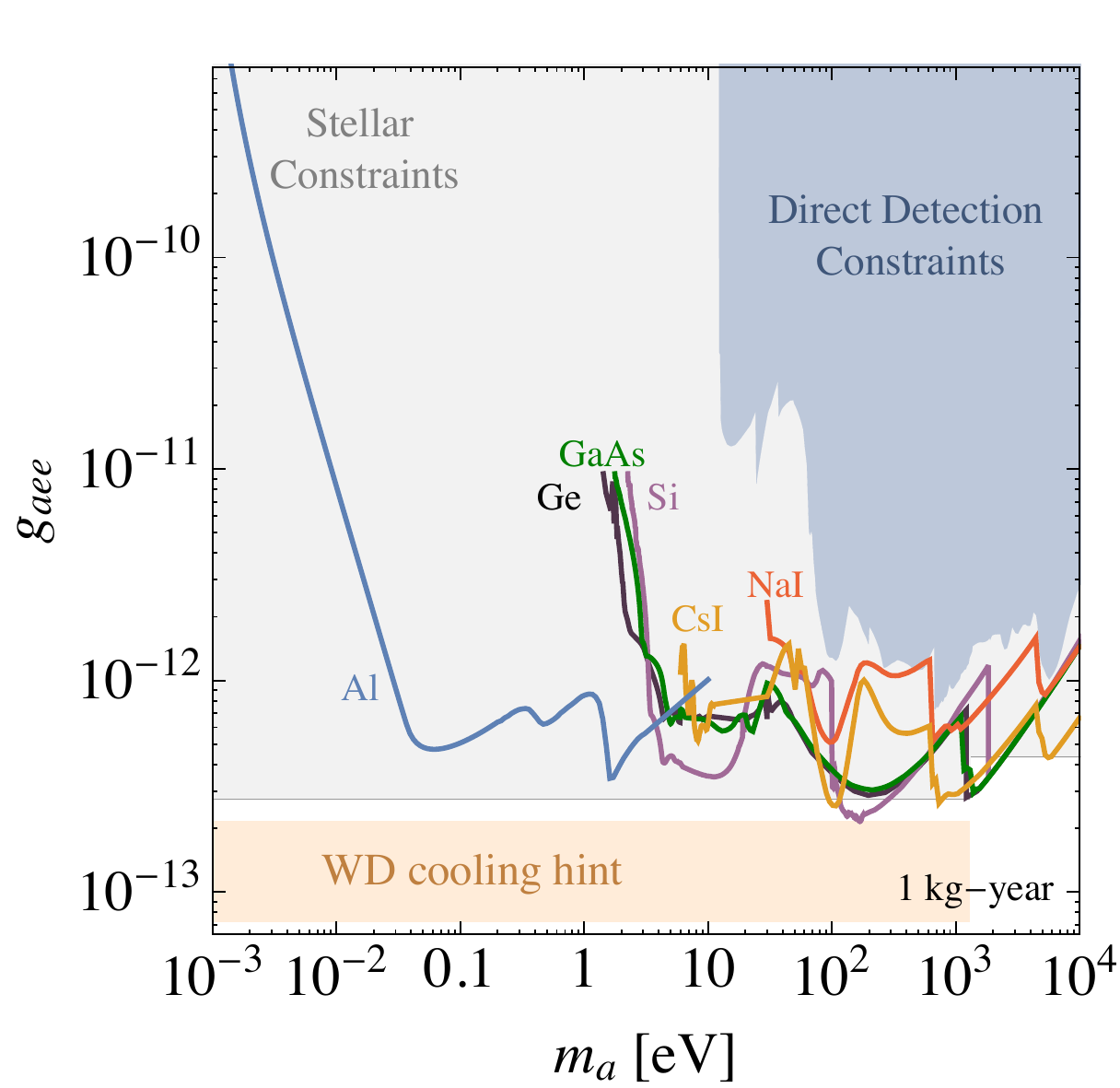}
\includegraphics[width=0.48\textwidth]{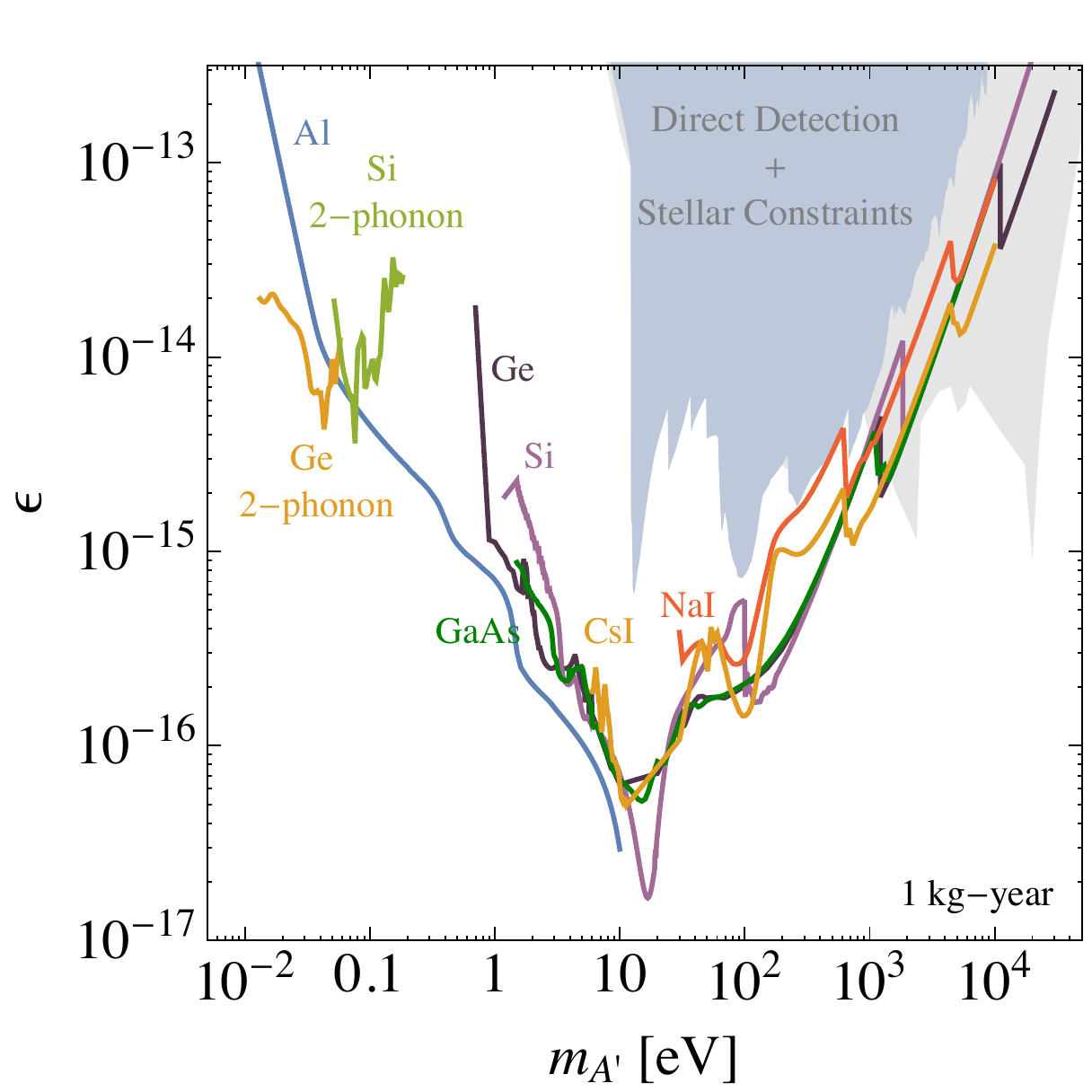}
\caption{
{\bf Event rates} for the {\bf absorption by an electron of axion-like particle (ALP) DM} ({\it left}) 
and {\bf dark-photon ($A'$) DM}  
({\it right}), assuming that the ALP/$A'$ constitutes all the DM~\cite{Bloch:2016sjj,Hochberg:2016sqx,Hochberg:2016ajh}.  
The {\it solid colored lines} show the ALP-electron coupling $g_{aee}$ or the kinetic-mixing parameter $\epsilon$ 
needed to produce 3 events for an exposure of 1 kg-year.  
{\it Blue regions} show constraints from WIMP direct-detection experiments 
\cite{Aprile:2014eoa,Armengaud:2013rta,Ahmed:2009ht,Aalseth:2008rx,An:2014twa,Bloch:2016sjj,Aguilar-Arevalo:2016zop}. 
Gray regions show stellar cooling constraints. 
In-medium effects are included for all $A'$ constraints and projections.   
Shaded orange region in left plot is consistent with an ALP possibly explaining the white dwarf luminosity function. 
}
\label{fig:absorption-WG1} 
\end{figure}
\begin{figure}[t!]
\includegraphics[width=0.48\textwidth]{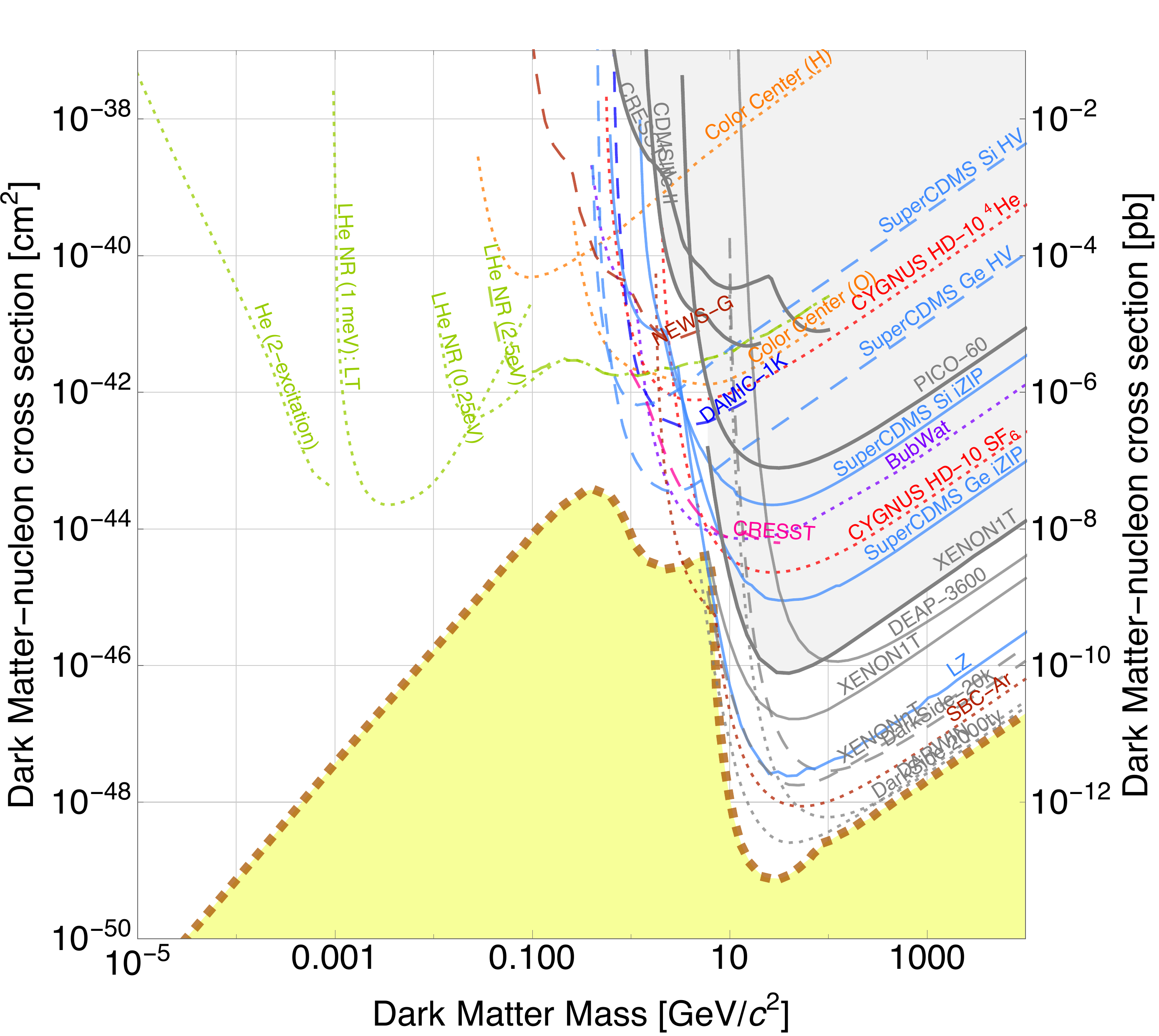}
~~\includegraphics[width=0.48\textwidth]{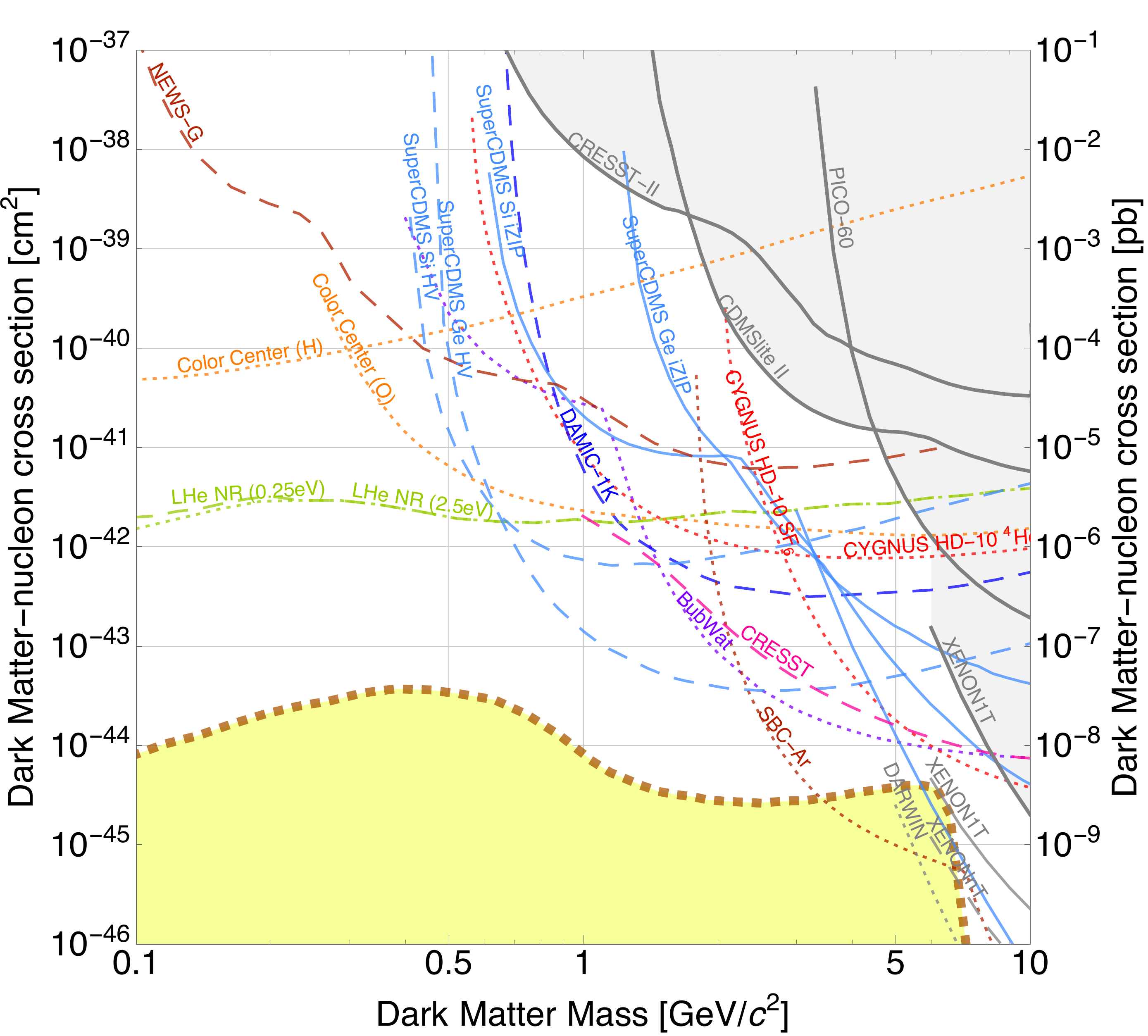}
\caption{
{\bf Left:} 
Constraints and projections (90\% c.l.) for the {\bf DM-nucleon scattering} cross section. 
Thick gray lines are current world-leading constraints~\cite{Aprile:2017iyp,Amole:2017dex,Agnese:2015nto,Angloher:2015ewa}. 
Projections are shown with solid/dashed/dotted lines indicating a short/medium/long timescale, respectively, 
with the same meaning as in Fig.~\ref{fig:scattering-ER}. 
Blue lines denote the DoE G2 experiment projections.  
Yellow region denotes the WIMP-discovery limit from~\cite{Ruppin:2014bra} 
extended to lower masses for He-based experiments.
{\bf Right:}
As in left plot, but focused on the 100~MeV to 10~GeV DM mass range.  
}
\label{fig:scattering-NR}
\end{figure}
\begin{figure}[b!]
\includegraphics[width=0.48\textwidth]{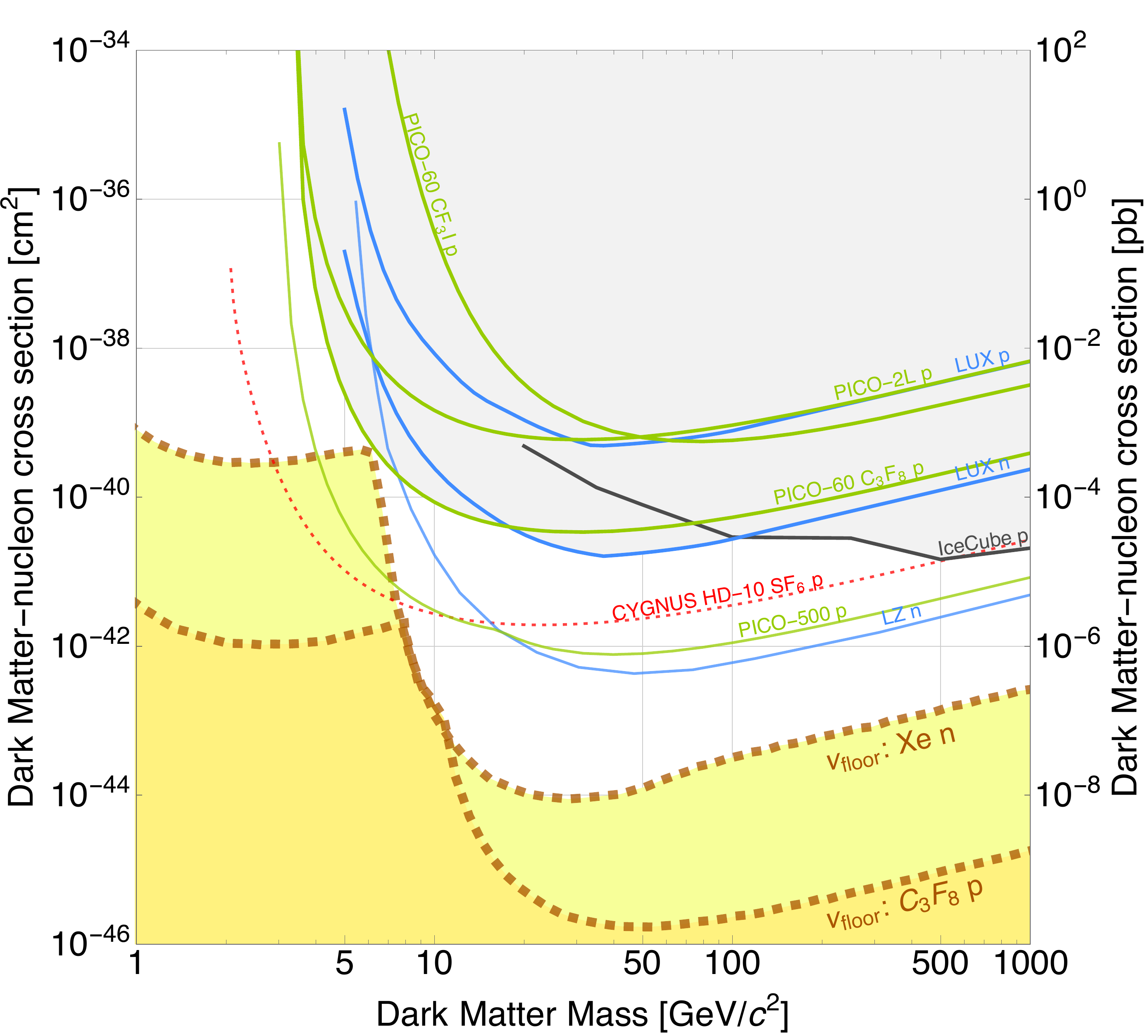}
\caption{
Constraints from direct-detection experiments (solid lines), colliders and indirect detection (labelled, dashed), and projections for new experiments (labelled, dashed/dotted lines) for the {\bf spin-dependent scattering cross section for protons or neutrons off nuclei}.  
Constraints are shown from PICO-60~\cite{Amole:2017dex}, LUX~\cite{Akerib:2017kat}, 
PICO-2L~\cite{PhysRevD.93.061101}, PICO-60 ${\mathrm{CF}}_{3}\mathrm{I}$~\cite{PhysRevD.93.052014}, 
and IceCube~\cite{Aartsen2017}.  Projections from PICO (proton) and LZ (neutron) are also shown~\cite{Akerib:2016lao}. 
The expected background from atmospheric, supernova and solar neutrinos in both xenon and C$_3$F$_8$ 
is shown by the shaded regions~\cite{Ruppin:2014bra}.
}
\label{fig:SD}
\end{figure}
\item Fig.~\ref{fig:absorption-WG1}: Event rates for an electron absorbing bosonic DM, such as an axion-like particle ({\bf left}) or 
a dark photon ({\bf right}). 
\item Fig.~\ref{fig:scattering-NR} shows projections for DM scattering off nuclei for a wide mass range ({\bf left}), 
and focused on the 100~MeV to 10~GeV mass range ({\bf right}).  
We note that the neutrino floor for low DM masses was calculated by assuming a liquid He-4 detector with 100\% recoil energy acceptance across the entire energy range, with coherent neutrino-nucleus scattering as the only background, and no nuisance parameters. Four combinations of exposure and energy threshold, which were chosen to represent an expected background rate of 40 events, were calculated and combined by choosing the lowest cross-section at each WIMP mass: (1 meV, 100 kg-yr), (90 eV, 350 kg-yr), (380 eV, 2000 kg-yr), (1500 eV, 3500 kg-yr).
\item Fig.~\ref{fig:SD}: Projections for DM scattering off nuclei through spin-dependent interactions.   
\end{itemize}

These figures contain various solid, dashed, or dotted lines, which show an estimate of the time to start taking data, 
corresponding approximately to a short, medium, and long timescale, respectively.  
A solid line indicates that the technology exists and that data taking has either already started or can begin in $\lesssim 2$ years.  
Moreover, the assumption of zero backgrounds (radioactive or dark currents) is reasonable for the indicated thresholds.  
A dashed line indicates experiments that require more R\&D, and if the R\&D is successful, data taking could start in $\sim 2-5$~years and 
potentially reach the sensitivity shown for the indicated target mass.  
A dotted line indicates longer-term R\&D efforts.  
We note that all exposures are approximate; experiments may run for more or less than 1-year, and may be deployed 
in stages with increasing target mass.

%%%%%%%%%%%%%%%%%%%%%%%%%%%%%%%%%%%%%%%%%%%%%%%%%%%%%%%%%%%%%%%%%%%%%
%%%%%%%%%%%%%%%%%%%%%%%%%%%%%%%%%%%%%%%%%%%%%%%%%%%%%%%%%%%%%%%%%%%%%
\subsection{Summary of Key Points}

\begin{itemize}
\item The direct detection of DM is a crucial experimental avenue to identify the nature of the DM particle. 
\item The direct-detection community is healthy and active, with several clear ideas to go beyond the funded G2 experiments. 
\item There are numerous science targets for searches for WIMPs and sub-GeV DM.  
These include {\it sharp targets} in parameter space from simple, predictive, and motivated DM candidates, as well as several general 
{\it regions of interest} in parameter space in which DM could hide.  In most cases, the proposed experimental ideas and experiments 
probe {\it several} sharp targets {\it and} general regions of interest.  
\item {\bf Several small projects, each with a cost of less than a few million dollars, can probe {\it orders of magnitude of new parameter space}, covering both sharp targets and general regions of interest for WIMPs as well as sub-GeV DM down to $\mathcal{O}$(MeV) masses, 
with project start-dates of FY19 and even earlier.}
\item Research and Development (R\&D) funding, in parallel to funds for small-scale projects, 
allows future projects to push below MeV masses and improve cross section sensitivities on a few-year timescale.  
\end{itemize}

Small-scale experiments with a few-million dollar price tag can explore vast areas of new parameter space beyond G2, since they use 
novel detection techniques and/or new target materials, and in many cases make use of advances in detector technology that 
allow for lower thresholds.  
Similarly to the Large Hadron Collider, which probed unexplored parameter space immediately when turning on due to its higher center-of-mass 
energy, new small-scale experiments can probe unexplored parameter space immediately when turning on due to lower thresholds.  
In this way, detectors with target masses as small as $\sim 1$~gram to $\sim 1$~kg can make enormous improvements over existing 
sensitivities in a parameter region complementary to that probed by the G2 experiments and for only a fraction of their cost.

\clearpage

%% file: WG2.tex
% !TEX root = WG2_wrapper.tex
\section{Detection of Ultra-Light (sub-milli-eV) Dark Matter}
\label{sec:WG2experiments}

\begin{figure}[h]
\begin{center}
    \includegraphics[width=0.9\textwidth]{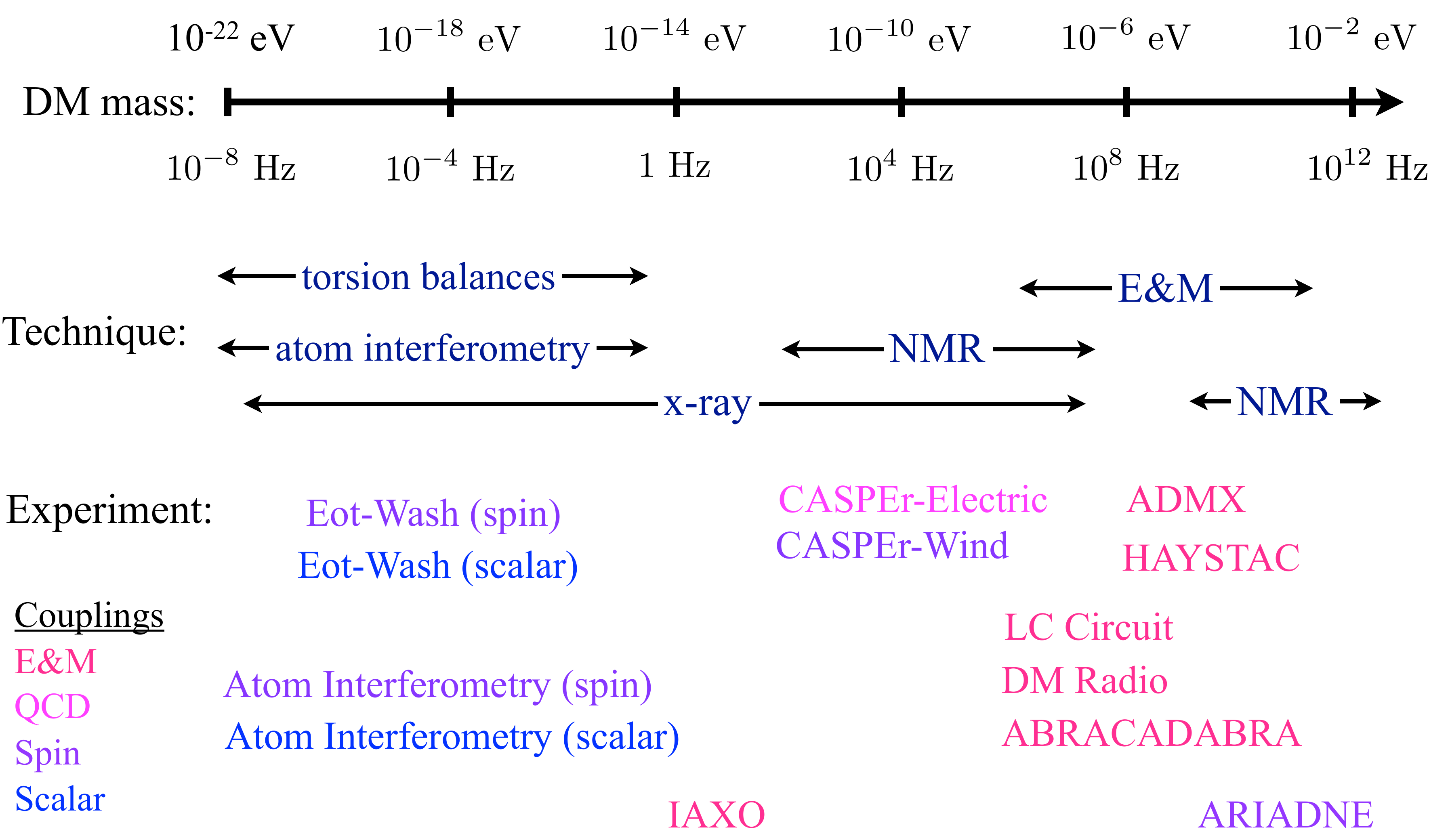}
    \caption{\label{fig:massrange} Mass range for ultralight dark matter.  Very rough optimal frequency ranges are shown for each experimental technique discussed in WG2.  Names of particular experiments and proposals discussed in this section are shown below their corresponding technique.  The names are color-coded by the DM coupling being searched for.  This is only meant as a cartoon --  for details of each experiment's sensitivity see the relevant discussion below.}    
\end{center}
\end{figure}

The axion and hidden photon are well-motivated dark matter candidates with models providing both viable production mechanisms and testable phenomenology.  To date, only a tiny fraction of the parameter space for such ultralight dark matter (as discussed in Section \ref{sec:Science-ultralight}) has been probed by existing experiments.  Excitingly, thanks to significant growth in interest in this area recently, there are now experiments or proposals which cover the entire viable mass range down to $10^{-22}$ eV.  These experiments are highly complementary in their mass reach as well as coupling type;  together they search for all four different possible types of couplings the dark matter can have (discussed in Section \ref{sec:Science-ultralight}).  Figure \ref{fig:massrange} is a rough cartoon of the complementary nature of these experiments, both in mass and coupling.  In particular, it now seems likely that a combination of these experiments can reach sensitivity to the QCD axion over a broad range of axion masses.

Searches for dark matter in this mass range use techniques which are very different than those used in traditional particle physics experiments.  In this range the dark matter can more usefully be thought of as a field (or wave) oscillating at a frequency equal to its mass.  Unlike a traditional particle detector (e.g. WIMP detection experiments) which looks for the energy deposited by a single hard collision, detectors searching for such light dark matter must look for the collective effect of all the dark matter particles in the wave.  This is analogous to gravitational wave detectors which search not for individual graviton scattering but for the semiclassical effect induced by the entire wave.  Thus, experiments searching for low mass bosonic dark matter utilize high precision sensors of continuous wave signals as opposed to the traditional impulse detectors used for single particle scattering.  Such sensors come from many areas of physics including condensed matter and atomic physics and are based on a wide range of techniques such as high-precision magnetometry, nuclear magnetic resonance (NMR), electromagnetic resonators, atomic clocks, and laser interferometry.

Indeed, the ability to reach pristine sensitivity to very weakly coupled bosonic dark matter with low cost experiments relies on the cross-disciplinary transfer of detector technology originally developed for other applications.  For example, some of these dark matter experiments rely on the high precision now achievable in magnetometers or atomic clocks developed by the quantum electronics and atomic physics communities.  Looking beyond the quick gains obtained from the initial technology transfer, the new dark matter application also provides vital and immediate motivation for further improvement of the sensitivities of these detector technologies which would also benefit science beyond just dark matter.    For example, torsion balances are also one of the best ways to search for new forces and equivalence principle violation.  Atomic interferometers allow searches for gravitational waves as well as sensitive equivalence principle tests and measurements of the fine structure constant, and have practical applications in geological mapping and inertial navigation.  Several of these technologies also have connections with work in quantum information and may be valuable for both fields.  As the above examples illustrate, improvement of these continuous wave detector technologies would have many broader impacts beyond just the bosonic dark matter search and cross-disciplinary collaborations should be encouraged.

%And in addition to searching for ultralight dark matter, networks of these sensors could also search for ultra heavy dark matter candidates through the correlated effects from long-range fields sourced by them.

The new direct dark matter detection projects discussed below are either already operating or constructing pathfinder experiments or in an advanced stage of hardware prototyping.  Future support would enable construction of full-sized detectors with signifiant reach into uncovered dark matter parameter space.  The timeline for these future experiments is around a few years in most cases.  Roughly ordered from high mass coverage to low mass coverage, these experiments include:
\begin{enumerate}
\item ADMX
\item HAYSTAC
\item LC Circuit
\item DM Radio
\item ABRACADABRA
\item CASPEr-electric
\item Torsion Balances
\item Atom interferometry
\end{enumerate}
Note that while ADMX is in fact a current DOE Generation 2 dark matter project, future funding would enable upgrades to the existing project which would expand the axion frequency range covered.  The first five experiments are electromagnetic detectors, with the first two using cavity resonators and the next three using lumped element (`LC circuit') resonators.  The first five experiments are all designed to reach sensitivity to the QCD axion as well as to simultaneously provide world-leading sensitivity to hidden photons.  CASPEr-electric searches specifically for the model-defining axion-gluon coupling, and the last two techniques are sensitive to ultralight scalar dark matter. 

There are also two experiments which are not directly searching for dark matter, but can cover interesting axion dark matter parameter space:
\begin{enumerate}
\item Mini-IAXO
\item ARIADNE
\end{enumerate}
And then additionally there are several areas of R\&D work which would enable significant future dark matter experiments.  These include work on high frequency electromagnetic resonators to allow detectors to push above cavity frequencies, high field magnet development for many axion experiments, and a full-scale IAXO project.

\subsection{Dark Matter Direct Detection Experiments}

\subsubsection{ADMX}

\noindent {\bf Ongoing search for QCD axions from 500~MHz-2~GHz and 2-10~GHz extension:}
The second generation Axion Dark Matter Experiment aims to discover or exclude QCD axion dark matter with sensitivity even to the most pessimistic axion-photon couplings predicted by the Dine-Fischler-Srednicki-Zhitnitsky (DFSZ) models at axion masses in the well-motivated 500~MHz-10~GHz range.  ADMX is an axion haloscope experiment that relies on dark matter axions converting into microwave photons in a strong magnetic field \cite{Sikivie:1983ip}.  This conversion is enhanced by a tuned high-Q cavity resonator, and the resulting photons are detected in a radiofrequency receiver.  ADMX at present uses a 50 cm bore 7.6 T magnet containing a single cylindrical microwave cavity, frequency-tuned by moving rods from the edge to the center of the cavity.  Previous versions of ADMX have already demonstrated sensitivity to the optimistically-coupled KSVZ axions.  Key to the sensitivity of ADMX G2 is the low thermal background obtained with a dilution refrigerator that cools the experiment to below 150 mK, and the quantum-limited SQUID and Josephson parametric amplifiers that are the first stage of the receiver chain.  At higher frequencies, the single cavity will be replaced with multiple power-combined cavities with higher resonant frequencies, but tuned in a similar manner.

Sited at the University of Washington, ADMX G2 is an approved and funded DOE Generation 2 project for its first stage operations which will cover axion masses up to 2~GHz.  The experimental collaboration includes around 30 members in 10 institutions.  The experiment is currently taking data with DFSZ sensitivity in the 600~MHz range.  The hardware to explore up to 1~GHz is constructed, and the hardware to reach 2~GHz is currently being built.  Operations in the 500~MHz-1 GHz will be completed in 2017, with 1-2 GHz covered the following year.  Designs are being prepared for the 2-10~GHz resonators and a proposal to continue the project in this higher frequency range will be submitted shortly.  Covering the entire frequency range under the current strategy is estimated to take 6 years.
 
 \bigskip
 
\noindent {\bf Longer-term R\&D for axion mass \textgreater 10~GHz:}
A subset of ADMX collaborators are engaged in longer term detector R\&D to enable dark matter axion searches at even higher masses.  This work includes high frequency resonators being developed at the University of Washington and LLNL, and novel single photon detectors being developed at FNAL and LLNL.
    The practice of cavity resonators with size matched to the photon wavelength will work up to 10 GHz, but at higher frequencies a number of issues need to be addressed.  Because of the smaller size of the individual cavities, the large number of combined resonators required to maintain large detection volume becomes intractable at higher frequencies.  Furthermore, the quality factor the high frequency cavities decreases, and  the quantum-limited noise of microwave amplifiers increases with frequency.  The ADMX R\&D addresses these issues with two separate thrusts.  The first is to develop sophisticated, tunable multiwavelength resonators that have good coupling to axions while maintaining  large volume and high Q.  One example of these is an open Fabry-Perot type resonator with strategically placed dielectrics to allow good coupling to the axion.   The second thrust is to develop single microwave photon detectors that evade the quantum-limited measurement noise by putting the backreaction into the unobserved phase quadrature.  The most promising direction for this thrust is to use the single photon manipulation hardware developed by the quantum computing community.  
    
The strategy is to develop the cavity and detector technologies separately in prototype experiments that probe previously unexplored axion-photon couplings, and when mature combine them to build an experiment sensitive to the weaker couplings predicted by QCD axion models.  Prototype tunable resonators have been built and operated at room temperature, and work is being done to developed higher-Q cryogenic prototypes.  Hardware for single frequency photon counters has been constructed and a prototype is under construction.  This R\&D is primarily funded by the Heising-Simons foundation.  In order to continue the US axion program at these frequencies, the right timescale to transform this R\&D into a full experimental proposal is 6 years, so it coincides with the end of the ADMX G2 program.

\subsubsection{HAYSTAC}

Another cavity-based haloscope experiment, the HAYSTAC (Haloscope At Yale Sensitive To Axion CDM) axion search \cite{Sikivie:1983ip}, supported by the NSF and the Heising-Simons Foundation is both an innovation works and a data pathfinder in the 2.5-12 GHz (~10-50 $\mu$eV) mass range.  A collaboration of Yale (S. Lamoreaux, PI), UC Berkeley (K. van Bibber, PI) and University of Colorado (K. Lehnert, PI) began design work in 2011, commissioned the experiment in 2015, and transitioned to data-taking in 2016.  First science results have recently been published \cite{Brubaker:2016ktl}, and has an instrumentation paper \cite{Kenany:2016tta}.  

%Thus the collaboration has established its bona fides to deliver results on spec, on budget, and on schedule.

The experiment as currently configured utilizes a superconducting 9 T magnet, a dilution refrigerator, and Josephson Parametric Amplifiers (JPA); the first experiment to do so.  Also for the first time, the experiment has achieved an operational system noise temperature only a factor of 2 above the Standard Quantum Limit, providing the best limits to date on the axion-photon coupling at these higher frequencies with a cavity volume of only 1.5 L.   

In the coming year, HAYSTAC will be validating in operations two critical technologies:  (i) A squeezed-state receiver developed at Colorado, which by evading the Standard Quantum Limit will dramatically improve the sensitivity and scan speed of the experiment.  The switch-over to this system will occur in Summer 2017.  (ii) A new cavity design that will enable the use of high harmonic TM$_{0n0}$ modes for higher frequency searches, and with the spectrum cleansed of interfering TE modes.  This will occur in early 2018.  Should both these technologies meet spec in operation, HAYSTAC will have validated a microwave cavity experiment as a full system with sensitivity between KSVZ and DFSZ models, up to 50 $\mu$eV in mass.  It should be noted that this development and demonstration requires no new resources; it is already funded and in progress.

An ultimate experiment that could reach up to 100 $\mu$eV, or beyond, with sensitivity better than DFSZ would require some additional R\&D for development of a resonator based on Photonic Band Gap (PBG) concepts and metamaterials; this work has already begun at Berkeley.   It will also require development of new JPA geometries to be developed at Colorado.  To integrate and commission an operational experiment based on these capabilities, Yale will alsoneed to procure a higher field Nb$_3$Sn magnet of order 10 L volume.  Nonetheless, including student/postdoc support during the R\&D phase, this project would fall into the small experiments category.   If the R\&D to purge the spectrum of TE mode-crossings at all frequencies is fully successful, the run time to cover up to 25 GHz (100 $\mu$eV) could be very short, of order 5 years.

\subsubsection{LC Circuit}

Another method to search for low-mass dark-matter axions is by using a lumped element $LC$ resonator instead of a cavity resonator in the strong magnet~\cite{sikivie2014proposal}. The premise of the method is that the axion field alters Maxwell's equations; in the presence of an external magnetic field $B_0$ there is an effective current parallel to the external field, $\vec j = -g\vec B_0{\partial a \over \partial t}$ where $a$ is the axion field and $g$ the axion coupling to two photons. The current oscillates at a frequency $\omega = (m_a c^2 +K)/\hbar$ where $m_a$ is the axion mass and $K\approx m_a c^2/10^6$ is the kinetic energy due to the axion orbital motion in our Galaxy. The current $\vec j$ produces by Ampere's law an AC magnetic field $\vec B_a$. A loop antenna occupying half the magnet bore with its plane perpendicular to $\vec B_a$ will have an emf induced in it by the time-varying flux associated with the field. 
As in the cavity detector exemplified by ADMX, , there is an enhancement in the antenna circuit by making it resonant at frequency $\omega$. The circuit is tuned by a variable capacitor and the output of the $LC$  circuit is brought to a low noise amplifier, mixed to audio frequencies, and detected. 

\begin{figure}[h!]
  \centering
  \includegraphics[width=0.9\textwidth]{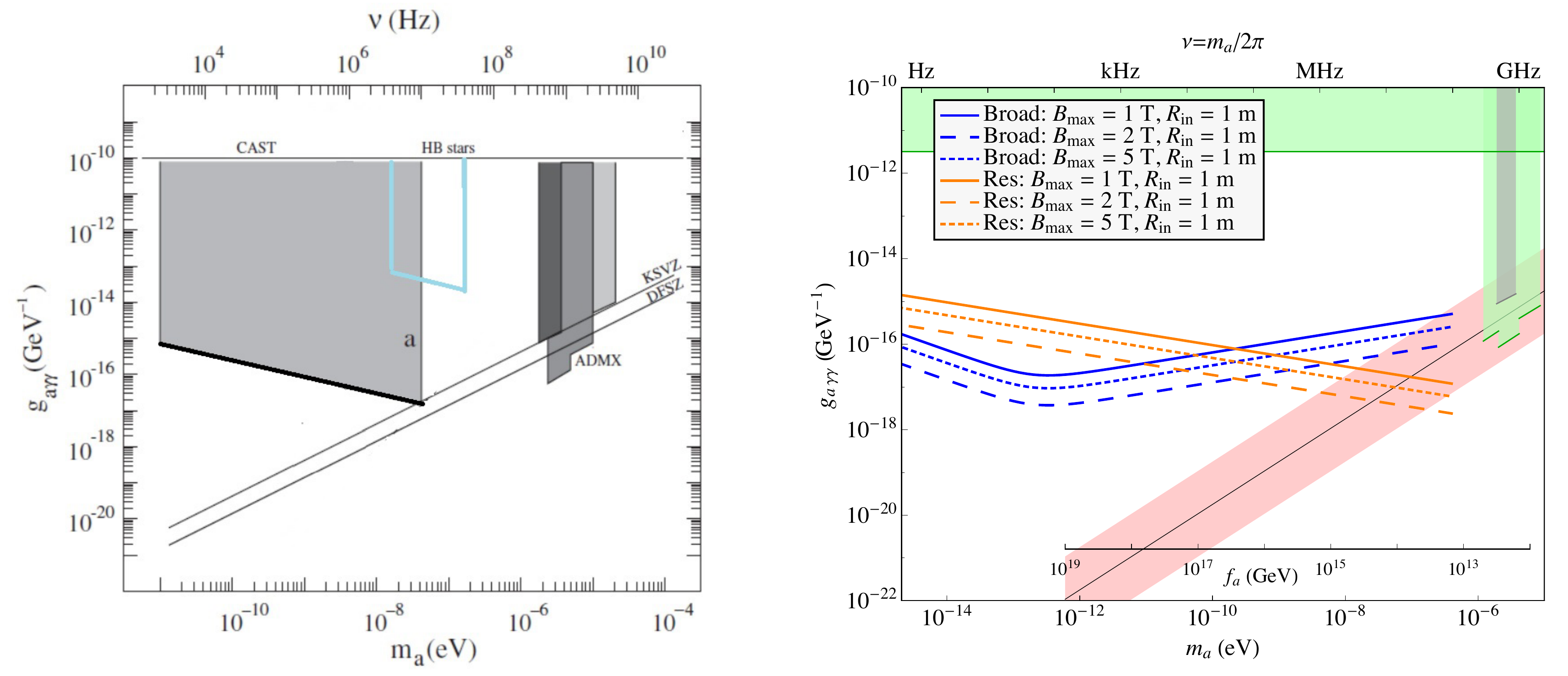}
  \caption{\label{fig:LCCircuit} (Left) Sensitivity of the $LC$ circuit in the ADMX solenoid magnet.  (Right) The SNR=1 sensitivity curves in the axion-photon coupling $g_{a\gamma\gamma}$ vs axion mass
    $m_a$ for ABRACADABRA in broadband only and resonant only readout
    modes after 1 year of data taking. The assumed aspect ratio for the toroidal magnet is
    $R_{\rm in}=\frac12R_{\rm out}=\frac13 h$. The QCD axion region is
    in shaded red.}
\end{figure}

The $LC$ circuit is sensitive to QCD axions and also to low-mass axion-like particles~\cite{sikivie2014proposal}. Using the ADMX magnet and operated at milliKelvin temperatures, the circuit would be able to scan the $g_{a\gamma\gamma}$ vs. $m_a$ region shown in grey in Figure \ref{fig:LCCircuit}.  Because this search involves a reuse of the ADMX magnet, the cost is not large. Capital costs are estimated to be under \$1M, in order to add another stage of cooling; the experiment should be operated at 1 mK or better, reachable with adiabatic demagnetization.   Operating costs will be similar to the operating costs of ADMX.

The University of Florida is planning a ``pilot experiment,'' building an LC circuit detector as a PhD student thesis project. It will employ a NbTi loop in series with parallel plate capacitor. The target $Q$ is 10,000 and a goal of the pilot experiment is to investigate the challenges of achieving this performance. It will use 8 T magnet (15 cm diameter, 45 cm length). The magnet volume is about 4\% of the ADMX magnet. We will scan 12 to 100 neV (3-30 MHz). The sensitivity goal of the pilot experiment, with loop at 0.4 K, is shown as the blue-outlined region in Figure \ref{fig:LCCircuit}.

%\begin{figure}
%\includegraphics[width=2.5in]{figs/LCsensitivitypilot}
%%  \epsfig{file=figs/LCsensitivitypilot.eps,width=3.4in}
%  \caption{\label{fig:loop1} Sensitivity of the $LC$ circuit in the ADMX magnet.}
%\end{figure}

\subsubsection{DM Radio}
\begin{figure}
\begin{center}
\includegraphics[width=3.1in]{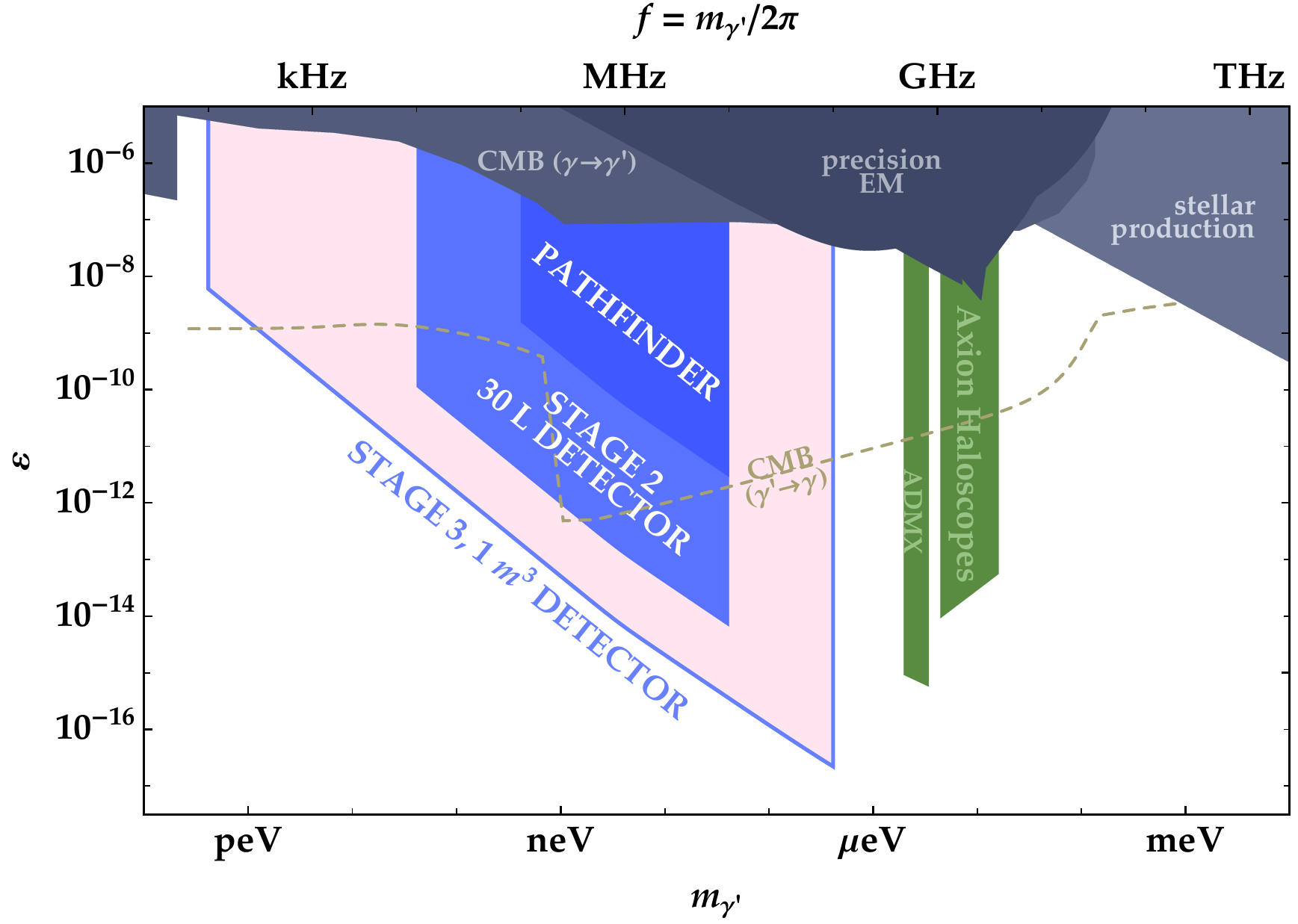} \ \ \ \includegraphics[width=3.1in]{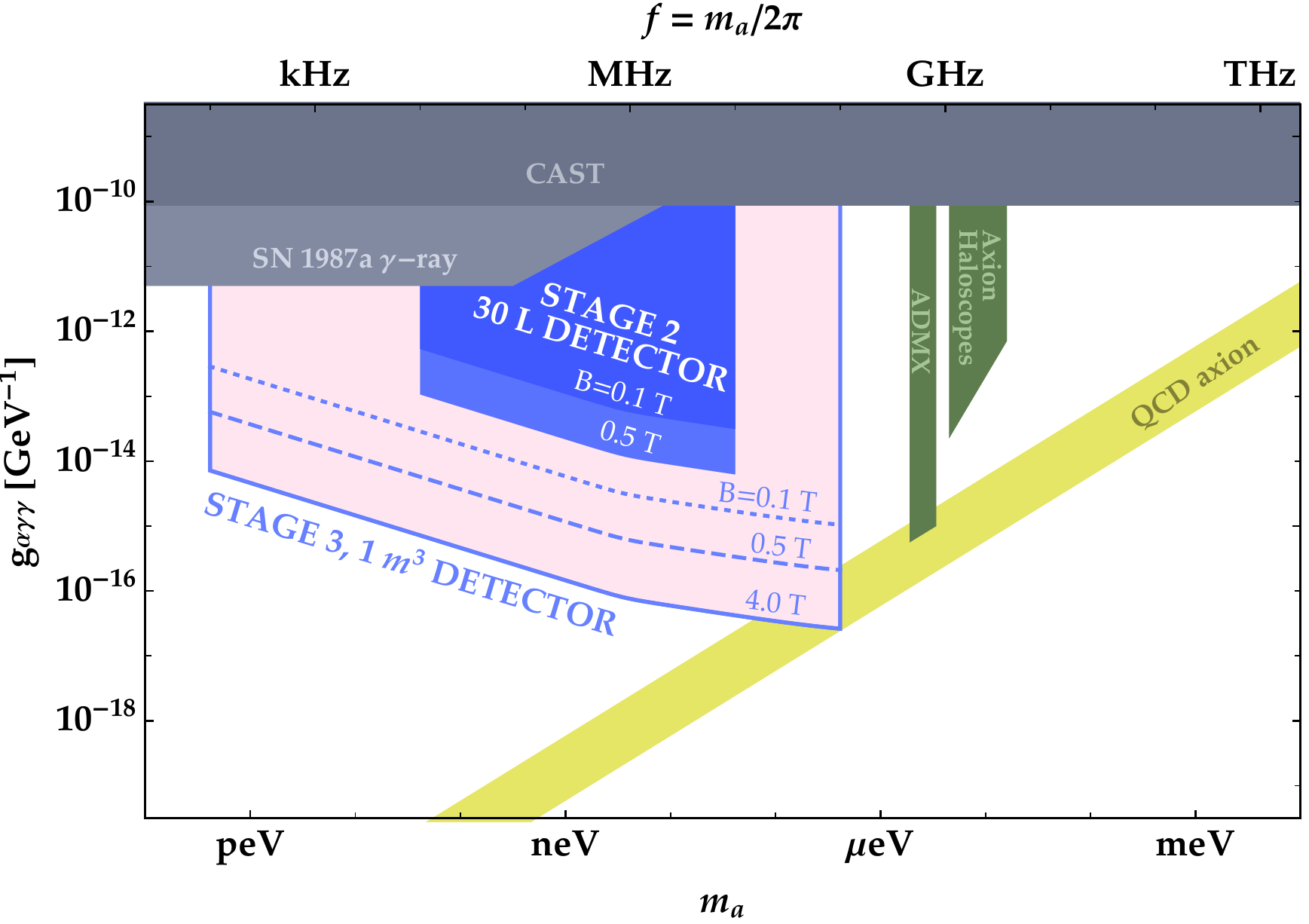}
\caption{\label{fig:dmradioscience}Left: DM Radio sensitivity to hidden photons: upper-level constraints on photon-hidden-photon coupling $\varepsilon$ as a function of hidden photon mass $m_{\gamma'}$ Right: DM Radio sensitivity to axions: upper-level constraints on axion-photon coupling $g_{\alpha\gamma\gamma}$ as a function of axion mass $m_a$ and applied DC magnetic field $B$.} 
\end{center}
\end{figure}

Dark Matter Radio (DM Radio) is a well-shielded, tunable lumped-element LC resonator for the detection of sub-eV hidden photon and axion dark matter\cite{dmradio2015prd}. A superconducting shield surrounding the resonator blocks ordinary electromagnetic signals, but is easily penetrated by hidden photons and axions. Hidden photons/axions generate an effective background current density within the shield that couples to the inductor. If the resonator is matched to the frequency (mass) of the hidden photon/axion field, it will ring up and generate a measurable voltage that may be sensed by a SQUID amplifier. The expected reach of the liter-scale DM Radio Pathfinder detector, a 30L Stage 2 detector, and 1m$^3$ Stage 3 detector are shown in Fig.~\ref{fig:dmradioscience}. The Pathfinder experiment is currently under construction at Stanford University and is expected to begin taking data Summer 2017, funded through the SLAC LDRD program\cite{silva2017design}. The development of the $\sim$\$1.3M Stage 2 experiment has received initial support from the Heising-Simons Foundation. A multi-site, multi-orientation implementation of the Stage 3 experiment would be within the purview of a new small experiments program.  The $\sim$15-person collaboration is a mixture of scientists from SLAC, Stanford, KIPAC, UC Berkeley, UC Davis, and Princeton.   

%\begin{figure}[h!]
%  \centering
%  \includegraphics[width=.4\textwidth]{figs/ABRASensitivity.pdf}
%  \caption{The SNR=1 sensitivity curves in $g_{a\gamma\gamma}$ vs
%    $m_a$ for ABRACADABRA in broadband only and resonant only readout
%    modes after 1 year of data taking. The assumed aspect ratio is
%    $R_{\rm in}=\frac12R_{\rm out}=\frac13 h$. The QCD axion region is
%    in shaded red.}
%\end{figure}

\subsubsection{ABRACADABRA}
In contrast, ABRACADABRA is a 1\,m scale {\it broadband} axion search designed to search
for axion and axion-like dark matter over the mass range
$10^{-12}\lesssim m_a\lesssim 10^{-6}$\,eV. The detector itself
consists of a toroidal magnet with a large magnetic field (of order
Tesla), with a superconducing pickup loop inside which is readout by a
SQUID current sensor.  The mass range to which ABRACADABRA is
sensitive corresponds to the frequency range
$1\lesssim 2\pi/m_a\lesssim 1\times10^8$\,Hz. In this range, the axion
wavelength is very long compared to the size of the detector and the
oscillating axion DM field generates \emph{effective currents} in the
toroid which in turn generate an oscillating magnetic field through
the center of the toroid. A sensitive magnetometer should be able to
detect this oscillating field \cite{Kahn:2016aff}.

The integrated flux through the center of the toroid is given by
\begin{equation}
  \Phi_a(t)=g_{a\gamma\gamma}B_{\rm max}\sqrt{2\rho_{\rm DM}}\cos(m_at)\mathcal{G}_V V.
\end{equation}
Where $B_{\rm max}$ is the maximum field in the toroid, $V$ is its
volume, $\mathcal{G}_V$ is a geometric factor that depends on the
aspect ratio and is typically $\sim$0.1.
% I assume that rho_DM and g_agg are defined elsewhere in the text.
The key advantage of this approach is that the field in the center
region of the toroid should ordinarily be zero. So with sufficient
shielding, ABRACADABRA will be searching for a small signal on top of
a nearly zero background.

Depending on the geometry, ABRACADABRA could be sensitive to the QCD
axion regime within a few years of continuous running. A $\sim$10\,cm prototype is being built at MIT and is expected to have data before the end of
2017. This prototype will not only give a better idea of the
challenges facing a larger 1\,m version, but will itself be sensitive
to unexplored regions of parameter space after only 1 month of data
taking.

\subsubsection{CASPEr}

The Cosmic Axion Spin Precession Experiment (CASPEr) searches for axion and axion-like dark matter via its interaction with nuclear spins. CASPEr-electric, the US-based experiment located at Boston University, searches for the axion-gluon coupling which gives rise to an oscillating nucleon electric dipole moment. The collaboration consists of two experimental groups and two theory groups. The primary physics goal is to reach experimental sensitivity that allows searching for axion and axion-like dark matter with couplings at the QCD axion level in a wide range of axion masses: approximately $10^{-12}$~eV to $10^{-6}$~eV, corresponding to frequencies $\sim$200~Hz to 200~MHz~\cite{Budker:2013hfa}.

%------------------------------------------------------------------ FIG
\begin{figure}[h!]
\begin{center}
    \includegraphics[height=6cm]{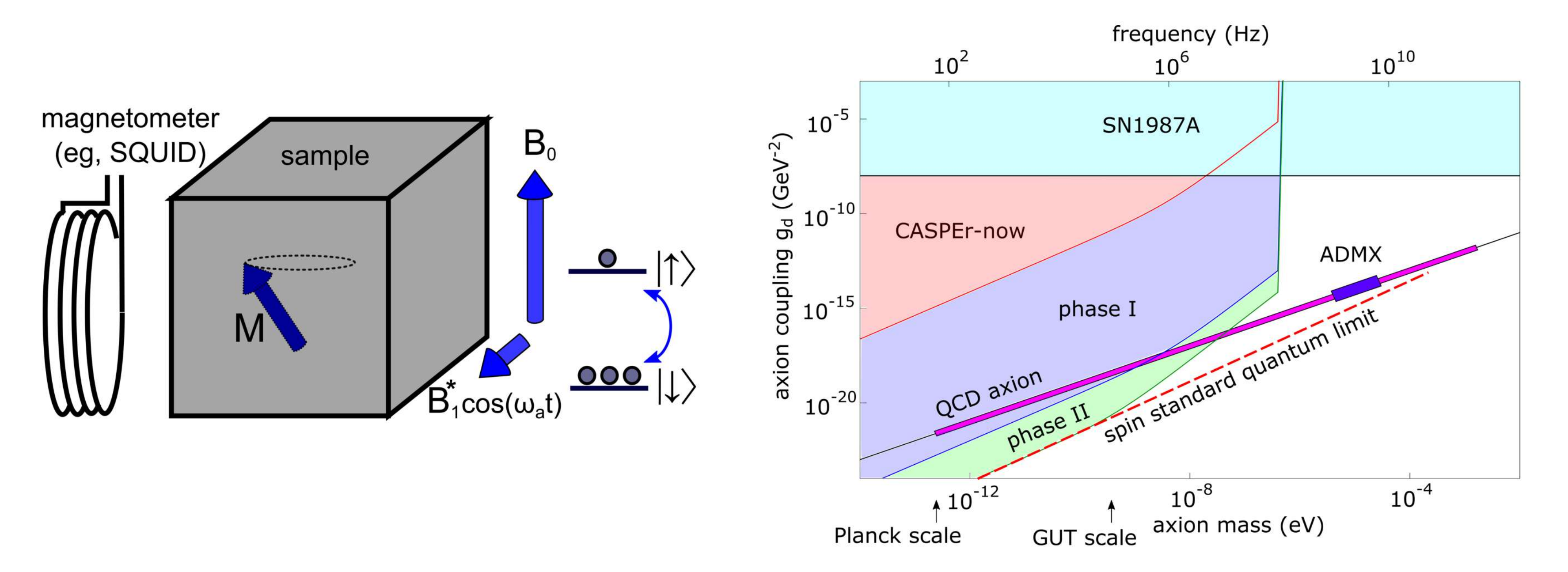}
    \caption{(Left) CASPEr experimental scheme for measuring an AC nucleon electric dipole moment $d_N$.  (Right) CASPEr projected sensitivity to the EDM coupling $g_d$ where $d_N = g_d \cdot a$ and $a$ is the local amplitude of the coherently oscillating dark matter wave.}
    \label{fig:CASPEr}
\end{center}
\end{figure}
%------------------------------------------------------------------ FIG

%\begin{wrapfigure}{r}{6.3cm}
%\includegraphics[width=6.3cm]{figs/CASPErSetup}
%\caption{\setstretch{0.8} \fontsize{9}{12}\selectfont
%CASPEr experimental scheme.
%}
%\label{fig:setup}
%\end{wrapfigure}
The experimental approach utilizes the existing technology of magnetic resonance and precision sensors. Briefly, the nuclear spins ($I=1/2$) in a solid ferroelectric sample experience an oscillating torque due to interaction with the axion dark matter field. If the constant bias magnetic field $B_0$ is such that the frequency of this torque (ie axion Compton frequency $\omega_a$) matches the nuclear spin Larmor frequency, the spins tilt and undergo Larmor precession, creating a transverse oscillating magnetization, detected by a sensitive magnetometer, such as a SQUID, see fig.~\ref{fig:CASPEr}. The search strategy is to sweep the value of the bias magnetic field, thus sweeping through a range of axion masses.

The pathfinder experiment (CASPEr-now) is currently under construction, this will prove feasibility and place limits on axion-like dark matter beyond current astrophysical constraints -- see fig.~\ref{fig:CASPEr}.  A full experiment would be similarly cost-effective and reach couplings at the QCD axion level over a wide range of masses over a 3-5 year time scale.

%------------------------------------------------------------------ FIG
%\begin{figure}[b!]
%\begin{center}
%    \includegraphics[height=6cm]{figs/CASPErSensitivity}
%    \caption{CASPEr projected sensitivity.}
%    \label{fig:sensitivity}
%\end{center}
%\end{figure}
%------------------------------------------------------------------ FIG

\subsubsection{Torsion Balances}

	The sensitivity range for ultra-light dark matter  can be dramatically extended by building two new state-of-the-art torsion balances.  One device will measure the time-dependent differential-acceleration signature of directly-coupled (hidden photon) dark matter \cite{Graham:2015ifn} and the other will probe the time-dependent spin-precession signature of spin-coupled (axion) dark matter \cite{Graham:2013gfa}.  Together these detectors will probe both classes of ultra-light dark matter over the lightest ~30\% of their possible log mass-ranges, from 10$^{-22}$ eV to 10$^{-15}$ eV, including the very well-motivated target of  10$^{-22}$ eV that would solve two outstanding discrepancies of conventional cold-dark-matter simulations\cite{Hu2000} \cite{Hui2017}.
	
	These devices will improve upon the very-high sensitivity rotating torsion-balance techniques developed by the Eot-Wash group that set the tightest bounds on DC  long-range forces \cite{Wagner2012}, \cite{Heckel2008}. The dark matter-induced twist on the pendulum will be triply-modulated, at the product of the turntable, the earth's rotation, and the DM Compton frequencies \cite{Graham:2015ifn}.  This distinctive signal will allow highly effective suppression of systematics and $1/f$ noise, allowing the new DM experiment to exploit fully the raw sensitivity of the torsion balance.  

The following innovations will increase the raw sensitivity relative to previous torsion balances:
\vspace{-0.6\topsep}	
\begin{enumerate}
\itemsep-0.2em
\item ultra-low-noise, high-stability fused-silica suspension fibers (40-80 reduction in the low-frequency side of the noise)
\item  longer optical-levers and interferometric twist-angle readout (10-100 times reduction in the high-frequency side of the noise)
\item  beryllium-polypropylene test bodies (4 times increase in sensitivity to B-L coupled DM) \cite{Adelberger2009}
\item very high stability turntables (to allow higher-frequency rotation)
\end{enumerate}
\vspace{-0.6\topsep}	
Combined, these will give a factor 100 or greater sensitivity to the DM coupling strength.

The R\&D and the construction of the two new balances can be accomplished in 2 years, by two FTE researchers collaborating with the Eot-Wash group.  Once operational and tested at the DOE lab at the Center for Experimental Nuclear Physics and Astrophysics at the University of Washington, the apparatus would be moved to the Sanford Underground Research Facility where environmental noise is 10-100 times smaller \cite{Harms2010}.  It should be noted that the sensitivity improvements and development-time estimates given above are conservative, being based on the decades of torsion balance experience of the Eot-Wash group.

\subsubsection{MAGIS-100: Atom interferometry for dark matter and gravitational waves}

\begin{figure}
\begin{center}
\includegraphics[width=2.9in]{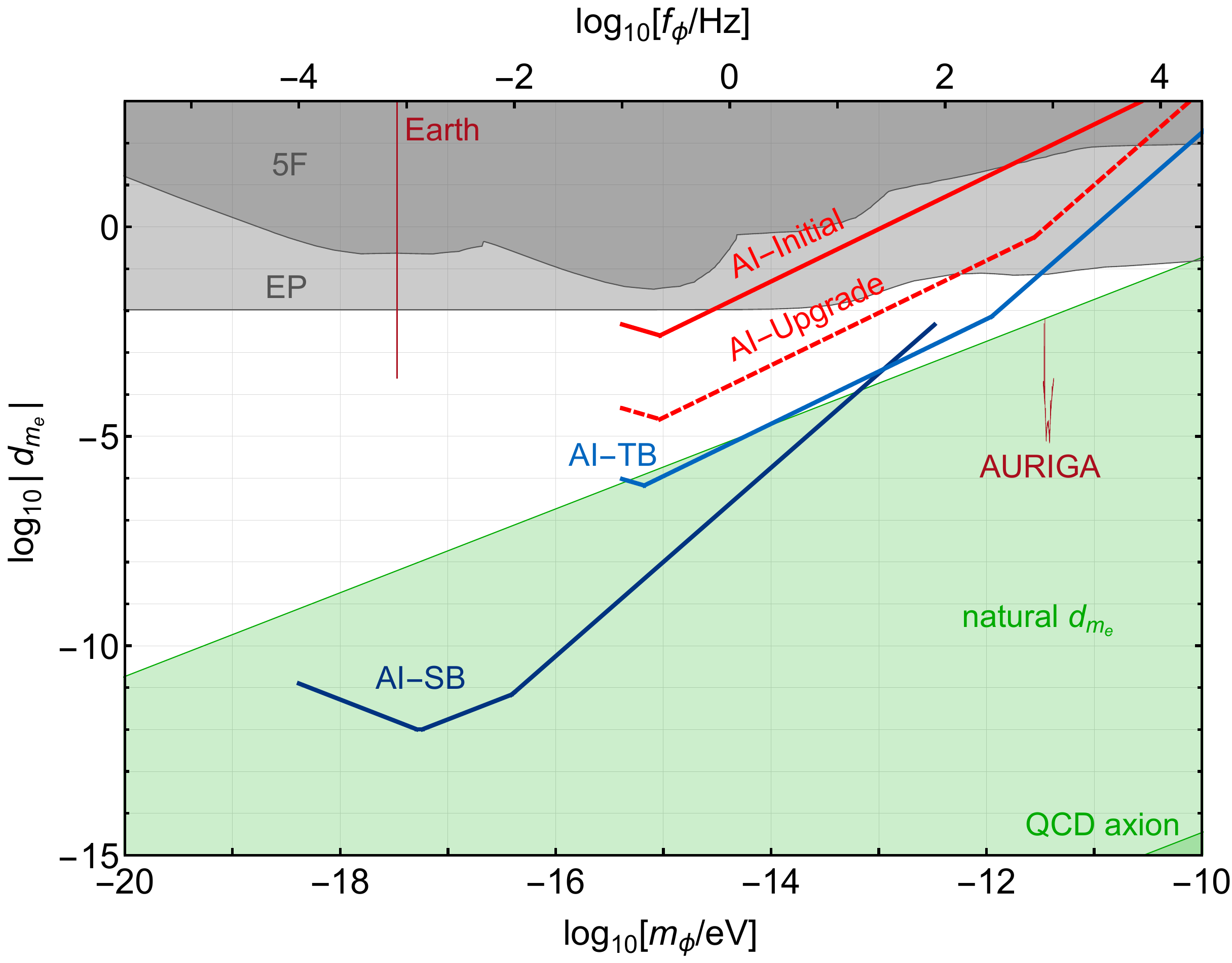} \ \ \ \includegraphics[width=3.3in]{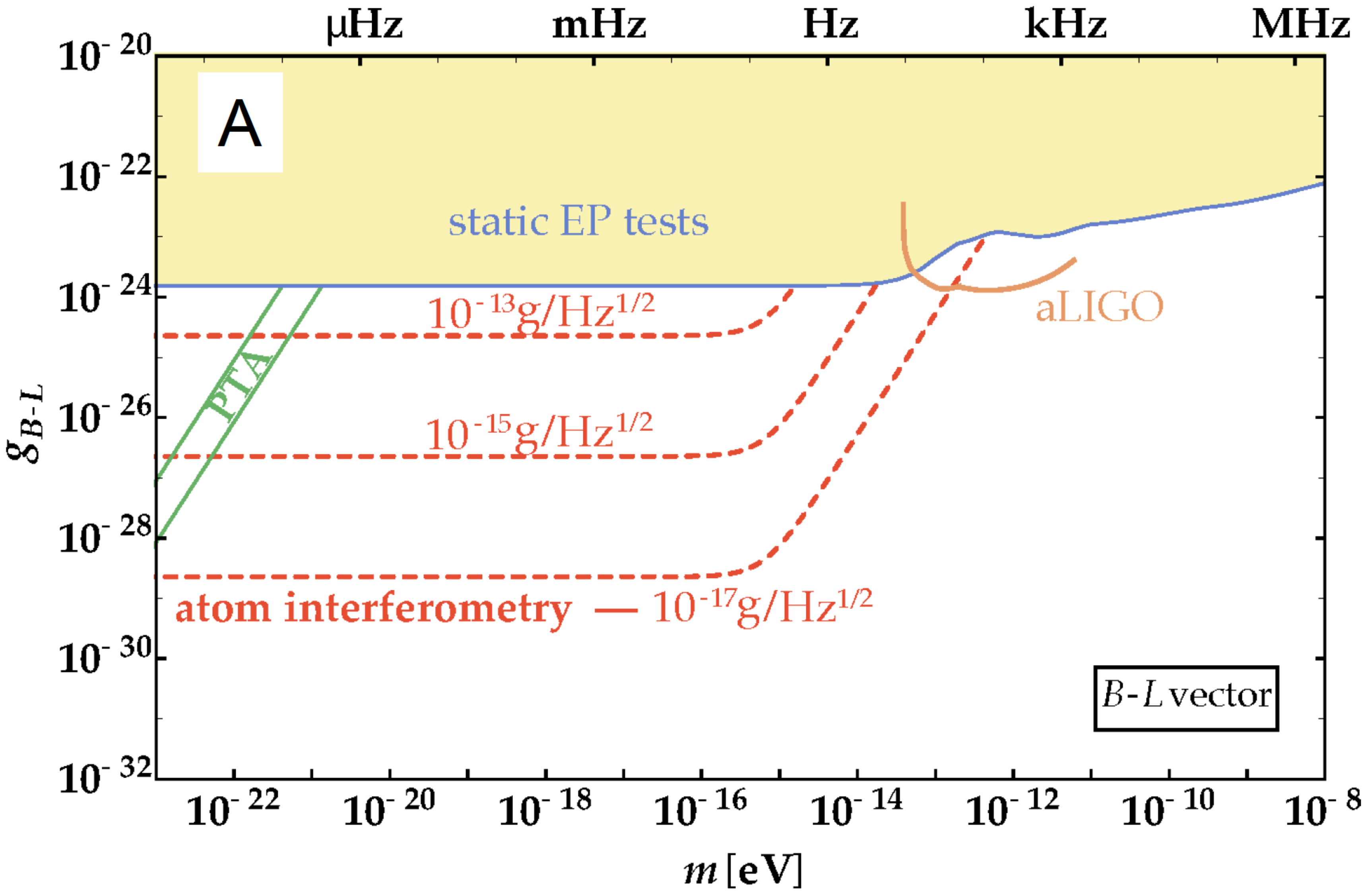}
\caption{Left:  Sensitivity of the MAGIS-100 atom interferometer to scalar DM-electron coupling $d_{m_e}$ measured relative to gravitational strength~\cite{Arvanitaki:2016fyj}.  Right: An initial BTBAI interferometer (dashed red) will reach coupling sensitivity of $10^{-15}\,g/$Hz$^{1/2}$ for $B-L$ dark matter \cite{Graham:2015ifn}.  %The $10^{-13}\,g/$Hz$^{1/2}$ (initially) and $10^{-15}\,g/$Hz$^{1/2}$ (final goal) will be reached a small scale experiment.  
The green curves give the range probed by the  EPTA and future SKA pulsar timing arrays.  
\label{fig:MAGIS-sensitivity} }
\end{center}
\end{figure}

MAGIS (Mid-band Atomic Gravitational wave Interferometric Sensor) is a new kind of atom interferometric sensor that aims to search for ultralight dark matter as well as gravitational waves.  Ultralight dark matter candidates with mass in the range $10^{-13}~\text{eV} - 10^{-16}~\text{eV}$ can cause time-varying atomic energy levels in the $0.1~\text{Hz}-10~\text{Hz}$ frequency range that can be detected with the proposed sensor \cite{Arvanitaki:2016fyj}.  The MAGIS detector would also provide unique sensitivity to gravitational waves in this mid-band frequency range -- between the frequency bands targeted by LIGO and LISA~\cite{graham2013new,hogan2015heterodyne}.  The discovery potential in this frequency band appears exciting, ranging from observation of new astrophysical sources (e.g. black hole and neutron star binaries) to searches for cosmological sources of stochastic gravitational radiation in addition to the searches for dark matter \cite{ResonantGW}.

%By operating in this mid-band, MAGIS can access an important frequency band in the gravitational wave spectrum that is otherwise not covered by existing and future detectors, and would thus be complementary to the LISA and LIGO detectors.  

The detector is based on a new atomic sensor concept that is a hybrid between an atomic clock and an atom interferometer.  Gravitational radiation is sensed through precise measurement of the light flight time between two distantly separated (atomic) inertial references \cite{graham2013new}.  Time is recorded by the accumulation of phase by these atoms, which also serve as precise differential clocks.  This same configuration is also sensitive to time-variations in the atomic energy levels caused by couplings to ultralight dark matter, since such energy level shifts change the phase accumulation by the separated atomic clocks \cite{Arvanitaki:2016fyj}.  Current work is focused on building a small-scale (10-meter) prototype detector to demonstrate required detector performance characteristics, including laser noise suppression.  Longer detector baselines are required to reach scientifically interesting strain sensitivity and dark matter couplings.  

The MAGIS-100 proposal is to build a 100-meter long detector to be located at Fermilab in an existing 100-meter vertical access shaft at the NuMI neutrino beam facility.   One atomic source would be located at the top of the shaft and one midway down, allowing for over 3 seconds of free-fall time and hence measurements at frequencies $<1~\text{Hz}$.  The initial detector would use state-of-the-art atom interferometry \cite{kovachy2015quantum,Asenbaum2016curvature} including $100 \hbar k$ enhanced atom optics and an atom flux of $10^6~\text{atoms/s}$ (``AI-Initial'' in Fig.~\ref{fig:MAGIS-sensitivity}).  Planned upgrades include larger atom optics ($1000 \hbar k$) and a larger atom flux of $10^8~\text{atoms/s}$ (``AI-future'' in Fig.~\ref{fig:MAGIS-sensitivity}).  This small experiment could be conducted over the course of 3 years.

\subsubsection{Berkeley Thick-Beam Atom Interferometer}

%\begin{figure}\centering
%\includegraphics[width=\textwidth]{figs/DoE_Figure_HMv2.pdf}
%\caption{(A) Reach of searches for $B-L$-DM \cite{Graham:2015ifn}. Dashed red: BTBAI. The $10^{-13}\,g/$Hz$^{1/2}$ (initially) and $10^{-15}\,g/$Hz$^{1/2}$ (final goal) will be reached here. We will develop technology for a future large-scale experiment, to reach even higher sensitivity. Green curves correspond to existing EPTA (upper) and upcoming SKA (lower) pulsar timing arrays.  (B) Sensitivity to photon-dark-photon coupling by measuring $\alpha$ with atom interferometers. Figure adapted from \cite{Curciarello:2016jbz}..} \label{fig:HMplot}
%\end{figure}

The Berkeley thick-beam atom interferometer (BTBAI) will be located at UC Berkeley's New Campbell Hall in a room with heavy acoustic and electromagnetic shielding. Control over systematic effects will be taken to the extreme by using a thick ($\sim 30\,$cm diameter), high-power (kW) laser beam that enables efficient atomic beam splitters, large atomic samples, and increases accuracy \cite{PhysRevLett.115.083002} and beam-splitter fidelity hundred- to several thousand-fold relative to atom interferometers with cm-sized beams. The experiment will look for dark matter and more general dark-sector constituents in two ways.   First, it will search for $B-L$, topological, and Higgs-portal dark matter by looking for oscillating accelerations on rubidium atoms in a differential measurement between rubidium isotopes at $10^{-15}\,g/$Hz$^{1/2}$ sensitivity ($g=9.8\,$m/s$^2$), see Figure~\ref{fig:MAGIS-sensitivity} \cite{Graham:2015ifn}. The high efficiency of large-momentum transfer beam splitters in BTBAI will be instrumental in reaching this goal. Second, it will search for dark photons in the MeV-GeV mass range %(Fig.~\ref{fig:MAGIS-sensitivity}) 
by measuring the fine structure constant at better than $10^{-11}$ accuracy and comparing with measurements of the electron's gyromagnetic anomaly $g_e-2$ \cite{PhysRevLett.100.120801} which now reach $2.2\times 10^{-10}$ accuracy but are expected to improve by an order of magnitude. 

The estimated time for this small experiment and R\&D effort is five years. BTBAI will enable broad progress in atom interferometry and laser technology, contributing to the development of gravitational wave detectors as well as to tools for geophysics.

%\begin{figure}\centering
%\includegraphics[width=0.5\textwidth]{figs/B-L_plot_2.pdf}
%\caption{Reach of searches for $B-L$-DM \cite{Graham:2015ifn}. Dashed red: BTBAI. The $10^{-13}\,g/$Hz$^{1/2}$ (initially) and $10^{-15}\,g/$Hz$^{1/2}$ (final goal) will be reached here. We will develop technology for a future large-scale experiment, to reach even higher sensitivity. Green curves correspond to existing EPTA (upper) and upcoming SKA (lower) pulsar timing arrays.} \label{fig:BLplot}
%\end{figure}

%\begin{wrapfigure}{r}{0.5\textwidth}\vspace{-1cm}
%\begin{figure}
%\begin{center}
%\includegraphics[width=0.4\textwidth]{figs/DarkPhotonLim.pdf}
%\end{center}
%\vspace{-0.5cm}
%\caption{\small Dark photon search by measuring $\alpha$ with atom interferometers. Figure adapted from \cite{Curciarello:2016jbz}.} 
%\label{fig:Dark_photons_alpha}
%\end{figure}

\subsection{Non-Direct Detection Experiments}

\subsubsection{Mini-IAXO}

\begin{figure}[b!]
\centering % \begin{center}/\end{center} takes some additional vertical space
\includegraphics[width=.5\textwidth]{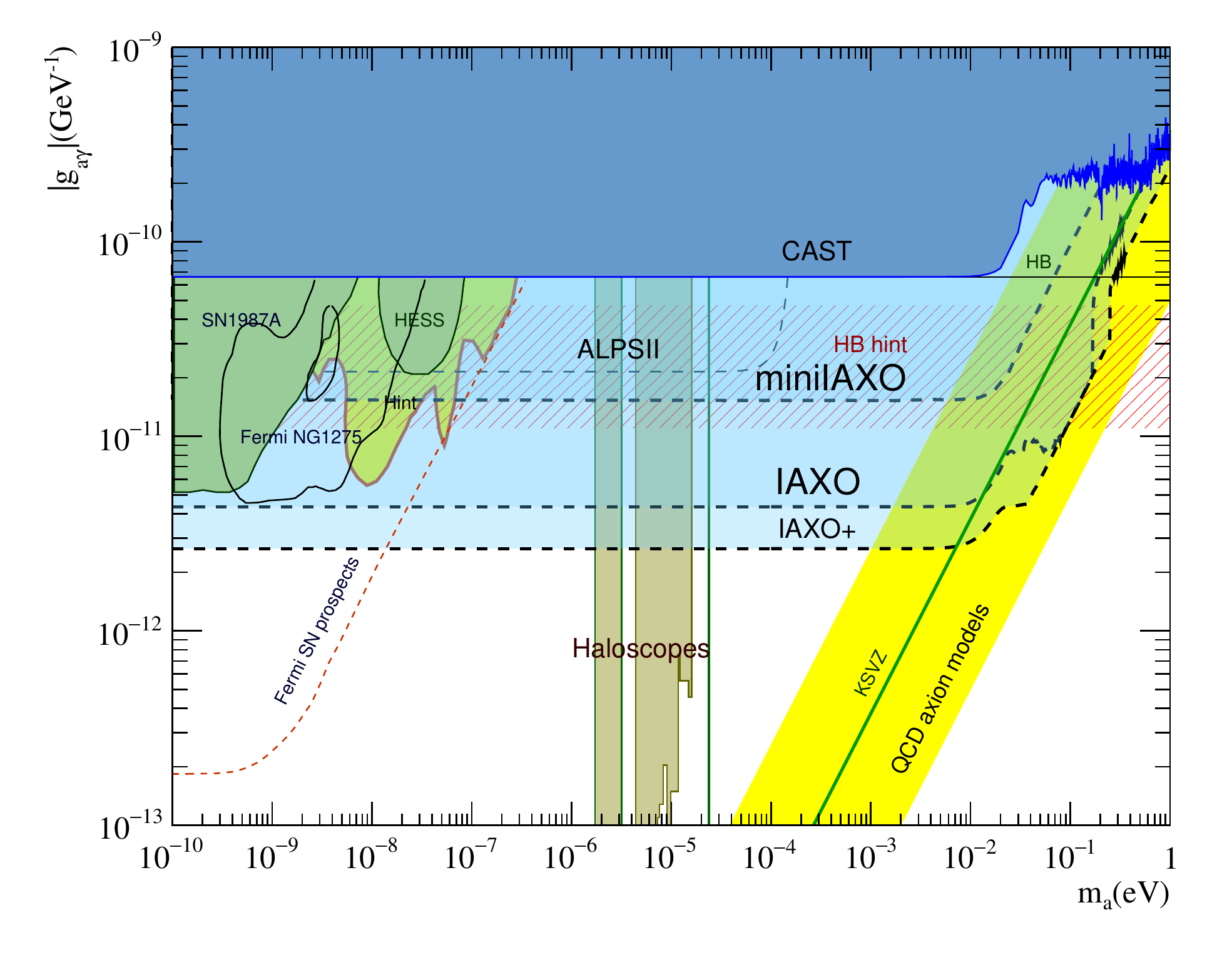}
\caption{\label{fig:i} Preliminary estimate of the mini-IAXO sensitivity in terms of axion coupling $g_{a \gamma\gamma}$ versus mass, indicating its complementarity to existing haloscope data.
%The solid blue and brown regions show the part of parameter space searched by CAST and haloscopes like ADMX. The yellow band indicates the region of QCD axion parameter space. The dashed lines indicate the sensitivity possible with mini-IAXO, different instantiations of IAXO and ALPS II. 
Mini-IAXO would be able to test models of axion-like particles favored by certain astrophysical observations.}
\end{figure}

The International Axion Observatory (IAXO)\cite{Irastorza:2011gs,Armengaud:2014gea,Irastorza:1567109} will be a fourth generation axion helioscope with the primary physics goal to look for axions or axion-like particles (ALPs) originating in the Sun via the Primakoff effect~\cite{Primakoff:1951}. Mini-IAXO is proposed as a small pilot experiment that will increase the sensitivity of the currently most powerful axion helioscope, CAST~\cite{Aune:2011rx, Arik:2008mq, Andriamonje:2007ew,Zioutas:1998cc,Arik:2013nya, Barth:2013sma}, reaching sensitivity to axion-photon couplings $g_{a \gamma\gamma}$ down to a few ${10^{-11}}$~GeV$^{-1}$ and thus probing new axion and ALP parameter space as shown in Fig.~\ref{fig:i}. %The experimen will also be sensitive to solar axions produced by mechanisms mediated by the axion-electron coupling $g_{ae}$ with sensitivity for the first time to values of $g_{ae}$ not previously excluded by astrophysics. \\

The preliminary design for mini-IAXO includes a single-bore, $10$~m superconducting magnet that will be instrumented with a focusing X-ray telescope and low-background X-ray detector to explore the above mentioned axion parameter space. Mini-IAXO will utilize existing infrastructure to produce the optics, consisting of multi-layer coated slumped-glass substrates~\cite{jakobsen2013},  and the low-background X-ray detector~\cite{Aune:2014}, consisting of gaseous Micromegas detectors. Such technologies have recently been developed for and demonstrated on CAST. This approach of combining very low-background detectors and efficient x-ray optics has led to a record-setting experimental sensitivity that has resulted in an upper limit on axion-photon coupling $g_{a \gamma\gamma}$ comparable to those obtained from astrophysical constraints~\cite{anastassopoulos:2017ftl}.

\subsubsection{ARIADNE}

\begin{figure}
\begin{center}
\includegraphics[width=0.8\columnwidth]{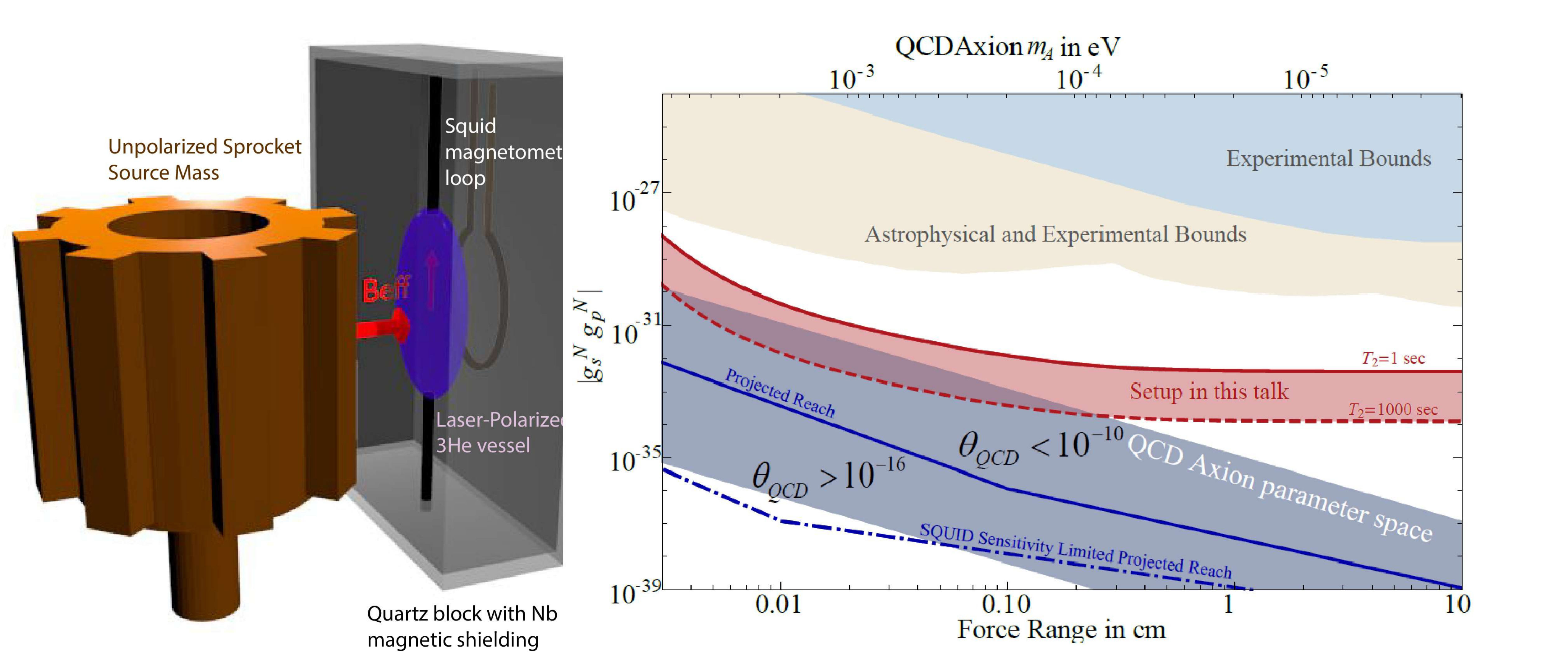}
\caption{\small{ (left) Setup: a sprocket-shaped source mass is rotated so its ``teeth'' pass near an NMR sample at its resonant frequency. (right) Projected reach for monopole-dipole axion mediated interactions. The band bounded by the red (dark) solid line and dashed line denotes the limit set by transverse magnetization noise, depending on achieved coherence time $T_2$. Current constraints and expectations for the QCD axion also are shown, adapted from Ref. \cite{Arvanitaki:2014dfa} \label{setup}.}}
\end{center}
\end{figure}

ARIADNE aims to detect axion-mediated spin-dependent interactions between an unpolarized source mass and a spin-polarized $^3$He low-temperature gas \cite{Arvanitaki:2014dfa}.  %Unlike all other direct axion search experiments which depend on the local Dark Matter density at the earth, with the axion being the particle that constitutes Dark Matter, \
%By \textbf{sourcing the axion locally in the lab}, the signal can be modulated in a controlled way.  The experiment probes QCD axion masses in the higher end of the traditionally allowed “axion window” of $1$ $\mu$eV to $6$ meV, which are \textbf{not accessible by any existing experiment} including dark matter “haloscopes” such as ADMX.  Thus ARIADNE \textbf{fills an important gap} in the search for the QCD axion in this important region of parameter space.
The axion can mediate a monopole-dipole (mass-spin) interaction between fermions (e.g. nucleons) with the scalar and dipole coupling constants $g_s^N$ and $g_p^N$, respectively. Since it couples to $\hat{\sigma}$ which is proportional to the nuclear magnetic moment, the axion coupling can be treated as a fictitious magnetic field $B_{\rm{eff}}$.  This fictitious field is used to resonantly drive spin precession in a laser-polarized cold $^3$He gas.  This is accomplished by spinning an unpolarized tungsten mass sprocket near the $^3$He vessel. As the teeth of the sprocket pass by the sample at the nuclear larmor precession frequency, the magnetization in the longitudinally polarized He gas begins to precess about the axis of an applied field. This precessing transverse magnetization is detected with a superconducting quantum interference device (SQUID). %The $^3$He sample acts as an amplifier to transduce the small fictitious magnetic field into a larger real magnetic field detectable by the SQUID. Superconducting shielding is needed around the sample to screen it from ordinary magnetic field noise which would otherwise limit the sensitivity of the measurement. The ultimate limit is set by spin projection noise in the sample itself \cite{minaandy}.
%\emph{Comparison with other experiments:} The QCD axion generically couples to photons as well as nucleons.  Similar in style to the light-shining-through-walls experiments searching for ALPS, we source the axion field in the lab by observing the axion as a force mediator.
The experiment sources the axion in the lab, and can explore all mass ranges in its sensitivity band simultaneously (Fig. \ref{setup}).
%, unlike experiments which must scan over the allowed axion oscillation frequencies (masses) by tuning a cavity or magnetic field. %In contrast to other magnetometry experiments \cite{explims}, the \textbf{resonant enhancement technique} allows orders of magnitude improvement, yielding sufficient sensitivity to detect the QCD axion (Fig. 1).
%\emph{Rough Timescale and Cost:} Design/construction/commissioning: \$$0.6$M/yr for 2-3 years.  Reaching QCD axion sensitivity: \$$0.6$M/yr for 3 more years, w/ R\&D for future upgrades.
A time scale of 2-3 years for construction followed by a 3-year operating phase is envisioned.

%% file: WG3.tex
% !TEX root = WG3_wrapper.tex
\section{Dark matter production at fixed target and collider experiments}\label{sec:WG3experiments}
Accelerator experiments are increasingly recognized as essential tools in probing the particle nature of dark 
matter (DM), and are especially suited to probing DM in the vicinity of known SM mass scales, roughly MeV - TeV. Indeed, a new 
generation of fixed target and collider experiments 
would be strongly positioned to test light thermal DM. The thermal paradigm is arguably one of the 
most compelling possibilities, and has driven much of the experimental DM program during the last decades. Among the 
thermal DM parameter space, light (sub-GeV) hidden sector thermal DM annihilating directly into Standard Model particles 
(the ``thermal relic target'') stands out for its predictiveness and testability. The virtue of an accelerator program relies on 
the fact that the rate of relativistic DM production is largely independent of the details of the dark sector and is predicted from 
freeze-out, whereas the rate of non-relativistic DM scattering is highly sensitive to the DM particle nature.  As a consequence, accelerators 
can probe {\it all} predictive direct annihilation scenarios, while the majority of these targets remain beyond the current capabilities 
of proposed direct 
detection experiments.

While a broad international program of accelerator experiments is currently focused on exploring light dark 
matter and any associated new forces, many of the theoretical milestones are only beginning to be probed. New, small-scale 
projects present an opportunity for the US DM program to play the leading role in light DM and dark sector 
physics during the next decade. By leveraging existing technologies and facilities, a high-impact program could 
be quickly deployed to achieve significant science in the next few years.

In addition to light, directly annihilating thermal DM, many of these proposals would also explore a wide parameter 
space for secluded thermal DM, as well as DM with quasi-thermal origins ({\it e.g.} asymmetric DM, SIMP/ELDER 
scenarios), and freeze-in models of light DM with a ``heavy'' mediator. More generally, they would offer 
sensitivity to any long-lived neutral particles below the GeV-scale, and provide a unique gateway to explore generic low-mass 
hidden sectors.

In the following, we briefly review the phenomenology of thermal DM, and underline the importance of the direct annihilation 
target. We discuss the need for an accelerator-based program, and provide a detailed discussion of future proposals to robustly 
test this scenario. 

\subsection{The Thermal Target at Accelerators}

The theoretical framework of hidden-sector thermal DM is summarized in Section~\ref{ssec:thermal}, including the definition 
of ``secluded'' and ``direct'' annihilation. We briefly review the science case for accelerators. The secluded annihilation rate 
is dictated by DM-mediator interaction strengths within the hidden sector~\cite{Finkbeiner:2007kk,Pospelov:2007mp}. Since arbitrarily 
small values of the SM-mediator coupling can be compatible with thermal light DM, this scenario is less predictive, although viable 
models, such as DM annihilation into two scalar or pseudo-scalar mediators, can still be probed with laboratory 
experiments~\cite{Dolan:2014ska, Krnjaic:2015mbs}. Direct annihilation, on the other hand, is controlled by the same couplings 
that are relevant for direct DM scattering, leading to well-defined predictions. In the case of scalar DM annihilating into leptons 
through the vector portal, the annihilation rate is given by:
\begin{eqnarray}
\langle \sigma v \rangle \approx \frac{1}{6\pi}\frac{g_{D}^2 \,  g_{\rm SM}^2  \,  m_{\rm DM}^2 \, v^2 }{(m_{\rm MED}^2 - 4 m_{\rm DM}^2)^2 + m_{\rm MED}^2 \Gamma_{\rm MED}^2  \!\!},
\end{eqnarray}
for $v \ll c$ and neglecting $m_e/m_{\rm DM}$ corrections. Sufficiently far away from the resonance region ($m_{\rm MED} = 2 m_{\rm DM}$) and 
assuming $m_{\rm MED} \gg \Gamma_{\rm MED}$, this cross-section depends on the dark sector parameters only through the DM mass $m_{\rm DM}$ 
and the dimensionless parameter

\begin{eqnarray}\label{eq:generic-thermal-target}
y\equiv\frac{g_D^2 \, g_{\rm SM}^2  }{16\pi^2} \left(\frac{m_{\rm DM}}{m_{\rm MED}}\right)^4
\end{eqnarray}

The observed DM abundance imposes a minimum bound on this cross-section, requiring $\langle \sigma v \rangle > \langle \sigma v \rangle_{\rm relic}$. 
Perturbativity and constraints on the mass ratio $m_{{\rm DM}}/m_{\rm MED}$, at most ${\cal O}(1)$ in this regime, imply in turn a {\it lower} bound 
on the magnitude of the SM-mediator and DM-mediator couplings to be compatible with a thermal history. In other words, a 
lower bound on the direct annihilation scenario. This constraint can be translated into a minimum value of $y$, which is qualitatively valid for 
every DM/mediator particle nature variation provided that $m_{\rm DM} < m_{\rm MED.}$. Larger values of $y$, which correspond to models where direct annihilation is not the dominant process that determines the DM abundance, could also be probed at accelerators. One caveat to the arguments above is the case where the DM mass is near the mediator's resonance region, {\it i.e.,} when $m_{\rm MED} \approx 2 m_{\rm DM}$. In this case, DM annihilation becomes extremely efficient, and thus freeze-out can be achieved with smaller couplings \cite{Feng:2017drg}.

The argument presented above is generic and equally applicable to any of the minimal portals between the SM and the DS. However, 
the vector/kinetic mixing portal is by far the most viable~\cite{Dolan:2014ska, Krnjaic:2015mbs,Alexander:2016aln} among the renormalizable operators. 
This portal should be viewed as both a concrete UV complete benchmark, as well as a simplified model, since it is representative of 
models where the mediator couples preferentially to baryonic (leptophobic DM), leptonic (leptophilic DM), or $(B - L)$ currents. In what 
follows, we denote the dark mediator by $A'$, its mixing with the SM by $\epsilon e$, and its coupling to DM by $g_d$. The terms 
mediator, dark photon or dark vector will be used interchangeably. Interestingly, the bottom-up values of the $\epsilon$ parameter in the kinetic 
mixing benchmark that are needed for a thermal target are well aligned with the top-down $\epsilon$ range motivated for all hidden 
sectors (Section~\ref{ssec:thermal}).

While we focus the remainder of the discussion on directly annihilating light thermal DM, since the scientific impact of reaching this sharp 
milestone is substantial and the opportunity to do so is timely, the scope of the accelerator-based program is much more extensive. 
The experimental approaches discussed below directly apply to many other important models, since analogous mappings allow constraints on 
the CMB and DM-SM scattering cross sections to be translated onto the $y$ parameter space. These models include asymmetric DM~\cite{Zurek:2013wia}, 
in which the DM abundance arises from a primordial asymmetry instead of from thermal freeze-out; SIMP DM~\cite{Hochberg:2015vrg}, which encompasses 
production of new DS resonances that can decay back to the SM directly or through hidden-valley-like topologies~\cite{Strassler:2006im}; models 
with different cosmological histories, {\it e.g.} ELDER DM~\cite{Kuflik:2015isi}; freeze-in models of sub-MeV DM with a ``heavy'' (GeV-scale) mediator (see Refs.~\cite{Giudice:2000dp, Gelmini:2004ah} for aspects on the cosmology of similar models); new force carriers decaying to SM particles~\cite{Alexander:2016aln} or searches for millicharged DS particles, either through missing energy signatures or through minimum 
ionizing signals~\cite{Prinz:1998ua,Haas:2014dda}. 

We finally emphasize that a comprehensive program, including both accelerator and direct detection experiments, would be most 
successful in exploring light dark matter. While accelerator-based experiments offers key advantages in probing several DM 
scenarios, some possibilities could only be explored with direct detection techniques, such as freeze-in models with an 
ultralight mediator, and models of ultralight bosonic DM. Other cases could be accessible to both accelerators and direct detection 
experiments, opening the exciting prospect of testing potential DM signal by different approaches.

\subsection{The Scientific Need for an Accelerator-based Program}

\begin{figure}[t!]
\center
\includegraphics[height=8.3cm]{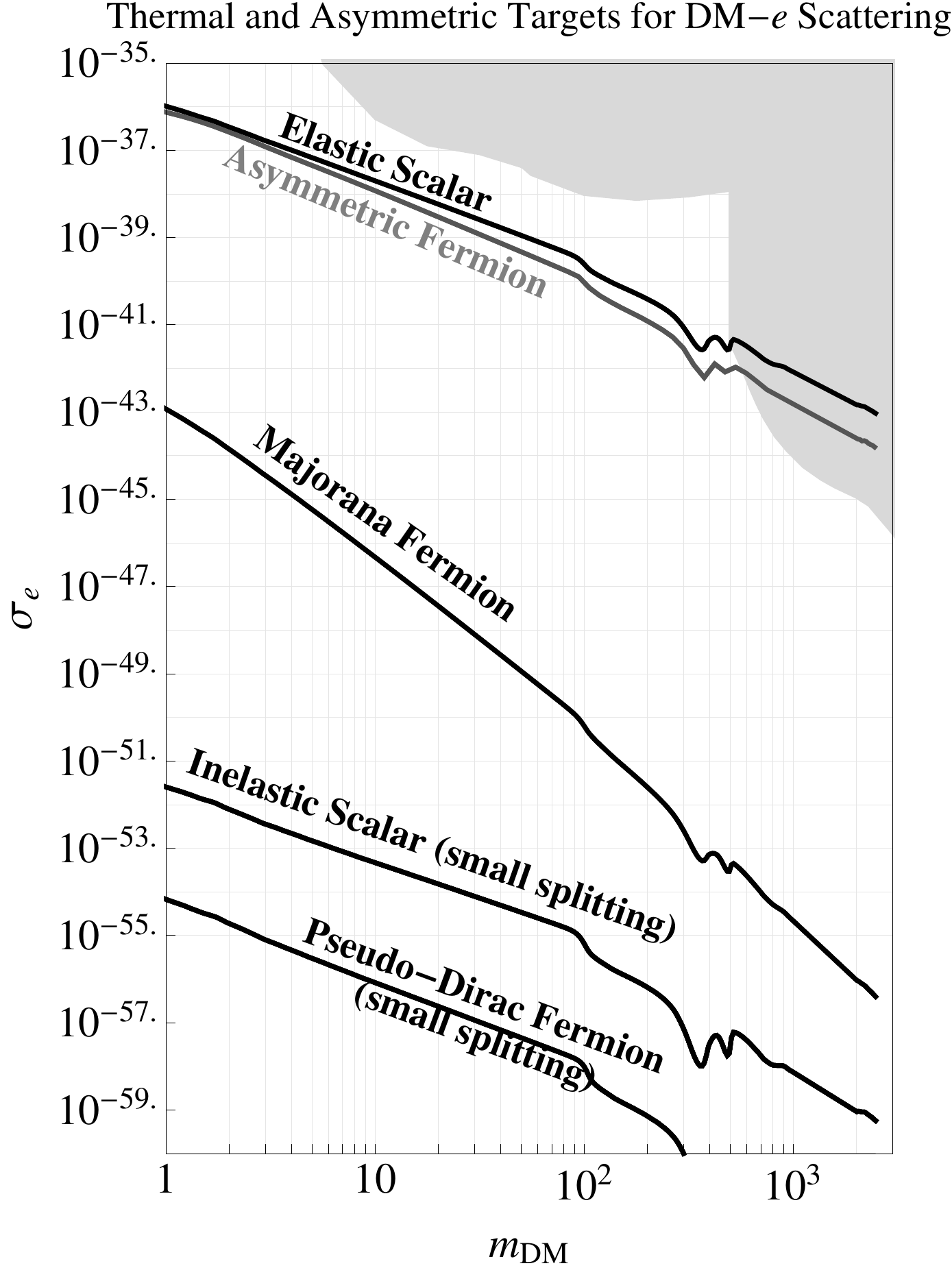} \hspace{0.5cm}
\includegraphics[height=6.3cm]{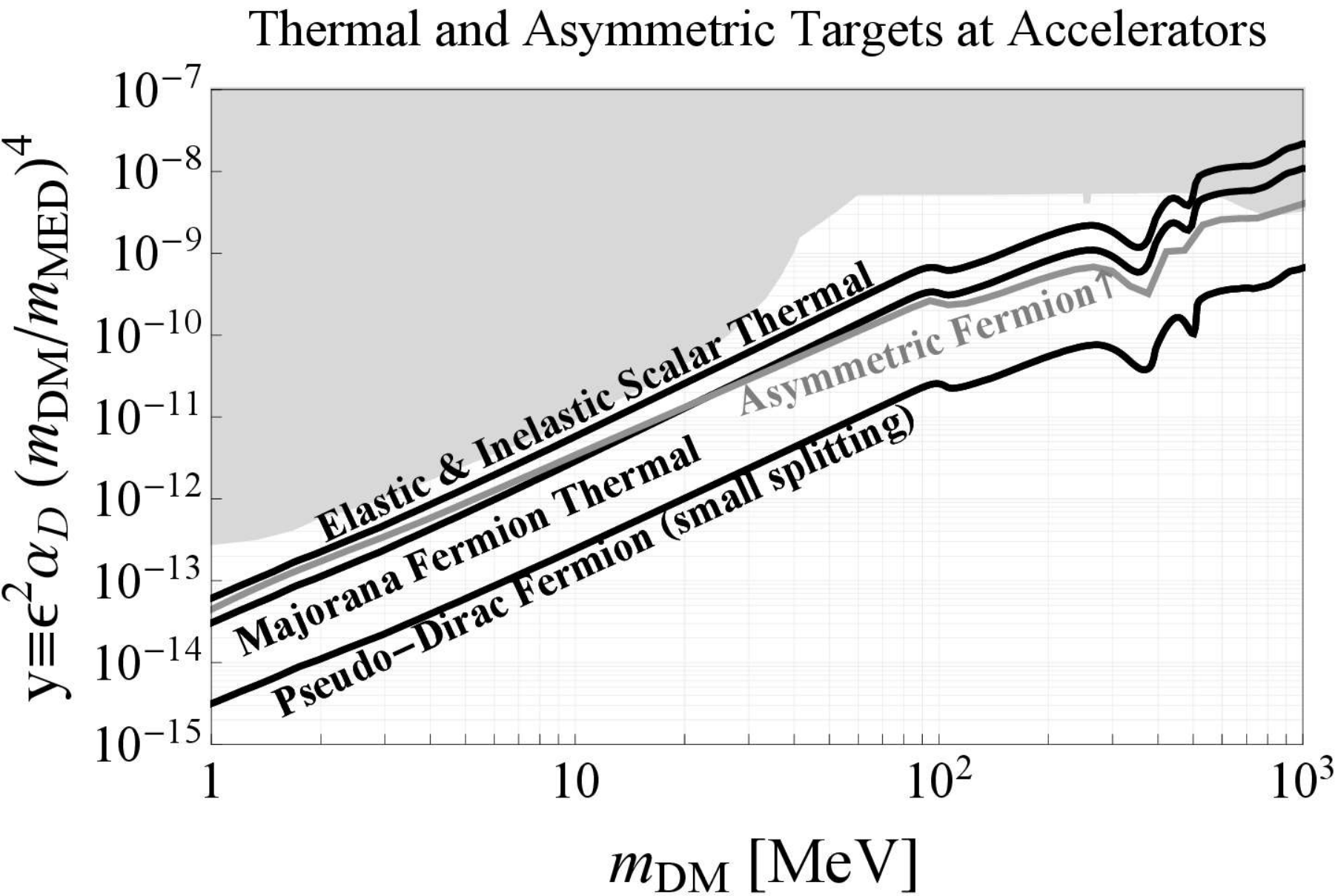} 
\caption{Direct annihilation thermal freeze-out targets and asymmetric DM target for (left) non-relativistic e-DM scattering 
probed by direct-detection experiments and (right) relativistic accelerator-based probes. The thermal targets include scalar, Majorana, 
inelastic, and pseudo-dirac DM annihilating through the vector portal. Current constraints are displayed as shaded areas. 
Both panels assume $m_{\rm{MED}} = 3 m_{\rm{DM}}$ and the dark fine structure constant $\alpha_D\equiv g_D^2/4\pi=0.5$. These choices 
correspond to a conservative presentation of the parameter space for accelerator-based experiments (see section~\ref{accproj}).}
\label{fig:accDDcomp}
\end{figure}

While important progress has been achieved from dedicated searches at current facilities or re-interpretation 
of previous results, new experiments are needed to cover decisive levels of sensitivity to the thermal-target 
parameter space. Compared to other approaches, accelerator experiments offer key advantages to robustly probe 
direct annihilation scenarios:

\begin{itemize}

\item {\bf Reduced dependence on DM particle nature:} Accelerator-based experiments are much less sensitive to 
the details of the DM particle nature than direct detection experiments, as illustrated in Fig.~\ref{fig:accDDcomp}. In some 
models, e.g. Majorana fermion DM, the direct detection cross-section is drastically reduced through its dependence 
on the halo DM velocity, well below current detection capabilities. On the other hand, DM is produced relativistically 
at accelerators, and the DM scattering cross section is only weakly dependent on the velocity. In missing energy/missing 
momentum experiments, the DM presence is inferred through energy/momentum imbalance, almost entirely insensitive to 
the DM velocity. 

\item {\bf Kinematic thresholds in the DS.} This effect can occur, for example, if the DM particle is part of a pseudo-Dirac 
fermion pair. In this scenario, DM (which we now label as $\chi_1$) is accompanied by a heavier excited state $\chi_2$. DM 
annihilation or scattering through the light mediator can feature dominantly off-diagonal couplings between the light 
mediator and the $\chi_1$ and $\chi_2$ particles, as opposed to diagonal mediator-$\chi_1$-$\chi_1$ couplings. The 
direct detection tree-level scattering can be reduced or altogether turned off whenever the DM's kinetic energy is 
insufficient to produce the excited state $\chi_2$~\cite{TuckerSmith:2001hy}. Instead, the dominant contribution at 
direct detection experiments could arise from exchange of two virtual light mediators. At accelerators, in contrast, 
the ground state can efficiently up-scatter into the excited state $\chi_2$ when detected through scattering off a 
SM target. Missing energy/momentum experiments yields are unaffected by kinematic thresholds as well. The 
heavier state $\chi_2$ may even exhibit macroscopic lifetimes that could be searched for at accelerator 
probes~\cite{Morrissey:2014yma,Izaguirre:2017bqb}.

\item {\bf Sensitivity to dark sector structure.} The mass of the mediator is not only accessible in scenarios where 
it decays dominantly into SM particles, but also in specific types of measurements for invisible decays. In the kinetic 
mixing portal, the dark photon can be resonantly produced and subsequently 
reconstructed as a dileptonic resonance or a peak in the $e^+e^- \rightarrow \gamma A'$ missing mass spectrum. The nature 
of the mediator-SM coupling, another fundamental property, can be investigated as well. In proton beam dump experiments, the 
mediator can be emitted by the incoming proton, or if kinematically allowed, from rare SM meson decays, while detection could 
proceed through DM-nucleon scattering. Thus, proton beam-dump experiments are uniquely sensitive to the coupling to quarks. On 
the other hand, leptonic couplings can be studied in electron beam-dump and fixed target experiments, where the 
mediator is radiated off the incoming electron beam. The DM is identified through its scattering off electrons at a 
downstream detector, or its presence is inferred as missing energy/momentum. 

\end{itemize}

\subsection{Experimental approaches and future opportunities}

The light DM paradigm has motivated extensive developments during the last few years, based on a combination of 
theoretical and proposed experimental work. As a broad organizing principle, these approaches can be grouped into 
the following generic categories:

\begin{itemize}

\item {\bf Missing mass:} The DM is produced in exclusive reactions, such as $e^+e^- \rightarrow \gamma (A' \rightarrow \chi \bar\chi)$ 
or $e^-p \rightarrow e^- p (A' \rightarrow \chi \bar\chi)$, and identified as a narrow resonance over a smooth background in the recoil 
mass distribution. This approach requires a well-known initial state and the reconstruction of all particles besides 
the DM. A large background usually arises from reactions in which particle(s) escape undetected, and detectors with good 
hermeticity are needed to limit their impact.

\item {\bf Missing momentum/energy:} The DM is produced in the fixed-target reaction $eZ \rightarrow eZ(A'\rightarrow \chi \bar\chi)$ 
and identified through the missing energy/momentum carried away by the escaping DM particles. This approach relies heavily on the 
detector hermeticity to achieve excellent background rejection, a critical aspect. In some implementations, the ability to measure 
the incoming electrons individually is also required. This method typically offers a better signal yield than beam dump experiments 
for a similar luminosity, as the DM particles are not required to scatter in the detector.

\item {\bf Electron and proton beam dump:} The DM is produced in $\pi^0/\eta^{(')} \rightarrow \gamma 
(A' \rightarrow \chi \bar\chi)$ decays, $p Z \rightarrow p Z A', A' \rightarrow \chi \bar\chi$ events 
(proton beam dump) or $e Z \rightarrow e Z (A'\rightarrow \chi \bar\chi)$ events (electron beam dump) and 
typically detected via $e \chi \rightarrow e \chi$ or $N \chi \rightarrow N \chi$ scattering in a 
downstream detector. This approach has the advantage of probing the DM interaction twice, providing 
sensitivity to the dark sector-mediator coupling, but requires a large proton/electron flux to compensate 
for the reduced yields. However, the signature is similar to that of neutrino interactions, which 
often constitutes the limiting factor on the sensitivity. Beam-dump experiments are also sensitive to 
the decay of excited states in the DS~\cite{Morrissey:2014yma,Izaguirre:2017bqb}, which can naturally 
occur in models where the DM is Pseudo-Dirac.

\item {\bf Direct dark photon searches:} focused on identifying the mediator through its decay into SM particles. This 
approach is of particular importance when $m_{A'} < 2m_\chi$, in which case the mediator decays visibly. The 
production mechanisms include among others $e^+e^- \rightarrow \gamma A'$, $eZ \rightarrow eZ A'$ or neutral 
meson decays, and the mediator is usually reconstructed through its leptonic decays $A' \rightarrow e^+e^-, 
\mu^+\mu^-$ as a narrow resonance over a wide background. The sensitivity is often limited by irreducible backgrounds, 
requiring large luminosities to extend the experimental reach. 

\end{itemize}

\subsection{Current constraints and on-going efforts}

Before discussing future experimental opportunities for DM searches, we briefly review the current constraints on the direct 
annihilation scenario. While the LHC experiments can search for invisible final states by looking for mediator 
or DM production in association with one or more visible objects, their sensitivities are limited to SM-mediator couplings 
$\epsilon$ of roughly $10^{-1}$~\cite{Izaguirre:2015yja}. Stronger bounds in the $\MeV-\GeV$ range are provided by the NA64 
experiment~\cite{Banerjee:2016tad}, mono-photon 
searches at \babar~\cite{Lees:2017lec}, and from a dedicated search for DM-nucleon scattering at MiniBooNE~\cite{Aguilar-Arevalo:2017mqx}. 
Estimated bounds from previous fixed target experiment in the low-mass range have been derived by reinterpreting prior measurements 
from LSND~\cite{deNiverville:2011it} and the E137 experiment~\cite{Batell:2014mga}. Finally, a 
re-analysis of electron-scattering data from direct detection experiments has led to constraints in the sub-$\GeV$ DM 
region~\cite{Essig:2012yx,Essig:2017kqs}. Combinations of these bounds are displayed in Fig.~\ref{DMAProj1}-\ref{DMAProj3}.

Constraints on visible decays are driven by dileptonic resonance searches~\cite{Lees:2014xha,Batley:2015lha,Merkel:2014avp} 
and re-interpretations of previous fixed target measurements in the low-mass region~\cite{Riordan:1987aw,Bjorken:1988as,Bross:1989mp}. 
A detailed discussion can be found in Ref.~\cite{Alexander:2016aln}.  

On-going experimental efforts are summarized in the following (non-exhaustive) list:
  
\begin{itemize}

\item {\bf APEX at JLab} (direct mediator search): search for prompt visible dark photon decays with fixed target experiment at 
Hall A at JLab using the CEBAF electron beam. Dark vectors are produced on a high-Z target and reconstructed with an existing 
high-resolution dual arm spectrometer.\\
Timeline: engineering run 2010 to demonstrate method, a one-month physics run is expected in 2018-2019.  \\
Sensitivity: $\epsilon^2 \gtrsim 10^{-7}$ for $60 < m_{A'} < 550~\MeV$. References:~\cite{Essig:2010xa,Abrahamyan:2011gv}.

\item {\bf HPS at JLab} (direct mediator search): visible dark vector decay searches with fixed target experiment installed within 
the Hall B at JLab using the CEBAF electron beam. Dark vectors are produced in a thin tungsten target and detected by a forward silicon 
tracker and a calorimeter. Sensitive to both prompt and displaced dark photon decays. 
Timeline: engineering runs in 2015 and 2016, physics run taking place 2018-2020. \\
Sensitivity: $\epsilon^2 \gtrsim 10^{-6}$ for prompt decays in the range $18 < m_{A'} < 500~\MeV$. Vertex reach still under evaluation.

\item {\bf MiniBooNE at FNAL} (proton beam-dump): DM scattering in a neutrino detector at the $8~\GeV$ Booster Neutrino 
Beamline at FNAL. MiniBooNE is a 800 ton mineral oil Cherenkov detector situated 490 m downstream of the beam dump. The DM production 
and detection mechanisms are similar to LSND. First results based on $1.8\times10^{20}$ POT have been published for DM-nucleon 
scattering, and on-going analyses of electron elastic scattering ($\chi e \rightarrow \chi e$) and inelastic production of the 
$\Delta$ resonance. \\
Timeline: on-going analyses. Sensitivity: $y \gtrsim 10^{-9}$ for $m_\chi < 400~\MeV$. Reference:~\cite{Aguilar-Arevalo:2017mqx}.

\item {\bf NA64 at CERN} (missing energy): missing energy experiment with a $100~\GeV$ secondary electrons beam from the SPS beam 
line at CERN. The detector consists of a magnetic spectrometer (tracker + bending magnet), followed by a calorimeter system composed 
of an ECAL, a charged track VETO, and a highly hermetic HCAL. The dark mediator is directly produced in the ECAL, and the signal is defined as 
a reconstructed track, an energy deposition in the ECAL below a certain threshold, and no activity in the VETO or the HCAL. 
The sensitivity is currently limited by the beam luminosity. By using a $\sim 150~\GeV$ muon beam instead of an electron beam, NA64 
could also search for a new mediator $Z'$ charged under the $L_\mu-L_\tau$ symmetry or leptophilic dark scalars.\\
Timeline: taking data, expect to collect $10^{11}$ EOT in 2017. Sensitivity: $\epsilon^2 \gtrsim 10^{-10}$ for $m_{A'} < 1~\GeV$.
Reference:~\cite{Banerjee:2016tad,Gninenko:2014pea,Chen:2017awl}.

\item {\bf TREK at J-PARC} (direct mediator search): visible dark vector decay searches at kaon decay experiments 
(TREK/E36 and TREK/E06) at J-PARC. A dark vector could be detectable in kaon and in pion decays, 
{\it e.g.} $K^+\rightarrow \mu^+ \nu (A' \rightarrow e^+e^-)$, $K^+\rightarrow \pi^+ (A' \rightarrow e^+e^-)$ or 
$\pi^0\rightarrow \gamma (A' \rightarrow e^+e^-)$. The experiment is currently analyzing its data, and TREK/E06 is 
planned upon realization of the Hadron Hall extension at J-PARC.\\
Timescale: E36: data currently analyzed, E06: planned. Sensitivity: N/A. Reference:~\cite{Albrow:2016ibs,TREKWeb}.
 
\end{itemize}

\subsection{Future experimental initiatives}

Future opportunities for DM searches are summarized in the following sections. We start by surveying international efforts before 
discussing new US-based proposals. While there has been a growing interest abroad to cover the thermal targets outlined in the 
introduction, relying on current constraints and international efforts only is {\it not} enough to robustly test the scientific 
goals outlined earlier in this chapter. The next generation of US-based experiments, such as BDX, LDMX, COHERENT 
and SBN, is needed to decisively test the direct annihilation scenarios. A key feature of these proposals is the ability 
to leverage existing technologies, enabling their rapid deployment.

\subsubsection{Future international initiatives}

\begin{itemize}

\item {\bf Belle-II at KEK:} missing mass and visible decay searches at the electron-positron 
collider at KEK. Belle-II is a multi-purpose detector with sensitivity to invisible $A'$ decays 
via mono-photon final state in the range $m_{A'} < 9.5~\GeV$. The sensitivity is limited by the 
calorimeter hermeticity and the tracker coverage, as well as the total luminosity. Belle-II can 
also search for visible $A'$ decays and more complex dark sector signatures ({\it e.g.} dark 
Higgs boson $h'$ in $e^+e^- \rightarrow A' (h' \rightarrow A'A')$). The large coupling between the 
SM Higgs boson and the $b$-quark also offers the opportunity to probe the scalar portal in 
$\Upsilon(nS) \rightarrow \chi \bar\chi (\gamma)$ decays. \\
Timeline: First data expected in 2018, and about 50 ab$^{-1}$ of data around 2025. Sensitivity: 
$\epsilon^2 \gtrsim 10^{-9}$ for $m_{A'} < 9.5~\GeV$ with full data set. Reference:~\cite{HeartyBelleII,Seong}.    

\item {\bf MAGIX at MESA:} visible dark photon decay searches with a dipole spectrometer at the $105~\MeV$ polarized 
electron beam at MAMI. The detector is a twin arm dipole spectrometer placed around a gas target. Production mechanism 
similar to HPS and identification through a di-electron resonance. The possibility of a beam dump setup similar to BDX is 
under study.\\
 Timeline: Proposal in 2017 with targeted operations in 2021-2022. 
Sensitivity: $\epsilon^2 \gtrsim 10^{-9}$ for $10 < m_{A'} < 60~\MeV$. Reference:~\cite{Denig:2016dqo}

\item {\bf PADME at LNF:} missing mass searches at the BTF in LNF. The principle is similar to the MMAPS 
experiment, using a $550~\MeV$ positron beam on a diamond target. In addition to invisible $A'$ decays, PADME 
is studying its sensitivity to diphoton decays of axion-like particles and dark Higgs decays.\\
Timeline: Expected to collect $10^{13}$ positron on target by end of 2018. Proposal to move PADME at Cornell 
if new positron beamline is approved. Sensitivity: $\epsilon^2 \gtrsim 10^{-7}$ 
in the range $m_{A'} < 24~\MeV$. Reference:~\cite{Raggi:2014zpa, Raggi:2015gza}.

\item {\bf SHIP at CERN:} DM scattering in neutrino detector at the $400~\GeV$ SPS beamline 
at CERN. The detector consists of OPERA-like bricks of laminated lead and emulsions placed in a magnetic 
field. The DM production mode is similar to that of MiniBooNE, and the detection occurs via electron elastic 
scattering ($\chi e \rightarrow \chi e$). The dominant backgrounds are expected to come from elastic, 
quasi-elastic, deep-inelastic and resonant neutrino scattering processes, and can be reduced using several 
topological and kinematical variables.\\
Expected to be able to deliver $10^{20}$ POT. Timeline: after 2026. 
Sensitivity: $y \gtrsim 10^{-12}$ for $m_\chi < few~\GeV$. Reference:~\cite{Alekhin:2015byh,Anelli:2015pba}.

\item {\bf VEPP3 at BINP:} missing mass and visible decay searches at BINP at Novosibirsk. Dark photons are 
produced by colliding a $500~\MeV$ positron beam on an internal gaseous hydrogen target, and both visible 
and invisible (via the missing mass mode) final state are identified. Elastic scattering will be used for a $17~\MeV$ 
signal search.\\
Timeline: First run is anticipated for 2019-2020. 
Sensitivity: $\epsilon^2 \gtrsim 10^{-8}$ in the range $5 < m_{A'} < 22 ~\MeV$. 
Reference:~\cite{Wojtsekhowski:2012zq}.

\end{itemize}

\subsubsection{Future US-based initiatives}

\begin{itemize}

\item {\bf BDX at JLab} (electron beam-dump): DM scattering in a scintillating crystal detector at the CEBAF(A) beam dump 
at JLab. The detector consists of 0.5 $\rm m^3$ of CsI(Tl) scintillating crystals situated 20 m downstream 
of the beam dump. The experiment is sensitive to elastic DM scattering $e \chi \rightarrow e \chi$ in the 
detector after production in $eZ \rightarrow eZ(A' \rightarrow \chi \bar\chi)$, as well as inelastic or pseudo-dirac 
DM scattering $\chi_1 (e/Z/N) \rightarrow \chi_2 (e/Z/N)$ or excited state decay-in-detector $\chi_2 \rightarrow \chi_1 e^+e^-$ following 
$eZ \rightarrow eZ(A' \rightarrow \chi_1 \chi_2)$ production. It seeks to improve upon E137 sensitivity by benefiting from the 
high intensity JLab beam, and by positioning the detector closer to the dump. 
The sensitivity is ultimately limited by the irreducible neutrino background, expected at the level of $\mathcal{O}(10)$ events for $10^{22}$ 
electrons on target. A different detection technique with directional capabilities based on a large drift chamber 
(BDX-DRIFT) is also explored.\\
Timeline: conditional approval at JLAB (PAC 44 in 2016). BDX-DRIFT 
at proposal stage. Sensitivity: $y \gtrsim 10^{-13}$ for $1 < m_\chi < 100~\MeV$ with $10^{22}$ 
EOT per year. Project cost within small-scale experiment guideline. Reference:~\cite{Battaglieri:2016ggd,Bondi:2017gul}.

\item {\bf COHERENT at ORNL} (proton beam-dump): DM scattering in scintillating crystals and liquid argon detectors 
at the Spallation Neutron Source at ORNL. The primary goal of the COHERENT experiment is to measure the coherent 
elastic neutrino nucleus scattering process. The current experimental setup includes ${\cal O}$(10kg) NaI(Tl) 
and CsI(Tl) detectors, and a 35 kg single-phase LAr scintillation detector. Possible upgrades to a 1-ton LAr 
or NaI detectors are envisioned. The dark matter is mainly produced via $\pi^0/\eta \rightarrow \gamma A'$ decays 
out of collisions from the primary proton beam, and identified through coherent scattering leading to a detectable 
nuclear recoil. The experiment seeks to exploit the large neutrino flux produced in the nearby target. Its sensitivity 
is limited by the DM flux and uncertainties on the neutrino-nucleon cross sections, and beam-unrelated backgrounds.\\
Timeline: currently taking data, upgrade after 2019. Sensitivity: $y \gtrsim 10^{-13}$ 
for $m_\chi < 60~\MeV$. Project cost within small-scale experiment guideline. Reference:~\cite{deNiverville:2015mwa,CoherentWeb}.

\item {\bf DarkLight at JLab} (missing mass) missing mass and visible decay searches at the Low Energy 
Recirculating Facility (LERF) at Jefferson Lab. Dark photons are produced in the reaction 
$e^-p \rightarrow e^-pA'$ colliding the $100~\MeV$ electron beam on a gaseous hydrogen target. 
The main advantage of this setup is the possibility to detect the scattered electron and recoil 
proton, enabling the reconstruction of invisible $A'$ decays via the missing mass technique, and 
providing a robust signature of visible $A'\rightarrow e^+e^-$ decays thanks to the fully reconstructible 
final state. The sensitivity is limited by the very large continuum QED background generated from the high-intensity 
beam. DarkLight is pursued in several stages to demonstrate the feasibility of the approach.\\
Timeline: phase I is currently 
taking data; on-going design studies for phase II. Sensitivity: $\epsilon^2 \gtrsim 10^{-6}$ 
in the range $10 < m_{A'} < 80~\MeV$. Project cost within small-scale experiment guideline. Reference:~\cite{Balewski:2014pxa}.

\item {\bf LDMX at SLAC or JLab} (missing momentum): missing momentum experiment at the DASEL beamline at SLAC or 
at the CEBAF facility at JLab. LDMX proposes to use a low current, high-repetition electron beam to achieve high statistics, 
with an energy in the few $\GeV$ range. DM is produced 
from interactions between a thin target and the electron beam via $eZ \rightarrow eZ (A' \rightarrow \chi \bar\chi)$. The 
experimental signature consists of a soft wide angle scattered electron, characteristic of DM production at an electron fixed-target 
reaction, and missing energy. The detector is composed of a tracker surrounding the target to measure each incoming and outgoing 
electron individually, and a fast hermetic calorimeter system capable of sustaining a $\mathcal{O}(100)$ MHz rate while vetoing 
few-multiplicity SM reactions that can mimic the DM signal.\\ 
Timeline: $>$ 2020. Sensitivity: $\epsilon^2 \gtrsim 10^{-12}$ (phase I) and 
$\epsilon^2 \gtrsim 10^{-14}$ (phase II) for $m_\chi < 400~\MeV$. Project cost within small-scale experiment guideline.
Reference:~\cite{LDMXWeb}.

\item {\bf MMAPS at Cornell} (missing mass): the principle of MMAPS consists of producing a dark vector in 
$e^+e^- \rightarrow \gamma A'$ reactions with a $5.3~\GeV$ positron beam on a fixed Be target. The beam is extracted 
in a slow spill from the Cornell synchrotron. The $A'$ mass is inferred by measuring the outgoing photon kinematics 
with a CsI calorimeter. This $A'$ search method provides sensitivity to all possible decay modes limited only by detector 
resolution and QED background from large-angle photon production, such as $e^+e^- \rightarrow \gamma \gamma$ or 
$e^+e^- \rightarrow \gamma e^+e^-$, where charged final particle(s) sometimes escape undetected.\\
Timeline: proposal stage, no starting date ($>$2020). 
Sensitivity: $\epsilon^2 \gtrsim 10^{-8}$ in the range $20 < m_{A'} < 75~\MeV$. Project cost within small-scale 
experiment guideline.
Reference:~\cite{cornell}.

\item {\bf SBN at FNAL} (proton beam-dump): DM scattering in liquid argon TPC detectors at the $8~\GeV$ Booster 
Neutrino Beamline at FNAL. The SBN program consists of three detectors of 112 ton (SBND), 89 ton (microBooNE), 
and 476 ton (ICARUS-T600) situated at 110 m, 470 m and 600 m downstream the beam dump, respectively. The 
dark matter beam is primarily produced via pion decays out of collisions from the 
primary proton beam, and identified via DM-nucleon or DM-electron elastic scattering in the detector. The neutrino-induced 
background could be significantly reduced by steering the proton beam around the production target in dedicated 
dark matter running modes. SBND is expected to yield the most sensitive results and could improve upon MiniBooNE 
by more than an order of magnitude with $6\times10^{20}$ POT. Further improvement could be achieved by replacing 
the neutrino horn with an iron target or building a new target station to allow simultaneous neutrino 
and dark matter running modes. 
Timeline: Detector commissioning and running in 2018. 
Sensitivity: $y \gtrsim 10^{-12}$ for $m_\chi < 400~\MeV$. Project cost within small-scale experiment 
guideline. Reference:~\cite{Antonello:2015lea,SBNRVW}.

\item {\bf SeaQuest} (direct mediator search): visible dark photon decay searches at the muon spectrometer at 
the $120~\GeV$ Main Injector beamline at FNAL. Parasitic run with the SeaQuest/E1039 polarized target 
experiment. Sensitive to prompt and long-lived dark photon dimuon decays, as well as more complex dark sector 
signatures ({\it e.g.} dark higgs, SIMP). \\
Timeline: Run with SeaQuest in 2017 and E1039 in 2018-2020 if funded. Potential upgrade to E1067 
in 2020-2025. Sensitivity: $\epsilon^2 \gtrsim 10^{-8}$ for $2m_\mu < m_{A'} < 9~\GeV$ (prompt decays), 
$\epsilon^2 \sim 10^{-14} - 10^{-8}$ for $m_{A'} < 2~\GeV$ (displaced decays). Project cost within small-scale 
experiment guideline. Reference:~\cite{doi:10.1142/S0217732317300087,Gardner:2015wea}.

\end{itemize}

\subsection{Facilities}

The facilities required to operate these experiments constitutes an important part of the accelerator 
program. A key aspect of several proposals is the possibility to leverage facilities which have already 
been operating successfully, or are close to starting operations, thus drastically reducing development 
time. A few proposals would require new beamlines at different US laboratories, which could be 
developed in the near future. We stress that these facilities would be used in parasitic/symbiotic mode, 
and current scientific activities would not be impacted by the proposed experiments. 

\begin{itemize}

\item {\bf CEBAF and LERF at JLAB:} The Continuous Electron Beam Accelerator Facility (CEBAF) provides an electron beam 
to four experimental areas with energies up to $12~\GeV$. The electron beam polarization is near 90\% and the  
Hall-D beamline includes a linearly polarized photon beam. Both the HPS and APEX experiments are approved to 
run at CEBAF. LERF is a one-pass energy recovery linac with a maximum electron beam energy of $170~\MeV$, currently 
hosting the DarkLight experiment. The BDX experiment has been recently conditionally approved at JLab. The M{\o}ller 
experiment, which will perform a precise measurement of the Weinberg angle $\theta_W$ and offer sensitivity to light 
DM, has received DOE approval. 
Reference:~\cite{Dudek:2012vr,Freyberger:2015rfv}.

\item {\bf CESR at Cornell:} Cornell University operates a high intensity positron source for the CESR storage ring 
which with the addition of an extraction beam line could provide a $6~\GeV$ extracted positron beam for the DM search. Such 
a beam line will operate parasitically to the CESR SR program. This is the only place in the world where a $\GeV$ energy 
large duty factor positron beam could be arranged at low cost. Reference:~\cite{cornell}.

\item {\bf DASEL at SLAC:} The Dark Sector Experiments at LCLS-II (DASEL) will deliver an almost CW beam of $4~\GeV$  
electrons using the LCLS-II superconducting RF (SCRF) linac in a parasitic mode. LCLS-II 
is under construction at SLAC as part of the photon science FEL program. The approach consists of 
filling unused RF buckets with a low current and diverting them to an experimental area without 
impacting the FEL program. LCLS-II is expected to operate more than 5000 hours per year, 
and an upgrade to increase the beam energy to $8~\GeV$ has received CD-0 from the DOE. 
Timeline: 2020+. Reference:~\cite{DASEL1,DASEL2}

\item {\bf SBN at FNAL:} The SBN facility features 8 GeV protons at the Booster Neutrino Beamline. Three Liquid Argon TPC 
detectors (LArTPC) of 112 ton, 89 ton, and 476 ton are situated 110 m, 470 m, and 600 m downstream the beam dump, 
respectively. Production and detection channels similar to MiniBooNE. The current plan is to collect 6$\times$10$^{20}$ POT, 
beginning in 2018, in on-target mode. Can be configured to collect 2$\times$10$^{20}$ POT in beam-dump mode after 
on-target run. Upgrades to BNB in 2016 will enable simultaneous on/off-target running. It would be possible to replace 
the neutrino horn with an iron target to improve sensitivity of future DM searches. Reference:~\cite{Antonello:2015lea,SBNRVW}.

\item {\bf Very asymmetric collider:} A design of a high-luminosity, very-asymmetric collider has been recently proposed. 
New advances in accelerator technology including the nano-beam scheme, high-current Energy Recovery Linacs, 
and magnetized beams have lead to the proposal of a very asymmetric collider capable or reaching luminosity 
greater than $10^{34}\rm \, cm^{-2}s^{-1}$ at a center-of-mass energy below $1~\GeV$. Such a machine could be deployed 
at any facility with a positron storage ring. Timeline: N/A. Project scale unknown. Reference:~\cite{Wojtsekhowski:2017szs}.

\item{\bf AWAKE at CERN:} The AWAKE experiment at CERN aims to demonstrate proton-driven plasma wakefield acceleration as a viable 
scheme to accelerate electrons to high energy.  By the late 2020s, it may be possible to construct a facility where electrons are 
accelerated up to about 100 GeV in about 100 m of plasma. Such an accelerator could improve by a factor 1000 the luminosity 
delivered to the NA64 experiment. This unique, high energy electron beam may have other applications. Timeline: late 2020s. 
Project scale unknown. Reference:~\cite{Caldwell:2015rkk, Adli:2016wah}.

\end{itemize}

\begin{sidewaystable}

\begin{tabular}{|l|c|c|c|c|c|c|c|c|}
\hline
Experiment & Machine              & Type                & $\rm E_{beam}$ ($\GeV$)   & Detection      & Mass range ($\GeV$)     & Sensitivity  & First beam  & Ref. \\\hline

\hline\multicolumn{9}{|c|}{\vspace{-0.2cm}}\\\multicolumn{9}{|c|}{\bf Future US initiatives}\\\multicolumn{9}{|c|}{\vspace{-0.2cm}}\\\hline

BDX        & CEBAF @ JLab         & electron BD 	& 2.1-11           & DM scatter     & $0.001 < m_\chi < 0.1$  & $y \gtrsim  10^{-13}$              & 2019+     &~\cite{Battaglieri:2016ggd,Bondi:2017gul}\\
COHERENT   & SNS @ ORNL           & proton BD           & 1                & DM scatter     & $m_\chi < 0.06$         & $y \gtrsim  10^{-13}$              & started   &~\cite{deNiverville:2015mwa,CoherentWeb}\\
DarkLight  & LERF @ JLab          & electron FT         & 0.17             & MMass (\& vis.)& $0.01 < m_{A'} < 0.08$  & $\epsilon^2 \gtrsim  10^{-6}$     & started   &~\cite{Balewski:2014pxa}  \\
LDMX       & DASEL @ SLAC         & electron FT         & 4 (8)*           & MMomentum      & $m_\chi < 0.4$          & $\epsilon^2 \gtrsim  10^{-14}$     & 2020+     &~\cite{LDMXWeb} \\
MMAPS      & Synchr @ Cornell     & positron FT         & 6                & MMass          & $0.02 < m_{A'} < 0.075$ & $\epsilon^2 \gtrsim  10^{-8} $    & 2020+     &~\cite{cornell}  \\
SBN        & BNB @ FNAL           & proton BD           & 8                & DM scatter     & $m_\chi < 0.4$          & $y \sim 10^{-12} $                & 2018+     &~\cite{Antonello:2015lea,SBNRVW} \\
SeaQuest   & MI @ FNAL            & proton FT           & 120              & vis. prompt    & $0.22 <m_{A'} < 9$      & $\epsilon^2 \gtrsim  10^{-8} $    & 2017      
&~\cite{doi:10.1142/S0217732317300087}  \\
           &                      &                     &                  & vis. disp.     & $      m_{A'} < 2$      & $\epsilon^2 \sim 10^{-14} - 10^{-8}$&         &  \\

\hline\multicolumn{9}{|c|}{\vspace{-0.2cm}}\\ \multicolumn{9}{|c|}{\bf Future international initiatives}\\\multicolumn{9}{|c|}{\vspace{-0.2cm}}\\\hline

Belle II   & SuperKEKB @ KEK      & $e^+e^-$ collider   & $\sim 5.3$       & MMass (\& vis.)& $0< m_\chi < 10$        & $\epsilon^2 \gtrsim  10^{-9}$     & 2018  &~\cite{HeartyBelleII} \\
MAGIX      & MESA @ Mami          & electron FT         & 0.105            & vis.           & $0.01 <m_{A'} < 0.060$  & $\epsilon^2 \gtrsim  10^{-9} $    & 2021-2022 &~\cite{Denig:2016dqo}  \\
PADME      & DA$\Phi$NE @ Frascati& positron FT         & 0.550            & MMass          & $m_{A'} < 0.024$        & $\epsilon^2 \gtrsim  10^{-7}$     & 2018    &~\cite{Raggi:2014zpa, Raggi:2015gza} \\
SHIP       & SPS @ CERN           & proton BD           & 400              & DM scatter     & $m_\chi < 0.4$          & $y \gtrsim  10^{-12}$             & 2026+     &~\cite{Alekhin:2015byh,Anelli:2015pba} \\
VEPP3      & VEPP3 @ BINP         & positron FT         & 0.500            & MMass          &  $0.005 < m_{A'} <0.022$& $\epsilon^2 \gtrsim  10^{-8}$     & 2019-2020 &~\cite{Wojtsekhowski:2012zq}  \\

\hline\multicolumn{9}{|c|}{\vspace{-0.2cm}}\\\multicolumn{9}{|c|}{\bf Current and completed initiatives}\\\multicolumn{9}{|c|}{\vspace{-0.2cm}}\\\hline

APEX         & CEBAF @ JLab	  & electron FT         & 1.1-4.5          & vis.           & $0.06 < m_{A'} < 0.55$  & $\epsilon^2 \gtrsim 10^{-7}$     & 2018-2019    &~\cite{Essig:2010xa,Abrahamyan:2011gv}\\
\babar\      & PEP-II @ SLAC 	  & $e^+e^-$ collider   & $\sim 5.3$       & vis.           & $0.02 < m_{A'} < 10$    & $\epsilon^2 \gtrsim 10^{-7}$     & done      &~\cite{Lees:2012ra,Lees:2014xha,TheBABAR:2016rlg}\\
Belle        & KEKB @ KEK 	  & $e^+e^-$ collider   & $\sim 5.3$       & vis.           & $0.1 < m_{A'} < 10.5$   & $\epsilon^2 \gtrsim 10^{-7}$     & done      &~\cite{TheBelle:2015mwa}\\
HPS         & CEBAF @ JLab         & electron FT        & 1.1-4.5          & vis.           & $0.015 <m_{A'} < 0.5$   & $\epsilon^2 \sim 10^{-7}$**      & 2018-2020 &~\cite{Battaglieri:2014hga} \\
NA/64      & SPS @ CERN           & electron FT         & 100              & MEnergy        & $m_{A'} < 1$            & $\epsilon^2 \gtrsim  10^{-10} $  & started   &~\cite{Banerjee:2016tad}  \\
MiniBooNE  & BNB @ FNAL           & proton BD           & 8                & DM scatter     & $m_\chi < 0.4$          & $y \gtrsim  10^{-9} $            & done      &~\cite{Aguilar-Arevalo:2017mqx} \\
TREK       & $K^+$ beam @ J-PARC  & $K$ decays          & 0.240            & vis.           &  N/A                    &  N/A                             & done      &~\cite{Albrow:2016ibs,TREKWeb} \\

\hline
\end{tabular}
\label{accTable}
\caption{Summary table of current light DM experiments and future proposals. The sensitivities are quoted either for the kinetic mixing or the variable $y$, whichever 
is most relevant (see the text and the corresponding figures for more detailed predictions). The range quoted for experiments sensitive to both 
visible and invisible decays refers to the invisible case. Starting dates are subject to variations. {\it Legend:} beam dump (BD), fixed target (FT), dark 
matter scattering (DM scatter), missing mass (MMass), missing momentum (MMomentum), missing energy (MEnergy), prompt/displaced visible decays (vis). 
{\it Notes:} *LDMX beam energy is $4~\GeV$ for phase I, and could be upgraded to $8~\GeV$ for phase II. **Sensitivity to displaced vertices under study.}
\end{sidewaystable}

\subsection{Projections}
\label{accproj}

In order to compare the reach of the different proposals, a few assumptions have to be made. When presented in the $m_\chi~\rm{vs}~y$ parameter space, 
the thermal target is an {\it invariant}, while the sensitivity of different experiments is usually not~\cite{Izaguirre:2015zva}. In particular, the DM 
signal yields at accelerator experiments are primarily sensitive to $\epsilon^2$ (missing mass and energy), or to $\epsilon^4 \alpha_D$ (beam dump) for 
a fixed $m_\chi/m_{A'}$ ratio, where $\alpha_D\equiv g_D^2 / 4\pi$ is the analogous dark sector fine structure constant in the vector portal. To express the 
reach of missing mass/energy/momentum experiments on the $m_\chi~\rm{vs}~y$ plane, one must choose a specific value of $\alpha_D$. In the following, we 
purposely adopt {\it a conservative approach} by setting the value of $\alpha_D$ near unity and a $\mathcal{O}(1)$ choice of the ratio $m_\chi/m_{A'}$ 
when required. This choice is conservative in the sense that smaller values of these parameters correspond to stronger experimental sensitivities, i.e. the 
quoted reach is the least constraining. One must also note that fixing the value of $\alpha_D$ near unity 
ensures perturbativity up to the Weak scale~\cite{Davoudiasl:2015hxa}.

\begin{figure}[htb]
\center
\includegraphics[height=6.4cm]{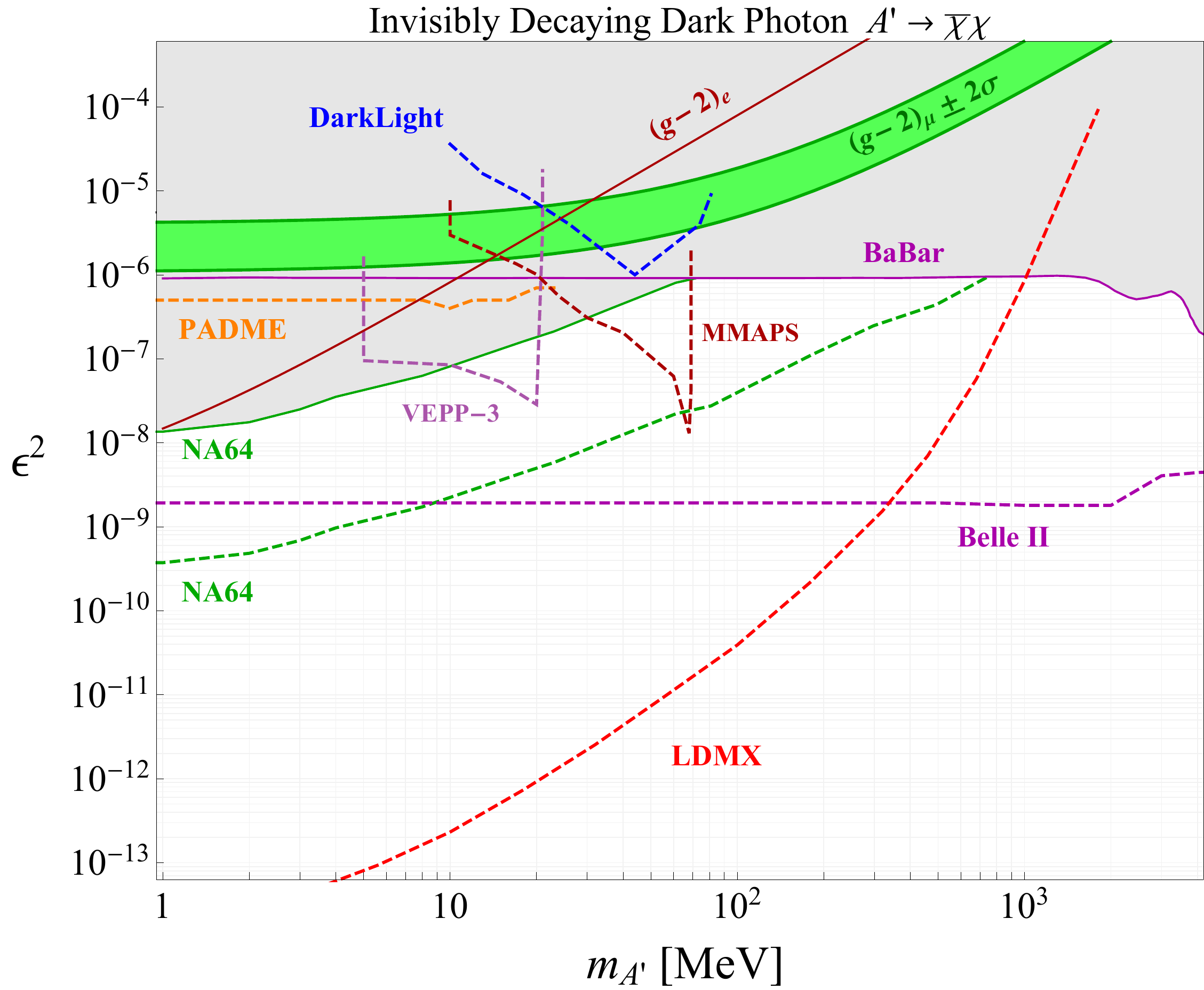} \hspace{0.5cm}
\includegraphics[height=6.4cm]{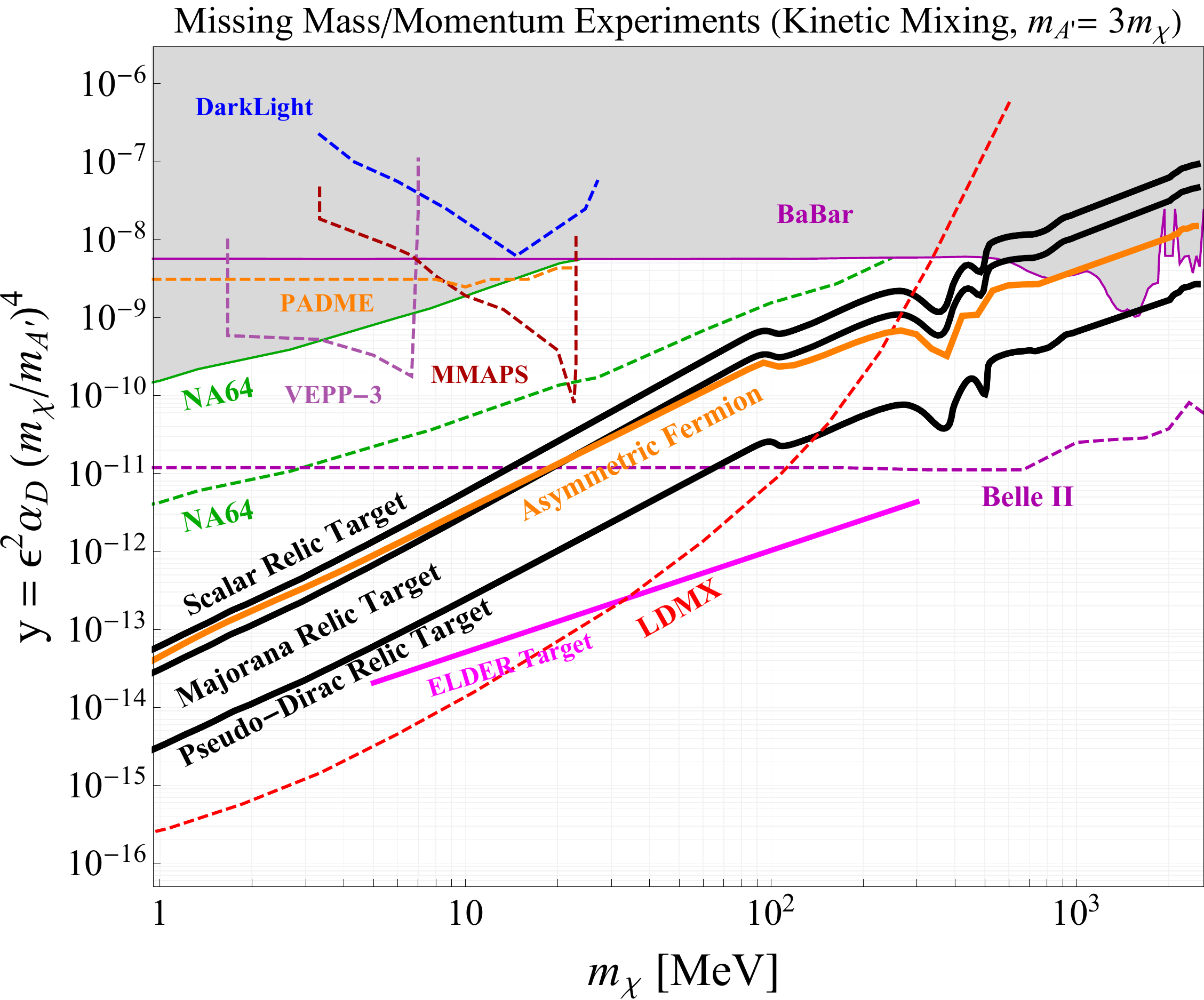}
\caption{Current constraints (shaded regions) and sensitivity estimates (dashed lines) on the SM-mediator coupling 
$\epsilon = g_{\rm{SM}}/e$, for various experiments based on the missing mass, missing energy and missing 
momentum approaches. The green band show the values required to explain the muon (g-2)$_\mu$ 
anomaly~\cite{Pospelov:2008zw}. Right: Corresponding curves on the parameter $y$, plotted alongside various 
thermal relic target. These curves assumes $m_{A'} = 3 m_\chi$ and $\alpha_D = 0.5$. For larger mass ratios or 
smaller values of $\alpha_D$, the experimental curves shift downward, but the thermal relic target remains 
invariant. The asymmetric DM and ELDER targets (see text) are also shown as solid orange and magenta lines, 
respectively. Courtesy G. Krnjaic.}
\label{DMAProj1}
\end{figure}

\begin{figure}[htb]
\center
\includegraphics[height=6.4cm]{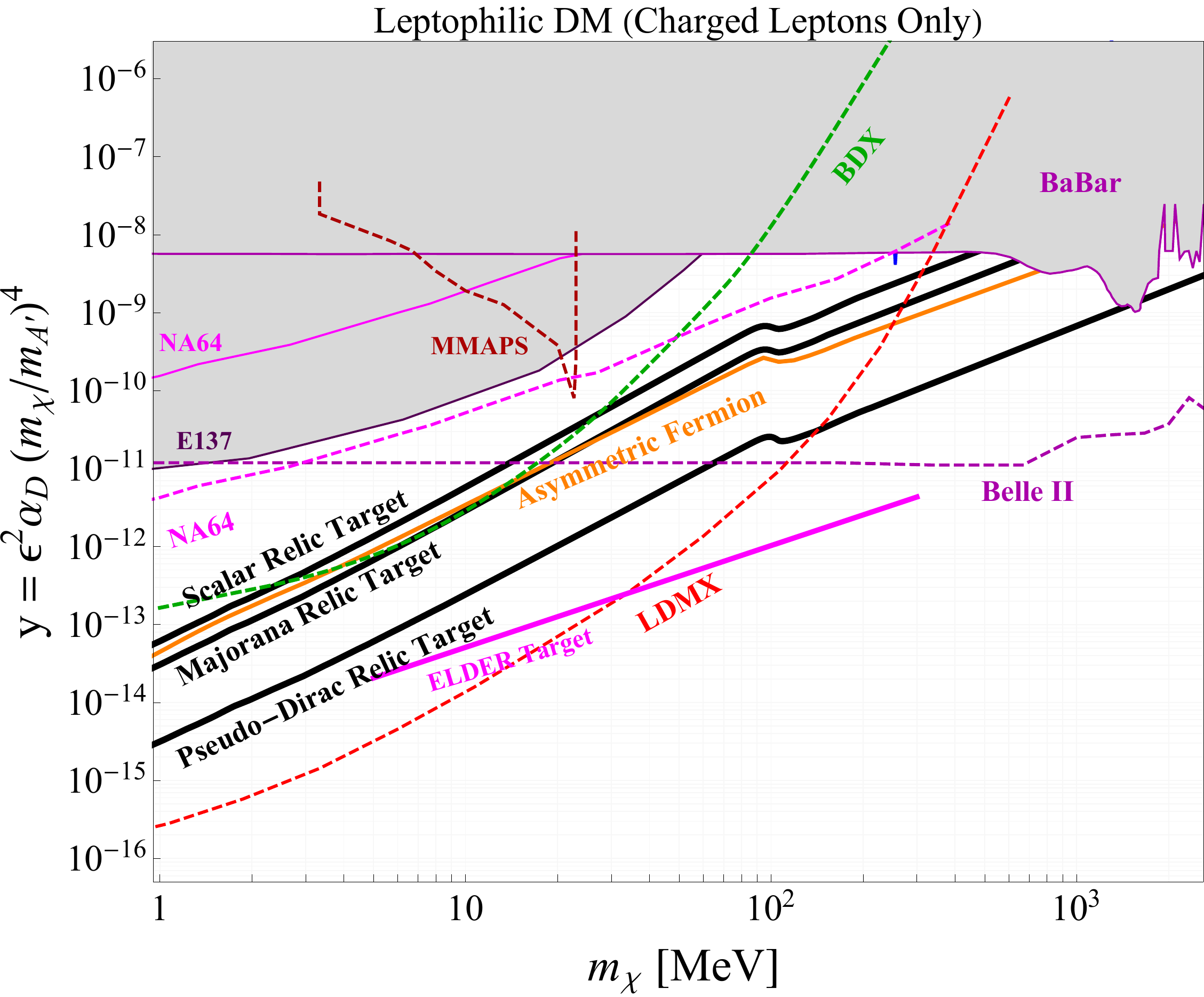} \hspace{0.5cm}
\includegraphics[height=6.4cm]{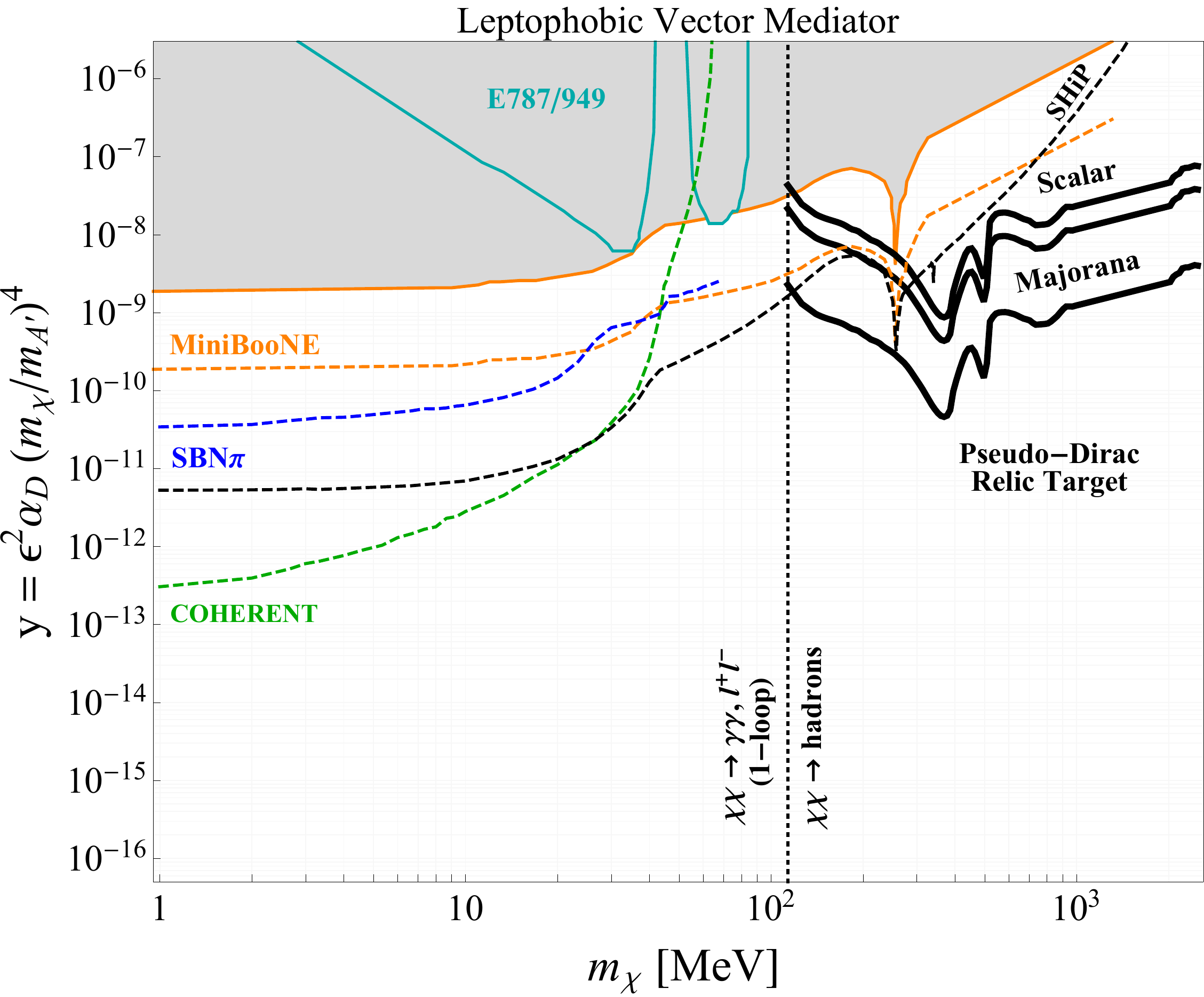}
\caption{Current constraints (shaded regions) and sensitivity estimates (dashed/solid lines) on the parameter $y$ 
for (left) leptophilic and (right) leptophobic dark forces coupled to dark matter for beam-dump experiments. The 
prescription $m_{A'} = 3 m_\chi$ and $\alpha_D = 0.5$ is adopted where applicable. The asymmetric DM and ELDER 
targets (see text) are also shown as solid orange and magenta lines, respectively. Courtesy G. Krnjaic, P. deNiverville.}
\label{DMAProj2}
\end{figure}

\begin{figure}[htb]
\center
\includegraphics[width=7.9cm]{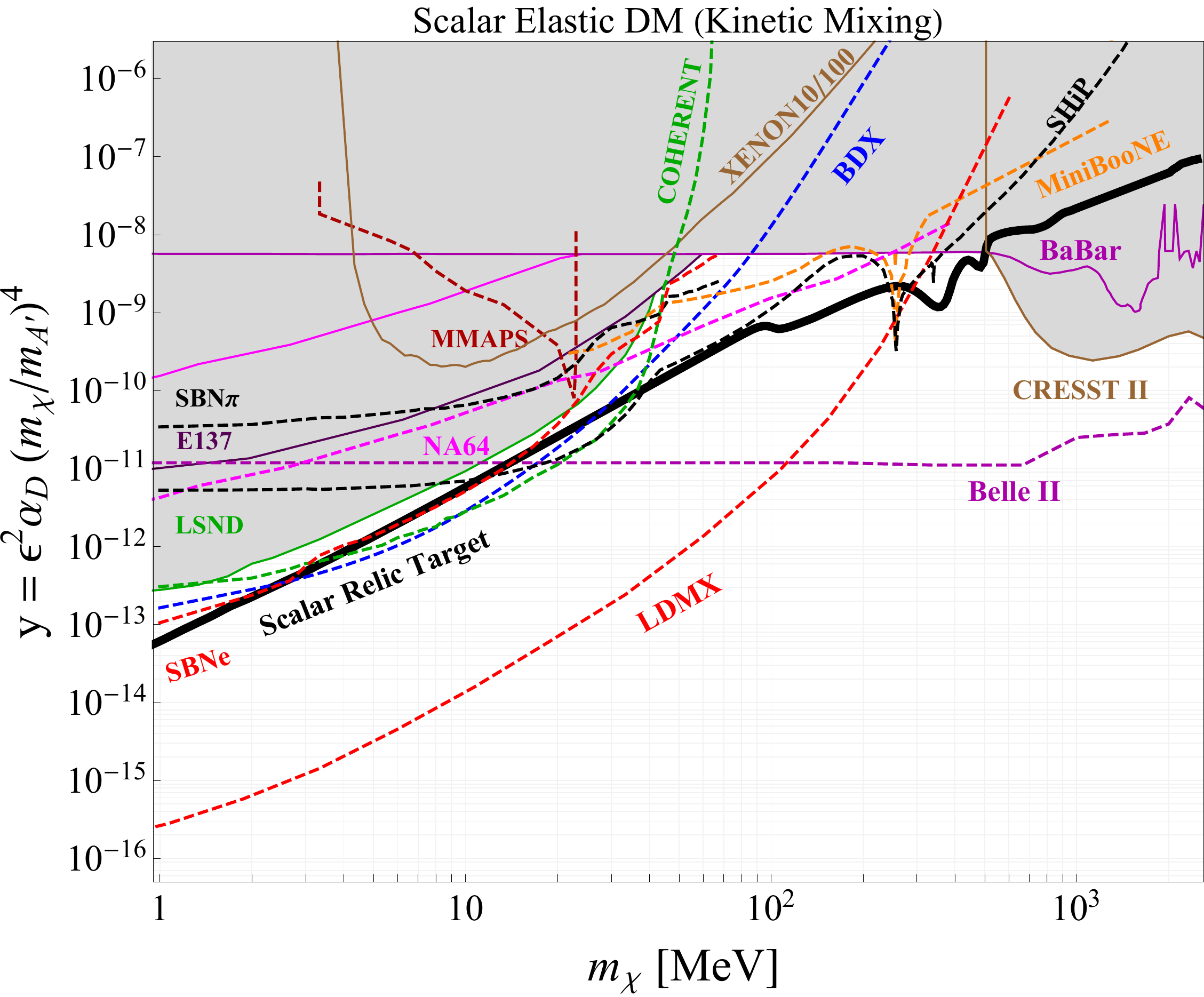}\hspace{0.5cm}
\includegraphics[width=7.9cm]{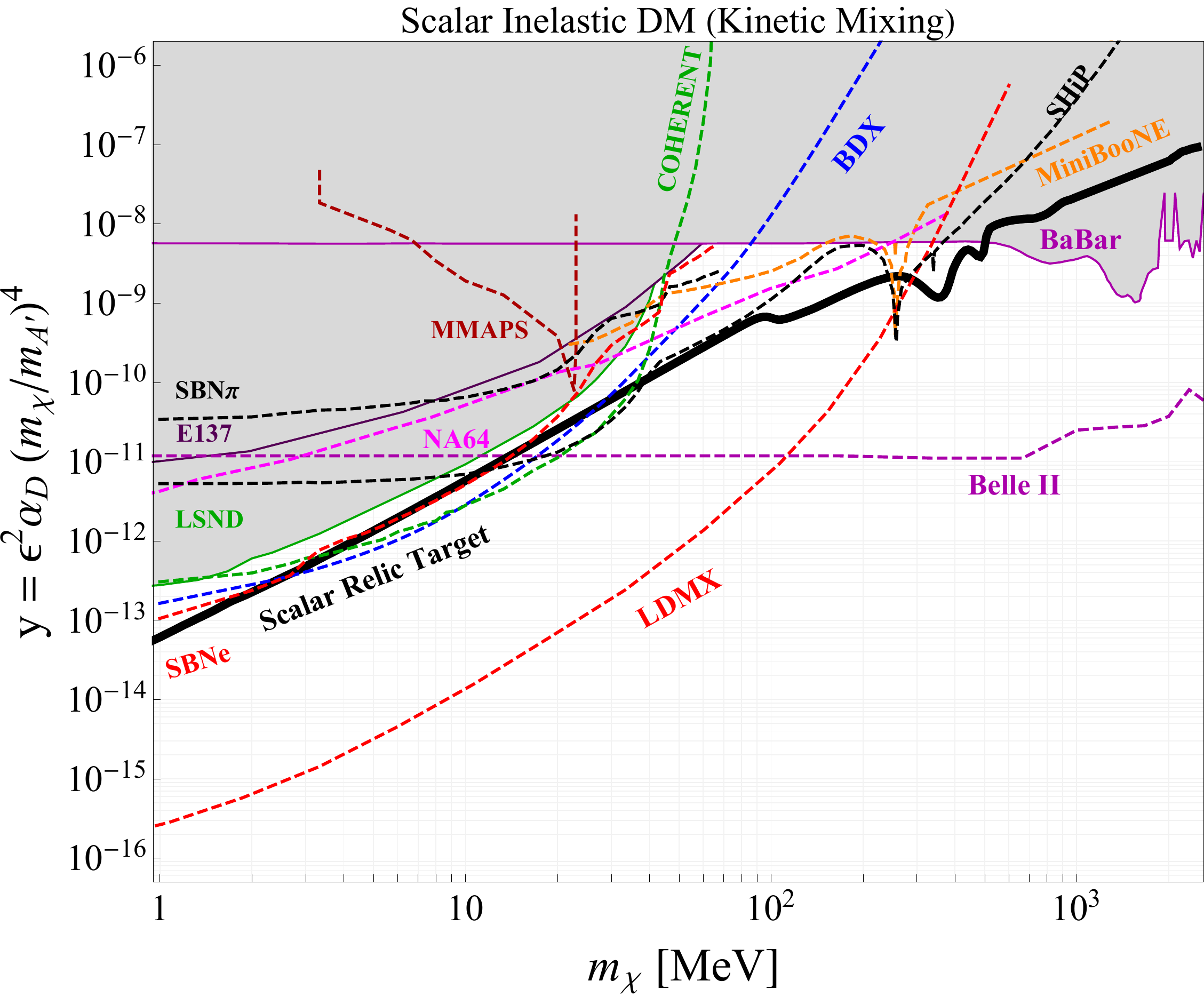} \vspace{0.0cm} \\
\includegraphics[width=7.9cm]{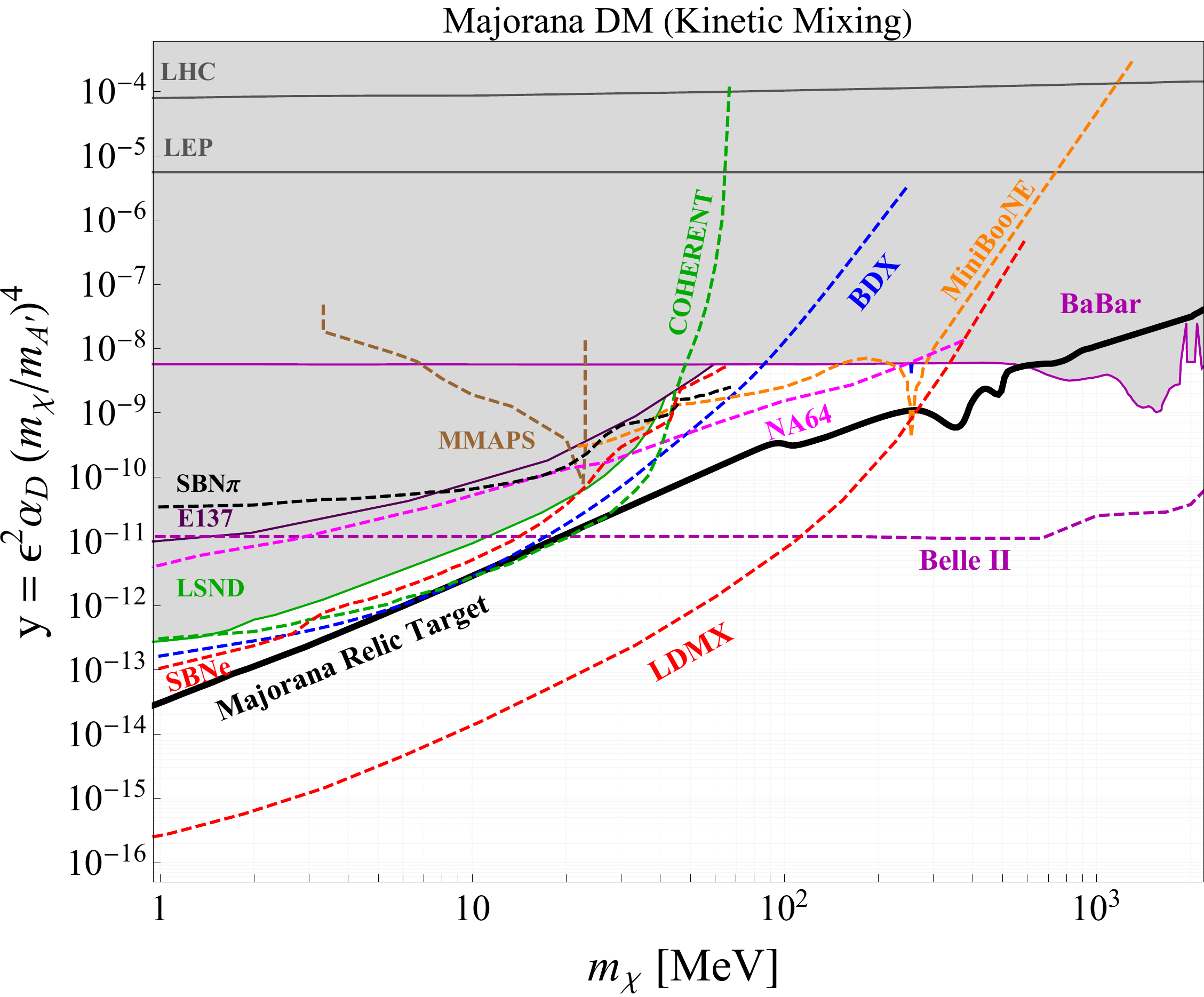}\hspace{0.5cm}
\includegraphics[width=7.9cm]{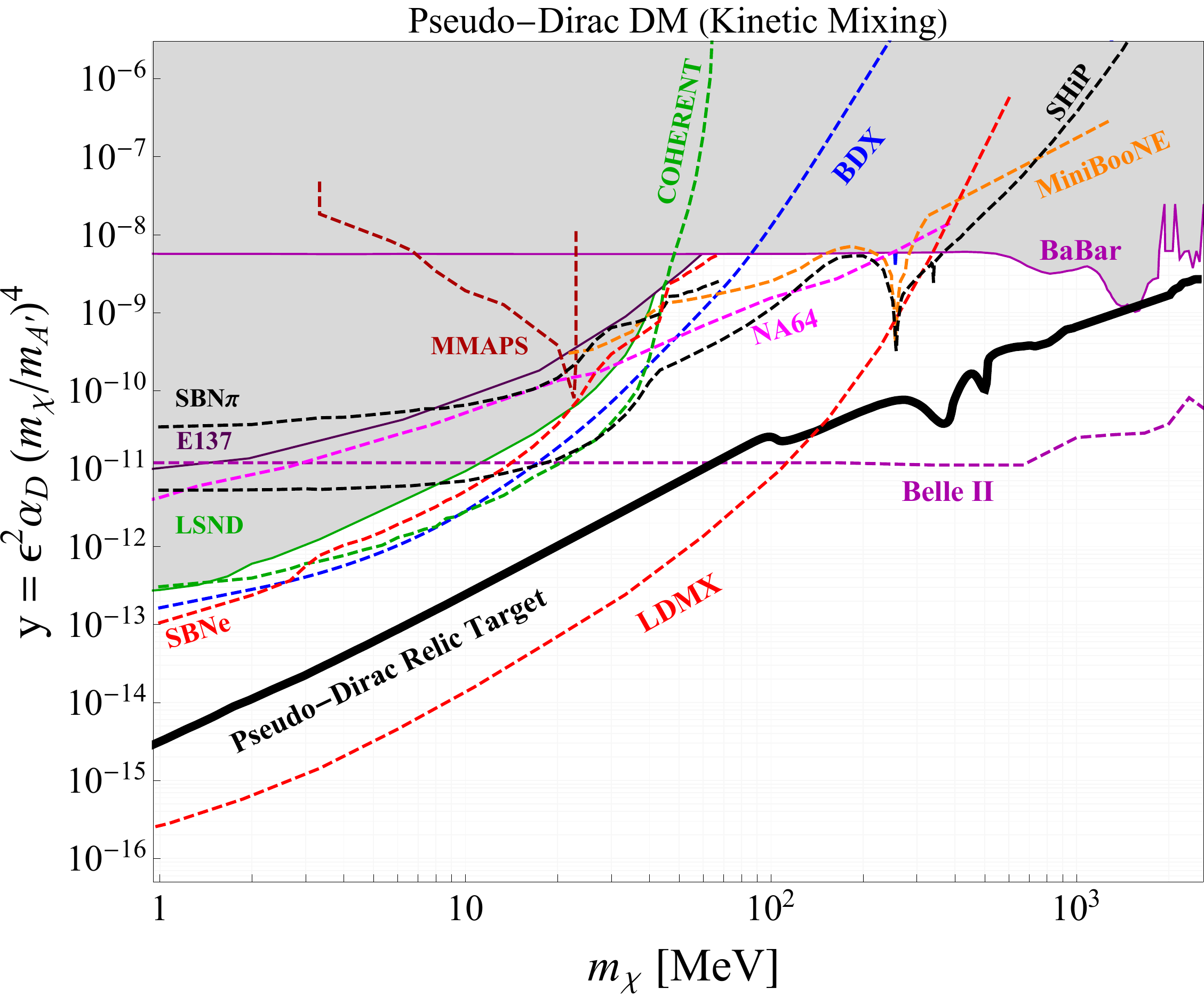} \\
\caption{Combined constraints (shaded regions) and sensitivity estimates (dashed/solid lines) on the parameter $y$  for scalar elastic, 
scalar inelastic, Majorana and pseudo-Dirac DM. The prescription $m_{A'} = 3 m_\chi$ and $\alpha_D = 0.5$ is adopted where applicable. For larger 
ratios or smaller values of $\alpha_D$, the accelerator-based experimental curves shift downward, but the thermal relic 
target remains invariant. See section~\ref{sec:WG2experiments} for sensitivity estimates for direct detection experiments. 
Courtesy G. Krnjaic.}
\label{DMAProj3}
\end{figure}

Current constraints and sensitivity estimates on the benchmark model of a dark photon kinetically mixed with hypercharge are shown in Fig.~\ref{DMAProj1} for 
various light DM experiments based on the missing mass, missing energy and missing momentum approaches. The corresponding curves on the parameter $y$ are 
also shown, assuming $m_{A'} = 3 m_\chi$ and $\alpha_D = 0.5$. Decreasing $\alpha_D$ pushes missing 
mass/energy/momentum curves downward. The analogous figures for electron and proton beam dump experiments are displayed in Fig.~\ref{DMAProj2}, assuming 
only electron (leptophilic) and nucleon (leptophobic) couplings, respectively. The combined projections and constraints are shown in Fig.~\ref{DMAProj3} as 
a function of the DM particle nature.  

\begin{figure}[t!]
\center
\includegraphics[width=16.5cm]{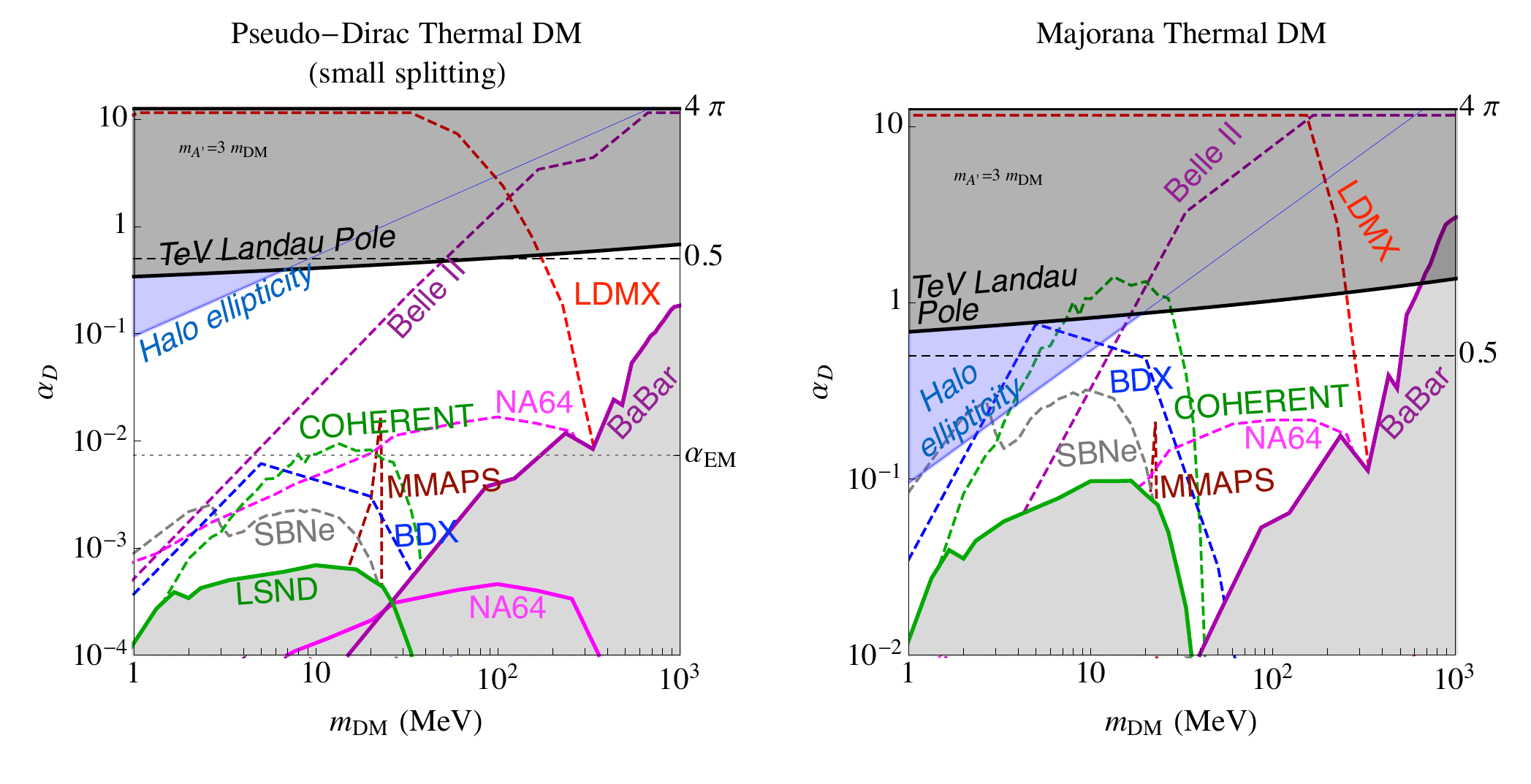}
\caption{Thermal ``area'' target in the $m_{\rm DM}$ vs $\alpha_D\equiv g_d^2/4\pi$ plane, for $m_{A'}/m_{\rm DM} = 3$, for 
pseudo-Dirac (left) and Majorana DM (right). The area in the white background is compatible with a thermal origin. Current 
constraints have largely ruled out the parameter space where the DM-mediator coupling constant is much smaller the $\mathcal{O}(1)$. The 
new proposals surveyed in this chapter aim to cover the region where the DM-mediator coupling is SM-like in strength. Courtesy N. Toro.}
\label{fig:alphadvsmass}
\end{figure}

For much smaller values of $\alpha_D$, the direct annihilation thermal scenarios are largely ruled out by previous experiments, namely 
LSND, \babar\, and E137. Fig.~\ref{fig:alphadvsmass} depicts the thermal-target area in the $(m_\chi,\alpha_D)$ parameter space still viable. Again, we 
fix the ratio $m_\chi = 3 m_{A'}$. Each point in the $(m_\chi,\alpha_D)$ plane corresponds to a SM-mediator coupling $\epsilon$ which has been fixed to 
the value needed to obtain the correct DM abundance. By and large, the open parameter space in these models corresponds to DM-mediator coupling strengths 
that are SM-like.

\begin{figure}[htb]
\center
\includegraphics[width=11cm]{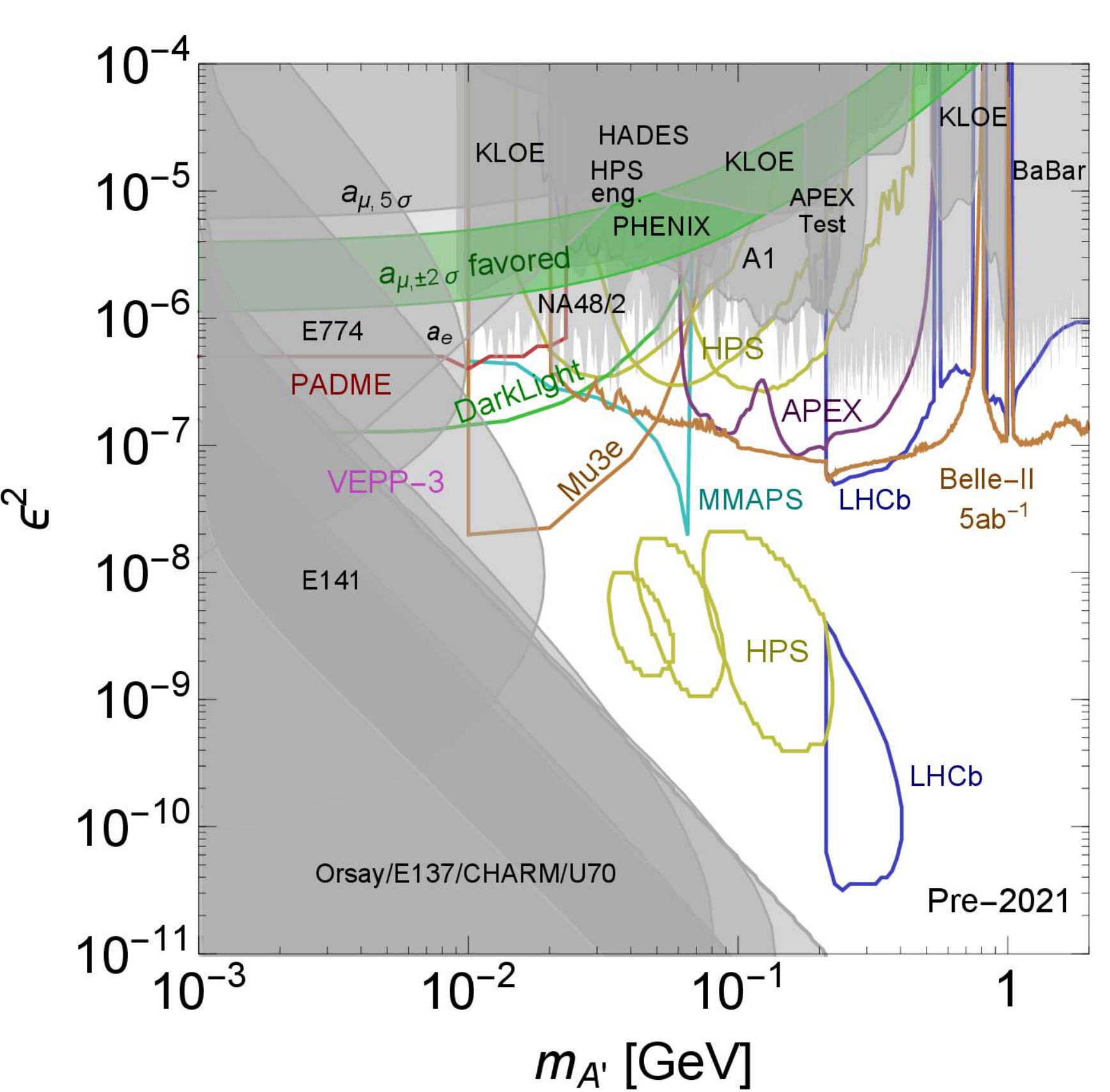}
\caption{Constraints on visibly-decaying mediators (shaded regions) and projected sensitivities of currently running or upcoming 
probes (solid lines). Visible decays of the mediator dominate in the $m_\chi > m_{A'}$ secluded annihilation regime.  Courtesy R. Essig.}
\label{fig:visible-decay}
\end{figure}

It is worth noting that the dimensionless variable $y$ is no longer a suitable parameter for presenting results when 
$m_\chi > m_{A'}$, as the DM annihilation proceeds trough $\chi \bar\chi \rightarrow A' A'$, independent of the kinetic mixing strength. However, 
accelerators can still probe interesting parameter space through off-shell DM production and through direct mediator searches, where the mediator 
decays back to Standard Model Final States. The present status and prospects for visibly-decaying $A'$ searches are shown in 
Fig.~\ref{fig:visible-decay}. These searches are set to continue testing the top-down motivated values of $\epsilon$ in the near future.

\subsection{Summary and key points}

This chapter has reviewed the science case for an accelerator-based program and outlined a path forward to reach 
decisive milestones in the paradigm of thermal light DM. The key points of the discussion could be summarized 
as follows:

\begin{itemize}

\item The scenario in which DM directly annihilates to the SM defines a series of {\bf predictive, well-motivated and bounded targets.} 
Exploring this possibility is an {\bf important scientific priority.}

\item A new generation of small-scale collider and fixed-target experiments {\bf is needed to robustly test this scenario}. The 
accelerator-based approach has the attractive feature of offering considerable model-independence in its sensitivity to the details 
of the dark sector, and can {\bf uniquely probe all predictive models.}

\item Most experimental proposals are based on {\bf existing, proven technology}, and {\bf could be operational in the near future.} 

\item {\bf A complementary approach is required to fully explore light, directly annihilating thermal DM.} Experiments relying on missing 
energy and missing momentum approaches generally offer the most favorable sensitivity. On the other hand, beam dump proposals enable access 
to the DM-mediator couplings and are especially well suited to probe the pseudo-Dirac DM scenario by looking for decays of the excited 
state inside the detector. Moreover, leptophobic DM models are best tested with proton beam-dump proposals. And finally, missing mass 
experiments offer a robust signature and a clean method to precisely determine the mass scale of the mediator in a largely model 
independent way.

\item The strategies discussed in this chapter can be readily applied {\bf to study other well-motivated scenarios 
with quasi-thermal production mechanisms}, such as asymmetric DM; as well as models in which the cosmological abundance is set by processes 
other than DM annihilation into SM particles ({\it e.g.} SIMPs and ELDERs); freeze-in models where the mediator is comparatively heavy 
with respect to DM; new light force carriers; and particles millicharged under electromagnetism. 

\end{itemize}

In summary, small-scale accelerator-based experiments could test important milestones of light DM parameter space. Among them, light 
thermal DM annihilating into SM final states, an important and well-motivated target, could be uniquely and robustly probed. By exploiting 
established detector technology and existing facilities, many proposals are ready for funding now and could achieve significant science 
in the next few years. Through a strong, vibrant contribution, the US dark matter program has the opportunity to play the leading role in 
light dark matter and dark sector physics during the next decade.

%% file: WG4.tex
\section{New Candidates, Targets, and Complementarity}
\label{WG4sec:WG4}

\subsection{Introduction}
\label{WG4sec:introduction}

For many years, research on the particle nature of dark matter focused on three classic candidates: WIMPs, the QCD axion, and sterile neutrinos.  These remain highly motivated particle candidates, and ongoing searches for them continue to be of great interest.  In the last few years, however, many new dark matter candidates have emerged.  This progress is striking for at least two reasons.  First, the motivations for these candidates are extremely varied, with some inspired by experimental discrepancies, others by theoretical considerations, and others representing ``lamppost physics,'' viable ideas that in many cases highlight broad swathes of parameter 
space that have not been experimentally investigated, but can be. Second, these developments have strikingly diverse implications for experiments and observations.  Some have strengthened the well-known, but still remarkable, synergy between particle physics and astrophysics, and others have generated completely new connections between dark matter and other subfields, including nuclear, atomic, and condensed matter physics.  

The recent progress in dark matter makes it difficult to form a coherent picture of the field and to chart future directions.  In this section, we give an overview of recent developments with the goal of providing some of the necessary background for the future prioritization of proposed experiments.  Our subjects include ``New Candidates,'' novel particle physics models and frameworks for dark matter, and ``Targets,'' particle candidates and regions of parameter space (for example, dark matter masses and couplings) that are of special interest, given compelling experimental puzzles or theoretical ideas.  We also discuss ``Complementarity,'' a breathtakingly broad catch-all term that includes the complementarity of proposed (small-scale) experiments with existing (large-scale) experiments; the complementarity of proposed experiments with each other; the complementarity of dark matter probes from the many relevant subfields of physics and astronomy; the complementarity of different experiments in their potential to discover dark matter; and the complementarity of experiments to precisely determine the properties of dark matter after the initial discovery.   

The number of recent developments and the complex inter-relationships between them make it impossible to comprehensively summarize all of them in neatly disjoint categories.  Below we organize our discussion into four broad and overlapping areas: Experimental Anomalies and Hints in~\secref{anomalies}, Cosmology and Astrophysics in~\secref{astrophysics}, Models and Relic Abundances in~\secref{models}, and Complementarity in~\secref{complementarity}.  We close in~\secref{conclusions} with some conclusions, including a few targets that seem especially ripe for experimental searches at this time.

%%%%%%%%%%%%%%%%%%%%%%%%%%
\subsection{Experimental Anomalies and Hints}
\label{WG4sec:anomalies}

Discrepancies between experimental results and SM predictions have been, and should be, among the prime motivations to search for new particles and forces. Examples include the GeV excess seen from the direction of the Galactic Center~\cite{Hooper:2010mq,TheFermi-LAT:2017vmf}, and the 3.5 keV X-ray line seen from galaxies and galactic clusters~\cite{Bulbul:2014sua,Boyarsky:2014jta,Abazajian:2017tcc}. In this section we discuss three leading experimental anomalies with relevance for the experiments discussed in this document: the anamalous magnetic moment of the muon, the proton radius, and the Beryllium-8 anomaly.

\ssection{Anomalous Magnetic Moment of the Muon}  Among the most longstanding puzzles is the $3.5\sigma$ discrepancy between experiment and theory in $(g-2)_{\mu}$, the anomalous magnetic moment of the muon~\cite{Miller:2012opa}. This may be resolved by weakly-interacting particles with milli-charged couplings to muons and 1 to 100 MeV masses~\cite{Pospelov:2008zw}.  Although dark photons with these properties are now excluded~\cite{Alexander:2016aln}, other light bosons remain viable solutions, as discussed below.  In the near future, the Muon $(g-2)$ Experiment at Fermilab is expected to reduce the experimental uncertainty in $(g-2)_{\mu}$ by a factor of four~\cite{Grange:2015fou}.   

\ssection{Proton Radius}  
%\contribution{Richard Hill}
Another muon-related anomaly is the proton radius puzzle, the $5.6 \sigma$ discrepancy between the proton electric charge radius $r_E^p = 0.8751(61)$ measured from a combination of electron scattering and (regular, electronic) hydrogen spectroscopy~\cite{Mohr:2015ccw}, and the radius $r_E^p=0.84087(26)(29)$ measured from muonic hydrogen spectroscopy~\cite{Antognini:1900ns}. (These numbers correspond to the CODATA 2014 adjustment of constants, and the updated 2013 CREMA analysis.)
%  The CODATA 2010 value from electron measurements~\cite{Mohr:2012tt}, $r_E^p = 0.8775(51)$, and the original 2010 CREMA analysis~\cite{Pohl:2010zza}, $r_E^p = 0.84184(36)(56)$, yielded a discrepancy of $6.9\sigma$.)
The large size of this discrepancy and its surprising appearance in seemingly well-known systems have motivated numerous theoretical and experimental efforts across particle, nuclear and atomic physics. The puzzle is likely to lead to a dramatic revision of the fundamental constants $r_E^p$ and the Rydberg constant, and it has forced a reexamination of lepton-nucleon scattering methodology, impacting in particular the long-baseline neutrino program~\cite{Hill:2017wzi}.

\ssection{Proton Radius: New Particle Physics}
Taking the data at face value, it is also interesting to consider potential implications for physics beyond the SM.  In its original form, the puzzle is a discrepancy between electron-based and muon-based measurements of the proton charge radius, $r_{\mu {\rm H}} < r_{e{\rm H}} \sim r_{e-p}$.  This hierarchy has been accommodated in phenomenological models involving $\sim$MeV force carriers with muon specific couplings.  New preliminary results for the hydrogen 2S-4P splitting have been reported by Beyer et al.~\cite{BeyerTalk}, with a ``small'' radius and error comparable to the existing hydrogen average.  The possible revision of the electronic hydrogen results, in agreement with muonic hydrogen, would leave a discrepancy between bound state and electron-proton scattering determinations of the radius: $r_{e{\rm H}} \sim r_{\mu {\rm H}} < r_{e-p}$.  Such a hierarchy would be predicted by an attractive Yukawa force mediated by a force carrier with a mass between the atomic Bohr momentum, $\sim m_\mu \alpha$, and momentum transfers probed in scattering experiments, $\sim 50\,{\rm MeV}$~\cite{Bernauer:2013tpr}.  One such interpretation is a dark photon model and the preferred parameter region is $\kappa/m_{A'} \sim \Delta r/\sqrt{6} \sim 10^{-4}~\text{MeV}^{-1}$, where $\kappa$ is the kinetic mixing parameter and $m_{A'}$ is the dark photon mass. 

%\contribution{Iftah Galon}
\ssection{$^8$Be Anomaly}  The $^8$Be anomaly is a 6.8$\sigma$ discrepancy reported by the ATOMKI group in their observations of the decays of excited $^8\rm{Be}$ nuclei to their ground state and an electron-positron pair, $^8\text{Be}^* \to {}^8\text{Be} \,  e^+ e^-$~\cite{Krasznahorkay:2015iga}.  A bump-shaped excess above the SM internal pair creation background appears in the distribution of $e^+e^-$ opening angles with a very high statistical significance. Its interpretation as a cascade decay of $^8\text{Be}^* \to {}^8\text{Be} \, X $ followed by $X \to e^+ e^-$, where $X$ is a new boson, fits with a $\chi^2/\text{dof} =1.07$ for milli-charged couplings and $m_X\approx 17~\text{MeV}$.  In contrast to the previous two anomalies, where new physics solutions involve virtual particles that can be as heavy as the weak scale, the $^8$Be anomaly, if taken as evidence for new particles, requires real particle production, and can only be resolved by light, weakly-coupled particles.

\ssection{$^8$Be Anomaly: New Particle Physics}  One may consider several spin-parity assignments for $X$ candidates that could account for the observed decay rate~\cite{Feng:2016jff,Feng:2016ysn}.  Scalar candidates are forbidden by parity conservation in nuclear decays~\cite{Feng:2016jff}, but pseudoscalars are a possibility~\cite{Ellwanger:2016wfe}.  Spin-1 bosons are also possible, but are constrained by null results from searches for $\pi^0\to \gamma X$~\cite{Batley:2015lha}; such constraints exclude, for example, dark photons as an explanation.  However, such $\pi^0$ decays are axial-anomaly driven~\cite{Sutherland:1967vf,Veltamn1967}, and so any particles that decouple from this decay, including protophobic gauge bosons~\cite{Feng:2016jff,Feng:2016ysn} and axial vectors~\cite{Kozaczuk:2016nma} are possible solutions.  We discuss these in turn.

\begin{figure}[tbp]
\begin{center}
\includegraphics[width=0.50\textwidth]{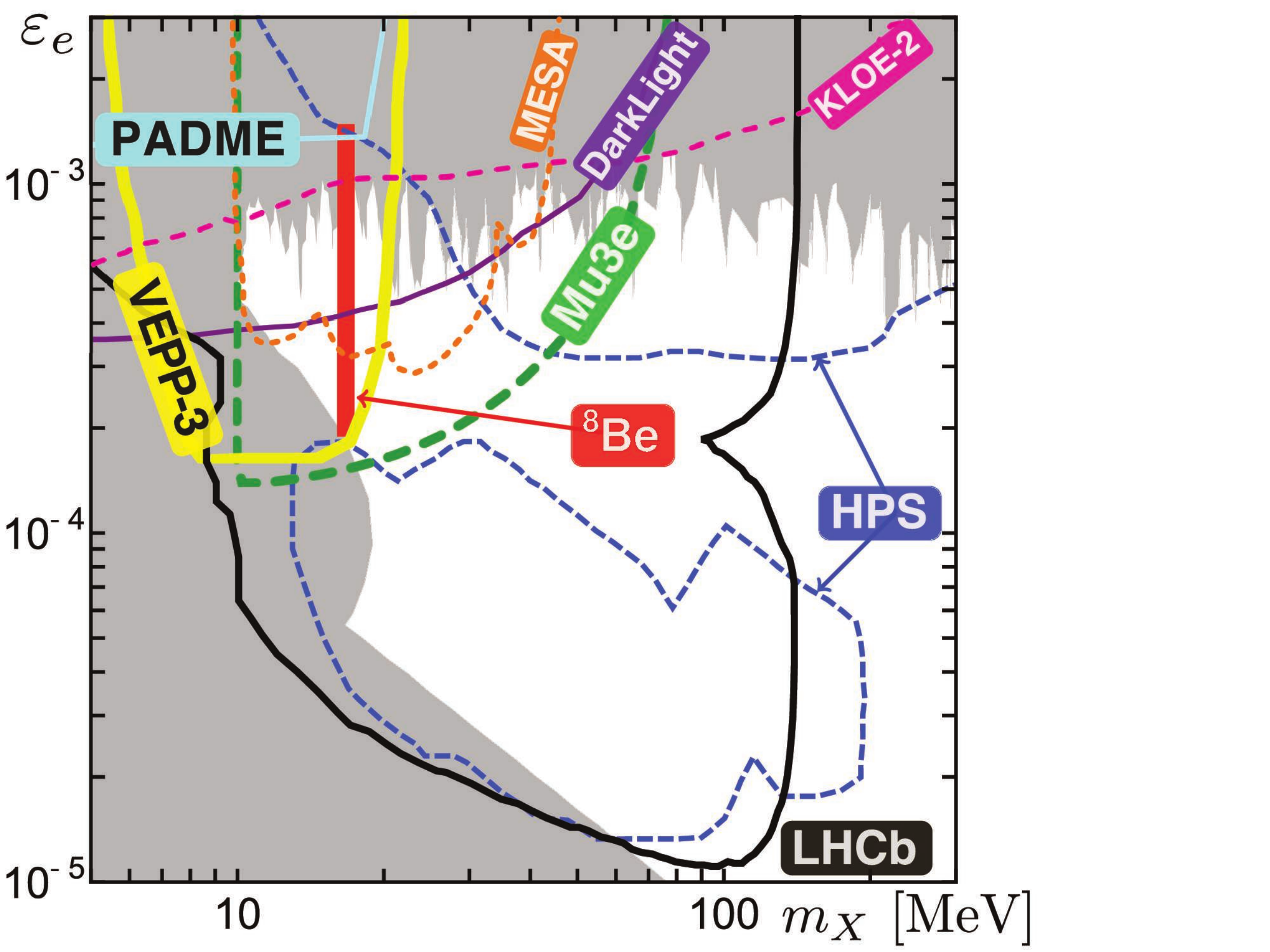}
\caption{The $^8$Be signal region in the $(m_X, \varepsilon_e)$ plane, along with current constraints (gray) and projected sensitivities of the indicated future experiments. From Ref.~\cite{Feng:2016ysn}.}
\label{fig:Be8searchregion}
\end{center}
\end{figure}

A viable protophobic vector candidate has milli-charged couplings to neutrons and electrons, and suppressed couplings to protons~\cite{Feng:2016jff}.  Such a particle can arise naturally as the force carrier of a spontaneously broken U(1)$_B$ or U(1)$_{B-L}$ symmetry that kinetically mixes with the photon~\cite{Feng:2016ysn}. In this case, the predicted leptonic couplings can be large enough to simultaneously ameliorate the discrepancy in $(g-2)_\mu$, providing an viable alternative to the now-excluded dark photon explanation.  These scenarios could be directly tested by repeating the experiment with $^8\rm{Be}$ or looking for similar decays in other nuclei (see below), or by testing the required electron couplings at $e^\pm$-beam-based experiments.  A number of accelerator experiments may probe the relevant couplings in the near future, Fig.~\ref{fig:Be8searchregion}.

%%%%%%%%%%%%%%%%%%%%%%%%%%%%%%%%%
%%%%%%%%%%%%%%%%%%%%%%%%%%%%%%%%%%%

%\contribution{Jonathan Kozaczuk}
An alternative explanation %of the ATOMKI $^8$Be anomaly 
is a light gauge boson that couples predominantly axially to quarks~\cite{Kozaczuk:2016nma}. In this case, the vector does not have to be protophobic, since the decay $\pi^0 \rightarrow \gamma X$ is forbidden in the chiral limit if $X$ has purely axial couplings, and so the constraints from NA48/2 on light vectors~\cite{Batley:2015lha} do not apply.  A light axial vector with mass $m_X\approx 17$ MeV can explain the ATOMKI result without violating existing experimental constraints, and such a particle can also arise from a self-consistent UV complete theory~\cite{Kozaczuk:2016nma}.  (For a related discussion of existing constraints and model-building, see Ref.~\cite{Kahn:2016vjr}.) The strongest constraints on the axial vector quark couplings in this scenario are from the non-observation of a corresponding bump in the predominantly isovector 17.64 MeV $^8$Be transition to the ground state. This illustrates the potential for nuclear decay experiments to provide experimental probes of light vectors that are complementary to those afforded by existing experiments. (Note that the potential for nuclear decay experiments to search for light, weakly coupled particles was pointed out some time ago~\cite{Donnelly:1978ty, Treiman:1978ge}.) Furthermore, both the axial- and protophobic vector interpretations of the $^8$Be anomaly highlight the importance of experimentally targeting both the leptonic and quark couplings of light hidden particles, since the relationship between them is model-dependent.

%\contribution{Xilin Zhang}
\ssection{$^8$Be Anomaly: Nuclear Theory} In investigating the $^8$Be anomaly, it is, of course, critical to know if the SM nuclear theory prediction is accurate.  The theoretical predictions for internal pair creation referred to in Ref.~\cite{Krasznahorkay:2015iga} are based on the classic work of Ref.~\cite{Rose:1949zz}.  This work are known to be incomplete, as it does not include pair emission anisotropy and  interference between the relevant EM multipole transitions.  The nuclear theory predictions have been improved recently~\cite{Zhang:2017zap} in work inspired by the nuclear-cluster-based effective field theory framework~\cite{Hammer:2017tjm, Zhang:2015ajn}.  In this work, the multipole interferences have been included, and the relative weights of the $\ee$ production have been constrained by photon production data, allowing a direct cross check with the weights extracted in (future) $\ee$ experiments. This work puts the nuclear theory expectations on solid footing.  The refined predictions cannot explain the observed $^8$Be anomaly, and the discrepancy with experiment remains.  Furthermore, the possible form factor associated with the M1 transition that would be needed to explain the anomaly requires an unrealistically large length scale (on the order of tens of fm) associated with the $^8\rm{Be}$ nucleus.  The refined model can be used for analyzing future experiments of this type and can also be adapted to study the interplay between the virtual-photon and new-boson decay mechanisms.   

%\contribution{Rafael Lang}
\ssection{$^8$Be Anomaly: Nuclear Experiments} Given the statistically significant discrepancy between experiment and refined nuclear theory predictions, as well as the existence of viable new physics explanations, it is clear that the original $^8$Be results should be followed up with dedicated optimized experiments. In searching for $X$, advantage can be taken of its long lifetime $\sim 10^{-13}$ s, meaning the spectrometer mass resolution dominates the measured mass width in the observation. In one proposed approach from Purdue~\cite{LangUSCosmicVisions}, 7 HPGe detectors with energy resolution $\delta E/E \sim 0.1\%$ are used to accurately measure the energy of the electron-positron pair. In addition, a magnetic field will be used to measure charge, reducing backgrounds due to Compton-scattered electron pairs. Additional particle ID will disentangle the proposed signal from various instrumental backgrounds. Tracks will be measured using two layers of silicon pixel detectors having measuring resolution of 25 micron and the track will be constrained by the production vertex. Because silicon detectors can be operated in vacuum, the spectrometer will not require a vacuum pipe between the detectors and the production target. This configuration will further reduce energy loss of the electron-positron pair in passing through the material of the spectrometer and will greatly reduce Compton-produced backgrounds within the structural elements supporting the charged particle detectors. With these improvements, the overall mass resolution should improve from 1.5 MeV, as observed in the ATOMKI experiment, to $<70$ keV, greatly improving the $\chi$ signal to noise ratio by more than an order of magnitude. Most of the equipment for this proposed spectrometer are already in hand at Purdue and can be installed at the Purdue Tandem facility, PRIME Lab, making the required funds~\cite{LangUSCosmicVisions} a small fraction of the small project threshold and allowing data taking to start rapidly.

%\contribution{Kyle Leach}
A similar proposal~\cite{LeachUSCosmicVisions} to address the $^{8}$Be anomaly, again with higher resolution and statistics, uses a different experimental technique: radiative proton capture on a $^7$Li target at the University of Notre Dame Nuclear Structure Laboratory (NSL). The NSL has a 5 MV single-end 5U accelerator, which is typically dedicated to similar radiative proton capture experiments and can provide proton beams with beam intensities of up to 200 $\mu A$, a factor of 200 increase relative to the original ATOMKI measurement~\cite{Krasznahorkay:2015iga}.  To provide the improvement in $e^+ e^-$ correlation detection,  a simple array of silicon strip detectors in a cubic configuration is used to provide high position resolution, followed by thick Si wafers to provide total energy information for the emitted leptons. The Si strip detectors have a very high granularity and will be configured to increase the angular resolution by up to an order of magnitude. If $\gamma$-ray tagging is required, there is access to high-resolution and high-efficiency HPGe clover detectors through the DOE clover-share program. The proposed setup is almost entirely achievable with existing equipment at the NSL, and the required funds area small fraction of the small project threshold; details are available in Ref.~\cite{LeachUSCosmicVisions}.  This project is currently in the design stage.  However, given the availability of equipment and facility (no program advisory committee is used at the NSL), we estimate that the equipment could be constructed, commissioned, and ready for the first physics run in less than 2 years after funding is available.  If this work is successful, and the measurement has been confirmed, additional light nuclei are planned to be studied with a more sophisticated setup.

%\contribution{Claudia Frugiele}
\ssection{Isotope Shift Spectroscopy}
There exist alternative ways to test for light gauge bosons.  One new frontier is to use precise tabletop atomic physics experiments to test for the existence of new light degrees of freedom.  To convert the high precision of atomic and molecular spectroscopy measurements into sensitivity to fundamental new physics, one has to either calculate atomic structure to high accuracy, or to find observables that are insensitive to theoretical uncertainties.  Recently Refs.~\cite{Delaunay:2016brc,Frugiuele:2016rii,Delaunay:2016zmu} have shown that precision isotope shift (IS) spectroscopy provides a probe of spin-independent couplings of light bosonic fields to electrons and neutrons that does not rely on precise theoretical predictions.  
Being data driven, this proposal has the advantage of not relying of any theoretical prediction for the background. On the other hand, new regions of the parameter space can be explored if and only if the SM background fits a particular (linear) shape in a so-called ``King plot''~\cite{ISKing}, a comparison of isotope shifts of two narrow transitions. This proposal is particularly interesting for the reported $^8$Be anomaly~\cite{Krasznahorkay:2015iga}, because it probes the coupling to neutrons and electrons for a spin-independent interaction, which are precisely the couplings predicted by the protophobic gauge boson interpretation of the data~\cite{Feng:2016ysn}. In the future, by looking at Yb$^+$ transitions with 1 Hz precision, IS measurements will probe all the couplings that could explain the $^8$Be anomaly~\cite{Berengut:2017zuo}, provided the data is compatible with King linearity.

%%%%%%%%%%%%%%%%%%%%%%%%%%
\subsection{Cosmology and Astrophysics}
\label{WG4sec:astrophysics}

At present, all evidence for dark matter is from the impact of its gravitational interactions on cosmological and astrophysical observations.  The astounding progress in cosmology in the past two decades has led to increasingly strong and varied evidence for dark matter and has now precisely determined the amount of dark matter in the Universe.  In recent years, however, advances in cosmology have begun to stringently constrain dark matter's particle properties and production mechanisms and even to motivate new ideas in particle physics, with the field on the threshold of even greater insights in the near future.  In this section, we illustrate the complementarity of astrophysics and cosmological probes with three topics: small scale structure, the cosmic microwave background, and supernovae.

%\contribution{Annika Peter}
\ssection{Small Scale Structure}   The microphysical properties of dark matter not only determine its detectability in the laboratory, but also dark matter's cosmological clustering.  Thus, astronomical probes of dark matter constitute a \emph{measurement} of dark matter's microphysical properties.  The key characteristics of the thermal WIMP dark matter candidate---its relative heaviness and electroweak-scale couplings with SM particles---lead to non-relativistic (``cold'') freeze-out with only minimal kinetic coupling to the SM after thermal decoupling, leaving the primordial inflationary perturbation spectrum untouched by free-streaming or interactions down to tiny scales.  In this ``cold dark matter'' (CDM) paradigm, the non-linear evolution is determined only by gravity.  On non-linear scales, the thermal WIMP/CDM paradigm makes its most striking prediction: the existence of a hierarchy of dense dark-matter halos down to free-streaming and kinetic decoupling ($\sim$Earth-mass) scales~\cite{diemand2005}.  

%%%%%%%%%%%%%%%%%%%%%%%%%%%%%%%%%
\begin{figure}[tb]
\begin{center}
\includegraphics[width=0.65\textwidth]{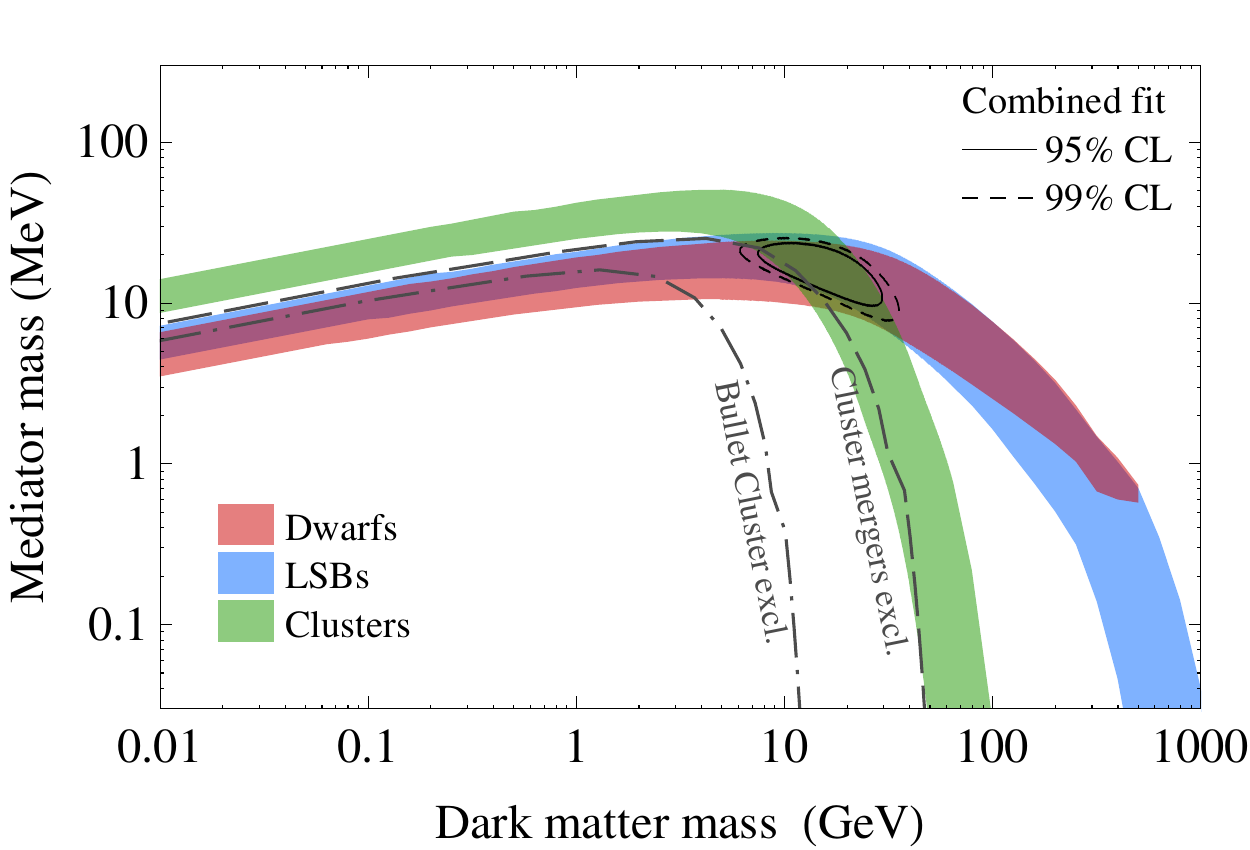}
\caption{The 95\% CL regions of dark matter-mediator parameter space preferred by lower-than-expected central dark matter density in dwarf galaxies (red), low surface brightness spiral galaxies (blue), and clusters (green).  The combined 95\% CL (99\% CL) region is enclosed by the solid (dashed) contours. The dark matter self-coupling strength is taken to be $\alpha'=\alpha$. The regions below the dot-dashed and long-dashed contours are excluded by the Bullet Cluster and the ensemble of merging clusters, respectively. From Ref.~\cite{Kaplinghat:2015aga}.}
\label{fig:DMselfinteractions}
\end{center}
\end{figure}
%%%%%%%%%%%%%%%%%%%%%%%%%%%%%%%%%%%

%\contribution{Manoj Kaplinghat}
The resulting hierarchical structure formation is an excellent description of the distribution and properties of galaxies on large scales. However, there are unexplained puzzles on scales much smaller than the virial radius of galaxies, such as the core-cusp problem, the missing satellites problem, and the too-big-to-fail problem~\cite{Tulin:2017ara}.  Dark matter with significant self-interactions has been argued to retain all the successes of CDM on large scales, while providing an economical solution to the small-scale puzzles~\cite{Tulin:2017ara}. The required ratio of scattering cross section to mass is a few barns per GeV, comparable to the strength for neutron-proton scattering. To be compatible with the measured dark matter densities in relaxed clusters of galaxies, this cross section must decrease with velocity, providing strong constraints on model building~\cite{Kaplinghat:2015aga,Kamada:2016euw}.  For example, for a simple model with dark matter interactions mediated by a single particle, the constraints on the dark matter and mediator masses are given in Fig.~\ref{fig:DMselfinteractions}; the favored mediator mass is $\sim 1 - 100$ MeV.  For direct searches, the implied momentum dependence in scattering off nucleons could be an important discriminant between WIMPs and SIDM candidates.

There are significant opportunities to leverage existing DOE investments in dark energy science for an enormous return on dark matter science.  The first step is to measure the halo mass function on small scales.  DES and LSST together can measure the galaxy luminosity function, and halo mass function, to unprecedented scales.  Halo masses $\lesssim 10^8-10^9 \, M_\odot$ are the domain of substructure lensing searches, which are on the cusp of realising their full potential.  DES is playing a major role in greatly increasing the sample of lens systems suitable for substructure studies~\cite{lin2017}.   DES and LSST will also open the door for novel astronomical probes of dark-matter physics~\cite{Bovy:2016irg,Kim:2016ujt}.  Using astronomy to measure dark-matter physics relies on a strong theory program, to make an accurate map between particle theory space and the astronomical observable and marginalize over uncertainties in the effects of galaxy formation.  A number of the tools required for this exercise are in place or are nearly so (high-resolution hydrodynamic simulations, semi-analytic codes), but require a modest increase in person-power to knit the tools together~\cite{vogelsberger2016,wetzel2016}.

%\contribution{Tracy Slatyer}
\ssection{Cosmic Microwave Background}  Measurements of the cosmic microwave background (CMB) can be used to set strong, robust, and largely model-independent constraints on annihilating or decaying dark matter. These constraints arise from the production of extra free electrons during the cosmic dark ages, by the cooling of electromagnetically interacting particles produced by dark matter annihilation/decay. Consequently, they can be evaded if annihilation is suppressed at late times or at low dark matter velocities (e.g., in the case of $p$-wave annihilation), or if the dark matter annihilates entirely to neutrinos or invisible particles; conversely, they become even stronger if the annihilation rate is enhanced at late times or low dark matter velocities (e.g., in the case of Sommerfeld enhancement due to a light dark mediator). Because the observable effect, to first order, depends only on the total ionizing energy liberated by dark matter annihilations/decays, the constraints only depend on details of the dark matter model via an overall efficiency factor (computed for general scenarios in Refs.~\cite{Slatyer:2015jla,Slatyer:2016qyl}) and are quite insensitive to the spectrum of annihilation/decay products. These bounds are particularly competitive for light dark matter with sub-GeV masses, where many direct dark matter searches lose sensitivity and other indirect searches may have difficulty detecting the products of annihilation. In particular, these limits exclude thermal relic dark matter annihilating to SM particles via $s$-wave processes for dark matter masses below $\sim 10$ GeV~\cite{Ade:2015xua,Slatyer:2015jla}, and decaying dark matter in the keV-TeV mass range with a lifetime shorter than $\sim 10^{{23}-{25}}$ seconds~\cite{Slatyer:2016qyl}, with the limit depending on the dark matter mass and decay channels.  The proposed CMB-S4 experiment~\cite{Abazajian:2016yjj} is expected to extend these constraints on the DM annihilation cross section by approximately a factor of 2 beyond the Planck results~\cite{Madhavacheril:2013cna}.

%\contribution{Sam McDermott}
\ssection{Supernovae}  Supernova 1987A created an environment of extremely high temperatures and nucleon densities during the core collapse supernova of a massive star in the Large Magellanic Cloud. The rough agreement between predictions of core collapse models and observations of  ``neutrino burst'' lasting  $\sim 10$ s has provided an opportunity to set bounds on a wide range of new physics models. In Ref.~\cite{Chang:2016ntp}, updated bounds on a dark sector model involving only a dark photon were presented. Among other novelties, these updates include finite-temperature effects on the production and trapping of the new particles for the first time.  They utilize a more realistic model of the high-mixing parameter space by including a fully energy-dependent differential optical depth, and they have investigated systematic uncertainties inherited from the wide range of progenitor models. Additional improvements include providing an exact calculation of the lifetime of dark photons below twice the electron mass, where derivative corrections to the Euler-Heisenberg Lagrangian are qualitatively important~\cite{McDermott:2017qcg}, and an investigation of the impact of invisible decays on the core collapse explosion~\cite{ChangEssigMcDermott:upcoming}.

%%%%%%%%%%%%%%%%%%%%%%%%%%
\subsection{Models and Relic Abundance}
\label{WG4sec:models}

Precise cosmological observations have not only provided overwhelming evidence for dark matter, but they have also determined the amount of dark matter in the Universe to the percent level.  The relic abundance of dark matter provides yet another criterion for selecting high-value targets: dark matter candidates and parameter regions that have thermal relic abundances in accord with observations merit special attention.  For this reason, axions with $\mu$eV masses have traditionally been viewed as the ideal target for axion searches, and WIMPs with TeV-scale masses have been a prime target of the worldwide effort to find dark matter.

In this subsection, we consider new particle candidates and the masses favored by their relic abundances.  If dark matter is confined to have SM interactions, the weak force is the only viable possibility, and TeV-scale particles are favored by relic density considerations; this is the coincidence known as the WIMP miracle.  Once one considers dark sectors, however, other mass scales may be preferred.  For example, dark matter can be lighter and more weakly interacting and still have the correct relic density, an alternative coincidence known as the WIMPless miracle~\cite{Feng:2008ya}. In the following sections, we review new progress that has found still other motivations for other mass scales and even models in which the mass of dark matter is not well defined. 

%\contribution{Kim Boddy}
\ssection{Non-Abelian Dark Sectors and Strongly Interacting Dark Matter} Dark sectors with Abelian symmetries, leading to dark photons, have become a standard reference model for light, weakly-interacting particles.  However, dark sectors with dark matter charged under a \emph{non-Abelian} SU($N$) gauge group is also perfectly viable. Non-Abelian models are particularly interesting, because they have the potential to undergo confinement at a scale $\Lambda$.  This naturally leads to strongly self-interacting dark matter, as may be indicated by the small scale structure issues discussed above, and more generally, allows dark matter to exhibit substantially different phenomenological features between the early and late universe.  

In a pure gauge theory, the hidden sector consists only of hidden gluons, which form into hidden glueballs below the confinement scale. There are $3\to 2$ scattering processes that keep the hidden glueballs in kinetic equilibrium and deplete their number density; thus, freeze-out is driven by this ``cannibalization'' mechanism rather than standard $2\to 2$ annihilations~\cite{Carlson:1992fn}.  If the hidden sector is secluded, the lightest hidden glueball state is dominant and stable with a mass $\sim\Lambda$, so cannibalization has little effect on the relic abundance~\cite{Boddy:2014yra}.
Alternatively, if the lightest state can efficiently decay into SM particles, the relic abundance of the heavier and presumably longer-lived states are dictated by the cannibalization process~\cite{Forestell:2016qhc}.  In a supersymmetric framework, the hidden sector also consists of hidden gluinos.
Under certain SUSY-breaking scenarios, the standard freeze-out process of gluinos can naturally produce a weak-scale dark matter relic abundance~\cite{Feng:2011ik}. Post-confinement, the self-interaction of hidden glueballinos can address small-scale structure anomalies~\cite{Boddy:2014yra}. As an additional consequence, the hidden glueballino spectrum has a hyperfine splitting of the right order to address the unexplained 3.5~keV line observed in the Perseus cluster~\cite{Boddy:2014qxa}.

%\contribution{Maxim Perelstein}
\ssection{SIMPs and ELDERs}  The possibility of significant $3 \to 2$ processes modifies freeze-out and provides an alternative to the ``WIMP miracle'' that has motivated so much of dark matter research to date.  For example, dark matter may appear with a mass not at the weak scale but near the QCD confinement scale, $\Lambda_{\rm QCD}\sim 100$ MeV. Such a dark matter particle could be a meson or a baryon of a ``mirror copy'' of the familiar QCD in the hidden sector, e.g., in twin Higgs models. In such a scenario, cannibalization processes naturally lead to a thermal relic abundance consistent with observations. This apparent coincidence, similar in spirit to the well-known ``WIMP miracle," was noted in Refs.~\cite{Hochberg:2014dra,Hochberg:2014kqa,Hochberg:2015vrg}, which dubbed such dark matter candidates Strongly Interacting Massive Particles (SIMPs). 

\begin{figure}[t]
\begin{center}
\includegraphics[width=0.65\textwidth]{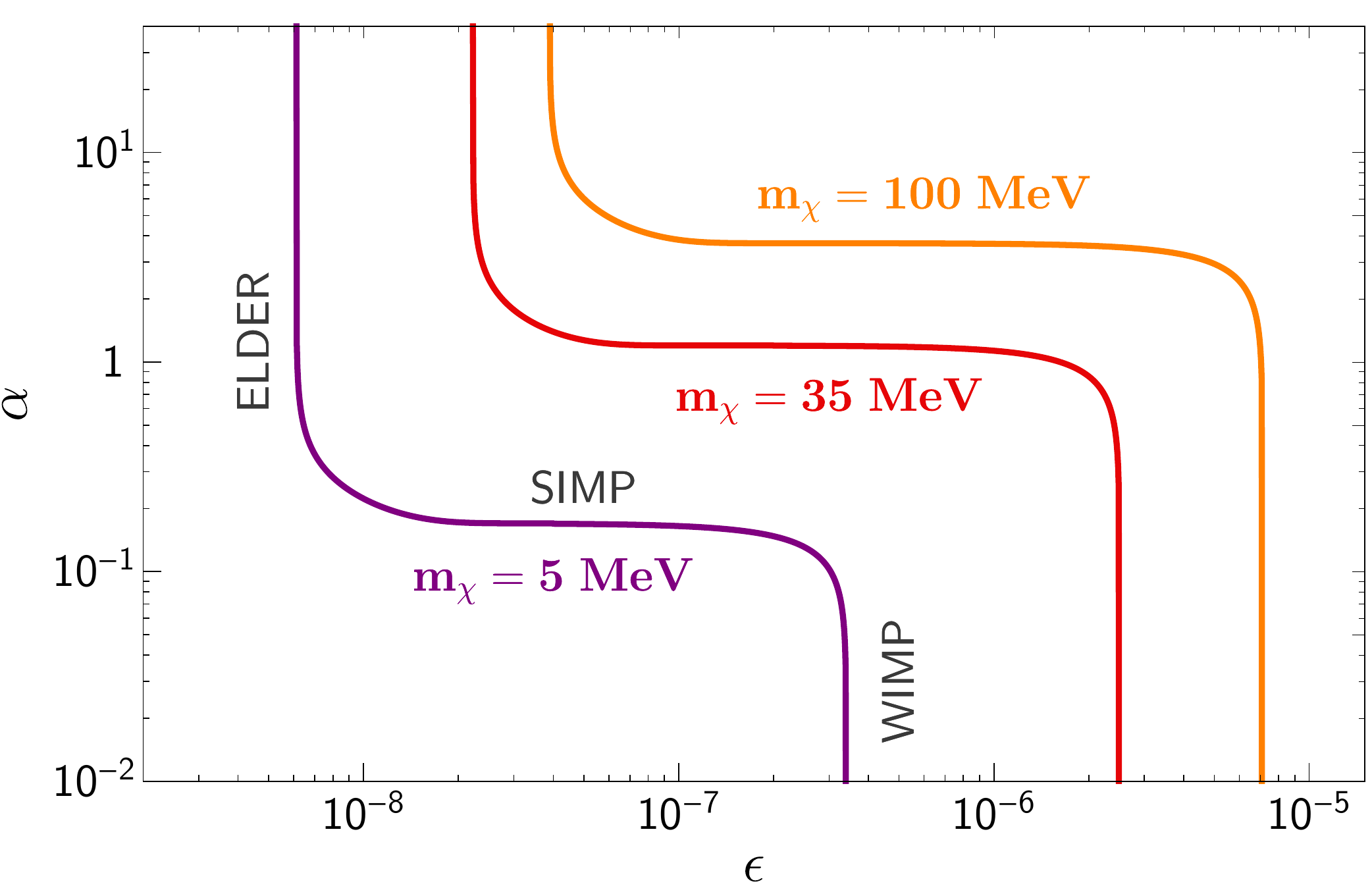}
\caption{Regions of parameters corresponding to the observed relic density, showing the smooth connection between the WIMP, SIMP and ELDER scenarios described in the text.  Dark matter elastic scattering and the cannibalization process scale as $\langle \sigma_{\mathrm{el}} v\rangle=\epsilon^2/m_\chi^2$ and $\langle \sigma_{3\rightarrow 2} v^2\rangle=\alpha^3/m_\chi^5$ respectively. From~Ref.~\cite{Kuflik:2015isi}.}
\label{fig:wimpsimpelders}
\end{center}
\end{figure}

A viable SIMP model requires elastic scattering between the SIMP and SM particles to keep the two sectors in kinetic equilibrium until the $3\to 2$ scattering freezes out. If the kinetic decoupling of the two sectors occurs before freeze-out, the dark matter sector will enter the cannibalization regime. In this case, the relic density is determined by the elastic scattering cross section, leading to the Elastically Decoupling Relic (ELDER) scenario~\cite{Kuflik:2015isi}. Elastic scattering between SIMP/ELDER and SM can be mediated by a dark photon. These scenarios make predictions for dark photon masses and couplings that are shown in Fig.~\ref{fig:wimpsimpelders} and will be probed by next-generation dark photon searches. The ELDER scenario also makes a robust prediction for the cross section of elastic scattering between $\chi$ and electrons, since it is precisely this process that sets the $\chi$ relic density. This prediction will be tested in future direct detection experiments that will search for electron recoils from interactions with ambient dark matter. 

%%%%%%%%%%%%%%%%%%%%%%%%%%%%%%%%%
%%%%%%%%%%%%%%%%%%%%%%%%%%%%%%%%%%%

%\contribution{Nikita Blinov}
\ssection{Non-Abelian Dark Sectors at Fixed Target Experiments} There are a number of ways to realize a dark matter scenario in which the dark matter is kept in kinetic equilibrium with the SM through a dark photon $A'$, but freezes out through the cannibalization process discussed above.  For example, the $3\rightarrow 2$ dark matter depletion mechanism can be realized in QCD-like confining dark sectors, where dark matter is composed of stable pions $\pi$ associated with the spontaneous breaking of a global chiral symmetry. In this case, the correct relic abundance is obtained in models where the vector mesons $V$ typically have $m_V \sim 2m_\pi$. When $m_V < 2m_\pi$, these vector mesons must decay to the SM via $V^0\rightarrow \bar f f$ or $V^\pm \rightarrow \pi^\pm \bar f f$. These dark sector states can be produced in fixed target collisions through the decays $A^\prime \to V\pi, \pi\pi, VV$. The vector mesons are unstable, but naturally long-lived. This gives rise to a displaced vertex signature accessible to present and future fixed target experiments. Different experimental baselines provide complementary coverage of the model parameter space. For example, the SLAC beam dump experiment E137 probed very long-lived vector mesons corresponding to small kinetic mixing parameter $\varepsilon$~\cite{Bjorken:1988as}. The currently operating HPS experiment can search for these displaced decays at larger $\varepsilon$~\cite{Battaglieri:2014hga}.  In particular, a near-future HPS run can probe theories where the hidden sector pions make up all of dark matter.  Future experiments, such as LDMX and long-baseline proton beam dumps like SeaQuest and SHiP are also sensitive to these signals.

%\contribution{Gordan Krnjaic}
\ssection{Co-annihilating Light Dark Matter} Dark sectors with coannihilating thermal relics are well motivated, easy to engineer, and arise in many extensions of the SM, yet they can be difficult to probe with traditional searches. In a representative class of such models, the dark matter abundance arises via $\chi_1 \chi_2 \to$ SM coannihilation, where $\chi_1$ is the stable dark matter candidate, and $\chi_2$ is a heavier unstable dark sector state.  After freeze-out, $\chi_2$ is depleted and annihilation shuts off, so these scenarios are safe from CMB bounds~\cite{Izaguirre:2014dua}, but the absence of $\chi_2$ in the halo also eliminates indirect detection observables. Furthermore, upscattering at direct detection requires $\chi_1 \to \chi_2$ transitions, which are kinematically forbidden for $m_2 - m_1 \gtrsim 100~\keV$, so this mechanism can only be tested with accelerator probes. 
 
 For dark matter masses in the few GeV to TeV range, the most powerful probes involve  searches for $\chi_2 \to \chi_1 + \text{SM}$, where the final state SM particles yield displaced vertices at BaBar and the LHC~\cite{Izaguirre:2015zva}. For masses in the MeV to few GeV range, it was shown in Ref.~\cite{Izaguirre:2017bqb} that powerful searches are possible through $\chi_{i} N \to \chi_j N$ scattering or $\chi_2 \to \chi_1+ \text{SM}$ processes observable in detectors positioned downstream of proton and electron beam dumps at MiniBooNE and BDX~\cite{Aguilar-Arevalo:2017mqx,Izaguirre:2013uxa,Battaglieri:2016ggd}. Similarly powerful probes involve missing energy and momentum at LDMX and NA64~\cite{Izaguirre:2014dua,Banerjee:2016tad}. The combined reach of these efforts can comprehensively test the thermal relic parameter space for $\chi_1\chi_2 \to$ SM coannihilation.

%\contribution{Glennys Farrar}
\ssection{Sexaquark Dark Matter} Without considering dark sectors, there exists a natural non-Abelian gauge group that could be involved in dark matter dynamics---QCD of the SM.  It has recently been pointed out that there may be an as-yet-undiscovered \emph{stable} particle in the SM, which would be an excellent dark matter candidate~\cite{fSDM17}.  The particle, called $S$, is a neutral, scalar sexaquark ($uuddss$) with baryon number $\text{B}=2$ and strangeness $\text{S}=-2$~\cite{fSableS17}.  Baryon number conservation implies that the $S$ is absolutely stable if its mass is $ \le 2 \, (m_p + m_e) = 1877.6$ MeV.   Its relic abundance can naturally be of the right order, and during nucleosynthesis it acts as inert relic dark matter, so nucleosynthesis constraints on baryons do not apply.   Simulations indicate that hadronic interactions with gas in the galaxy can bring the dark matter in the solar neighborhood into co-rotation, so it has too little energy to have been detected so far~\cite{fSDM17,wfCoRotation17}.    

Two accelerator experiments are proposed to discover the $S$ if it exists through the processes $K^-  p \rightarrow S \bar{\Lambda}$, and $\Upsilon [\rightarrow {\rm gluons}] \rightarrow S \, \bar{\Lambda} \, \bar{\Lambda}$ (or charge conjugate).  The $\Lambda \, (\bar{\Lambda})$'s can be reconstructed with high efficiency, and their 4-momenta well measured.   If all final particles but the $S$ are detected, missing mass gives the mass of the $S$; B and S conservation establish its distinctive quantum numbers.  The $K^-  p \rightarrow S \bar{\Lambda}$ experiment can be done with the NA61 detector and beam;  all that is needed are simulations to optimize background rejection and a dedicated run with appropriate beam and target.  Rate estimates for $\Upsilon$ decay suggest that events with $S$ or $\bar{S}$ in the final state have already been collected; only the resources for their analysis is needed.  The two approaches are complementary and can be completed quickly and at very low cost.

%\contribution{Keith Dienes}
\ssection{Dynamical Dark Matter}  Although many non-standard dark matter scenarios transcend the traditional WIMP or axion frameworks and involve new regions of dark matter parameter space, perhaps none do so as dramatically as those that arise within the Dynamical Dark Matter (DDM) framework~\cite{DDM1,DDM2}.  In DDM scenarios, the requirement of dark matter stability is replaced by a balancing of lifetimes against cosmological abundances across a large ensemble of individual dark matter species that exhibit a broad spectrum of masses, lifetimes, and abundances.  Thus, in such scenarios, it is the entire DDM ensemble that collectively serves as the dark matter ``candidate,'' albeit one which cannot be characterized by a single mass or cross section.

This change in the nature of the dark-matter candidate has numerous consequences for experimental dark matter searches.  First, there are fundamental differences in the search strategies best suited for discovering DDM ensembles at traditional dark matter experiments.  For example, in collider experiments, the  distributions of relevant kinematic variables can be significantly modified~\cite{DDMLHC}, with major changes for standard experimental handles (such as the ``mass edge'' often apparent in the invariant-mass distributions of particles produced alongside the dark matter constituents).   Likewise, at direct detection experiments, dark-matter recoil energy distributions can also be modified in dramatic ways~\cite{DDMDD}.  Second, and perhaps even more interestingly, entirely new experimental techniques may also now become relevant for probing the DDM dark sector.  For example, a proposed experiment such as MATHUSLA~\cite{MATHUSLA}---a surface detector designed to detect long-lived (but not cosmologically stable) particles at the  LHC by searching for displaced vertices on $\mathcal{O}(10^2\mathrm{~m})$ length scales---can serve as a probe of certain otherwise inaccessible regions within the DDM ensemble.  This leads to additional complementarities between existing and new probes of the dark sector. 

%%%%%%%%%%%%%%%%%%%%%%%%%%
\subsection{Complementarity}
\label{WG4sec:complementarity}

As noted in \secref{introduction}, classic dark matter candidates and existing experiments remain interesting, because ongoing and planned searches continue to probe highly motivated regions of parameter space, these programs will continue to dominate the funding profile in dark matter for the foreseeable future, and they provide the context in which new experiments and complementarity are to be evaluated.  In addition, there is continual progress in both theory and experiment in these areas.  In this section, we discuss recent developments in this area, as well as a few ``exotic'' dark matter candidates where novel experimental searches have been proposed.

%\contribution{Howie Baer}
\ssection{Mixed WIMP/Axion Dark Matter}
Models with supersymmetry yield neutralino WIMP dark matter, and models with Peccei-Quinn symmetry~\cite{Peccei:1977hh} yield axion dark matter. An attractive possibility is to consider models with both supersymmetry and PQ symmetry that simultaneously solve the gauge hierarchy and strong CP problems~\cite{Bae:2013hma}.  Natural models of supersymmetry have a light Higgsino ($\sim 100-200\GeV$), and the Higgsino may be the only supersymmetric state at the weak scale.  Such a Higgsino-like WIMP would be the LSP and, if thermally produced, would only contribute as a sub-dominant part of the dark matter.  However, in supersymmetric models with PQ symmetry, the axion can contribute as the remainder of the dark matter abundance.  When combined with supersymmetry to form the SUSY DFSZ solution~\cite{Kim:1983dt}, the breaking of PQ symmetry generates the Higgsino $\mu$-term, $\mu\sim f_a^2/M_p$, and the SUSY spectrum admits a natural little hierarchy $\mu\sim f_a^2/M_p \sim 100-200 \GeV \ll m_{\text{SUSY}}\sim m_{3/2} \sim 1-20 \TeV$.  Some experimental consequences of this setup are that Higgsino dark matter could be detectable by multi-ton-scale noble liquid detectors, the axion lies in the range $3\times 10^{-7}\eV \lsim m_a \lsim 3\times 10^{-4}\eV$, and the presence of Higgsinos in the loop generating the $g_{a\gamma\gamma}$ coupling may further suppress this coupling, requiring deeper axion probes~\cite{Bae:2017hlp}.

%\contribution{Mark Boulay/Dark Side}
\ssection{Future Argon Direct Detection Experiments}  In addition to new regions of dark matter parameter space, we must continue to search for ``conventional" dark matter using ever more precise techniques.
Researchers from four collaborations who have pioneered the development of argon dark matter searches are in the process of forming a joint collaboration towards a coordinated global future dark matter program.  Argon is unique in that it allows excellent pulse-shape discrimination using scintillation light in a single-phase detector, and a TPC exploiting the ratio of primary scintillation and ionization (S1/S2) can be used to increase background rejection.  The collaboration, numbering over 350 researchers, brings together complementary expertise from miniCLEAN, DEAP-3600, ArDM and DarkSide.  

The new collaboration will develop and operate DarkSide-20k at LNGS (DS-20k).  The DS-20k program will enhance our sensitivity to WIMPs, particularly at high WIMP mass, using 20 tonnes of UAr and will also be the first large-scale detector to make use of Silicon Photomultipliers (SiPMs) for light readout.    DS-20k has a dark matter sensitivity competing with that of future searches using xenon, to which it is complementary with a ``background-free'' technology (detection in both targets allows better determination of mass and cross section).  The DS-20k program also complements LHC searches, with the direct argon search sensitive to a higher mass range than accessible with colliders.  DS-20k is designed to collect an exposure of 100 tonne-years completely free of neutron-induced nuclear recoil background and all electron recoil background. DS-20k is set to start operating by 2021 and, for 1 TeV WIMPs, will probe WIMP-nucleon spin-independent cross sections of $1.2\times 10^{-47}$ cm$^2$ ($7.4\times 10^{-48}$ cm$^2$) after 5 (10) years.  The collaboration is also targeting a longer-term multi-hundred tonne LAr detector that will follow DS-20k, which reaches down to the neutrino floor and is immune to the solar $pp$ elastic scattering neutrino background, which is a concern for xenon detectors with 1/2 event per tonne-year after recoil discrimination. The program includes further development of underground argon, SiPM photosensors, and low background materials screening.  DOE support in these areas will be extremely valuable to this effort.

\ssection{Future Two-phase Xenon Experiments}  As well as proposed new targets and technologies, existing approaches are being strengthened and extended.  For WIMP masses above 4 GeV, searches for spin-independent and neutron-spin-dependent dark matter interactions are being led by two-phase xenon experiments, such as XENON~\cite{Aprile:2015uzo}, LUX~\cite{Akerib:2016vxi}, and PandaX~\cite{Tan:2016zwf}. At a WIMP mass of 50 GeV$/c^2$, WIMP-nucleon spin-independent cross sections above $1.1\times 10^{-46}\mathrm{cm}^2$ are excluded by LUX, which had a 250 kg active xenon mass.  PandaX-II is currently operational with a 500 kg active xenon mass, and the PandaX collaboration is designing a new experiment at CJPL, PandaX-nT. XENON1T is beginning operations with a 2000 kg active mass, has a projected sensitivity of $1.6\times 10^{-47}\mathrm{cm}^2$ at 50 GeV, and is planned to be upgraded to XENONnT, which will use the same infrastructure at LNGS.  Currently under construction, the LUX-ZEPLIN (LZ) experiment~\cite{Mount:2017qzi} will use a 7000 kg dual-phase xenon time projection chamber for the direct detection of dark matter.  Suppression of backgrounds is achieved through fiducialization and a veto strategy involving anti-coincidence between the main time projection chamber and outer detectors (an instrumented xenon ``skin'' and liquid scintillator detector). LZ is projected to have a baseline sensitivity of  $2.3\times 10^{-48}$ $\mathrm{cm}^2$ for a 40 GeV/$c^2$ WIMP mass. Operation of LZ will begin in 2020 at the Sanford Underground Research Facility (SURF).

%\contribution{Brian Humensky}
\ssection{Cherenkov Telescope Array} The Cherenkov Telescope Array (CTA)~\cite{2013APh....43....3A} will provide an order-of-magnitude improvement in sensitivity over current imaging air Cherenkov telescopes in the 100 GeV -- 10 TeV energy range, along with new sensitivity from 20 GeV up to 300 TeV. An 8$^\circ$ field of view combined with 2-3 arcmin angular resolution enables efficient surveys and studies of extended and diffuse emission. Better than 10\% energy resolution above 100 GeV enables resolution of spectral features. For deep observations ($\sim$500 hrs) of the Galactic Center, CTA will have the sensitivity to reach the thermal relic cross section for a broad range of WIMP particle masses and annihilation channels~\cite{2015PhRvD..91l2003L}. CTA is especially powerful in searching for WIMPs with masses at the TeV scale and higher, making it a necessary complement to other techniques to span the full dark matter discovery parameter space~\cite{2013arXiv1305.0302W,2015PhRvD..91e5011C}. CTA will contribute more broadly to dark matter science via measurements of the cosmic-ray electron spectrum to several 10's of TeV or higher, depending on whether local sources or more exotic production mechanisms (such as dark matter) contribute, and via searches for axion-like particles through studies of the $\gamma$-ray opacity of the universe~\cite{2013APh....43..189D}.

%\contribution{Stefano Profumo}
\ssection{MeV Gamma-Ray Detectors} Gamma-ray observations in the MeV energy range with future telescopes, such as e-ASTROGAM~\cite{DeAngelis:2016slk}, offer opportunities to search for signals from the annihilation or decay of particle dark matter. Intriguingly, MeV ``excesses'' have been identified, both in the MeV diffuse extragalactic gamma-ray background, as well as in the Galactic MeV emission compared to expected astrophysical background~\cite{Strong:2004ry,Strong:2004de,Lacki:2012si}.  Such excesses could be associated, for example, with dark matter decay~\cite{Cembranos:2006gt}. Dark matter particles with masses in the MeV range can generically produce detectable MeV gamma-ray signals that are compatible with existing constraints from BBN and CMB, as recently studied in, for example, Refs.~\cite{Boddy:2015efa, Bartels:2017dpb, Gonzalez-Morales:2017jkx} and in Ref.~\cite{Boddy:2016fds} in the context of dynamical dark matter models.

%\contribution{Marco Trovato}
\ssection{ATLAS and CMS Searches} The LHC provides stringent limits on dark matter production via spin-0 and spin-1 mediators~\cite{Abercrombie:2015wmb,CMS:DP2016057,CMS:2017xrr,ATLAS:PUBPAGE}. Reinterpretations of CMS and ATLAS results in terms of dark matter-nucleon cross sections demonstrate the complementarity between collider and direct detection measurements, and the assumptions involved in both approaches. In particular, collider limits typically become the most stringent for spin-dependent cross sections and for light dark matter masses $\lesssim$ 100 GeV, assuming the mediators involved in DM production are not light.

The High-Luminosity LHC (HL-LHC) will provide ATLAS and CMS with 3000 fb$^{-1}$, a factor of 100 times more data than currently collected. Projections for dark matter searches at the HL-LHC indicate that the reach of collider experiments will extend below the coherent neutrino scattering limit~\cite{Buchmueller:2014yoa}, where direct detection experiments have little sensitivity. Thanks to the large amount of available data and plans to broaden the dark matter program beyond the currently explored signatures, HL-LHC will also extend its discovery reach to weaker coupling scenarios.  For this program to be successful, it is imperative that the performance of the ATLAS and CMS detectors is improved beyond their present levels. One of the primary challenges for the HL-LHC experiments will be the extremely large number of interactions per beam crossing (pile-up). Pile-up mitigation can be achieved by introducing tracking information in the hardware triggers of the experiments, allowing information from fully reconstructed events to be leveraged early in the process of data selection. A remarkable background rate reduction will be achieved, while keeping good efficiency and measurement resolutions for dark matter signals.

%\contribution{Philip Ilten}
\ssection{LHCb Searches} Many light dark sector candidates, e.g., the dark photon, are expected to be produced in rare meson decays. The high luminosity environment of the LHC creates copious numbers of mesons, which provide a complementary and already existing avenue to dark sector searches. However, data acquisition of rare meson decays in high pile-up LHC events is experimentally challenging. The LHC beauty experiment (LHCb) was purpose-built for detecting rare $B$-hadron decays and, consequently, is ideally suited for rare meson decay searches. LHCb has a flexible trigger system where real-time detector calibration, low transverse-momentum thresholds, and full event reconstruction at trigger level allow for the acquisition of high statistic data samples with rare meson decays. During Run 3 of LHC operation, the LHCb data acquisition system will be upgraded to a triggerless readout with full software reconstruction, enabling even more efficient data collection.

LHCb searches have already been performed for dark bosons produced in $B^0(B^+) \to K^{*0}(K^+) \mu^+ \mu-$ decays~\cite{Aaij:2015tna,Aaij:2016qsm} and Majorana neutrinos in $B^- \to \pi^+ \mu^- \mu^-$ decays~\cite{Aaij:2014aba}. Two dark photon searches using inclusive di-muon production~\cite{Ilten:2016tkc} and $D^{*0} \to D^0 \, e^+ e^-$ decays~\cite{Ilten:2015hya} were recently proposed. Because LHCb has an excellent lifetime resolution of $\approx 50~\mathrm{fs}$, prompt and displaced searches can be performed simultaneously, allowing much of the open parameter space in the kinetic mixing parameter $\varepsilon$ between prompt and beam-dump limits to be covered with the full Run 3 LHCb dataset. The $D^{*0}$ search will cover dark photon masses from the di-electron mass threshold up to $1.9~\gev$, and the inclusive di-muon search will cover masses from the di-muon threshold upwards. Further channels are being considered to cover the gap between these two channels. Although the $D^{*0}$ search requires Run 3 triggers, the inclusive di-muon search is already possible with the current Run 2 LHCb data. 

%\contribution{Claudia Frugiele}
\ssection{Light Dark Matter at Neutrino Facilities} Neutrino facilities can probe light dark matter-nucleon coupling in a fashion complementary to present and future direct detection experiments~\cite{Batell:2009di}. If dark matter interacts with quarks via a light mediator, a dark matter beam is produced at these facilities along with the neutrino beam, and the dark matter particles then enter the near detector and scatter with the nucleons inside, just as for neutrinos. Hence, the challenge of this program is the suppression of the neutrino background.  

The MiniBooNE (MB) collaboration has already carried out the first dedicated search for light dark matter at a neutrino facility~\cite{Aguilar-Arevalo:2017mqx}.  They placed  strong bounds on the 1 MeV--1 GeV mediator mass region, covering a significant part of unexplored territory.  Building on the success of MB, the authors of Refs.~\cite{Dobrescu:2014ita,Coloma:2015pih,Frugiuele:2017zvx} investigated the reach of facilities near the 120 GeV proton beam at the Main Injector facility and the future LBNF facility, both at Fermilab.  These higher energy proton beams offer the possibility to extend the reach up to 7--8 GeV mediator masses. The signal in this case is given by deep inelastic scattering events.  The neutrino background, which presents a problem for quasi-on-axis detectors like MINOS or NOvA, can be sufficiently suppressed by going far off-axis.  The ideal position to maximize signal over background is $ 6.5^\circ$ off-axis and 200 m away~\cite{Coloma:2015pih}, which is, coincidentally, very close to the location of the MB detector relative to the Main Injector ($6.5^\circ$ and 750 m).  Therefore, by analyzing existing data coming from the Main Injector, MB can extend its reach for light dark matter and also can get similar sensitivity for sub-GeV mediators~\cite{Frugiuele:2017zvx}. This proposal is that it is completely symbiotic to the neutrino program.

%\contribution{Jeff Martoff}
\ssection{Trapped Atom Search for Sterile Neutrino Dark Matter} An alternative, and perhaps equally well-motivated, candidate to WIMP dark matter is the sterile neutrino~\cite{Shi:1999}.  The HUNTER Experiment (Heavy Unseen Neutrinos from Total Energy-momentum Reconstruction)~\cite{Smith:2016vku} would achieve this in a non-accelerator experiment using a medical isotope and existing technologies.  High-precision, kinematically complete measurements of K-capture decays are made, using a sample of $>10^8$ radioactive $^{131}$Cs atoms contained in a laser atom trap. Any sterile neutrino would be produced in the decay as an addition to the electron neutrino mixture of $\nu_1$, $\nu_2$, $\nu_3$.  Momenta of the recoiling $^{131}$Xe atom and the X-ray and Auger electron(s) produced in the decay are all measured with the requisite accuracy using the MOTRIMS spectroscopic method~\cite{Ullrich:2003}, and the neutrino 4-momentum and mass are reconstructed.  A sterile neutrino signal appears as a separated population of events at nonzero neutrino mass.   The initial implementation would probe sterile neutrino masses in the range of ten to a few hundred keV and coupling constants in the high 10$^{-5}$ range, requiring a three-year program and funding at a small fraction of the small projects portfolio.  Subsequent upgrades of the trap and the detection system would probe coupling constants down to the 10$^{-11}$ range.  

%\contribution{Leah Broussard/Ben Rybolt}
\ssection{Mirror Neutron Searches}
A novel approach to probe the nature of dark matter is to search for neutron oscillations into a hidden dark sector.  Neutron oscillations are predicted by theories that postulate a parallel sector with identical particles and interactions as in the SM, such as mirror matter~\cite{Berezhiani:2005hv}.  Big Bang nucleosynthesis and cosmological limits imply mirror matter should be colder than ordinary matter and therefore helium-dominated with faster cosmological evolution. Mirror matter could explain baryogenesis and some or all of dark matter~\cite{Berezhiani:2003xm}.

Ultracold neutron storage measurements place limits on the oscillation time of $\tau<448$~s, assuming no mirror magnetic field $B'$~\cite{Serebrov:2008hw}. When a nonzero $B'$ is considered, some conflicting results have been reported for $\tau<10$~s~\cite{Berezhiani:2012rq, Altarev:2009tg}. To clarify the situation, a disappearance-regeneration ``beam-dump'' type experiment has been proposed~\cite{Berezhiani:2017azg}. This type of experiment is uniquely sensitive to a certain class of dark matter.  Existing cold neutron beamlines such as at the High Flux Isotope Reactor (HFIR) at Oak Ridge National Laboratory or the National Institute of Standards and Technology Center for Neutron Research (NIST NCNR)  could produce results in a few years for a small fraction of the small project threshold~\cite{BroussardRyboltUSCosmicVisions}, which will either exclude those controversial results or discover a new phenomenon that will inform us about the nature of dark matter. 

%%%%%%%%%%%%%%%%%%%%%%%%%%%%%%%%%
\begin{figure}[t]
\begin{center}
\includegraphics[width=0.55\textwidth]{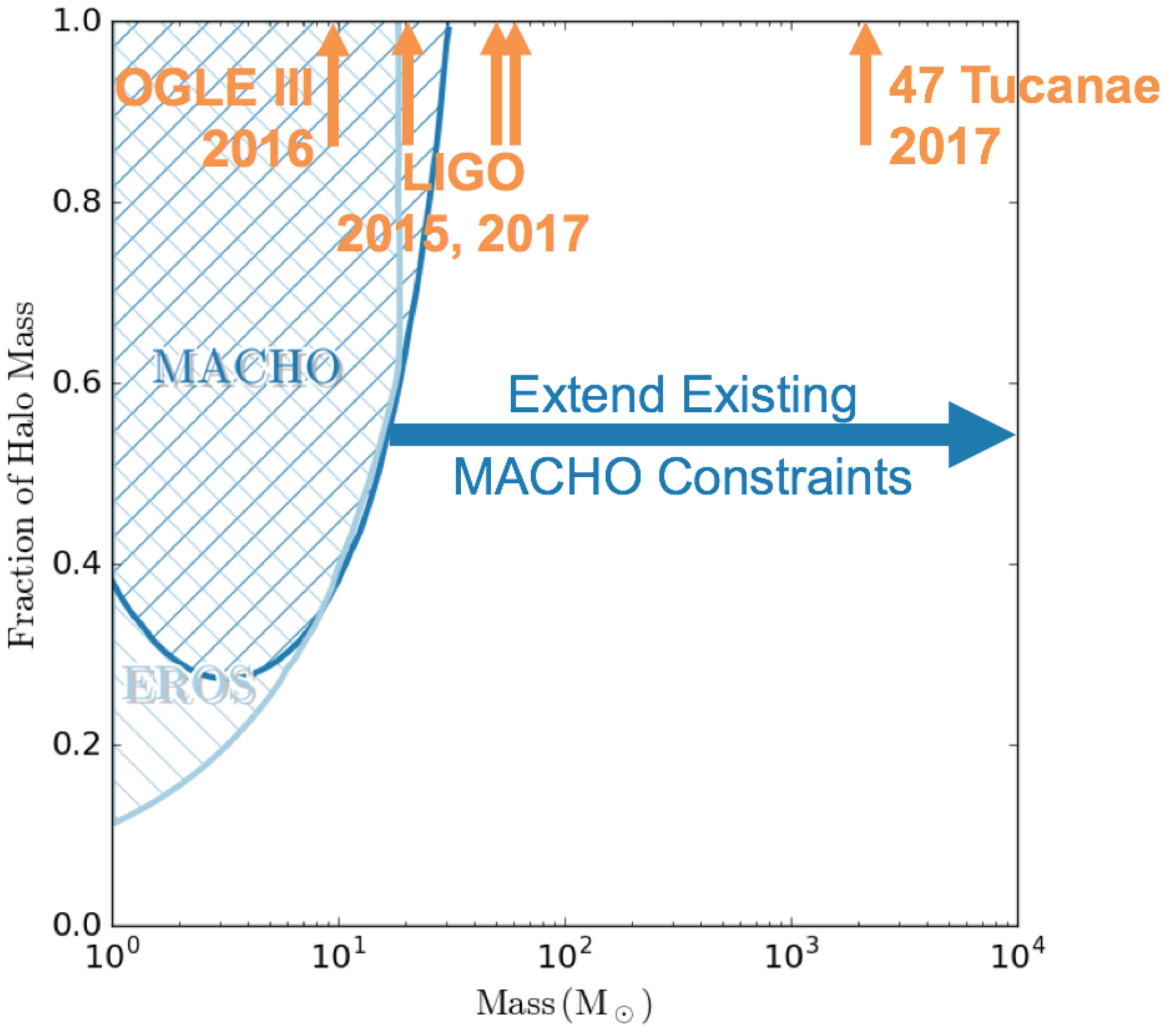}
\caption{Existing microlensing constraints on the fraction of the mass in the Milky Way halo that can be composed of intermediate mass MACHO dark matter.  The masses of detected black holes \cite{Abbott:2016blz,Wyrzykowski:2015ppa,LoebNature} are indicated in orange. From Ref.~\cite{DawsonUSCosmicVisions}.}
\label{fig:PBHchimney}
\end{center}
\end{figure}
%%%%%%%%%%%%%%%%%%%%%%%%%%%%%%%%%%%

%\contribution{Will Dawson}
\ssection{Microlensing Searches for Black Hole Dark Matter}  LIGO's recent discovery of $30M_\odot$ black holes have renewed interest in the possibility that dark matter consists entirely of intermediate mass black holes formed less than one second after the Big Bang. 
Massive compact halo object (MACHO) searches from the 1990's constrained the MACHO content of the universe for MACHO masses below $15M_\odot$.  While the original CMB and wide-binary constraints from the 2000's appeared to exclude the intermediate mass (IM) MACHO parameter space above $2M_\odot$, recent studies have shown that the complex and poorly constrained astrophysical assumptions that the CMB and wide-binary constraints relied on were incorrect.  Once again opening the parameter space above $15M_\odot$ where black holes like those detected by LIGO may account for the majority of dark matter (Fig.~\ref{fig:PBHchimney}).  Rather than rely on probes with complex astrophysical assumptions and associated systematics, one can exploit direct detection microlensing probes of the IM MACHO population to determine if they comprise all of the dark matter.  

With a 5-year, 700-square-degree, multi-band survey of the Galactic bulge with DECam (4 nights/month, 8 months/year, resulting in $\sim 60$ measurements/year of $\sim 500$ million stars) one expects $\sim 100$ microlensing events by intermediate mass MACHOs in the $15-10^4 M_\odot$ range. LSST also has potential to be the ideal survey for this science (with $\sim 1000$ expected microlensing events); however, microlensing is currently outside its purview. Additionally, current survey plan options, which only observe the Milky Way in the first year, will unnecessarily preclude microlensing dark matter science. It is pressing that we begin this effort now while there is still time to influence the LSST survey strategy.  By leveraging existing DOE investments in DECam and LSST, LLNL and FNAL computing, and LLNL LDRD staff support, this survey can be carried out for a small fraction of the small project threshold~\cite{DawsonUSCosmicVisions}.  

%%%%%%%%%%%%%%%%%%%%%%%%%%
\subsection{New Candidates, Targets, and Complementarity--Summary}
\label{WG4sec:conclusions}

Dark matter has long been one of the leading scientific open questions of our time, but the field has entered a new era.  Simply put, dark matter has been transformed in recent years by innovative cross talk across many fields of physics.  Previously confined to the cosmic frontier of particle physics, it now spans the cosmic, energy, and intensity frontiers, has become an incredibly fertile field for creative ideas about new particles and forces, and is the source of fascinating new connections between particle physics, astrophysics, and many other subfields, including nuclear, atomic, and condensed matter physics.  Most notably, these innovative ideas have opened up completely new directions that can be explored by inexpensive experiments, creating opportunities not seen since the early days of particle searches for dark matter in the 1980's.  

In this chapter we have briefly reviewed the new candidates, target regions of parameter space, and complementarity between different approaches that have emerged.  There are many exciting opportunities for high-value investments in dark matter research.  We highlight a few of them here:
\begin{itemize}

\item {\em Importance of Investment in Theory.} Theory has played a particularly important role in these recent developments, motivating new models and regions of parameter space, suggesting new experiments, and drawing connections between disparate phenomena.  {\bf Healthy support for theory is essential to maintaining the flow of creative and cross-disciplinary ideas that have been seen in recent years, and which may finally unmask the particle identity of dark matter.}

\item {\em Nuclear and Accelerator Tests of the $^8$Be Anomaly.} The $6.8\sigma$ anomaly in $^8$Be nuclear decays may be evidence for a new 17 MeV boson and provides a well-defined target for future experiments.  The experimental results disagree with new and refined nuclear theory predictions, and several viable new particle explanations exist.  Additional experiments are needed to confirm or exclude the anomaly.  {\bf The $^8$Be anomaly strongly motivates proposed followup nuclear experiments that are fast (under 2 years) and cheap (a small fraction of the small projects threshold), as well as isotope shift spectroscopy experiments and accelerator searches for new bosons with masses $\sim 10$ MeV and electron couplings $\varepsilon \sim 10^{-4} - 10^{-3}$.}

\item {\em Synergy with Cosmology and Astrophysics.}  Precision cosmology now probes the microscopic particle properties of dark matter.  Observations of the CMB and supernovae constrain regions of parameter space inaccessible to particle experiments, and small-scale structure has motivated self-interacting dark matter models with implications for particle experiments.  {\bf Small investments in simulations and astroparticle theory can leverage the enormous amount of cosmological data already being collected and are certain to provide not just {\em constraints} on dark matter {\em candidates}, but {\em measurements} of the properties of {\em dark matter}.} 

\item {\em Importance of the 1 to 100 MeV Mass Scale.}  The WIMP miracle has motivated searches for weak-scale dark matter, but recent developments have broadened the range of interesting masses for both dark matter particles and the particles mediating their interactions with the standard model. For example, motivations for sub-GeV dark matter from hidden thermal relics and asymmetric dark matter are discussed above.  In the work discussed in this section, a diverse set of new considerations motivate the 1 to 100 MeV mass scale for dark sector particles.  These considerations include thermal relic abundances in SIMP and ELDER models with significant $3\to2$ interactions, small-scale structure puzzles that motivate QCD-like self-interactions induced by 1 to 100 MeV mediators, and existing anomalies, such as the muon $(g-2)$, the proton radius and $^8$Be anomalies.  {\bf New models, astrophysical observations, and existing experimental anomalies point to the 1 to 100 MeV mass scale as a high-value target region for dark matter and dark mediator searches.}  

\item {\em Microlensing Searches for Solar Mass Black Hole Dark Matter.} The LIGO observation of colliding $\sim 30 M_{\odot}$ mass black holes has renewed interest in the possibility that such black holes make up some or all of the dark matter.  {\bf The LIGO discovery of gravitational waves from colliding black holes strongly motivate a proposed microlensing search that can confirm or exclude the possibility of intermediate mass black hole dark matter using existing facilities with minimal funding.}
\end{itemize}

%% file: conclusions.tex
\section{Conclusions}
This whitepaper has summarized the science opportunities and experimental ideas presented at the workshop ``US Cosmic Visions: New Ideas in Dark Matter''.   The possibilities for dark matter explored by small projects span a wide range of possibilities for the nature of dark matter, extending from the observational lower bound of $10^{-22} \eV$ particles up to $30 M_\odot$ black holes.  Two particular areas of focus are \emph{ultralight (sub-keV) dark matter}, which behaves as a coherent (bosonic) field, and of which the QCD axion is a particularly motivated and well-known example; and \emph{hidden-sector dark matter}, neutral under Standard Model forces but interacting through a new force, with testable sharp targets in parameter space motivated by DM production mechanisms and anomalies in data.  These two broad scenarios stand out as simultaneously well-motivated and accessible to small-scale experiments.

There is a broad and active community of physicists pursuing several new experimental directions: new direct detection experiments, ultralight (sub-eV) DM searches, accelerator-based searches for DM production, and small-scale experiments exploring anomalies in existing data that suggest new forces.  Each of these techniques covers broad parameter regions with great sensitivity and decisively explores high-priority science targets; any one of them could revolutionize our understanding of Nature's dark sector.   In addition, theory has played a notably essential role in developing new small-scale experiments and the connections among sub-fields on which many of these experimental techniques are based.  The experimental approaches presented at the workshop are highly complementary --- each of the working groups has identified models for which particular techniques are uniquely sensitive, while in many other cases a combination of different experimental approaches is required to move from discovery to a physical understanding of dark matter.  The breadth and importance of dark matter science therefore strongly motivate a portfolio of small experiments spanning all of the above techniques, as well as continued investment in theory.